\documentclass{aa}

  \usepackage{graphicx}
  \usepackage[utf8]{inputenc}
  \usepackage{txfonts}
  \usepackage{mathtools}
  \usepackage{xcolor}
  \usepackage[inline]{enumitem}
  \definecolor{colurl}{rgb}{0.,0.5,0.}
  \definecolor{colcite}{rgb}{0.2,0.2,1}
  \definecolor{collink}{rgb}{0.7,0.1,0.}
  \definecolor{colsec}{rgb}{0.7,0.1,0.}
  \definecolor{colhead}{rgb}{0.6,0.6,0.6}
  \usepackage[colorlinks=true,
              breaklinks=true,
              pdfstartview=FitV, 
              linkcolor=collink, 
              citecolor=colcite, 
              urlcolor=colurl]{hyperref}
  \usepackage{thumbpdf}
  \usepackage{tabularx}
  \newcolumntype{R}{>{\raggedleft\arraybackslash}X}
  \usepackage{rotating}
  \usepackage{aalongtable}
  \usepackage{pdflscape}
  \usepackage{amsmath}
              
  \graphicspath{{Figures/}}
  \DeclareGraphicsExtensions{.pdf,.png}
  \usepackage{float}

\newcommand{\citext}[1]{\textit{#1}}
\newcommand{\expression}[1]{\textit{#1}}
\newcommand{\familyname}[1]{\textsc{#1}}

\newcommand{\sms}[1]{{\mbox{{\scriptsize #1}}}}

\newcommand{\dd}{\textnormal{d}}
\newcommand{\ddiff}{\,\textnormal{d}}

\newcommand{\E}[1]{\times10^{#1}}

\newcommand{\vect}[1]{\protect\overrightarrow{#1}}
\newcommand{\mat}[1]{\protect\overleftrightarrow{#1}}

\providecommand{\refeq}[1]{\hyperref[#1]{Eq.~(\ref{#1})}}
\newcommand{\refeqp}[1]{\hyperref[#1]{(Eq.~\ref{#1})}}
\newcommand{\refeqnp}[1]{\hyperref[#1]{Eq.~\ref{#1}}}
\newcommand{\refeqs}[2]{\ifthenelse{\equal{#2}{}}%
                      {\hyperref[#1]{Eqs.~(\ref{#1})}}%
                      {\hyperref[#1]{Eqs.~(\ref{#1})}~--~\hyperref[#2]{(\ref{#2})}}}
\newcommand{\refeqsnp}[2]{\ifthenelse{\equal{#2}{}}%
                        {\hyperref[#1]{Eqs.~\ref{#1}}}%
                        {\hyperref[#1]{Eqs.~\ref{#1}}~--~\hyperref[#2]{\ref{#2}}}}

\newcommand{\reftab}[1]{\hyperref[#1]{Table~\ref{#1}}}
\newcommand{\reftabs}[2]{\ifthenelse{\equal{#2}{}}%
                       {\hyperref[#1]{Tables~\ref{#1}}}%
                       {\hyperref[#1]{Tables~\ref{#1}}~--~\hyperref[#2]{\ref{#2}}}}

\newcommand{\reffig}[1]{\hyperref[#1]{Fig.~\ref{#1}}}
\newcommand{\reffigs}[2]{\ifthenelse{\equal{#2}{}}%
                       {\hyperref[#1]{Figs.~\ref{#1}}}%
                       {\hyperref[#1]{Figs.~\ref{#1}}~--~\hyperref[#2]{\ref{#2}}}}
\newcommand{\refsubfig}[2]{\hyperref[#1]{Fig.~\ref{#1}.#2}}

\newcommand{\refchap}[1]{\hyperref[#1]{Chap.~\ref{#1}}}
\newcommand{\refchaps}[2]{\ifthenelse{\equal{#2}{}}%
                       {\hyperref[#1]{Chaps.~\ref{#1}}}%
                       {\hyperref[#1]{Chaps.~\ref{#1}}~--~\hyperref[#2]{\ref{#2}}}}

\newcommand{\refsec}[1]{\hyperref[#1]{Sect.~\ref{#1}}}
\newcommand{\refsecs}[2]{\ifthenelse{\equal{#2}{}}%
                       {\hyperref[#1]{Sects.~\ref{#1}}}%
                       {\hyperref[#1]{Sects.~\ref{#1}}~--~\hyperref[#2]{\ref{#2}}}}
\newcommand{\refS}[1]{\hyperref[#1]{\S\ref{#1}}}

\newcommand{\refapp}[1]{\hyperref[#1]{Appendix~\ref{#1}}}
\newcommand{\refapps}[2]{\ifthenelse{\equal{#2}{}}%
                       {\hyperref[#1]{Appendices~\ref{#1}}}%
                       {\hyperref[#1]{Appendices~\ref{#1}}~--~\hyperref[#2]{\ref{#2}}}}

\newcounter{textlistctr}

\newcommand{\squishlist}{
   \begin{list}{$\bullet$}
    { \setlength{\itemsep}{0pt}      \setlength{\parsep}{3pt}
      \setlength{\topsep}{3pt}       \setlength{\partopsep}{0pt}
      \setlength{\leftmargin}{1.5em} \setlength{\labelwidth}{1em}
      \setlength{\labelsep}{0.5em} } }
\newcommand{\squishlisttwo}{
   \begin{list}{$\bullet$}
    { \setlength{\itemsep}{0pt}    \setlength{\parsep}{0pt}
      \setlength{\topsep}{0pt}     \setlength{\partopsep}{0pt}
      \setlength{\leftmargin}{2em} \setlength{\labelwidth}{1.5em}
      \setlength{\labelsep}{0.5em} } }
\newcommand{\squishend}{
    \end{list}  }

\newcounter{obsrefctr}
\newcommand{\eg}{\textit{e.g.}}
\newcommand{\ie}{\textit{i.e.}}
\newcommand{\cf}{\textit{cf.}}

\newcommand{\etc}{\textit{etc.}}

\newcommand{\ncode}[1]{\texttt{#1}}

\newcommand{\mic}{\mu\textnormal{m}}

\newcommand{\Lsun}{\textnormal{L}_\odot}
\newcommand{\Msun}{\textnormal{M}_\odot}
\newcommand{\Zsun}{\textnormal{Z}_\odot}

\newcommand{\emic}{\;\mic}

\newcommand{\eMsun}{\;\Msun}
\newcommand{\eZsun}{\;\Zsun}

\newcommand{\tmic}{$\mic$}

\newcommand{\tLsun}{$\Lsun$}
\newcommand{\tMsun}{$\Msun$}

\newcommand{\cii}{C$\,$\textsc{ii}}

\newcommand{\ha}{H$\alpha$}

\newcommand{\hi}{H$\,$\textsc{i}}
\newcommand{\hii}{H$\,$\textsc{ii}}

\newcommand{\neiii}{Ne$\,$\textsc{iii}}
\newcommand{\neii}{Ne$\,$\textsc{ii}}

\newcommand{\oi}{O$\,$\textsc{i}}

\newcommand{\hmol}{H$_\textnormal{2}$}

\newcommand{\hiline}{[\hi]$_{21\,\textnormal{cm}}$}

\newcommand{\ciiline}{[\cii]$_{158\mu\textnormal{m}}$}

\newcommand{\haline}{\ha$_{656.3\textnormal{nm}}$}

\newcommand{\neiiiline}{[\neiii]$_{15.56\mu\textnormal{m}}$}
\newcommand{\neiiline}{[\neii]$_{12.81\mu\textnormal{m}}$}

\newcommand{\oiline}{[\oi]$_{63\mu\textnormal{m}}$}
\newcommand{\oilineb}{[\oi]$_{145\mu\textnormal{m}}$}
\newcommand{\oiiiline}{[O$\,$\textsc{iii}]$_{88\mu\textnormal{m}}$}

\newcommand{\COio}{$^{12}$CO(J$=$1$\rightarrow$0)$_{2.6\textnormal{mm}}$}

\newcommand{\COiiitoii}{$^{12}$CO(J$=$3$\rightarrow$2)$_{867\mu\textnormal{m}}$}

\newcommand{\met}{12+\log(\textnormal{O/H})}
\newcommand{\tmet}{$\met$}

\newcommand{\IC}[1]{IC$\;$#1}

\newcommand{\M}[1]{M$\;$#1}
\newcommand{\mrk}[1]{Mrk$\;$#1}
\newcommand{\ngc}[1]{NGC$\;$#1}
\newcommand{\ugc}[1]{UGC$\;$#1}
\newcommand{\pgc}[1]{PGC$\;$#1}
\newcommand{\ic}[1]{IC$\;$#1}
\newcommand{\eso}[1]{ESO$\;$#1}

\newcommand{\hs}[1]{HS$\;$#1}

\newcommand{\sbs}[1]{SBS$\;$#1}

\newcommand{\izw}{I$\;$Zw$\;$18}

\newcommand{\viizw}{VII$\;$Zw$\;$403}

\newcommand{\snii}{SN$\,$\textsc{ii}}

\newcommand{\iras}{\textit{IRAS}}

\newcommand{\wise}{\textit{WISE}}
\newcommand{\spitz}{\textit{Spitzer}}

\newcommand{\hersc}{\textit{Herschel}}
\newcommand{\planck}{\textit{Planck}}

\newcommand{\spica}{\textit{SPICA}}
\newcommand{\cobe}{\textit{COBE}}

\newcommand{\irasiv}{\iras$_\sms{100$\mu m$}$}

\newcommand{\citeprep}[1]%
           {\citeauthor{#1}, \textcolor{colcite}{\textit{in prep.}}}
\newcommand{\citetprep}[1]%
           {\citeauthor{#1} \textcolor{colcite}{(\textit{in prep.})}}
\newcommand{\citepprep}[1]%
            {(\citeauthor{#1}, \textcolor{colcite}{\textit{in prep.}})}
\newcommand{\citesubm}[1]%
           {\citeauthor{#1}, \textcolor{colcite}{\textit{submitted}}}
\newcommand{\citetsubm}[1]%
           {\citeauthor{#1} \textcolor{colcite}{(\textit{submitted})}}
\newcommand{\citepsubm}[1]%
           {(\citeauthor{#1}, \textcolor{colcite}{\textit{submitted}})}
\newcommand{\citepress}[1]%
           {\citeauthor{#1}, \textcolor{colcite}{\textit{in press}}}
\newcommand{\citetpress}[1]%
           {\citeauthor{#1} \textcolor{colcite}{(\textit{in press})}}
\newcommand{\citeppress}[1]%
           {(\citeauthor{#1}, \textcolor{colcite}{\textit{in press}})}

  \newcommand{\HB}{\ncode{HerBIE}}
  \newcommand{\THE}{\ncode{THEMIS}}
  \newcommand{\CIG}{\ncode{CIGALE}}

  \newcommand{\powerU}{\textsl{powerU}}
  \newcommand{\starBB}{\textsl{starBB}}

  \newcommand{\ugca}{UGCA$\,$20}

  \newcommand{\Md}{M_\sms{dust}}
  \newcommand{\tMd}{$\Md$}
  \newcommand{\Mdref}{M_\sms{dust}^\sms{ref}}
  \newcommand{\tMdref}{$\Md^\sms{ref}$}
  \newcommand{\Uav}{\langle U\rangle}
  \newcommand{\tUav}{$\Uav$}
  \newcommand{\qAF}{q_\sms{AF}}
  \newcommand{\tqAF}{$\qAF$}
  \newcommand{\qAFref}{q_\sms{AF}^\sms{ref}}
  \newcommand{\tqAFref}{$\qAFref$}
  \newcommand{\qPAH}{q_\sms{PAH}}
  \newcommand{\tqPAH}{$\qPAH$}
  \newcommand{\Mg}{M_\sms{gas}}
  \newcommand{\tMg}{$\Mg$}
  \newcommand{\Ms}{M_\star}
  \newcommand{\tMs}{$\Ms$}
  \newcommand{\sMd}{\textnormal{s}\Md}
  \newcommand{\tsMd}{$\sMd$}
  \newcommand{\sMg}{\textnormal{s}\Mg}
  \newcommand{\tsMg}{$\sMg$}
  \newcommand{\Umin}{U_\sms{min}}
  
  \newcommand{\DU}{\Delta U}
  
  \newcommand{\DtoB}{f_\sms{dust}}
  \newcommand{\tDtoB}{$\DtoB$}
  \newcommand{\fg}{f_\sms{gas}}
  
  \newcommand{\tZ}{$Z$}
  \newcommand{\Zd}{Z_\sms{dust}}
  \newcommand{\tZd}{$\Zd$}
  \newcommand{\ZOH}{12+\log(\textnormal{O}/\textnormal{H})}
  \newcommand{\tZOH}{$\ZOH$}
  \newcommand{\CR}[1]{\textnormal{CR}_{95\%}(#1)}

  \defcitealias{draine07}{DL07}
  \defcitealias{clark18}{C18}
  \defcitealias{galliano18a}{G18}
  \defcitealias{lianou19}{L19}
  \defcitealias{de-looze20}{DL20}
  \defcitealias{nanni20}{N20}
  \renewcommand{\refeq}[1]{\textcolor{collink}{Eq.~(\ref{#1})}}
  \newcommand{\DPar}{\href{http://dustpedia.astro.noa.gr}%
                        {http://dustpedia.astro.noa.gr}}

  \setlist[description]{font=\bfseries}
  \newlist{inlinelist}{enumerate*}{1}
  \setlist*[inlinelist,1]{label=\textit{(\roman*)},}

\begin{document}

  \title{A Nearby Galaxy Perspective on Dust Evolution} 
  \subtitle{Scaling relations and constraints on the dust build-up in 
            galaxies with the DustPedia\thanks{DustPedia is a
         collaborative focused research project supported by the European 
         Union under the Seventh Framework Programme (2007–2013) call 
         (proposal no.~606847, PI J.~I.~Davies). 
         The data used in this work is publicly available at 
         \DPar.} and DGS samples}
  \authorrunning{F. Galliano et al.}
  \author{Fr\'ed\'eric~\familyname{Galliano}\inst{1}
          \and 
          Angelos~\familyname{Nersesian}\inst{2,4}
          \and
          Simone~\familyname{Bianchi}\inst{3}
          \and 
          Ilse~\familyname{De~Looze}\inst{4,5}
          \and 
          Sambit~\familyname{Roychowdhury}\inst{6,7,8}
          \and
          Maarten~\familyname{Baes}\inst{4}
          \and 
          Viviana~\familyname{Casasola}\inst{9}
          \and 
          Letizia, P.~\familyname{Cassar\'a}\inst{10}
          \and 
          Wouter~\familyname{Dobbels}\inst{4}
          \and
          Jacopo~\familyname{Fritz}\inst{11}
          \and
          Maud~\familyname{Galametz}\inst{1}
          \and 
          Anthony~P.~\familyname{Jones}\inst{6}
          \and 
          Suzanne~C.~\familyname{Madden}\inst{1}
          \and
          Aleksandr~\familyname{Mosenkov}\inst{12,13}
          \and
          Emmanuel~M.~\familyname{Xilouris}\inst{2}
          \and
          Nathalie~\familyname{Ysard}\inst{6}}
  \institute{
             AIM, CEA, CNRS, Universit\'e Paris-Saclay, Universit\'e Paris 
             Diderot, Sorbonne Paris Cit\'e, F-91191 Gif-sur-Yvette, France \\
             \email{frederic.galliano@cea.fr} 
             \and
             National Observatory of Athens, Institute for Astronomy, 
             Astrophysics, Space Applications and Remote Sensing, Ioannou 
             Metaxa and Vasileos Pavlou GR-15236, Athens, Greece
             \and
             INAF - Osservatorio Astrofisico di Arcetri, Largo E. Fermi 5, 
             I-50125, Florence, Italy
             \and
             Sterrenkundig Observatorium, Ghent University, Krijgslaan 281 - 
             S9, 9000 Gent, Belgium
             \and
             Dept.\ of Physics \& Astronomy, University College London, Gower 
             Street, London WC1E 6BT, UK
             \and 
             Université Paris-Saclay, CNRS, Institut d’Astrophysique Spatiale, 
             91405 Orsay, France
             \and
             International Centre for Radio Astronomy Research (ICRAR), M468, 
             University of Western Australia, 35 Stirling Hwy, Crawley, 
             WA 6009, Australia
             \and
             ARC Centre of Excellence for All Sky Astrophysics in 3 Dimensions 
             (ASTRO 3D), Australia
             \and
             INAF --~Istituto di Radioastronomia, Via P.~Gobetti 101, 40129 
             Bologna, Italy
             \and
             INAF - Istituto di Astrofisica Spaziale e Fisica Cosmica, Via 
             Alfonso Corti 12, 20133, Milan, Italy
             \and
             Instituto de Radioastronomía y Astrofísica, UNAM, Antigua 
             Carretera a Pátzcuaro \#~8701, Ex-Hda. San José de la Huerta, 
             58089 Morelia, Michoacán, Mexico
             \and
             Central Astronomical Observatory of RAS, Pulkovskoye Chaussee 
             65/1, 196140 St.\ Petersburg, Russia
             \and
             St.\ Petersburg State University, Universitetskij Pr.\ 28, 198504 
             St.\ Petersburg, Stary Peterhof, Russia}
  \date{Received 16 October 2020; accepted 23 December 2020}
  \abstract
    {The efficiency of the different processes responsible for the evolution of 
     interstellar dust on the scale of a galaxy, are to date 
     very uncertain, spanning several orders of magnitude in the literature.
     Yet, a precise knowledge of the grain properties is the key to addressing 
     numerous open questions about the physics of the interstellar medium and 
     galaxy evolution.}
    {This article presents an empirical statistical study, aimed at quantifying
     the timescales of the main cosmic dust evolution processes, as a function
     of the global properties of a galaxy.}
    {We have modelled a sample of $\simeq800$ nearby galaxies, spanning a 
     wide range of metallicity, gas fraction, specific star formation rate
     and Hubble stage.
     We have derived the dust properties of each object from its spectral 
     energy distribution.
     Through an additional level of analysis, we have inferred the 
     timescales of dust condensation 
     in core-collapse supernova ejecta, grain growth in cold clouds and dust 
     destruction by shock waves.
     Throughout this paper, we have adopted a hierarchical Bayesian approach, 
     resulting in a single 
     large probability distribution of all the parameters of all the galaxies, 
     to ensure the most rigorous interpretation of our data.}
    {We confirm the drastic evolution with metallicity of the dust-to-metal 
     mass ratio (by two orders of magnitude), found by previous studies.
     We show that dust production by core-collapse supernovae is efficient only
     at very low-metallicity, a single supernova producing on average less than
     $\simeq0.03\eMsun$/SN of dust.
     Our data indicate that grain growth is the dominant formation mechanism
     at metallicity above $\simeq1/5$ solar, with a grain growth timescale 
     shorter than $\simeq50$~Myr at solar metallicity.
     Shock destruction is relatively efficient, a single supernova clearing 
     dust on average in at least $\simeq1200\eMsun$/SN of gas.
     These results are robust when assuming different stellar initial mass 
     functions.
     In addition, we show that early-type galaxies are outliers in several
     scaling relations.
     This feature could result from grain thermal sputtering in hot X-ray 
     emitting gas,
     an hypothesis supported by a negative correlation between the 
     dust-to-stellar mass ratio and the X-ray photon rate per grain.
     Finally, we confirm the well-known evolution of the 
     aromatic-feature-emitting grain mass fraction as a function of metallicity
     and interstellar radiation field intensity.
     Our data indicate the relation with metallicity is significantly 
     stronger.}
    {Our results provide valuable constraints for simulations of galaxies.
     They imply that grain growth is the likely dust production mechanism in
     dusty high-redshift objects.
     We also emphasize the determinant role of local low-metallicity systems to 
     address these questions.}
  \maketitle

  \section{INTRODUCTION}
  \label{sec:intro}

Characterizing the dust properties across different galactic environments is an important milestone towards understanding the physics of the \expression{InterStellar Medium} (ISM) and galaxy evolution \citep[][ for a review]{galliano18}.
Interstellar grains have a nefarious role of obscuring our direct view of star formation \citep[\eg\ ][]{bianchi18}.
The subsequent unreddening of \expression{Ultraviolet}-(UV)-visible observations relies on assumptions about the constitution and size distribution of the grains, as well as on the relative star-dust geometry \citep[\eg\ ][]{witt92,witt00,baes01}.
Dust is also involved in several important physical processes, such as the photoelectric heating of the gas in \expression{PhotoDissociation Regions} \citep[PDR; \eg][]{draine78,kimura16} or \hmol\ formation \citep[\eg\ ][]{gould63,bron14}.
The uncertainties about the grain properties have dramatic consequences on the rate of these processes and are the main cause of discrepancy between PDR models \citep{rollig07}.
Finally, state-of-the-art numerical simulations of galaxy evolution are now post-processed to incorporate the full treatment of dust radiative transfer in order to reproduce realistic observables \citep[\eg\ ][]{camps16,camps18,trayford17,rodriguez-gomez19,trcka20}.
These simulations rely on assumptions about the dust emissivity, absorption and scattering properties, which changes from galaxy to galaxy \citep[\eg\ ][]{clark19,bianchi19}.

We are far from being able to predict the dust composition, structure and size distribution, in a given ISM condition.
\textit{Ab initio} methods are currently impractical because of the great complexity of the problem.
Empirical approaches face challenges, too.
The last four decades have shown that each time we were investigating a particular observable with the hope of constraining the grain properties, these properties were proving themselves more elusive because of their evolution,
as illustrated in the following.
\begin{itemize}
  \item The diversity of extinction curve shapes within the Milky Way   
    \citep{fitzpatrick86,cardelli89} and the Magellanic Clouds \citep{gordon03} 
    can be explained by variations of the size distribution and 
    carbon-to-silicate grain ratio \citep{pei92,kim94,clayton03,cartledge05}.
  \item Elemental depletion patterns strongly depend on the density of the 
    medium, suggesting that dust is partly destroyed and reformed in 
    the ISM \citep{savage79,crinklaw94,jenkins09,parvathi12}.
  \item Variations of the \expression{InfraRed} (IR) to 
    \expression{submillimeter} (submm) emissivity as a function 
    of the density of the medium, whether in the Milky Way 
    \citep{stepnik03,ysard15} or in the 
    Magellanic clouds \citep{roman-duval17}, are interpreted as variations of 
    the grain mantle thickness and composition, and grain-grain coagulation
    \citep[\eg\ ][]{kohler15,ysard16}.
  \item The wide variability of the relative intensity and band-to-band ratio
    of the mid-IR aromatic feature spectrum, witnessed within the Milky Way and 
    nearby galaxies, is the evidence of the variation of the abundance, 
    charge and size distribution of the band carriers
    \citep[\eg\ ][]{boulanger98,madden06,galliano08b,schirmer20}.
  \item The wavelength-dependent polarization fraction in extinction and 
    emission provides additional constraints on the composition of the grains,
    as well as on their shape and alignment 
    \citep[\eg\ ][]{andersson15,fanciullo17}.
\end{itemize}

To understand the evolution of the global dust content of a galaxy, one must be able to quantify the timescales of the processes responsible for dust formation and destruction.
These processes can be categorized as follows.
\begin{description}
  \item[Stardust production] takes place primarily in the ejecta of:
    \begin{inlinelist}
      \item \expression{Asymptotic Giant Branch} (AGB) stars; and
      \item \expression{type~II SuperNovae} (SN~II; core-collapse SN).
    \end{inlinelist}
    SNe~II potentially dominate the net stardust injection rate  
    \citep[\eg\ ][ for a review]{draine09}, but their
    actual yield is the subject of an intense debate.
    Most of the controversy lies in the fact that, while large amounts could 
    form in SN~II ejecta \citep[\eg\ ][]{matsuura15,temim17}, a large fraction 
    of freshly formed grains could not survive the reverse shock 
    \citep[\eg\ ][]{nozawa06,micelotta16,kirchschlager19}.
    Estimates of the net dust yield of a single SN ranges in the literature from
    $\simeq10^{-3}$ to $1\eMsun/\textnormal{SN}$ 
    \citep[\eg\ ][]{todini01,ercolano07,bianchi07b,bocchio16b,marassi19}.
  \item[Grain growth] in the ISM refers to the addition of gas atoms 
    onto pre-existing dust seeds \citep[\eg\ ][]{hirashita12}.
    It could be the main grain formation process, happening on timescales 
    $\lesssim1$~Myr \citep[\eg\ ][]{draine09}.
    It is however challenged due to the lack of direct constraints and because
    we currently lack a proven dust formation mechanism, at cold temperatures.
  \item[Grain destruction] can be attributed to:
    \begin{inlinelist}
      \item astration, \ie\ their incorporation into stellar interiors 
        during star formation;
      \item sputtering and shattering by SN blast waves.
    \end{inlinelist}
    The second process is the most debated, although it is less controversial 
    than stardust production and grain growth efficiencies.
    Timescales for grain destruction in the Milky Way range from $\simeq 
    200$~Myr to $\simeq 2$~Gyr \citep{jones94,slavin15}.
\end{description}
In the Milky Way, there is a convergent set of evidence pointing toward a scenario where:
\begin{inlinelist}
  \item stardust accounts for at most $\simeq10\,\%$ of ISM dust;
  \item most grains are therefore grown in the ISM
\end{inlinelist}
\citep[\eg\ ][]{draine79,dwek80,jones94,tielens98,draine09}.
The goal of this paper is to explore how particular environments could affect these conclusions, by modelling nearby extragalactic systems.
By studying nearby galaxies, we also aim to provide a different perspective on these questions.

Several attempts have been made in the past to quantify the evolution of the dust content of galaxies \textit{via} fitting their \expression{Spectral Energy Distribution} \citep[SED; \eg ][]{lisenfeld98,morgan03,draine07b,galliano08a,remy-ruyer14,remy-ruyer15,de-vis17,nersesian19,de-looze20,nanni20}.
Although these studies provided important benchmarks, most of them were limited by the following issues:
\begin{inlinelist}
  \item their coverage of the parameter space was often incomplete, especially 
    in the low-metallicity regime, which is crucial to quantify stardust 
    production (\cf~\refsec{sec:dustvol});
  \item potential systematic effects, originating either in the SED fitting
    or in the ancillary data, were questioning some of the conclusions;
  \item dust evolution models were most of the time not fit to each galaxy,
    but simply visually compared\footnote{Among the cited studies, only 
    \citet{nanni20} and \citet{de-looze20} perform an actual fit of individual 
    galaxies.}, which can lead to some inconsistencies 
    (\cf~\refsec{sec:tracks}).
\end{inlinelist}

The present paper is an attempt at addressing these limitations.
We rely on the homogeneous multi-wavelength observations and ancillary data of the DustPedia project \citep{davies17} and of the \expression{Dwarf Galaxy Sample} \citep[DGS;][]{madden13}.
We adopt a \expression{Hierarchical Bayesian}\footnote{In principle, we should say \expression{Bayesian-Laplacian}, in place of \expression{Bayesian}, as Pierre-Simon \familyname{Laplace} is the true pioneer in the development of statistics using the formula found by Thomas \familyname{Bayes} \citep{hahn05,mcgrayne11}.} (HB; \cf~\refsec{sec:HerBIE}) approach when comparing models to observations, in order to limit the impact of systematic effects on our results \citep[][ hereafter \citetalias{galliano18a}]{galliano18a}.
Finally, we perform a rigorous dust evolution modelling of individual objects in our sample, in order to unambiguously constrain the dust evolution timescales.
\refsec{sec:data} presents the data we have used.
\refsec{sec:model} presents our model and discusses the robustness of the derived dust parameters.
\refsec{sec:trends} provides a qualitative discussion of the derived dust evolution trends.
\refsec{sec:dustvol} describes the quantitative modelling of the main dust evolution processes.
\refsec{sec:summary} summarizes our results.
Several technical arguments are detailed in the appendices, so that they do not alter the flow of the discussion.

  \section{THE GALAXY SAMPLE}
  \label{sec:data}

This study focuses on global properties of galaxies.
Integrating the whole emission of a galaxy complicates the interpretation of the trends, as we will discuss in \refsec{sec:trends}.
However, it also presents some advantages: 
\begin{inlinelist}
  \item we can include galaxies unresolved at infrared wavelengths; and
  \item we can provide benchmarks for comparisons to unresolved studies of the 
    distant Universe or to one-zone dust evolution models.
\end{inlinelist}
Several upcoming studies on a subsample of resolved DustPedia galaxies will discuss the improvements that the spatial distribution of the dust properties provides (\citeprep{roychowdhury20}; \citeprep{casasola20b}).

    \subsection{The Infrared Photometry}
    \label{sec:photometry}

We present here the integrated, multi-wavelength photometry of our sample, used to constrain the global dust properties of each galaxy.
We focus on the mid-IR-to-submm regime, as it is where dust emits.

      \subsubsection{The DustPedia Aperture Photometry}
      \label{sec:DustPedia}

We use the photometry of the 875 galaxies of the DustPedia sample presented by \citet[][ hereafter \citetalias{clark18}]{clark18}\footnote{Available at \href{http://dustpedia.astro.noa.gr/Photometry}{http://dustpedia.astro.noa.gr/Photometry}.}.
Since we focus on the mid-IR-to-submm regime, we restrain the wavelength range to photometric bands centered between 3~\tmic\ and 1~mm.
The list of photometric bands we have used is given in \reftab{tab:photometry}.
\citetalias{clark18} has provided a dedicated reduction of the \hersc\ broadband data and an homogenization of the \spitz, \wise\ and \planck\ observations.
Foreground stars have been masked.
Aperture-matched photometry has been performed on each image and a local background has been subtracted from each flux.
A consistent noise uncertainty was estimated for each measurement.
\iras\ fluxes from \citet{wheelock94} were added to the catalog.
We refer the reader to \citetalias{clark18} for more details about the data reduction and photometric measurement.
\begin{table}[htbp]
  \caption{Number of galaxies observed through each photometric band.
           During the inference process (\refsec{sec:ref}), the observed fluxes are compared to the SED model integrated within the transmission of these filters, with the appropriate flux convention.}
  \label{tab:photometry}
  \begin{tabularx}{\linewidth}{lrrrR}
    \hline\hline
      Instrument & Wavelength & Label  & 
        \multicolumn{2}{c}{Number of galaxies} \\
                 & (central)  &  & 3$\sigma$ detection & Total \\
    \hline
WISE & $3.4\emic$ & WISE1 & 725 & 751 \\
 & $4.6\emic$ & WISE2 & 663 & 739 \\
 & $11.6\emic$ & WISE3 & 554 & 728 \\
 & $22.1\emic$ & WISE4 & 438 & 694 \\
    \hline
IRAC & $3.5\emic$ & IRAC1 & 277 & 292 \\
(\spitz) & $4.5\emic$ & IRAC2 & 359 & 390 \\
 & $5.7\emic$ & IRAC3 & 100 & 113 \\
 & $7.8\emic$ & IRAC4 & 116 & 130 \\
    \hline
MIPS & $23.7\emic$ & MIPS1 & 125 & 178 \\
(\spitz) & $71\emic$ & MIPS2 & 32 & 41 \\
 & $156\emic$ & MIPS3 & 18 & 25 \\
    \hline
PACS & $70\emic$ & PACS1 & 108 & 144 \\
(\hersc) & $100\emic$ & PACS2 & 273 & 456 \\
 & $160\emic$ & PACS3 & 296 & 493 \\
    \hline
SPIRE & $250\emic$ & SPIRE1 & 481 & 674 \\
(\hersc) & $350\emic$ & SPIRE2 & 446 & 658 \\
 & $500\emic$ & SPIRE3 & 404 & 634 \\
    \hline
IRAS & $60\emic$ & IRAS3 & 282 & 360 \\
 & $100\emic$ & IRAS4 & 391 & 501 \\
    \hline
HFI & $350\emic$ & HFI1 & 217 & 275 \\
(\planck) & $550\emic$ & HFI2 & 127 & 182 \\
 & $850\emic$ & HFI3 & 93 & 125 \\
    \hline
  \end{tabularx}
\end{table}

The SED model (\refsec{sec:HerBIE}), that we have applied to our data, performs a complex statistical treatment, allowing us to analyze even poorly sampled SEDs.
However, the efficiency of such a model can be affected by the presence of systematic effects not properly accounted for by the uncertainties.
In order to be conservative, we have therefore excluded a series of fluxes, based on the following criteria.
\begin{enumerate}
  \item\label{item:flag}
    The \citetalias{clark18} catalog flags about $22\,\%$ of its fluxes, 
    for different reasons: artefacts, contamination by nearby sources, 
    incomplete extended emission, \etc\
    We have excluded all the fluxes that are flagged.
    For 89 galaxies, all IR fluxes end up flagged.
    These galaxies have therefore been excluded.
  \item\label{item:AGN}
    Several galaxies contain a significant emission from their 
    \expression{Active Galactic Nucleus} (AGN).
    This emission is characterized by a prominent synchrotron continuum and 
    copious amounts of hot dust ($T_\sms{dust}\gtrsim300$~K), resulting in a 
    rather flat mid-IR continuum.
    Our dust model (\refsec{sec:themis}) is optimized for regular interstellar 
    dust.
    Our distribution of starlight intensities (\refsec{sec:dale}) is usually 
    not flexible enough to account for the hot emission from the torus.
    We have therefore excluded 19 sources presenting such an emission
    at a significant level, following \citet{bianchi18}, who used the criterion 
    of \citet{assef18} based on the WISE1 and WISE2 fluxes to identify 
    AGNs\footnote{Those are:
    \eso{434-040}, \ic{0691}, \ic{3430}, \ngc{1068}, \ngc{1320}, \ngc{1377}, 
    \ngc{3256}, \ngc{3516}, \ngc{4151}, \ngc{4194}, \ngc{4355}, \ngc{5347}, 
    \ngc{5496}, \ngc{5506}, \ngc{7172}, \ngc{7582}, \ugc{05692}, \ugc{06728}, 
    \ugc{12690}.}.
  \item\label{item:bad}
    We have performed a preliminary least-squares fit with our reference
    SED model (\refsec{sec:robust}), in order to identify where the largest 
    residuals are.
    \begin{enumerate}
      \item A few short wavelength bands present a large deviation from
        their SED model and their adjacent fluxes\footnote{Those are:
        WISE2 and IRAC2 for \eso{0358-006}, \eso{0116-012} and 
        \eso{0358-006}; and IRAC3 for \ngc{3794}.}.
        After inspection of the images, the discrepancies are likely due to 
        residual starlight contamination.
      \item Several long-wavelength \iras\ fluxes are significantly deviant 
        from their nearby MIPS and PACS fluxes\footnote{Those are: 
        IRAS4 for \ngc{254}, \ngc{4270}, \ngc{4322}, \ngc{5569} 
        and \ugc{12313}; and both IRAS3 and IRAS4 for \ic{1613}, 
        \ngc{584}, \pgc{090942}, \ic{2574}, \ugc{06016}, \ngc{3454}, 
        \ngc{4281}, \ngc{4633}, \ngc{5023}, \ngc{7715}.}.
        The reason of these discrepancies is obscure. 
        However, \refsec{sec:robust} will present a test comparing SED results 
        with and without \iras\ data, showing they are not crucial to our study.
      \item Three additional galaxies are not properly 
        fitted\footnote{Those are: \ngc{1052}, \ngc{2110} and \ngc{4486}.}.
        These objects present the characteristics of an AGN and are indeed
        classified as such \citep{liu02}.
        They were not accounted for by the \citet{assef18} criterion.
        It is however consistent with the fact that this criterion has a 
        $90\,\%$ confidence level.
    \end{enumerate}
    We have excluded all these problematic fluxes.
    Criterion \ref{item:bad} is more qualitative than \ref{item:flag} and 
    \ref{item:AGN}, but it allows us to identify potential systematics that 
    were missed.
\end{enumerate}
In total, we are left with 764 DustPedia galaxies.

The monochromatic fluxes (in Jy) were converted to monochromatic luminosities (in \tLsun/Hz), using the distances from the HyperLeda database \citep{makarov14}.

      \subsubsection{The Dwarf Galaxy Sample}
      \label{sec:DGS}

Metallicity is a crucial parameter to study dust evolution \citep[\cf~\refsec{sec:dustvol} and ][]{remy-ruyer14,remy-ruyer15}.
In particular, the low-metallicity regime, represented by dwarf galaxies, provides unique constraints \citep[\eg\ ][]{galliano18}, yet the DustPedia sample selected sources larger than 1\arcmin.
In order to improve the sampling of the low-metallicity regime, we have thus included additional galaxies from the \expression{Dwarf Galaxy Survey} \citep[DGS;][]{madden13}.

Among the 48 sources of the DGS, 35 are not in the DustPedia sample.
We have added these sources to our sample.
These galaxies have been observed with \spitz, \wise\ and \hersc.
We use the aperture photometry presented by \citet{remy-ruyer13,remy-ruyer15}.
We do not expect systematic differences between the DustPedia and DGS aperture fluxes.
\refapp{app:phot} compares the photometry of the DGS sources that are in DustPedia.
Both are indeed in very good agreement.
Similarly to the DustPedia sample, we apply the three exclusion criteria of \refsec{sec:DustPedia}.
\begin{enumerate}
  \item We exclude the fluxes that have been flagged.
  \item No significant AGN contribution is present in this sample.
  \item \citet{remy-ruyer15} advise to not trust the PACS photometry for 
    \hs{0822+3542}. 
    Since this source is neither detected with SPIRE, we simply exclude it.
\end{enumerate}
In total, we are left with 34 DGS galaxies, in addition to those already included in DustPedia.

      \subsubsection{Photometric Uncertainties}
      \label{sec:calunc}

Our combined sample contains 798 galaxies.
For each of them, we consider the following two sources of photometric uncertainty.
\begin{description}
  \item[The noise:]%
    it includes statistical fluctuations of the detectors
    and residual background subtraction.
    It has been thoroughly estimated by \citetalias{clark18} and 
    \citet{remy-ruyer15}.
    We assume that the noise of each waveband of each galaxy follows an 
    independent normal distribution\footnote{We note here that the background 
    subtraction introduces an uncertainty which is independent between galaxies.
    Indeed, we estimate the background in each waveband for each galaxy 
    separately.
    The resulting biases are thus randomly distributed across the sample.
    It would not have been the case, if we had considered
    individual pixels inside a galaxy.}.
  \item[Calibration uncertainties:]%
    they are systematics, \ie\ fully correlated
    between different galaxies, and partially correlated between wavebands.
    We assume they follow a joint, multivariate normal distribution, whose 
    covariance matrix is given in \refapp{app:calunc}.
\end{description}
\reftab{tab:photometry} gives the number of galaxies observed though each waveband, and the number of detections (flux greater than $3\sigma$, where $\sigma$ refers solely to the noise uncertainty).
The number of available filters per galaxy ranges between 1 and 19, and its median is 11.
There is a median number of 2 upper limits per galaxy.

    \subsection{The Ancillary Data}
    \label{sec:ancillary}

We present here the ancillary data gathered in order to characterize the ISM conditions in each galaxy.

      \subsubsection{The Stellar Mass}
      \label{sec:Mstar}

For the DustPedia sample, we adopt the stellar masses of \citet{nersesian19}.
Theses masses were derived from UV-to-mm SED fitting, using the code \CIG\ \citep{boquien19}, with two stellar populations.
For the DGS, the stellar masses are given by \citet{madden14}, using the \citet{eskew12} relation, based on the IRAC1 and IRAC2 fluxes.
\citet{eskew12} emphasize that the largest source of systematic uncertainties in the stellar mass is the \expression{Initial Mass Function} (IMF).
Both \citet{nersesian19} and \citet{eskew12} adopt a \citet{salpeter55} IMF, therefore limiting potential biases between the two samples.
Other potential biases, such as the form of the assumed star formation history, should not be an issue with our sample.
For instance, both \citet{mitchell13} and \citet{laigle19} tested the reliability of stellar mass estimates using numerical simulations of galaxies, 
and showed that they gave consistent results at low redshift.
We use $M_\star$ to denote the stellar mass.

\citet[][ hereafter \citetalias{nanni20}]{nanni20} have reestimated the stellar masses of the DGS, with \CIG.
They report systematically lower values, compared to \citet{madden14}, sometimes by an order of magnitude.
We discuss the possible reasons of this discrepancy in \refapp{app:Mstar}, concluding that our estimates are likely more reliable.

      \subsubsection{The Metallicity}
      \label{sec:metal}
 
For DustPedia galaxies, we use the metallicities derived by \citet{de-vis19}, using the $S$ calibration of \citet[][ hereafter PG16\_S]{pilyugin16}.
For the DGS, we use the metallicities derived by \citet{de-vis17}, using the same PG16\_S calibration.
\citet{de-vis17} show that this particular calibration is the most reliable at low-metallicity.

We adopt the solar elemental abundances of \citet{asplund09}:
the hydrogen mass fraction is $X_\odot=0.7381$, the helium mass fraction, $Y_\odot=0.2485$, the heavy element mass fraction, $Z_\odot=0.0134$, and the oxygen-to-hydrogen number ratio, $\met_\odot=8.69\pm0.05$.
Throughout this study, we assume a fixed elemental abundance pattern.
It implies that the total metallicity, $Z$, scales with the oxygen-to-hydrogen number ratio as: 
\begin{equation}
  Z\simeq2.04\E{-9}\times10^{\met}\eZsun.
  \label{eq:metal}
\end{equation}
For that reason, in the rest of the paper, we refer to both $Z$ and $\met$, as \expression{metallicity}.

      \subsubsection{The Gas Mass}    
      \label{sec:Mgas}

\paragraph{Neutral gas.}
The neutral hydrogen masses, derived from \hiline, have been compiled by \citet{de-vis19}, for DustPedia, and \citet{madden13}, for the DGS.
Integrating the \hi\ mass in the aperture used for the photometry is not always possible for the smallest sources, as they are not resolved by single dish observations.
This is particularly important for dwarf galaxies, where the \hi\ disk tends to be significantly larger than the optical/IR radius \citep[\eg\ ][]{begum08}, as we will discuss in \refsec{sec:D2M}.
\citetprep{roychowdhury20} recently obtained interferometric, spatially-resolved \hiline\ observations of 20 of the lowest-metallicity galaxies in our sample ($12+\log(\textnormal{O/H})\le8.0$)\footnote{Those are:
\ugc{00300}, \ngc{625}, \eso{358-060}, \ngc{1569}, \ugc{04305}, \ugc{04483}, \ugc{05139}, \ugc{05373}, \pgc{029653}, \ic{3105}, \ic{3355}, \ngc{4656}, \pgc{044532}, \ugc{08333}, \eso{471-006}, \mrk{209}, \ngc{2366}, \viizw, \izw\ and \sbs{1415+437}.}.
They integrated these \hi\ masses in the aperture used for the photometry in order to provide a better estimate of the gas mass associated with the dust emission.
We have adopted these more accurate values, for these 20 objects.
We have multiplied the \hi\ mass of each galaxy by $1/(1-Y_\odot-Z)\simeq1.35$ to account for helium (assumed independent of metallicity) and heavier elements.
We use $M_\sms{\hi}$ to denote the \emph{total} atomic gas mass probed by \hiline.

\paragraph{Molecular gas.}
\citet{casasola20} compiled observations of $^{12}$CO lines for 245 late-type DustPedia galaxies.
They converted these observations in molecular masses, assuming a constant $X_\sms{CO}$ conversion factor \citep{bolatto13}, and corrected them for aperture effects.
The uncertainties are the quadratic sum of the $^{12}$CO line noise measurement and $30\,\%$, corresponding to the uncertainty on the $X_\sms{CO}$. 
We adopt these values when available.
For the other DustPedia and DGS galaxies, we infer the \hmol\ mass, with the approximation used by \citet[][ Eq.~7]{de-vis19}.
It uses the scaling relation between $M_\sms{\hi}/\Ms$ and $M_\sms{\hmol}/M_\sms{\hi}$, derived by \citet{casasola20}.
The scatter of this relation is propagated and accounted for in the uncertainties.
These molecular gas masses are also corrected for helium and heavy elements.
We use $M_\sms{\hmol}^\sms{CO}$ to denote the \emph{total} molecular gas mass derived from actual $^{12}$CO line observations, and $M_\sms{\hmol}$, to denote the method-independent total molecular mass, either derived from $^{12}$CO lines or from the \citet{de-vis19} approximation.

      \subsubsection{The Star Formation Rate}
      \label{sec:SFR}

For DustPedia, the \expression{Star Formation Rate} (SFR), has been derived by \citet{nersesian19}.
The SFR was a free parameter of the SED fit they performed with \CIG.
For the DGS, we used the SFR estimated by \citet{remy-ruyer15}, using a combination of the \haline\ line and of the \expression{Total InfraRed luminosity} (TIR).
Although the two estimators are different, we do not expect significant biases
between DustPedia and the DGS.
Indeed, both account for:
\begin{inlinelist}
  \item the escaping power from young ionizing stars, with the UV SED or
    the \haline\ line; and
  \item the power re-radiated by dust, with the IR SED or the TIR.
\end{inlinelist}
In addition, both assume a \citet{salpeter55} IMF, which, similarly to the stellar mass (\refsec{sec:Mstar}), is the main source of uncertainty.
We use SFR to denote the numerical value of the SFR, expressed in \tMsun/yr and, $\textnormal{sSFR}\equiv\textnormal{SFR}/M_\star$, to denote the \expression{specific SFR}.

      \subsubsection{Uncertainties of the Ancillary Data}

The SED analysis we will present in \refsec{sec:HerBIE} includes the ancillary data, as a \expression{prior}.
For that purpose, it is preferable to consider the logarithm of quantities that can span several orders of magnitude.
The independent ancillary data parameters we use as a constraint are listed in the first five lines of \reftab{tab:ancillary}.
The total gas mass is $\Mg=M_\sms{\hi}+M_\sms{\hmol}$ and the $^{12}$CO-line-estimated molecular fraction, $f_\sms{\hmol}^\sms{CO}=M_\sms{\hmol}^\sms{CO}/\Mg$.
We also list, in the second part of \reftab{tab:ancillary}, parameters derived from these quantities, including the method-independent molecular fraction, $f_\sms{\hmol}=M_\sms{\hmol}/\Mg$.
All the extensive quantities (masses and SFR) have been homogenized to the distances adopted in \refsec{sec:photometry}.
\begin{table}[htbp]
  \caption{Number of galaxies per ancillary data constraint.}
  \label{tab:ancillary}
  \begin{tabularx}{\linewidth}{Xrr}
    \hline\hline
      Ancillary data     & Parameter    & Number of galaxies \\
    \hline
Total gas mass & $\ln\Mg$ & 518 \\
Molecular fraction & $\ln f_\sms{\hmol}^\sms{CO}$ & 179 \\
Stellar mass & $\ln\Ms$ & 785 \\
Star formation rate & $\ln\textnormal{SFR}$ & 749 \\
Metallicity & \tmet & 376 \\
    \hline
Molecular fraction & $\ln f_\sms{\hmol}$ & 514 \\
Gas-to-star ratio & $\ln (\Mg/\Ms)$ & 514 \\
Gas fraction & $\ln f_\sms{gas}$ & 514 \\
Specific SFR & $\ln\textnormal{sSFR}$ & 745 \\
    \hline
  \end{tabularx}
\end{table}

We assume that the uncertainty of the 5 independent ancillary parameters (first part of \reftab{tab:ancillary}) follows a split-normal distribution \citepalias[][ Sect.~3.2.3]{galliano18a}.
We propagate the uncertainties and their correlations in the derived parameters (second part of \reftab{tab:ancillary}).

  \section{THE SED MODELLING APPROACH}
  \label{sec:model}

We now describe our modelling approach, as well as the consistency tests we have performed to assess the robustness of our results.

    \subsection{Our In-House Hierarchical Bayesian SED Model}
    \label{sec:HerBIE}

\HB\ \citepalias[\expression{HiERarchical Bayesian Inference for dust Emission}; ][]{galliano18a} is a hierarchical Bayesian model aimed at inferring the \expression{Probability Density Functions} (hereafter PDF) of dust parameters (dust mass, \etc), from their SED, rigorously accounting for the different sources of uncertainties.
As any Bayesian model, \HB\ computes a \expression{posterior} PDF as the product of a classical \expression{likelihood} and a \expression{prior} PDF.
What makes this model \expression{hierarchical} is that the prior depends on a set of \expression{hyperparameters}.
These hyperparameters are:
\begin{inlinelist}
  \item the average of each physical parameter; and 
  \item their covariance matrix.
\end{inlinelist}
The hyperparameters are inferred, together with the parameters of each individual galaxy.
We are therefore sampling a single, large-dimension\footnote{The dimension of the parameter space is approximately the product of the number of galaxies and the number of model and ancillary parameters \citepalias[][ Sect.~3.3]{galliano18a}. In the present case, it has $\simeq798\times(7+5)\simeq10\,000$ dimensions.}, joint PDF.
Since the shape of the prior is inferred from the data, the information of the whole sample is used, in this process, to refine our knowledge of each individual galaxy.
In particular, it is relevant to keep in our sample even poorly-constrained SEDs, with upper limits or missing fluxes.
Such a model is also efficient at suppressing the numerous noise-induced, false correlations between parameters, encountered when fitting SEDs with least-squares or non-hierarchical Bayesian methods \citep{shetty09,kelly12,galliano18a,lamperti19}.
It has recently been used to study the anomalous microwave emission in the Milky Way \citep{bell19}.

From a technical point of view, \HB\ uses a \expression{Markov Chain Monte-Carlo} (hereafter MCMC), with Gibbs sampling \citep{geman84}.
As mentioned in \refsec{sec:ancillary}, we include the ancillary data in our prior.
This process is fully demonstrated in Sect.~5.3 of \citetalias{galliano18a}.
These ancillary data do not enter the dust model, but they provide information that helps to better constrain the hyperparameters, in a \expression{holistic} way.
In other words, the information provided by the gas mass or the metallicity helps to better constrain the dust SED fit.
It is a Bayesian implementation of \expression{Stein's paradox} 
\citep{stein56,efron77}.

    \subsubsection{Parametrization of the Dust Model}
    \label{sec:themis}

To infer dust parameters with \HB, we adopt the framework provided by the grain properties of the \THE\footnote{\href{http://www.ias.u-psud.fr/themis/}{www.ias.u-psud.fr/themis/}} model \citep{jones17}.
\THE\ is built, as much as possible, on laboratory data, and reproduces dust observables of the Galactic ISM.
One of its originalities is that it accounts for the aromatic and aliphatic mid-IR features with a single population of small, partially hydrogenated, amorphous carbons, noted a-C(:H).
Consequently, it does not include \expression{Polycyclic Aromatic Hydrocarbons} (PAH) \textit{per se}. 
Although largely dehydrogenated, small a-C(:H) are very similar to PAHs.
The other main component of \THE\ is a population of large, a-C(:H)-coated, amorphous silicates, with Fe and FeS nano-inclusions.

\THE\ is designed to model the evolution of:
\begin{inlinelist}
  \item the size distribution, 
  \item the a-C(:H) hydrogenation, and 
  \item the mantle thickness,
\end{inlinelist}
with the \expression{InterStellar Radiation Field} (ISRF) and gas density.
With the observational constraints of \refsec{sec:photometry}, we can not reliably constrain the mantle thickness, as its effect on the shape of the far-IR SED is too subtle.
Nor can we constrain the a-C(:H) hydrogenation, as broadband fluxes do not provide unambiguous constraints on the 3.4~\tmic\ feature.
We can however study the variations of the size distribution of small a-C(:H), from galaxy to galaxy, as it will affect the strength of the bright mid-IR aromatic features.

\begin{figure}[htbp]
  \includegraphics[width=\linewidth]{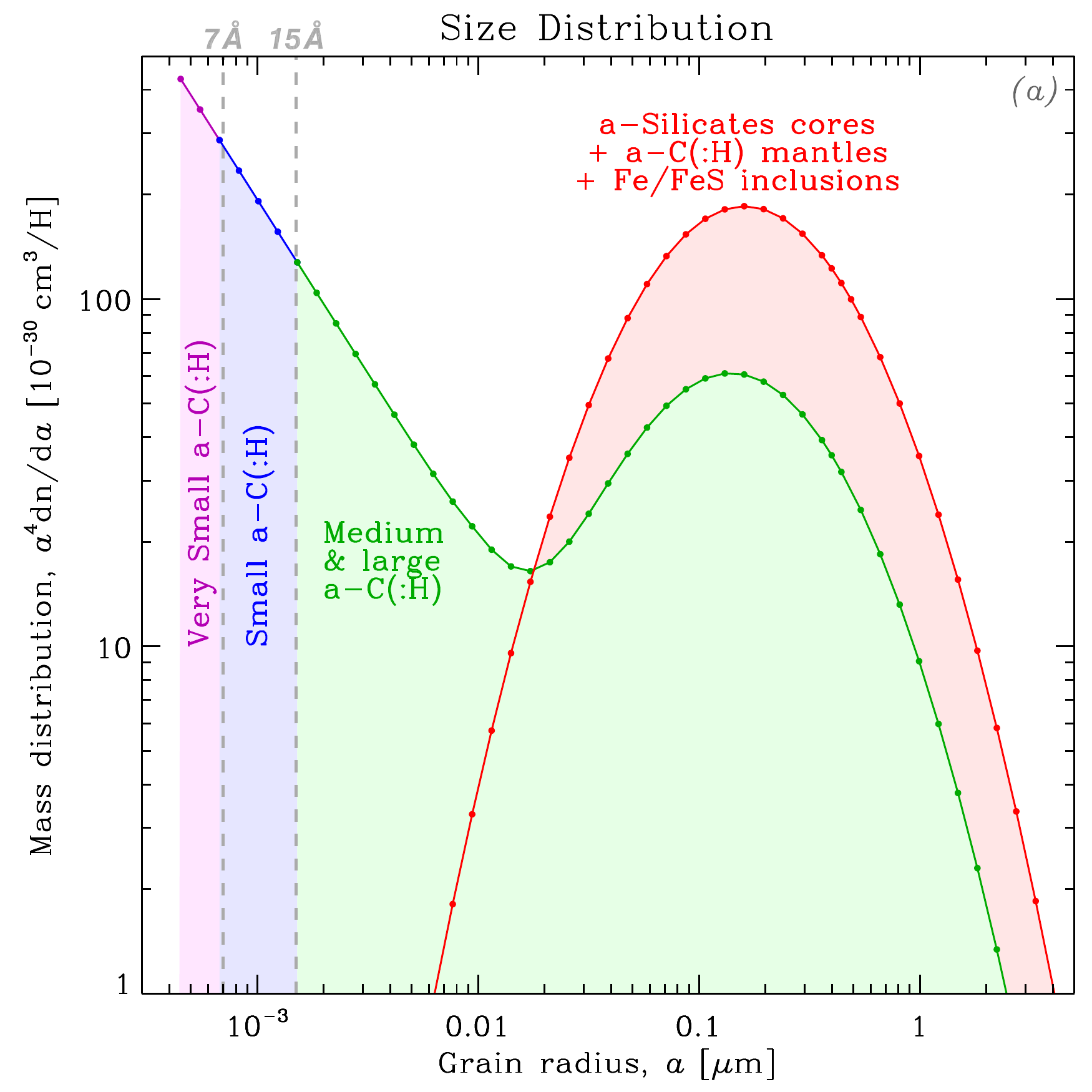} \\
  \includegraphics[width=\linewidth]{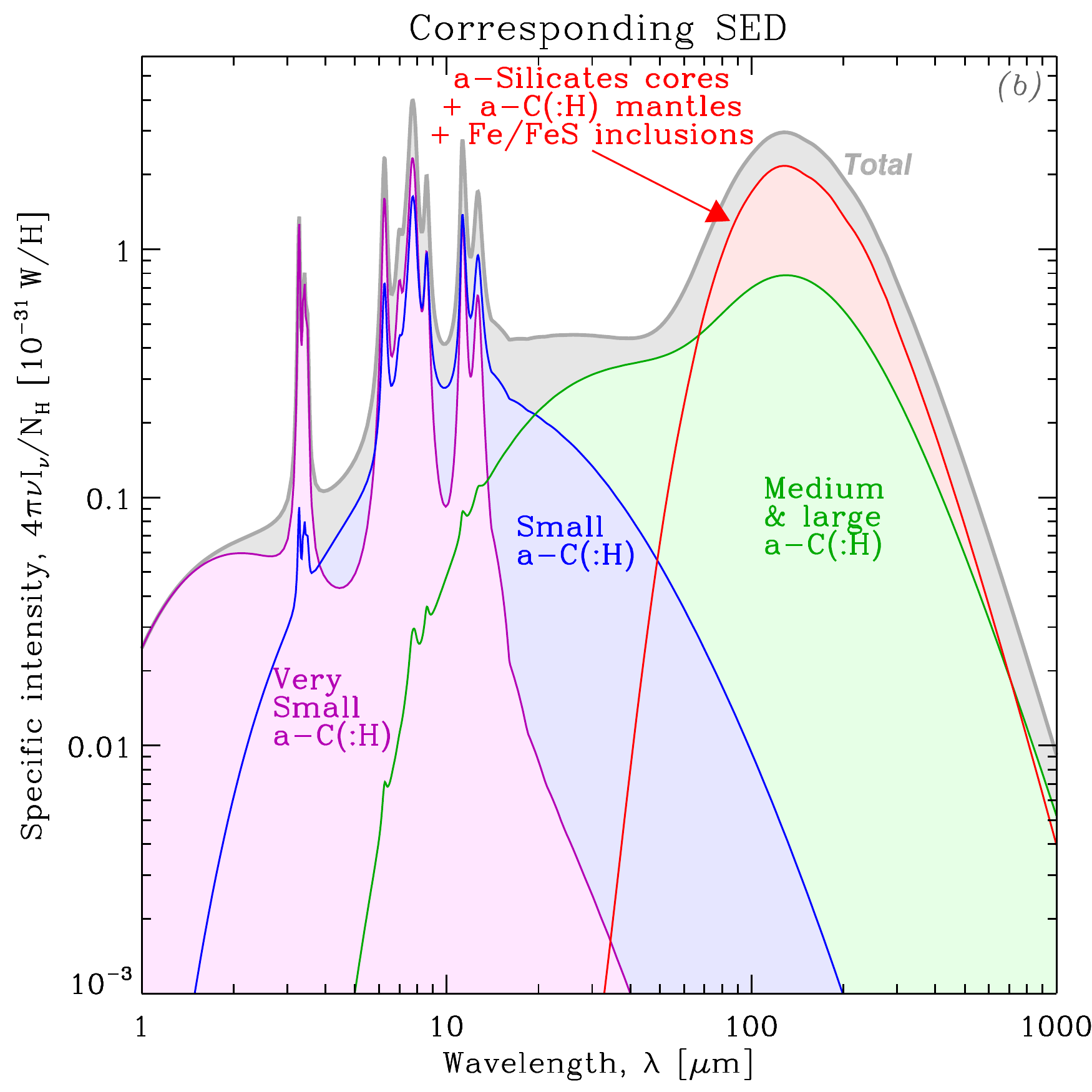}
  \caption{\textsl{Parametrization of the \THE\ model.}
           Panel \textit{(a)} shows the size distribution of the two main 
           components of \THE: amorphous carbons and silicates.
           We show how we divide the a-C(:H) size distribution into three 
           independent components.
           Panel \textit{(b)} shows the SED corresponding to each 
           component (same color code as panel \textit{a}).
           The SED is shown for the ISRF of the solar neighborhood 
           \citep{mathis83}.}
  \label{fig:themis}
\end{figure}
We have therefore parametrized the size distribution in the following way.
\begin{enumerate}
  \item We have divided the a-C(:H) component into (\cf~\reffig{fig:themis}):
    \begin{description}
      \item[Very small a-C(:H)]%
        (denoted VSAC)
        of radius smaller than 7~\AA, responsible for the mid-IR 
        feature emission, with more weight in the short wavelength 
        bands;
      \item[Small a-C(:H)]%
        (denoted SAC)
        of radius between 7~\AA\ and 15~\AA, responsible for the mid-IR
        feature emission, with more weight in the long wavelength bands;
      \item[Medium and large a-C(:H)]%
        (noted MLAC)
        of radius larger than 15~\AA,
        carrying the featureless mid-IR continuum and a fraction of the far-IR 
        peak.
    \end{description}
  \item  The silicate-to-MLAC ratio is kept constant.
\end{enumerate}
Using $q_X$ to denote the mass fraction of component $X$, the size distribution is controlled by two parameters:
\begin{inlinelist}
  \item the mass fraction of aromatic-feature-carrying grains, 
    $\qAF\equiv q_\sms{VSAC}+q_\sms{SAC}$; and
  \item the very-small-a-C(:H)-to-aromatic-feature-emitting-grain
    ratio, $f_\sms{VSAC}\equiv q_\sms{VSAC}/\qAF$.
\end{inlinelist}
Comparing these parameters to dust models with PAHs \citep[\eg\ ][]{zubko04,draine07,compiegne11}, \tqAF\ is the analogue of the PAH mass fraction, $q_\sms{PAH}$ introduced by \citet[][ hereafter \citetalias{draine07}]{draine07}.
The difference here is that, $\qAF=17\,\%$, while $q_\sms{PAH}=4.6\,\%$, in the diffuse Galactic ISM\footnote{We note here that the estimated mass fraction of PAHs in the diffuse Galatic ISM depends on the set of mid-IR observations used to constrain it. 
For instance, \citetalias{draine07} use \cobe/DIRBE broadbands, while \citet{compiegne11} use a scaled \spitz/IRS spectrum.
It results in different levels of mid-IR emission and PAH fraction: \citet{compiegne11} find $q_\sms{PAH}=7.7\,\%$.
Since \THE\ is calibrated with the same observations as \citet{compiegne11}, it is possible to estimate the value of $q_\sms{PAH}$ that would result in the same level of mid-IR emission as \tqAF: $q_\sms{PAH}\simeq0.45\times\qAF$ (ratio of $7.7\,\%$ to $17\,\%$).}.
This is because a-C(:H) have a smaller fraction of aromatic bonds per C atom.
More mass is therefore needed to produce the feature strength of a PAH.

This parametrization has the great advantage of being linear.
A simpler version, fixing $f_\sms{VSAC}$, has been implemented in \CIG\ by \citet{nersesian19}.
A more physical way to vary the size distribution would be to vary the minimum cut-off size and the index of the power-law size distribution \citep[\cf\ ][ for the description of the size distribution of \THE]{jones13}.
However, this alternate parametrization would be CPU time consuming, and would 
not produce noticeable differences in the resulting broadband SED (\cf\ \refapp{app:themis}).

    \subsubsection{The Mixing of Physical Conditions}
    \label{sec:dale}

A model, such as \THE, is not directly applicable to observations of galaxies.
Indeed, the monochromatic flux of a whole galaxy comes from the superposition of grains exposed to a range of physical conditions.
This well-known problem can be addressed assuming a distribution of starlight intensity inside each galaxy.
We adopt the prescription proposed by \citet{dale01}, where
the dust mass, \tMd, follows a power-law distribution, as a function of the starlight intensity, $U$: 
\begin{eqnarray}
  \dd\Md\propto U^{-\alpha}\ddiff U 
    & \mbox{ for } & \Umin < U < \Umin+\DU.
  \label{eq:dale}
\end{eqnarray}
The parameter $U$ is the frequency-integrated monochromatic mean intensity of the \citet{mathis83} diffuse ISRF.
It is normalized so that $U=1$ in the solar neighborhood \citep[\eg\ Eq.~2 of ][]{galliano11}.
This phenomenological composite SED is the \powerU\ model component of \citetalias{galliano18a} (Sect.~2.2.5).
We add to the emission the \starBB\ component of \citetalias{galliano18a} (Sect.~2.2.6), in order to account for the residual diffuse stellar emission at short wavelengths.

The list of free SED model parameters, that \HB\ infers, is thus the following.
Similarly to the ancillary parameters (\reftab{tab:ancillary}), we use the natural logarithm of quantities varying over more than an order of magnitude.
\begin{enumerate}
  \item $\ln\Md$, the dust mass, scales with the whole dust SED. 
  \item $\ln\Umin\in[\ln0.01,\ln10^3]$ is the minimum cut-off in 
    \refeq{eq:dale}.
  \item $\ln\DU\in[\ln1,\ln10^7]$ controls the range in \refeq{eq:dale}.
  \item $\alpha\in[1,2.5]$ is the power-law index in \refeq{eq:dale}.
  \item $\ln\qAF\in[\ln10^{-5},\ln0.9]$ is defined in \refsec{sec:themis}.
  \item $f_\sms{VSAC}\in[0,1]$ is defined in \refsec{sec:themis}.
  \item $\ln L_\star$  is the luminosity of the residual stellar emission
    \citepalias[\cf\ ][ Sect.~2.2.6]{galliano18a}.
\end{enumerate}
These parameters are however not the most physically relevant.
In the rest of the present article, we focus our discussion on the following three parameters, marginalizing over the other ones:
\begin{inlinelist}
  \item \tMd;
  \item \tUav; and
  \item \tqAF,
\end{inlinelist}
where \tUav\ \citepalias[Eq.~9 of][]{galliano18a} is the mean of the distribution in \refeq{eq:dale}.
It quantifies the mass-averaged starlight intensity illuminating the grains.
It is a function of the three parameters $U_\sms{min}$, $\Delta U$ and $\alpha$
\citepalias[Eq.~10 of ][]{galliano18a}.
It can be related to an equivalent large grain equilibrium temperature, $T_\sms{eq}$, through: $\Uav\simeq(T_\sms{eq}/18\;\textnormal{K})^{5.8}$ \citep[\eg\ ][]{nersesian19}.

      \subsubsection{Questionable Assumptions and Residual 
                           Contaminations}
      \label{sec:contaminations}

In our experience, the model of \refsec{sec:dale} is the most appropriate for galaxies observed with the typical spectral coverage of \reftab{tab:photometry}.
It presents however the following limitations.

\paragraph{Shape of the ISRF.}
We assume that dust is heated by the \citet{mathis83} ISRF, scaled by a factor $U$.
We assume in the model that the shape of this ISRF does not vary between galaxies nor within regions inside galaxies.
It is obvious that this assumption is not correct. 
However, its consequences are minimal on the parameters we are interested in, for the following reasons.
\begin{itemize}
  \item \tMd\ and \tUav\ depend mainly on the far-IR peak emission, which is 
    dominated by large grains.
    These grains are at thermal equilibrium.
    Their emission therefore does not depend on the shape of the ISRF, only on 
    the absorbed power.
  \item \tqAF\ controls the fraction of small, stochastically-heated grains.
    The emission from these grains depends on the shape of the ISRF 
    \citep[\eg\ ][]{camps15b}.
    However, since small a-C(:H) are destroyed in \hii\ regions 
    \citep[\cf\ Sect.~4.2.3 of ][ and \refsec{sec:qAFev}]{galliano18}, they are 
    effectively heated by a rather narrow spectral range 
    ($4\;\textnormal{eV}\lesssim h\nu<13.6\;\textnormal{eV}$), where they exist.
    This effect is demonstrated on Fig.~7 of \citet{draine11}, with PAHs.
    We thus do not expect that actual variations of the ISRF shape will 
    significantly bias our estimate of \tqAF.
\end{itemize}

\paragraph{Evolution of small a-C(:H).}
The abundance of small a-C(:H) and their properties evolve with the ISRF and the gas density.
These grains are dehydrogenated and destroyed by intense ISRFs;
they are also accreted onto large grains, in dense regions \citep[\eg\ ][]{jones13,kohler15}.
Our parametrization (\refsec{sec:themis}) allows us to explore variations of \tqAF\ between galaxies, but we assume that \tqAF\ is constant, for all $U$, within each galaxy.
However, this assumption will not bias our global estimate of \tqAF, as this parameter is merely a way to give a physical meaning to the observed $L_\sms{AF}/\textnormal{TIR}$ ratio ($L_\sms{AF}$ denoting the power emitted by the aromatic features).
This assumption would only be problematic if we were trying to estimate the local value of \tqAF\ in PDRs, for instance.
Indeed, we would, in this case, underestimate \tqAF, by assuming that a fraction of the aromatic feature emission comes from \hii\ regions, which are generally hotter than PDRs, and thus more emissive.

\paragraph{Grain opacity.}
We assume that the grain opacity does not vary between galaxies, nor within regions inside galaxies.
There are several indications that this hypothesis is not correct \citep[\eg\ Sect.~4.2.1 of ][]{galliano18}.
In particular, the far-IR opacity can typically vary by a factor of $\simeq3$ with the accretion/removal of mantles and grain-grain coagulation \citep{kohler15}.
Such variations will bias our estimate of the dust mass.
In addition, the submm/mm silicate emissivity implemented in \THE\ has an unphysical power-law dependence at long wavelengths.
Accounting for a more realistic opacity, based on laboratory measurements such as those of \citet{demyk17a,demyk17b}, will likely affect our dust mass estimates.
This is an important effect that we can not avoid.
We discuss its consequences in \refsec{sec:D2M}, when interpreting our results.

\paragraph{Residual contaminations.}
The following emission processes, not accounted for in our model, could be present in our observations.
\begin{itemize}
  \item Several gas lines contribute to the emission in the photometric bands 
    of \reftab{tab:photometry}. 
    The brightest are: \oiline, \oiiiline, \ciiline, \oilineb\ and 
    \COiiitoii.
    Without dedicated spectroscopic observations, the subtraction of these 
    lines is hazardous.
    Luckily enough, their intensities are, to first order, proportional 
    to TIR \citep[\eg\ ][]{cormier15}.
    They constitute therefore a bias proportional to the flux,
    that can be taken into account together with the calibration bias, as
    we will show in \refsec{sec:ref}.
  \item The \expression{submm excess} is an excess emission of debated origin, 
    appearing longward 500~\tmic\ 
    \citep[\cf\ Sect.~3.5.5 of ][ for a review]{galliano18}.
    Since it is not accounted for in our model, it could bias our submm SED.
    This effect is however probably negligible in our sample, for the following 
    reasons.
    \begin{enumerate}
      \item This submm excess is model-dependent.
        \THE\ has a flatter submm opacity than models based on the 
        \citetalias{draine07} optical properties 
        \citep[\eg\ Fig.~4 of ][]{galliano18}.
        It is therefore more emissive in the submm regime and minimizes the
        contribution of the excess.
      \item This excess is observed in dwarf galaxies, but is rarely detected
        in higher metallicity systems
        \citep[\eg\ ][]{galliano03,galliano05,galametz09,remy-ruyer15,dale17}.
        Yet, data longward 500~\tmic, in our sample, are available only for 
        large objects, due to the large \planck/HFI beam.
      \item Residuals of our model run (\refsec{sec:ref}) do not show 
        significant excesses in the submm bands.
    \end{enumerate}
  \item Residual stellar emission could contaminate the shortest 
    wavelength bands.
    Improper stellar subtraction could lead to positive or negative offsets,
    independently in the different bands.
\end{itemize}

    \subsection{The Reference Run}
    \label{sec:ref}

\begin{figure*}[htbp]
  \includegraphics[width=\textwidth]{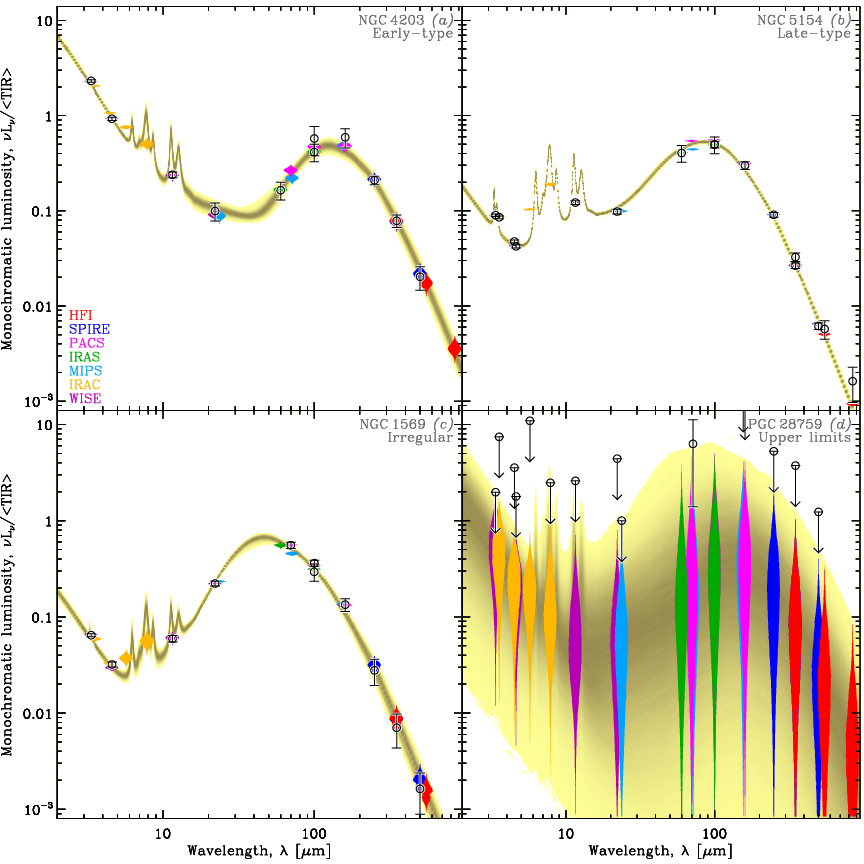}
  \caption{\textsl{Select SED fits.}
           Each panel shows the inferred SED PDF, in yellow intensity scale,
           normalized by the MCMC-averaged TIR.
           The color violin plots represent the synthetic photometry PDF,
           for each waveband.
           The black error bars are the observations.
           Panels \textit{(a)}, \textit{(b)} and \textit{(c)} show arbitrary 
           chosen galaxies, representative of their  classes (early-type, 
           late-type and irregulars, respectively).
           Panel \textit{(d)} shows the case, where most constraints are only 
           upper limits.}
  \label{fig:SEDs}
\end{figure*}

The results we will present in \refsec{sec:trends} are obtained with the model described in \refsec{sec:HerBIE}.
We call it our \expression{reference} run.
\reffig{fig:SEDs} shows the SED fit of four representative objects, obtained with this run.
In panels \textit{(a)}, \textit{(b)} and \textit{(c)}, SEDs of three arbitrarily chosen galaxies, representative of \expression{Early-Type Galaxies} (ETG), \expression{Late-Type Galaxies} (LTG) and irregulars, respectively, are displayed.
The PDF of the SED is shown as a yellow density plot.
We also show the PDF of the synthetic photometry (violin plots\footnote{violin plots are $90^\circ$-rotated histograms.} of different colors).
The comparison of this synthetic photometry to the observed flux (circles with 
an error bar) reflects the quality of the fit. 
We discuss a more thorough and technical fit quality test in \refapp{app:ppp}.
Panel \textit{(d)} shows the example of a galaxy where most of the fluxes are $3\sigma$-upper limits (displayed when $F_\nu<\sigma_\nu^\sms{noise}$).
When the evidence provided by the data is weak, which is the case when few detections are available, the posterior distribution becomes dominated by the prior.
This is what we see here.
The range spanned by this SED PDF is the extent of the prior.
The global scaling of this SED is very uncertain, but its shape is realistic.
This would not be the case with a non-hierarchical model.

\begin{figure}[htbp]
  \includegraphics[width=\linewidth]{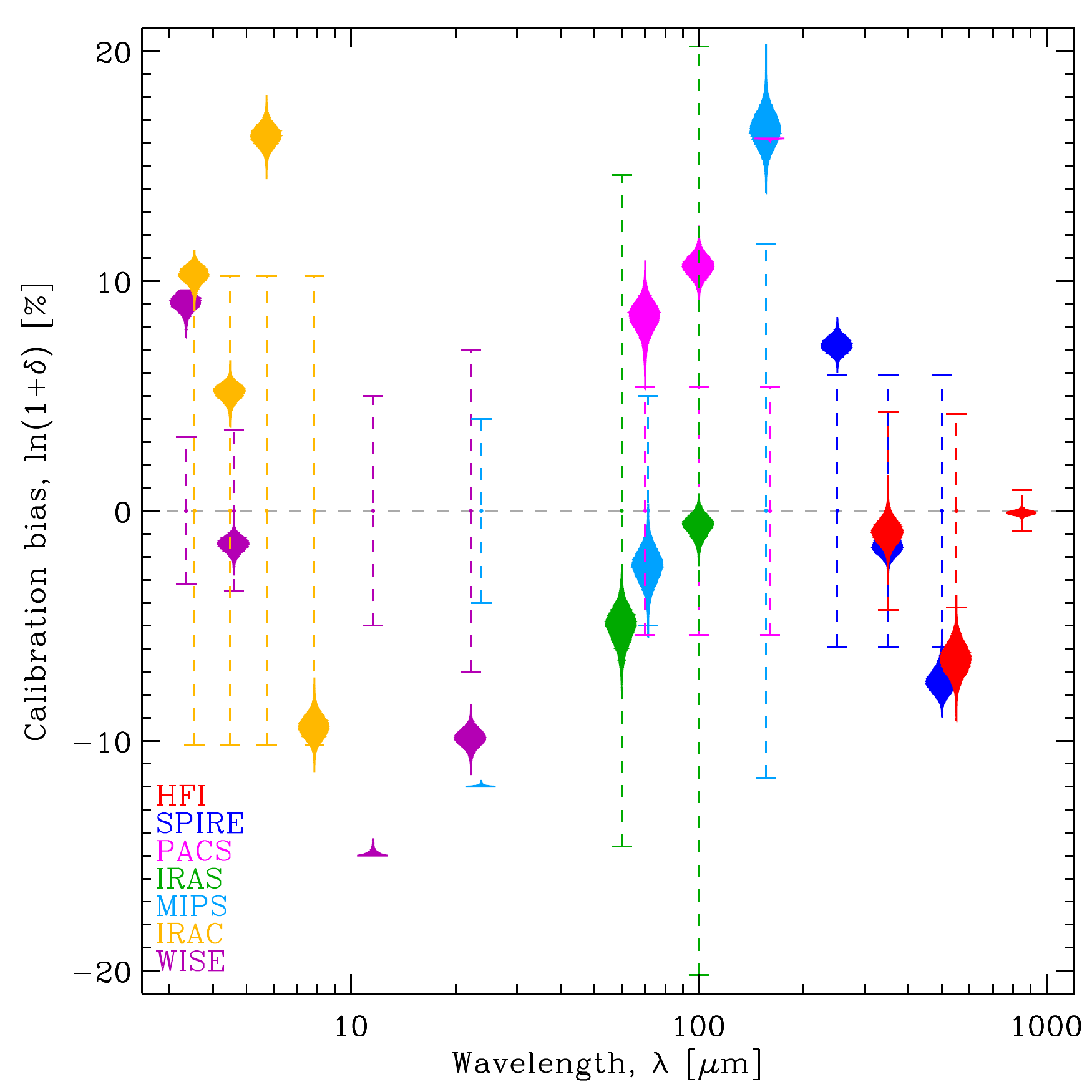}
  \caption{\textsl{Inferred calibration biases.}
           This figure shows our inference of the calibration bias, $\delta$, 
           defined in Sect.~3.2.2 of \citetalias{galliano18a},
           for each of the photometric filters in \reftab{tab:photometry}.
           Each instrument is color-coded.
           The vertical dashed error bars represent the calibration prior, 
           \ie\ the $\pm1\sigma$ range given in \refapp{app:calunc}.
           The \expression{violin plots} show the actual posterior PDFs of 
           $\delta$, for each broadband.
           The width of a single violin plot is proportional to the PDF as a 
           function of $\delta$ (vertical axis).}
  \label{fig:delta}
\end{figure}
\reffig{fig:delta} shows the inference of the calibration biases, $\delta(\lambda)$, of each waveband (\refsec{sec:calunc}).
Basically, the model fluxes are corrected by factors $1+\delta(\lambda)$.
For each waveband, the dashed-line error bar represents the $\pm1\sigma$ interval of the prior on the calibration bias.
This prior is a multivariate normal distribution, centered on 0, with the covariance matrix of \refeq{eq:Vcal}.
As discussed in \refsec{sec:contaminations}, besides accounting for the uncertainty in the calibration of each instrument, these coefficients can account for external contaminations and fitting residuals.
We can interpret the most deviant values of \reffig{fig:delta} in this light.
The posterior values of $\delta$ lying outside their $\pm2\sigma$ prior are the following.
\begin{description}
  \item[160~$\pmb{\mu}$m:] both overlapping bands PACS3 and MIPS3 
    indicate an excess emission of about $15\,\%$.
    A significant fraction of this excess is likely due here to contamination 
    by the brightest line of the ISM, \ciiline.
    The \ciiline\ intensity is about $\simeq1\,\%$ the FIR (far-IR; 
    60-200~\tmic) power 
    \citep{malhotra97,malhotra01,brauher08,cormier15}.
    Its contribution to the PACS3 and MIPS3 is thus expected to be a few 
    percent, as these bands are narrower than the FIR range.
  \item[12~$\pmb{\mu}$m and 22~$\pmb{\mu}$m:] the deficit could be 
    due to a systematic difference between the size distribution of medium 
    a-C(:H) grains implemented in \THE\ and in the diffuse ISM of 
    these galaxies.
    Although the model was different, \citet[][ Appendix~A.2]{galliano11} had 
    to decrease the abundance of the grains carrying the 24~\tmic\ continuum in 
    the \expression{Large Magellanic Cloud} (LMC), for a similar reason.
\end{description}
Apart from these deviations, the two overlapping bands WISE1 and IRAC1 indicate that the model, on average, underestimates the observations by $\simeq10\,\%$ around 3.5~\tmic.
This could be due to either the continuum or the 3.3/3.4~\tmic\ features.
The constraint of the size distribution of \THE\ by the SED of the Milky Way  \citep[Fig.~3 of][]{jones13}, has been performed with a medium resolution IRS spectrum for all a-C(:H) features, except the two in question here.
The global emissivity of these features is the most uncertain of the model.
We also notice that the SPIRE calibration biases go in opposite directions, while they should be partially correlated (\refapp{app:SPIRE}).
The HFI1 and HFI2 calibration biases agree very well with SPIRE.
This is the sign that there is a systematic residual proportional to the flux in this wavelength range.
It could mean that the slope of \THE\ is not steep enough in the $200-500\emic$ range\footnote{We do not implement the evolution of the grain properties with accretion and coagulation that could account for a steeper far-IR slope.}.
The fact that the HFI3 is almost 0 would mean that the slope flattens again
longward $\simeq800\emic$.
This is very speculative, but we note that this behaviour is qualitatively consistent with the optical properties measured in the laboratory by \citet{demyk17b}.

The parameters inferred with this run, for each galaxy, are given in \reftab{tab:resref} (electronic version only).
We report only the parameter values and their uncertainties (mean and standard-deviation of the posterior PDF).
The skewness and correlation coefficients can be retrieved from the DustPedia archive (\DPar).
The length of the MCMC of this run is $1\,000\,000$, where we have excluded the first $100\,000$ steps, to account for burn-in.
The maximum integrated autocorrelation time \citepalias[Eq.~43 of][]{galliano18a} is $65\,000$.

    \subsection{Additional Runs: Robustness Assessment}
    \label{sec:robust}

\begin{figure*}[htbp]
  \includegraphics[width=\textwidth]{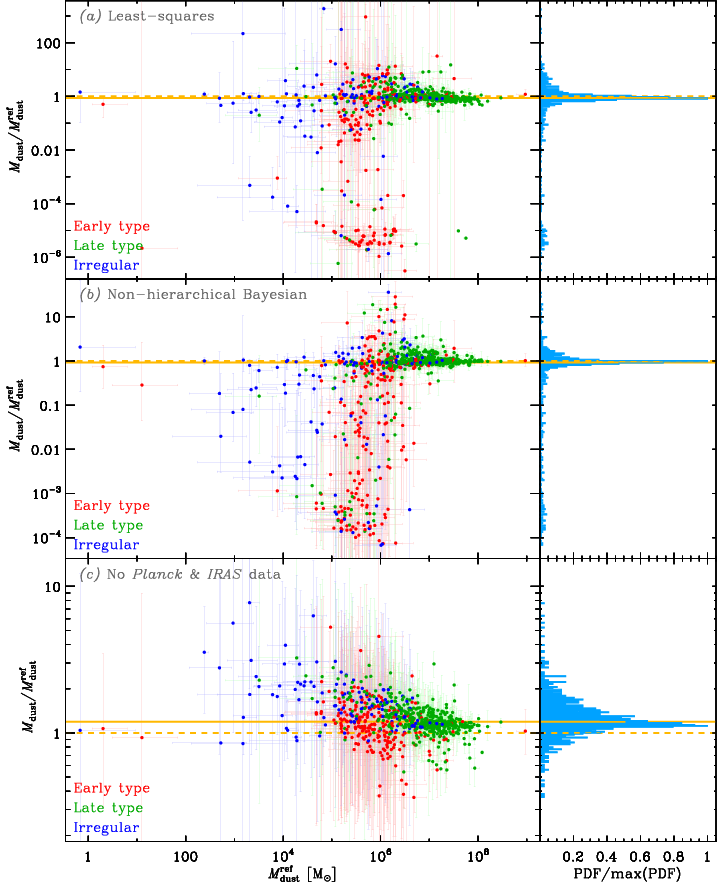}
  \caption{\textsl{Robustness assessment.}
           Each of the three horizontal panels of Figs.~\ref{fig:comp1} to 
           \ref{fig:comp3} display the comparison of the various tests of
           \refsec{sec:robust} to our reference run (\refsec{sec:ref}).
           The left column panels show the ratio of either \tMd\ or \tqPAH, 
           derived from the test (grey label in the top left corner), to its 
           equivalent with the reference run, as a function of \tMdref. 
           Galaxies are color-coded according to their type.
           The right column plot shows the PDF (normalized) of the distribution
           of the ratio.
           The median of the ratio is displayed as an orange solid line.
           The 1:1 ratio is highlighted as an orange dashed line.
           This figure shows the influence of various fitting methods.}
  \label{fig:comp1}
\end{figure*}

In order to assess the robustness of our results, we have performed several additional runs, with different assumptions.
We now discuss the comparison of these tests with our reference run.
We focus on the derived dust mass and do not display the redundant comparisons with $\Uav$.
Indeed, TIR is usually well constrained and, by model construction, $\textnormal{TIR}\propto\Md\times\Uav$.
We discuss the difference in \tqAF, when relevant.
\reffigs{fig:comp1}{fig:comp3} compares \tMd\ or \tqAF\ derived by these tests to the same quantity derived by the reference run, \tMdref\ and \tqAFref, respectively.
\reftab{tab:comp} reports some statistics on this comparison.
We color code the galaxies according to their Hubble stage, $T$:
\begin{description}
  \item[Early-type:] $T\le 0$ (red);
  \item[Late-type:] $0<T<9$ (green);
  \item[Irregulars:] $T\ge9$ (blue).
\end{description}

\begin{figure*}[htbp]
  \includegraphics[width=\textwidth]{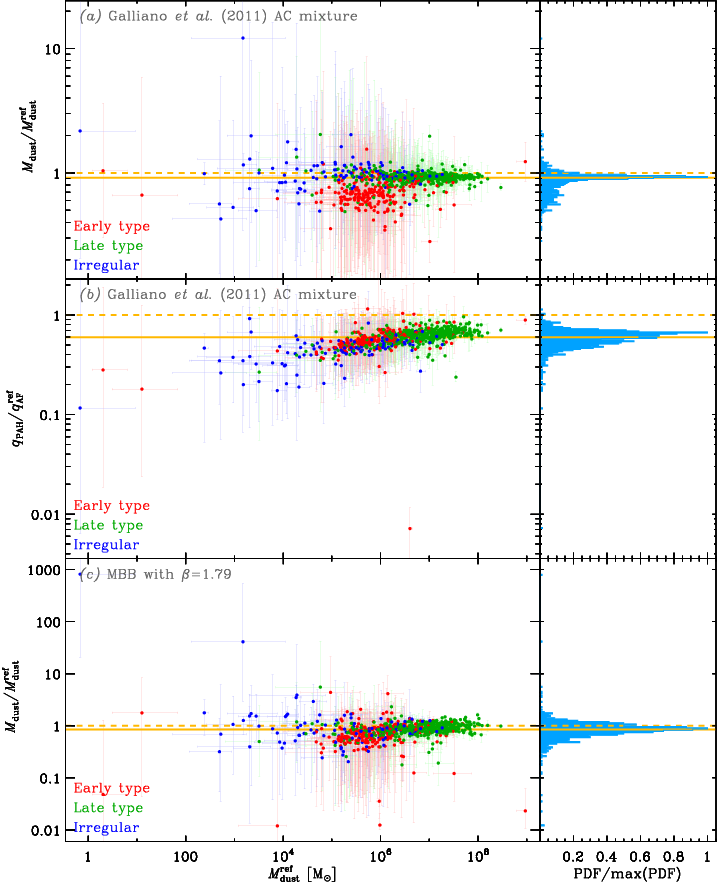}
  \caption{Same as \reffig{fig:comp1}.
           This figure shows the influence of various fitting methods.}
  \label{fig:comp2}
\end{figure*}
\begin{figure*}[htbp]
  \includegraphics[width=\textwidth]{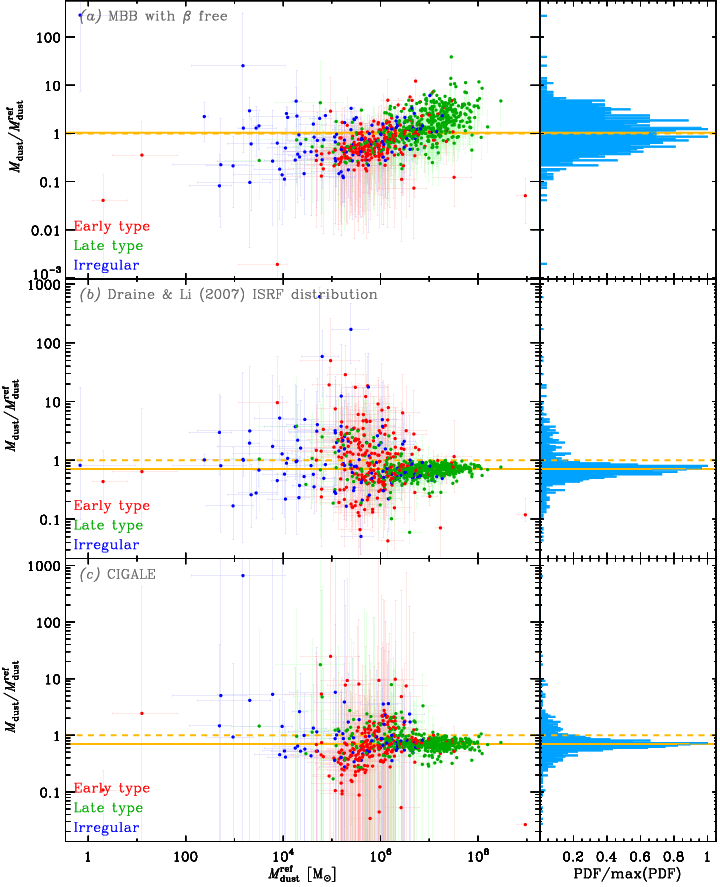}
  \caption{Same as \reffig{fig:comp1}.
           The first two panels continue to show the influence of various dust 
           model assumptions as in \reffig{fig:comp1}. 
           The other panel is a comparison with results from a previous work 
           done using the same sample but different method.}
  \label{fig:comp3}
\end{figure*}
\begin{table*}[htbp]
  \caption{\textsl{Robustness assessment.}
         Statistics of the comparison
         between the reference run and the 
         various tests of Figs.~\ref{fig:comp1} to \ref{fig:comp3}.
         The median, $68\,\%$ and $95\,\%$ intervals refer to the
         distribution of the quantity in the first column.
         See \refsec{sec:robust} for more details.}
  \label{tab:comp}
  \centering\begin{tabular}{llrrrr}
    \hline\hline
      Quantity & Run & Median & $68\,\%$ interval 
      & $95\,\%$ interval & Number of sources\\
    \hline
$\Md/\Mdref$ & Least-squares & $0.86$ & $0.128$ -- $1.50$ & $2.83\E{-6}$ -- $7.0$ & 783 \\
$\Md/\Mdref$ & Non-hierarchical Bayesian & $0.92$ & $0.0223$ -- $1.22$ & $0.000164$ -- $3.7$ & 798 \\
$\Md/\Mdref$ & No \planck\ \&\ \iras\ data & $1.19$ & $1.00$ -- $1.60$ & $0.69$ -- $2.44$ & 783 \\
$\Md/\Mdref$ & \citet{galliano11} AC mixture & $0.91$ & $0.69$ -- $0.97$ & $0.51$ -- $1.15$ & 798 \\
$\qPAH/\qAFref$ & \citet{galliano11} AC mixture & $0.60$ & $0.48$ -- $0.68$ & $0.37$ -- $0.80$ & 798 \\
$\Md/\Mdref$ & MBB with $\beta=1.79$ & $0.85$ & $0.61$ -- $1.08$ & $0.35$ -- $1.39$ & 798 \\
$\Md/\Mdref$ & MBB with $\beta$ free & $1.00$ & $0.43$ -- $2.75$ & $0.183$ -- $5.3$ & 798 \\
$\Md/\Mdref$ & \citet{draine07} ISRF distribution & $0.71$ & $0.49$ -- $1.31$ & $0.186$ -- $4.6$ & 798 \\
$\Md/\Mdref$ & CIGALE & $0.70$ & $0.50$ -- $1.06$ & $0.227$ -- $3.6$ & 756 \\

    \hline
  \end{tabular}
\end{table*}

      \subsubsection{Influence of the Fitting Method}
      \label{sec:robustfit}

\paragraph{Least-squares.}
We have fit the full sample of \refsec{sec:data} with the physical model of \refsec{sec:HerBIE}, using a standard least-squares method 
(\cf\ Appendix~C of \citetalias{galliano18a}).
The goal of this run is to demonstrate the importance of the fitting method.
The result is shown in panel $(a)$ of \reffig{fig:comp1}.
The results for sources with high signal-to-noise ratio ($\Mdref\gtrsim10^6\eMsun$; mainly LTGs) are in very good agreement with the reference run.
However, there is a significant scatter for sources with $\Mdref\lesssim10^6\eMsun$ (mainly ETGs and irregulars).
Furthermore, there are more drastic underestimates than overestimates.
The galaxies with $\Md/\Mdref\simeq10^{-5}$ are indeed essentially sources with only far-IR upper limits, having a negligible weight in the chi-squared.
These SEDs are thus wrongly fit a with very high \tUav, and scaled on the mid-IR fluxes, leading to a drastic underestimate of the dust mass.

\paragraph{Non-hierarchical Bayesian.}
We have fit the full sample of \refsec{sec:data} with the physical model of \refsec{sec:HerBIE}, replacing the hyperparameter distribution by a flat prior
(\cf\ Sect.~4.2.3 of \citetalias{galliano18a}).
This flat prior is bounded, with limits well beyond the displayed parameter range.
This run thus samples the likelihood of each galaxy, independently, but does not benefit from an informative prior, inferred from the whole sample.
The result is shown in panel~\textit{(b)} of \reffig{fig:comp1}.
Similarly to the least-squares example above, there is no bias for the well sampled sources ($\Mdref\gtrsim10^6\eMsun$), but there is a significant scatter for sources with $\Mdref\lesssim10^6\eMsun$.
The scattered sources are underestimates, for similar reasons as the least-squares example.
The amplitude of the scatter is however smaller than for the least-squares run.
In addition, even the most extremely scattered sources are $2\sigma$-consistent
with the 1:1 relation.
This is because a Bayesian method samples the likelihood as a function of the parameters, but stays conditional on the data, while a frequentist approach (\eg\ least-squares) samples the likelihood as a function of the data, thus considering data that have not actually been observed, leading to biases \citep[\eg\ ][]{jaynes76,wagenmakers08}.

\paragraph{No \planck\ \&\ \iras\ data.}
To understand the role of the far-IR coverage, we have fit the sample of \refsec{sec:data} without the \planck\ and \iras\ data, using the physical model of \refsec{sec:HerBIE}.
In this case, the far-IR coverage is provided by the sole \hersc\ data.
There is thus no data longward 500~\tmic.
The result is shown in panel~\textit{(c)} of \reffig{fig:comp1}.
The absence of this data does not bias the fit, it simply introduces some scatter, consistent with the 1:1 relation (\cf\ Sect.~5.2 of \citetalias{galliano18a} for more discussion on this effect).
The irregulars are marginally overestimated, due to the lack of sufficient SPIRE detections. 
Some of these sources do not have any submm detections, higher dust masses are therefore allowed.

      \subsubsection{Influence of the Dust Model Assumptions}
      \label{sec:robustmod}

\paragraph{With the \citet{galliano11} dust mixture.}
We have fit the full sample of \refsec{sec:data} with the HB model of \refsec{sec:HerBIE}, replacing the \THE\ grain properties by those of the \expression{Amorphous Carbon} (AC) model of \citet{galliano11}.
The far-IR opacity of the two grain mixtures are comparable \citep[\cf\ Fig.~4 of ][]{galliano18}.
The only fundamental difference is that the aromatic features are accounted for by PAHs, not by a-C(:H).
Panel~\textit{(a)} of \reffig{fig:comp2} compares \tMd\ to our reference model.
As expected, the two values are in good agreement.
The moderate scatter is due to the mild difference between the two far-IR opacities.
However, most of the ratios are $1\sigma$-consistent with the 1:1 relation.
Panel~\textit{(b)} of \reffig{fig:comp2} compares the $q_\sms{PAH}$ of this run to the \tqAF\ of the reference run.
In \refsec{sec:themis}, we noted that we should have $q_\sms{PAH}/\qAF\simeq0.45$.
In the present case, we have a ratio of $\simeq0.6$.
A possible explanation of this difference is the following.
The parametrization of the \THE's aromatic spectrum shape is controlled by $f_\sms{VSAC}$ (\refsec{sec:themis}), which alters the a-C(:H) mass.
On the contrary, for the \citet{galliano11} AC model, the shape of the aromatic spectrum is controlled by the fraction of ionized PAHs, which does not alter the PAH mass.
A systematically lower 8-to-12~\tmic\ ratio compared to the Galaxy's diffuse ISM would explain a $q_\sms{PAH}/\qAF$ ratio higher than expected.

\paragraph{Modified black body with $\beta=1.79$.}
We have fit the photometry of \refsec{sec:data}, longward 100~\tmic, with a Modified Black Body (MBB; \eg\ Sect.~2.2.2 of \citetalias{galliano18a}).
In this first test, we fix the emissivity index $\beta=1.79$ and the level of the opacity to mimic the far-IR opacity of \THE:
$\kappa(\lambda)=0.64\;\textnormal{m}^2\,\textnormal{kg}^{-1}\times(250\emic/\lambda)^{1.79}$ \citep[\eg\ Sect.~3.1.1 of ][]{galliano18}.
Panel~\textit{(c)} of \reffig{fig:comp2} compares \tMd\ to our reference run.
We can see that \tMd\ is about 0.8 times lower.
This value is consistent with what \citet[][ Appendix~C.2]{galliano11} found in the LMC.
This difference is due to the fact that a MBB is an isothermal approximation.
Since a SED fit is roughly luminosity weighted, the MBB does not account for the coldest, less emissive, but massive regions in the galaxy.
It thus systematically underestimates the mass.

\paragraph{Modified black body with free $\beta$.}
Similarly to the previous test, we have fit the photometry of \refsec{sec:data}, longward 100~\tmic, with a MBB, but letting $\beta$ free, this time.
Such a model can potentially infer the grain optical properties through the value of $\beta$ and the grain physical conditions through the temperature, $T_\sms{d}$.
This potentiality is however limited by the mixing of physical conditions \citep[\eg\ Sect.~2.3.1 of ][]{galliano18}.
Panel~\textit{(a)} of \reffig{fig:comp3} compares \tMd\ to our reference run.
We can see the dust mass is not extremelly biased, although there is some scatter.
The $\beta-T_\sms{d}$ relation of this run is presented in \refapp{app:MBB}.

\paragraph{With the \citetalias{draine07} ISRF distribution.}
We have fit the full sample of \refsec{sec:data} with the physical model of \refsec{sec:HerBIE}, replacing the ISRF distribution of \refeq{eq:dale} by the \citetalias{draine07} prescription:
\begin{equation}
  \frac{\dd\Md}{\dd U}=\Md\left[(1-\gamma)\delta(U-U_\sms{min})
    + \frac{\gamma(\alpha-1)U^{-\alpha}}{U_\sms{min}^{1-\alpha}
           -(U_\sms{min}+\Delta U)^{1-\alpha}}
    \right].
  \label{eq:draine}
\end{equation}
This prescription is our \refeq{eq:dale} plus a uniformly illuminated component at $U=U_\sms{min}$, with the parameter $\gamma$ controlling the weight of the two components.
We follow \citetalias{draine07} by fixing $\alpha=2$ and $U_\sms{min}+\Delta U=10^6$.
Consequently, the far-IR/submm slope, beyond the large grain peak emission ($\simeq100\emic$), is the slope of the large grain emission at $U=U_\sms{min}$, making this model very similar to a fixed-$\beta$ MBB in the far-IR/submm range.
In comparison, the ISRF distribution of our reference run allows us to account for a flattening of the far-IR/submm slope, by mixing different ISRF intensities, down to lower temperatures \citep[Sect.~2.3.1 of][]{galliano18}.
Apart from the ISRF distribution, the other features are similar to our reference run: THEMIS grain mixture and HB method.
Panel~\textit{(b)} of \reffig{fig:comp3} compares \tMd\ to our reference run.
We notice, as expected, the same systematic shift, here by a factor $\simeq0.7$, as for the fixed-$\beta$ MBB (\reftab{tab:comp}).

      \subsubsection{Comparison to \CIG\ Results}
      \label{sec:robustCIG}

\citet{nersesian19} have modelled the DustPedia sample, using the code \CIG\ \citep{boquien19}, which fits the dust SED with:
\begin{inlinelist} 
  \item the \THE\ grain properties;
  \item the \citetalias{draine07} ISRF distribution; and
  \item a non-hierarchical Bayesian method.
\end{inlinelist}
Panel~\textit{(c)} of \reffig{fig:comp3} shows its comparison to our reference run.
As expected, it is very similar to the \citetalias{draine07} ISRF distribution test in panel~\textit{(b)} of the same figure.
In particular, the mass is shifted by the same $\simeq0.7$ factor (\reftab{tab:comp}). 
Note that \citet{nersesian19} did not analyze the sources from the DGS, which are mostly the scattered values (in blue) at low \tMdref.
Concerning the fraction of small a-C(:H), the values of \citet{nersesian19} are in very good agreement with our reference run.

In summary, the comparisons of \refsecs{sec:robustfit}{} to \ref{sec:robustCIG} have demonstrated that the discrepancies induced by different fitting methods or model assumptions can be well understood.
We are therefore confident that our \expression{reference} run is the most robust among the diversity of approaches we have tested.

    \subsection{Comparison with \citet{lianou19}}
    \label{sec:sophiasco}

\begin{figure*}[htbp]
  \includegraphics[width=\textwidth]{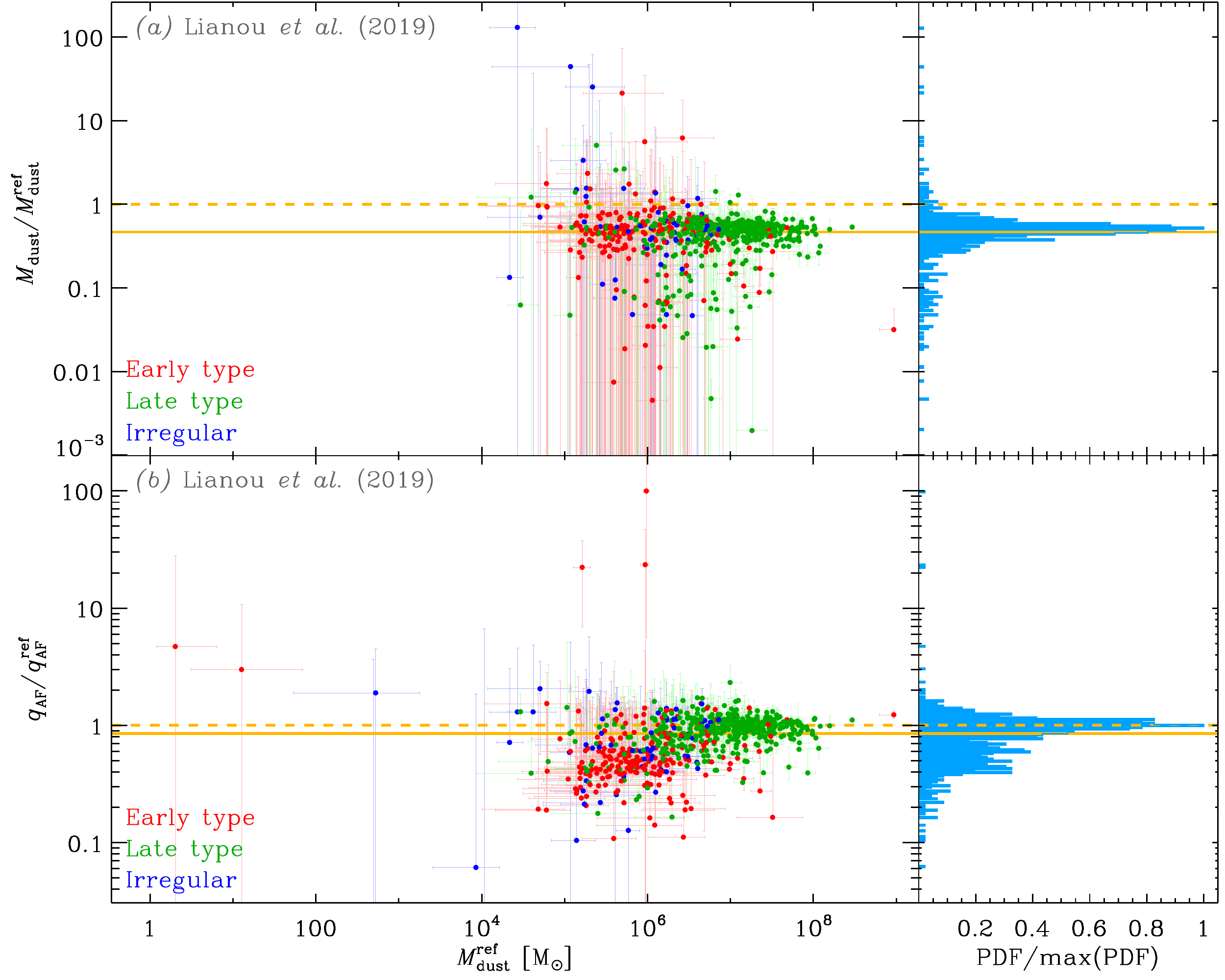}
  \caption{\textsl{Comparison to \citet{lianou19}.}
           Same conventions as \reffig{fig:comp1}.}
  \label{fig:compL19}
\end{figure*}
\citet[][ hereafter \citetalias{lianou19}]{lianou19} recently used our model, \HB, to analyze the DustPedia photometry.
There are three main technical differences between our analysis and theirs.
\begin{enumerate}
  \item They used the implementation of the 2013 version of the \THE\ model 
    \citep{jones13}, while we use the revised 2017 version \citep{jones17}.
  \item They did not profit from the possibility to include ancillary data in 
    the prior (\refsec{sec:HerBIE}).
  \item They did not include the remaining sources from the DGS sample.
\end{enumerate}
The comparisons of \tMd\ and \tqAF\ are shown in \reffig{fig:compL19}.

Panel~\textit{(a)} of \reffig{fig:compL19} shows that their dust mass is about a factor of 2 lower than ours.
This can be partly understood in the light of the difference in grain mixtures.
Indeed, \citet{jones17} revised the \THE\ model by including the more realistic \citet{kohler14} optical properties.
To fit the same observational constraints, they compensated this update by changing the mantle thickness, as well as the dust-to-gas mass ratio: $\Md/M_\sms{H}=8.6\E{-3}$ according to Table~2 of \citet{jones13} and
$\Md/M_\sms{H}=7.4\E{-3}$ according to Table~1 of \citet{jones17}.
This sole modification explains why a mass derived with the 2013 \THE\ version would be a factor of $\simeq0.86$ lower than with the 2017 version.
However, this does not explain the whole extent of the discrepancy.
The comparison of our dust masses with those derived with \CIG\ (\refsec{sec:robustCIG}) certainly excludes such a large error, on our side.
It can be noted that there is also some scatter around the median ratio of $\Md/\Mdref$.
A part of this discrepancy can naturally be explained by the fact that several galaxies have a very poor spectral coverage.
As demonstrated in \refsec{sec:ref}, the prior becomes dominant in this case.
In our case, the prior contains the information provided by the ancillary data, thus helping to reduce the dust parameter range.

Panel~\textit{(b)} of \reffig{fig:compL19} shows the comparison of \tqAF.
The two quantities are in good agreement.
There is some scatter around the median, for the same reason as mentioned for \tMd.
However, the problem with this quantity is the way \citetalias{lianou19} discuss it.
They improperly report the meaning of \tqAF\ that they call ``\textit{`QPAH'}''
\citepalias[][ Sect.~3, $5^{\textnormal{th}}$ item]{lianou19}.
They write it represents \citext{``the mass fraction of hydrogenated amorphous carbon dust grains with sizes between 0.7~nm and 1.5~nm''}, while it actually is the mass fraction of a-C(:H) with sizes between 0.4~nm and 1.5~nm.
Furthermore, they claim the Galactic value of this parameter is $7.1\,\%$, while it is $\qAF^\sms{Gal}=18.6\,\%$ for the 2013 version and $\qAF^\sms{Gal}=17.0\,\%$ for the 2017 version.
The value of $7.1\,\%$ is the mass fraction of a-C(:H) between 4~\AA\ and 7~\AA.
Consequently, they mistakenly show that most of the DustPedia sample has a higher fraction of small a-C(:H) than the Milky Way, while it is not the case (\cf\ \refsec{sec:qAFev}).

In summary, we are confident that our derived parameters are both more precise and more accurate than \citetalias{lianou19}'s.

  \section{THE DERIVED DUST EVOLUTION TRENDS}
  \label{sec:trends}

In this section, we present the main dust evolution trends derived from the \expression{reference} run (\refsec{sec:ref}).
These results are displayed as correlations between two inferred parameters, for each source in the sample.
Displaying the full posterior PDF of each galaxy as density contours is visually impractical.
Instead, we display its extent as a \expression{Skewed Uncertainty Ellipse} (SUE; \refapp{app:unc}).
SUEs approximately represent the $1\sigma$ contour of the PDF, retaining the information about the correlation and the skewness of the posterior, with a dot at the maximum \textit{a posteriori}.
When discussing parameter values in the text, we often quote the \expression{95$\,\%$ Credible Range} (CR$_{95\%}$), which is the parameter range excluding the $2.5\,\%$ lowest and $2.5\,\%$ highest values of the PDF.
We also adopt the following terminology.
\begin{itemize}
  \item We call \expression{Extremelly Low-Metallicity Galaxy} (ELMG), a system 
    with $Z\lesssim Z_\odot/10$.
  \item To simplify the discussion, since the heavy-element-to-gas mass ratio, 
    $Z$, is usually called \expression{metallicity}, we introduce the term 
    \expression{dustiness} to exclusively denote the dust-to-gas mass ratio: 
    \begin{equation}
      Z_\sms{dust}\equiv\frac{\Md}{\Mg}.
    \end{equation}
  \item We denote by \expression{specific}, quantities per unit stellar mass
    (similar to sSFR):
    \begin{inlinelist}
      \item the \expression{specific dust mass} is $\sMd\equiv\Md/\Ms$;
      \item the \expression{specific gas mass} is $\sMg\equiv\Mg/\Ms$.
    \end{inlinelist}
\end{itemize}
Note that, in all the displayed relations, the number of objects depends on the availability of the ancillary data (\reftab{tab:ancillary}).

    \subsection{Evolution of the Total Dust Budget}
    \label{sec:Mdev}

We first focus on scaling relations involving the total dust mass, \tMd, with respect to the gas and stellar contents, the metallicity and the star formation activity.
\citet{casasola20} have explored additional scaling relations, focussing on DustPedia LTGs.

      \subsubsection{Qualitative Discussion}
      \label{sec:scaldust}

\begin{figure*}[htbp]
  \begin{tabular}{cc}
    \includegraphics[width=0.48\linewidth]{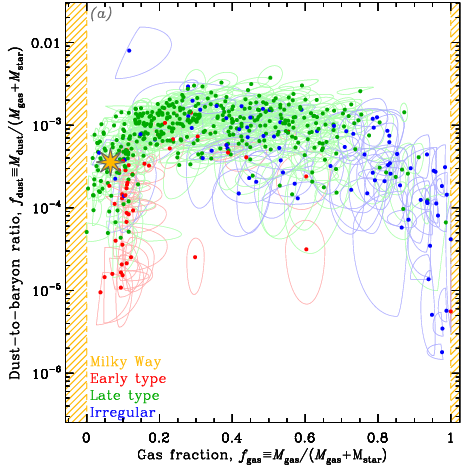} &
    \includegraphics[width=0.48\linewidth]{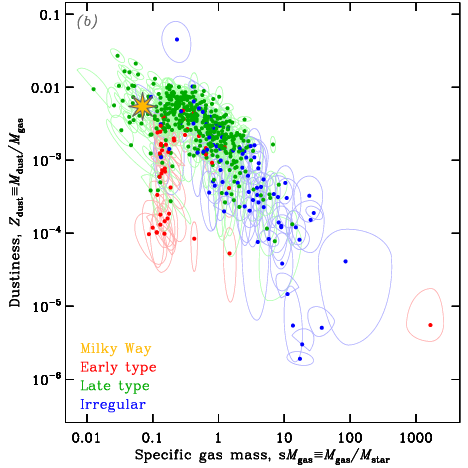} \\
    \includegraphics[width=0.48\linewidth]{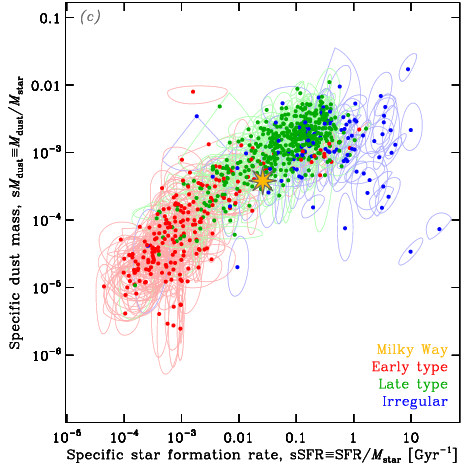} &
    \includegraphics[width=0.48\linewidth]{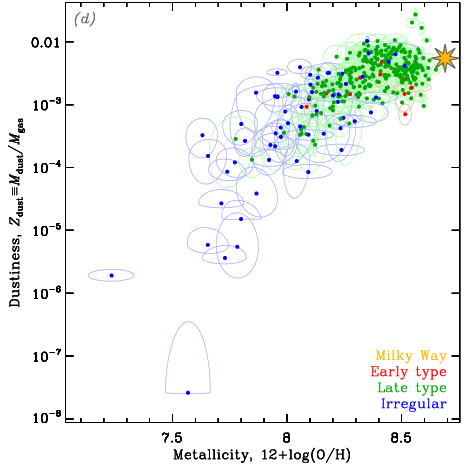} \\
  \end{tabular}
  \caption{\textsl{Dust-related scaling relations.}
           In the four panels, the SUEs represent the posterior of each galaxy,
           using the \expression{reference} run (\refsec{sec:ref}).
           The SUEs are color-coded according to the Hubble stage 
           of the object (\refsec{sec:robust}).
           We show the Milky Way values, as a yellow star, for comparison.
           We emphasize that, although we are showing different parameters of 
           the same sample, the different panels do not exactly contain the same
           number of objects (\reftab{tab:ancillary}).
           In particular, there are fewer reliable metallicity measurements, 
           especially for ETGs.}
  \label{fig:scaldust}
\end{figure*}
\reffig{fig:scaldust} presents four important scaling relations.
Panel~\textit{(a)} shows the evolution of the dust-to-baryon mass ratio:
\begin{equation}
  \DtoB\equiv\frac{\Md}{\Mg+\Ms}, 
\end{equation}
as a function of the gas fraction:
\begin{equation}
  \fg\equiv\frac{\Mg}{\Mg+\Ms}.
\end{equation}
This well-known relation was previously presented by \citet{clark15}, \citet{de-vis17a} and \citet{davies19}.
It shows that:
\begin{inlinelist}
  \item at early stages ($\fg\gtrsim0.7$; mainly irregulars), there is a net
    dust build-up;
  \item it then reaches a plateau ($0.2\lesssim\fg\lesssim0.7$; mainly LTGs) 
    where the dust production is counterbalanced by astration;
  \item at later stages ($\fg\lesssim0.2$; mainly ETGs), there is a net dust 
    removal.
\end{inlinelist}
Several sources have peculiar positions relative to the above mentioned trend.
\begin{enumerate}
  \item The irregular (blue SUE) with $\DtoB\simeq0.01$ is \pgc{166077}.
    It is however technically not an outlier, as
    $\CR{\DtoB}=[1.7\E{-3},2.8\E{-2}]$, overlapping with the rest 
    of the sample.
  \item The two ETGs (red SUEs) with a low \tDtoB, at 
    $\fg\simeq0.3$ and $\fg\simeq0.6$, are \ngc{5355} and 
    \ngc{4322}, respectively.
    For \ngc{4322}, $\CR{\DtoB}=[1.2\E{-6},1.4\E{-4}]$, marginally 
    overlapping with the rest of the sample.
    For \ngc{5355}, however, $\CR{\DtoB}=[2.7\E{-6},6.8\E{-5}]$, making 
    it a true outlier.
  \item The ETG (red SUE) with $\fg\simeq1$ is \eso{351--030}.
    It is the Sculptor dwarf elliptical galaxy.
    This is not an outlier to the relation, but it indisputably lies outside of 
    its group.
\end{enumerate}

Panel~\textit{(b)} of \reffig{fig:scaldust} presents the relation between the specific gas mass, and the dustiness.
Such a relation was previously presented by \citet{cortese12}, \citet{clark15} and \citet{de-vis17a}.
There is a clear negative correlation between these two quantities, showing that when a galaxy evolves, its ISM gets progressively enriched in dust and its gas content gets converted to stars. 
There is one notable feature deviating from this \expression{main sequence}:
a vertical branch at $\sMg\simeq0.1$, exhibiting a systematically lower dustiness.
This branch is mostly populated by ETGs and contains most of the ETGs of the relation.
We will discuss the likely origin of this branch in \refsec{sec:Lx}.
Finally, the peculiar sources of panel~\textit{(a)} logically stand out in this panel too.
\begin{enumerate}
  \item \pgc{166077} is the blue SUE at $\Zd\simeq0.05$.
    It is not a clear outlier, as $\CR{\Zd}=[0.01,0.13]$.
  \item \ngc{5355} and \ngc{4322} are the red SUEs at $\sMg\simeq0.4$ and 
    $\sMg\simeq1.5$, respectively.
  \item The bottom red SUE, at $\sMg\simeq1600$, is \eso{351-030}.
\end{enumerate}

Panel~\textit{(c)} of \reffig{fig:scaldust} shows the relation between the specific dust mass and the specific star formation rate, sSFR.
This relation was previously presented by \citet{remy-ruyer15} and \citet{de-vis17a}.
It was also discussed in \citet{cortese12} with sSFR replaced by its proxy NUV-r. 
In the same way, but on resolved 140-pc scales, it was shown in \M{31} by \citet{viaene14}.
There is a clear positive correlation between the two quantities. 
The different galaxy types are grouped in distinct locations:
\begin{inlinelist}
  \item ETGs are clustered around 
    $10^{-4}\lesssim\textnormal{sSFR}\lesssim0.01\;\textnormal{Gyr}^{-1}$;
  \item LTGs are clustered around
    $0.01\lesssim\textnormal{sSFR}\lesssim1\;\textnormal{Gyr}^{-1}$;
  \item Irregulars tend to lie around 
    $0.1\lesssim\textnormal{sSFR}\lesssim10\;\textnormal{Gyr}^{-1}$, but are 
    more scattered 
    than the other types, with a systematically lower $\sMd$ than what the 
    extrapolation of the trend would suggest.
\end{inlinelist}
This scaling relation provides an interesting approximation to derive the dust content from SFR and \tMs, two quantities which usually are easy to estimate.
In particular, the relation is quasi-linear in the low sSFR regime, with:
\begin{equation}
  \frac{\Md}{\Msun}\simeq7.3_{-6.1}^{+10.4}\E{7}
    \times\frac{\textnormal{SFR}}{\Msun/\textnormal{yr}}
    \;\;\mbox{ for }\;\; \textnormal{sSFR}\lesssim0.1\;\textnormal{Gyr}^{-1},
\end{equation}
with $\CR{\Md/\textnormal{SFR}}=[0.12,53]\times10^7$~yr.

Panel~\textit{(d)} of \reffig{fig:scaldust} shows the variation of the dustiness as a function of the metallicity \refeqp{eq:metal}.
Such a relation is one of the most common benchmarks for global dust evolution models (\cf~\refsec{sec:dustvol}), and has been presented by numerous studies
\citep[\eg\
][]{lisenfeld98,james02,draine07,galliano08a,galametz11,remy-ruyer14,de-vis17}.
There is a clear correlation between the two quantities, reflecting the progressive dust enrichment of the ISM, built from heavy elements injected by stars at the end of their lifetime.
Few ETGs are present, due to the lack of reliable gas metallicity estimates for these objects.
These would occupy the high-metallicity regime.
Compared to \citet{remy-ruyer14}, who presented a similar trend for the DGS sources, we notice that a few sources are missing and some metallicities have been updated.
This comes from the fact that we have adopted the more robust DGS metallicity reestimates by \citet{de-vis17}, with published nebular lines and using the PG16\_S calibration (\refsec{sec:metal}).
The lowest metallicity source, at $\ZOH\simeq7.15$, is \izw.
The most dust-deficient source of the panel, at $\Zd\simeq1.9\E{-6}$, is \ugca.
It is a clear outlier to the trend with $\CR{\Zd}=[1.3\E{-6},3.0\E{-6}]$.

      \subsubsection{Comparison to X-Ray Luminosities}
      \label{sec:Lx}

\begin{figure}[htbp]
  \includegraphics[width=\linewidth]{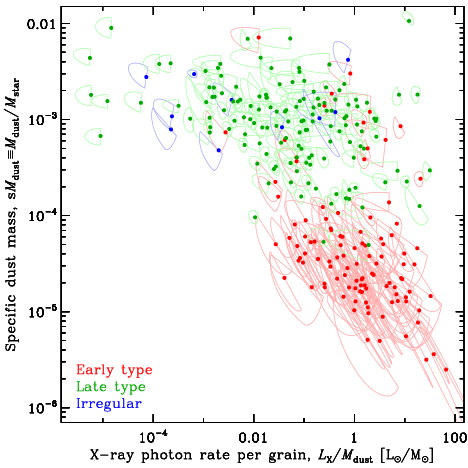}
  \caption{\textsl{Relation of dust mass to X-ray luminosity.}
           This figure shows the variation of the dust-to-stellar mass ratio
           as a function of the X-ray photon rate per dust grain, for the 
           sample of \reftab{tab:Lx}. 
           This is a subset of the \expression{reference} run 
           (\refsec{sec:ref}).
           The color convention is similar to \reffig{fig:scaldust}.
           The Bayesian correlation coefficient of this relation is 
           $\rho=-0.596_{-0.015}^{+0.017}$, with $\CR{\rho}=[-0.62,-0.56]$.}
  \label{fig:Lx}
\end{figure}

The ETG outliers, visible in panel~\textit{(b)} of \reffig{fig:scaldust} and discussed in \refsec{sec:scaldust}, are likely due to enhanced dust destruction by thermal sputtering in their hot, X-ray emitting gas \citep[\eg\ ][]{bocchio12,smith12}, as suggested by \citet{de-vis17a}.
We could conceive a reverse causality where it is the low dust abundance that allows a higher fraction of escaping X-ray photons.
However, this is unrealistic, as X-ray observations clearly indicate that ETGs are permeated by a coronal gas that is usually not found in later types of galaxies \citep[][ for a review]{mathews03}.
In order to support the likeliness of the thermal sputtering scenario, we have compiled X-ray luminosities from the literature (\reftab{tab:Lx}).
\reffig{fig:Lx} displays the dust-to-stellar mass ratio as a function of the X-ray photon rate per dust grain, $L_\sms{X}/\Md$, for the 256 galaxies in our sample with published X-ray luminosities.
This figure shows a mild negative correlation between the two quantities, consistent with the enhanced dust destruction in X-ray-bright galaxies.
ETGs clearly lie in the bottom right quadrant of this figure.
It is reminiscent of Figure~10 in \citet{smith12}, displaying $L_\sms{FIR}/L_\sms{B}$ as a function of $L_\sms{X}/L_\sms{B}$.

We note there is however a significant intrinsic scatter in this relation.
This could be due to the fact that most of the studies listed in \reftab{tab:Lx} quote a total X-ray luminosity, including both:
\begin{inlinelist}
  \item point sources (AGNs, binary systems); and 
  \item the diffuse thermal emission that is sole relevant to our 
    case\footnote{\citet{david06}, \citet{diehl07}, \citet{rosa-gonzalez09} 
    and \citet{kim19}
    extract the thermal emission of the gas. We use these values for the 
    sources in these catalogs.}.
\end{inlinelist}
In addition, the spectral range used to compute the X-ray luminosity varies from one instrument to the other (\reftab{tab:Lx}), leading to systematic differences.
Nonetheless, accounting for these differences, which is beyond the scope of this paper, would likely not change the plausibility that dust grains are significantly depleted in ETGs due to thermal sputtering.
\begin{table}[htbp]
  \caption{\textsl{X-ray luminosity references.}
           The total number of sources (last line) is fewer than 
           the sum of each sample size (last column).
           This is because the same sources
           were observed by different instruments.
           When it is the case, we keep the most recent estimate.}
  \label{tab:Lx}
  \begin{tabularx}{\linewidth}{Xllr}
    \hline\hline
      Bibliographic & Telescope & Range    & Size \\
      reference     &           & [keV] &    \\
    \hline
\citet{fabbiano92} & \textit{Einstein} & $0.2-4$ & 172 \\
\citet{brinkman94} & \textit{ROSAT} & $0.1-2.4$ & 7 \\
\citet{osullivan01} & \textit{ROSAT} & $0.1-17$ & 83 \\
\citet{tajer05} & \textit{ROSAT} & $0.1-2$ & 10 \\
\citet{cappi06} & \textit{XMM} & $0.2-10$ & 16 \\
\citet{david06} & \textit{Chandra} & $0.5-2$ & 8 \\
\citet{diehl07} & \textit{Chandra} & $0.3-5$ & 7 \\
\citet{rosa-gonzalez09} & \textit{XMM} & $0.5-2$ & 1 \\
\citet{gonzalez-martin09} & \textit{Chandra} & $0.5-10$ & 30 \\
\citet{akylas09} & \textit{XMM} & $2-10$ & 19 \\
\citet{grier11} & \textit{Chandra} & $0.3-8$ & 29 \\
\citet{brightman11} & \textit{XMM} & $2-10$ & 37 \\
\citet{liu11} & \textit{Chandra} & $0.3-8$ & 147 \\
\citet{kim19} & \textit{Chandra} & $0.5-5$ & 12 \\
    \hline
\multicolumn{3}{r}{Total} & 256 \\
    \hline
  \end{tabularx}
\end{table}

      \subsubsection{On the variation of the Dust-to-Metal Mass 
                           Ratio}
      \label{sec:D2M}

Panel~\textit{(d)} of \reffig{fig:scaldust} shows that the dustiness-metallicity relation is non-linear.
\citet{remy-ruyer14} first argued, relying on insights from the models of \citet{asano13} and \citet{zhukovska14}, that such a trend was the result of different dust production regimes:
\begin{inlinelist}
  \item at low \tZOH, dust production is dominated by condensation in type~II  
    Supernova (SN) ejecta, with a low yield;
  \item around a \expression{critical metallicity}\footnote{This is a concept
    introduced by \citet{asano13}. Its exact value depends on the star 
    formation history of each galaxy.} of $\ZOH\simeq8$, 
    grain growth in the ISM becomes dominant, causing a rapid increase of \tZd;
  \item at high \tZOH, the dust production is dominated by grain growth in the 
    ISM, with a yield about two orders of magnitude higher than \snii, and is 
    counterbalanced by \snii\ blast wave dust destruction.
\end{inlinelist}
\citet{remy-ruyer14} argued that the intrinsic scatter of the relation, which could not be explained by SED fitting uncertainties, was due to the fact that each galaxy has a particular \expression{Star Formation History} (SFH).
We will explore this aspect in \refsec{sec:dustvol}.
In the following paragraphs, we discuss the different biases that could have induced an artificial non-linearity in our empirical trend.

\begin{figure}[htbp]
  \includegraphics[width=\linewidth]{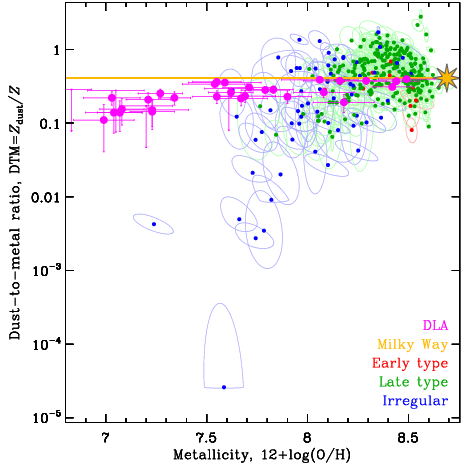}
  \caption{\textsl{Comparison to DLAs.}
           This figure shows the relation between the metallicity and the
           dust-to-metal mass ratio, for the \expression{reference} run 
           (\refsec{sec:ref}).
           This figure is very similar to panel~\textit{(d)} of 
           \reffig{fig:scaldust}: the only difference is that the $y$-axis has
           been divided by \tZ\ (related to the $x$-axis through 
           \refeqnp{eq:metal}).
           We have overlaid in magenta the DLA measures from Table~6 of 
           \citet{de-cia16}.
           The horizontal yellow line corresponds to the Galactic dust-to-metal 
           mass ratio.
           The Bayesian correlation coefficient of the nearby galaxy sample is
           $\rho=0.636_{-0.023}^{+0.021}$, with $\CR{\rho}=[0.59,0.68]$.}
  \label{fig:DLA}
\end{figure}

\paragraph{Comparison to DLAs.}
\reffig{fig:DLA} shows the evolution of the \expression{dust-to-metal mass ratio} (DTM) as a function of metallicity.
It is another way to look at the data in panel~\textit{(d)} of \reffig{fig:scaldust}.
A constant DTM corresponds to a linear dustiness-metallicity trend.
The SUEs in \reffig{fig:DLA} are not consistent with a constant DTM (horizontal yellow line).
However, there are reports in the literature of objects exhibiting an approximately constant DTM, down to very low metallicities: \expression{Damped Lyman-$\alpha$ Absorbers} (DLA).
We have overlaid the DLA sample of \citet{de-cia16}.
These measures are performed in absorption on redshifted systems, along the sightline of distant quasars.
The metallicity and DTM are derived from a combination of atomic lines of volatile and refractory elements.
We can see that the DLA DTMs vary significantly less than in our nearby galaxy sample\footnote{We note that metallicities measured in absorption tend to always be lower than metallicities measured in emission \citep[\eg\ ][]{hamanowicz20}}.
Such an evolutionary behaviour requires a high SN-dust condensation efficiency coupled to a weak grain growth rate in the ISM \citep[\eg\ ][]{de-vis17}.
Alternatively, the DLA estimates could be biased.
It is indeed not impossible that the hydrogen column density of most DLAs includes dust- and element-free circumgalactic clouds in the same velocity range.
It would result in a typical solar metallicity object appearing to have a solar DTM, and at the same time, an artificially lower metallicity, diluted by the additional pristine gas along the line of sight.

\paragraph{Accounting for the Gas Halo.}
The contamination of the gas mass estimate by external, dust- and element-poor gas is also a potential issue for our nearby galaxy sample.
This is particularly important for low-metallicity, dwarf galaxies, where the IR-emitting region is usually small compared to the whole \hi\ halo \citep[\eg\ ][]{walter07,begum08}.
For instance, \citet{draine07b} hinted that the trend between \tZ\ and \tZd\ was close to linear, when the gas mass used to estimate \tZd\ was integrated in the same region as the dust mass.
They only had 9 galaxies below $\ZOH<8.1$.
We have addressed this issue by adopting the interferometric \hiline\ observations of 20 of the lowest metallicity galaxies in our sample, including the lowest metallicity system, \izw\ (\citeprep{roychowdhury20}; \refsec{sec:Mgas}).
We have integrated the gas mass within the photometric aperture for these 20 objects.
\izw\ still lies two orders of magnitude below the Galactic DTM, despite this correction.
On the contrary, it is possible that the DTM of \ugca, the lowest SUE in \reffig{fig:DLA}, has been underestimated, as it has not been resolved in \hi.
We have to admit that there is still room for improvement as several of the ELMGs are barely resolved in the IR.
Thus, although we corrected for a large fraction of the \hi\ halo, there might still be residual gas not associated with the star forming region within our aperture.
We might therefore be underestimating the dustiness of our ELMGs.
This will have consequences in \refsec{sec:dustvol}.
The amplitude of this underestimation is not quantifiable as these sources are not resolved in the far-IR.
It is however difficult to imagine that there would still be $99\,\%$ of gas not associated with IR emission within our aperture.
Indeed, for the most extreme case, \izw, our aperture is only 1.5 times the optical radius \citep{remy-ruyer13}, which should be comparable to the IR radius.
The overall rising DTM with \tZOH\ of \reffig{fig:DLA} is therefore unlikely due to an improper correction of the \hi\ envelopes of ELMGs.

\paragraph{Variation of the Grain Opacities.}
We have noted in \refsec{sec:contaminations} that our dust mass estimates depend on the rather arbitrary grain opacity we have adopted.
This grain opacity has been designed to account for the emission, extinction
and depletions of the diffuse Galactic ISM (\cf~\refsec{sec:themis}).
A systematic variation of the overall grain opacity with metallicity could change the slope of the trend in \reffig{fig:DLA}. 
In order to move \izw\ up to the Galactic DTM, we would need to adopt a grain emissivity diminished by about two orders of magnitude\footnote{For \izw, $\CR{\textnormal{DTM}}=[2.3\E{-3},7.9\E{-3}]$, while $\textnormal{DTM}_\odot\simeq0.5$.}.
Qualitatively, the ISM of an ELMG, such as \izw, is
\citep[\eg\ ][]{cormier19}:
\begin{inlinelist}
  \item permeated by hard UV photons;
  \item very clumpy with a low cloud filling factor.
\end{inlinelist}
With more UV photons to evaporate the mantles and less clouds to grow them back, we could assume that the grains in such a system would be reduced to their cores \citep[\cf~Figure~16 of ][]{jones13}.
Demantled, crystalline, compact grains are indeed among the least emissive grains.
However, to our knowledge, there are no interstellar dust analogs having a far-IR opacity two orders of magnitude lower than those used in the \THE\ model.
The \citet{draine07} mixture, which resembles compact bare grains, is only a factor of $\simeq2$ less emissive than \THE\ \citep[\eg\ Figure~4 of ][]{galliano18}.
We could also imagine that the composition of the dust mixture itself changes.
In particular, the fraction of silicates could be higher in ELMGs, as C/O and O/H are correlated \citep{garnett95}.
This would have a limited effect, since:
\begin{inlinelist}
  \item a decrease of the large a-C(:H) abundance by $50\,\%$ would decrease 
    the global emissivity of \THE\ in the SPIRE bands by $15-30\,\%$;
  \item a variation of the forsterite-to-enstatite ratio would change the 
    emissivity by less than $30\,\%$.
\end{inlinelist}
It is therefore very unlikely that the dependence of DTM with metallicity is artificially induced by our grain opacity assumption.

\begin{figure}[htbp]
  \includegraphics[width=\linewidth]{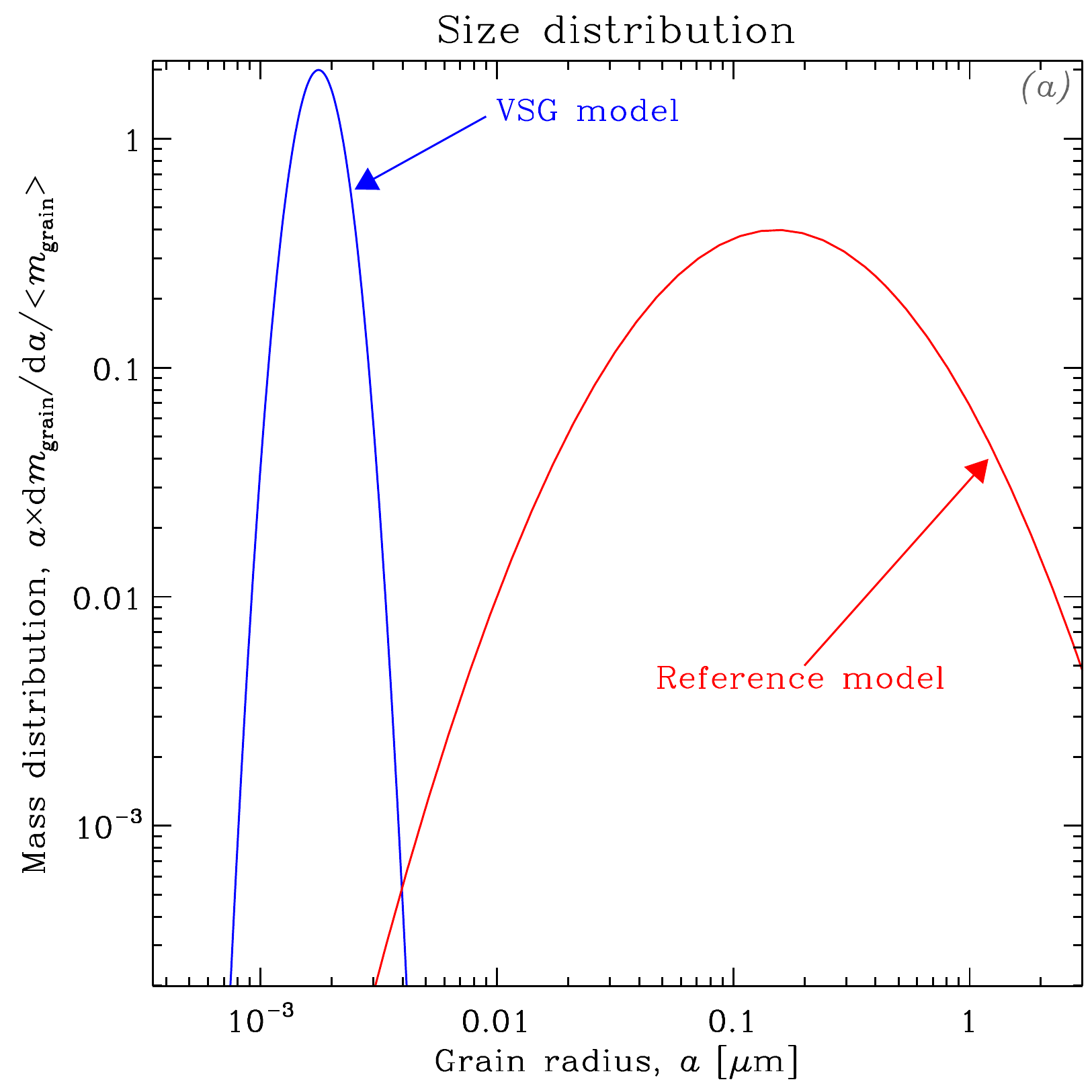} \\
  \includegraphics[width=\linewidth]{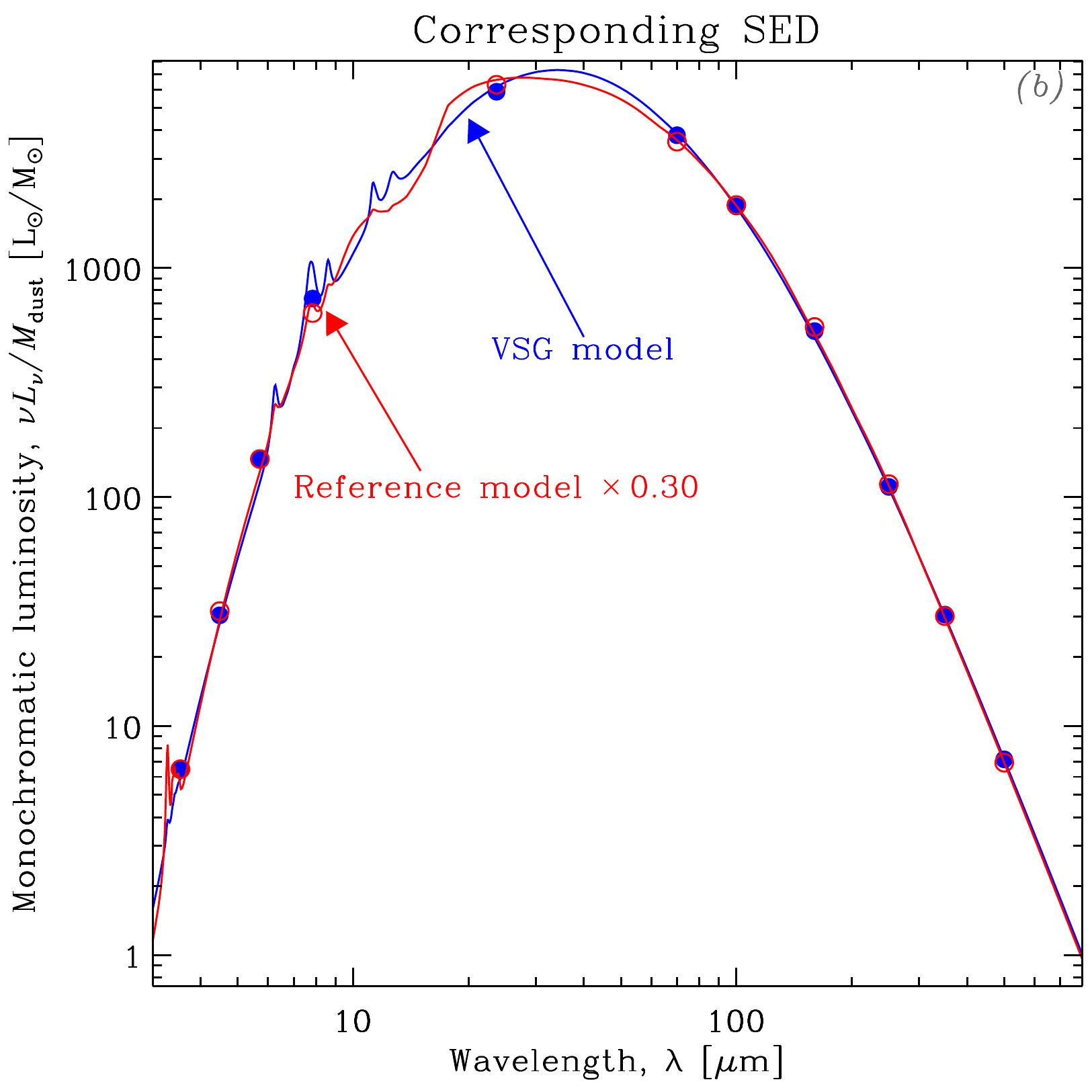}
  \caption{\textsl{Demonstration of a VSG-dominated SED.}
           Panel~\textit{(a)} shows the size distribution of the two models
           we are comparing here: the \expression{reference} model in red, 
           which is simply the \THE\ model, with the small grains scaled down
           to mimic the SED of a typical ELMG;
           and a VSG model, in blue, which is a log-normal size distribution, 
           peaking at $1.5$~nm with a log-width of $0.2$.
           The optical properties of the VSG model are those of \THE, and we
           have kept the same silicate-to-carbon grain mass ratio.
           In both cases, the curves represent the size distributions of 
           a-C(:H) and silicates.
           Panel~\textit{(b)} shows the corresponding SEDs for the two models.
           The SED of the \expression{reference} model has been scaled down
           by a factor of $0.30$, to fit the VSG SED.
           It means the VSG SED reproduces the reference model with a dust mass
           $3.33$ times higher.
           The VSG SED is uniformly illuminated by an ISRF with $U=10$, while 
           the \expression{reference} model has a distribution of ISRFs with 
           $\langle U\rangle=200$.
           The blue filled and red empty circles show the synthetic photometry
           of each model in the IRAC, MIPS1, PACS and SPIRE bands.}
  \label{fig:VSG}
\end{figure}
\paragraph{Variation of the Size Distribution.}
Another factor that could affect the trend of \reffig{fig:DLA} is that we have fixed the size distribution of the large grains.
In principle, relaxing this assumption and allowing the size distribution to be dominated by \expression{Very Small Grains} (VSG) in low-metallicity sources, would raise the DTM of objects such as \izw.
Indeed, VSGs are stochastically heated.
They spend most of their time at very low temperatures between successive photon absorptions.
Their excursion at $T\gtrsim20$~K will span only a fraction of their time.
A mixture of VSGs would thus appear less emissive than a larger grain at an equilibrium temperature close to the high end of their temperature distribution.
We have demonstrated this effect in \reffig{fig:VSG}.
We have simulated a VSG-dominated SED mimicking a typical ELMG, peaking around $\lambda\simeq40\emic$, and with very weak aromatic feature emission (\reffig{fig:VSG}; panel~\textit{b}; blue curve).
The size distribution needed to produce such a SED is made almost exclusively of $a\simeq1.5$~nm radius grains (\reffig{fig:VSG}; panel~\textit{a}; blue curve).
We have fit this synthetic SED with our \expression{reference} model (\refsec{sec:dale}), keeping the large grain size distribution fixed (\reffig{fig:VSG}; panel~\textit{a}; red curve), but varying the ISRF distribution.
This model reproduces very well the photometric fluxes (\reffig{fig:VSG}; panel~\textit{b}; red circles), but requires a total dust mass a factor of $\simeq3$ lower than the VSG model.
This effect goes in the right direction but is not enough to explain the 2 orders of magnitude required to account for a constant DTM in \reffig{fig:DLA}.
We could further decrease the emissivity of the VSG model by lowering the size of the grains.
However, it would result in a SED peaking shortward $\lambda\simeq30\emic$, inconsistent with our observations.
Finally, VSG-dominated ELMGs would not be in agreement with theoretical dust size distribution evolution models \citep[\eg\ ][]{hou17,hirashita19,aoyama20}.
At early stages, these models predict the ISM being populated of large grains.
The reason is that the dust production at these stages is thought to be dominated by SN-\textsc{ii}-condensed dust, and grains condensed in core-collapse SN ejecta are essentially large as the small ones are kinetically sputtered \citep[\eg\ ][]{nozawa06}.

\paragraph{Very Cold Dust.}
A significant fraction of the dust mass could have been overlooked, hidden in the form of very cold dust ($T_\sms{dust}\lesssim10$~K).
The emission of this component would manifest as a weakly emissive continuum at submm wavelengths, that we have not accounted for.
Very cold dust has been invoked to explain the \expression{submm excess} in dwarf galaxies (\cf~\refsec{sec:contaminations}).
It could thus be invoked to flatten our dust-to-metal mass ratio trend, in principle.
However, while the excess has been reported by numerous studies, predominantly in dwarf galaxies \citep[\eg\ ][]{galliano03,galliano05,dumke04,bendo06,galametz09,bot10}, very cold dust does not appear as one of its viable explanations.
Indeed, to reach such a low temperature, very cold dust should be shielded from the general ISRF in massive dense clumps.
In the LMC, \citet{galliano11} showed that the excess emission at 10~pc scale was diffuse and negatively correlated with the gas surface density, inconsistent with the picture of a few dense clumps.
Similarly, \citet{galametz14} and \citet{hunt15} showed that this excess was more prominent in the outskirts of late-type disk galaxies, where the medium is less dense.
In addition, other realistic physical processes to explain this excess have been proposed \citep{meny07,draine12}.
Dust analogs also exhibit a flatter submm slope that could partly account for this excess \citep{demyk17b}.
More qualitatively, it is difficult to conceive of the presence of massive dense clumps in ELMGs, where the dust-poor ISM is permeated by intense UV photons from the young stellar populations that are forming.
In addition, this dust would likely be associated with dense gas that we have not accounted for, limiting its impact on the estimated dustiness.

\paragraph{Systematic Uncertainty on the ELMG's DTM.}
Overall, it appears that none of our model assumptions could be demonstrated to be responsible for artificially inducing the observed variation of the DTM.
The trend of \reffig{fig:DLA}, for nearby galaxies, is therefore likely real.
It is however possible that the dustiness of the ELMGs has been underestimated.
We can quote a rough systematic uncertainty for an object such as \izw, the following way.
\begin{itemize}
  \item Since the \hi\ aperture is a factor of $\simeq1.5$ times the optical 
    radius, $Z_\sms{dust}$ might be underestimated by a factor at most 
    $\simeq1.5^2=2.25$.
  \item We have discussed that the grain opacity could have been overestimated
    by a factor of $\simeq2$.
    The systematic uncertainty on $Z_\sms{dust}$ is thus at most a factor of 
    $\simeq2$.
  \item We have discussed that our grain size distribution assumption could
    lead to an underestimate of $Z_\sms{dust}$ by a factor of at most $\simeq3$.
\end{itemize}
Overall, we can consider that the dustiness of ELMGs could have been underestimated by a factor of $\simeq\sqrt{2.25^2+2^2+3^2}=4.25$.

    \subsection{Evolution of the Aromatic Feature Emitters}
    \label{sec:qAFev}

We now focus our discussion on another important dust parameter, the mass fraction of aromatic feature emitting grains, \tqAF\ (\refsec{sec:themis}).
Note that, in the \THE\ model, all grains are either pure a-C(:H) or a-C(:H)-coated.
All of them therefore potentially carry aromatic features.
However, only the smallest ones ($a\lesssim10$~nm) will fluctuate to high enough temperatures ($T\gtrsim300$~K) to emit these features.
Finally, we remind the reader that, assuming aromatic features are carried by PAHs, \tqAF\ is formally equivalent to $\simeq2.2\times\qPAH$ (\refsec{sec:themis}).

\subsubsection{The Competing Small a-C(:H) Evolution Scenarios}
\label{sec:qAFscenario}

The strong spatial variability of the aromatic feature strength has been known for several decades.
Dramatic variations have been observed within Galactic and Magellanic \hii\ regions \citep[\eg\ ][]{boulanger98,madden06,galametz13}, as well as among nearby galaxies \citep[\eg ][]{galliano03,galliano05,engelbracht05,madden06}.
The following scenarios have been proposed.

\paragraph{Photodestruction.}
The aromatic feature strength appears to be negatively correlated with the presence of intense/hard UV fields.
For instance, \citet{boulanger98} showed that their strength in Galactic PDRs is decreasing when the intensity of the ISRF is increasing.
\citet{madden06} showed that their strength, within Galactic and Magellanic regions and within nearby galaxies, is negatively correlated with \neiiiline/\neiiline, an ISRF hardness indicator.
Numerous studies confirmed these trends, showing the depletion of aromatic features around massive stars, as a result of the photodissociation and photosublimation of small a-C(:H).
At the same time, it was also shown that the aromatic feature strength is empirically correlated with metallicity.
This is clear in nearby galaxies, as a whole \citep[\eg\ ][]{engelbracht05,madden06}, as well as within extragalactic \hii\ regions \citep[\eg\ ][]{gordon08,khramtsova13}.
This correlation is not inconsistent with photodestruction, as most low-metallicity galaxies detected at IR wavelengths to date are actively forming stars.
Their ISM, less opaque due to a lower \tZd, is permeated by an intense UV field, potentially destroying small a-C(:H) on wider scales.
It is also consistent with the increasing strength of the 2175~\AA\ bump with metallicity, comparing sightlines in the \expression{Small Magellanic Cloud} (SMC), LMC and Milky Way \citep[\eg\ ][]{gordon03}.
The extinction bump is indeed thought to be carried by small carbon grains with aromatic bonds \citep{joblin92}.

\paragraph{Inhibited formation.}
Alternatively, several studies have explored the possibility that small a-C(:H) are less efficiently formed at low-metallicity.
\citet{galliano08a} showed that, assuming small a-C(:H) are formed from the carbon produced by AGB stars (on timescales of $\simeq400$~Myr) and the rest of the grains are mostly made out of the elements produced by massive stars (on timescales of $\simeq10$~Myr), the aromatic feature emitters are under-abundant at early stages of galaxy evolution.
Another scenario, developed by \citet{seok14}, reproduces the trend of aromatic feature strength with metallicity, assuming small a-C(:H) are the product of the shattering of larger carbon grains \citep[see also][]{rau19,hirashita20}.
Finally, \citet{greenberg00} have proposed that aromatic feature carriers could form on grain surfaces in molecular clouds and be photoprocessed in the diffuse ISM.
\citet{sandstrom10} and \citet{chastenet19} show that the spatial distribution of $q_\sms{PAH}$ in the Magellanic clouds is consistent with this scenario.
This process would also explain the fact that these grains are under-abundant in ELMGs, as the molecular gas fraction rises with metallicity.

\subsubsection{Empirical Correlations within Our Sample}

\begin{figure*}[htbp]
  \includegraphics[width=\textwidth]{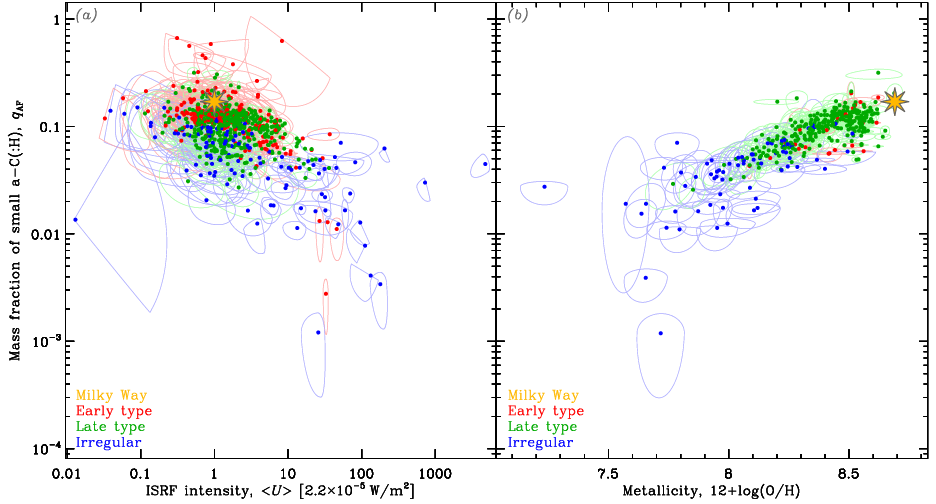}
  \caption{\textsl{Evolution of small a-C(:H) grains.}
           The SUEs are color-coded according to the type of galaxy (\cf\ 
           \refsec{sec:robust}).
           The Milky Way is shown as a yellow star.
           There are fewer objects with metallicity measurements 
           (376; \reftab{tab:ancillary}), especially among ETGs 
           (panel~\textit{b}).}
  \label{fig:scalAF}
\end{figure*}
The different processes we have just listed are not exclusive and could very well compete within the ISM.
It is however important to understand which one controls the overall abundance of small a-C(:H), at galaxy-wide scales.

\reffig{fig:scalAF} shows \tqAF\ as a function of the average starlight intensity, \tUav\ (\refsec{sec:dale}), and metallicity, in our sample.
Similar relations were previously shown by \citet{draine07b}, \citet{galliano08a}, \citet{khramtsova13} and \citet{remy-ruyer15}.
The two SUEs at high \tUav\ (panel~\textit{a}) are \izw\ and \sbs{0335-052}.
Their SEDs indeed peaks around 30~\tmic.
However, their \tqAF\ appear higher than the extrapolation of the general trend, with $\CR{\qAF}=[0.012,0.035]$ for \izw\ and $\CR{\qAF}=[0.024,0.064]$ for \sbs{0335-052}.
Mid-IR \spitz-IRS spectroscopy did not detect aromatic features in these galaxies \citep{wu07,houck04}.
The true value of \tqAF\ is likely lower for these two objects.
The reason of this overestimation lies in the difficulty to estimate aromatic feature strengths solely with broadband fluxes, when the feature-to-continuum ratio is weak.
Indeed, in the weak aromatic feature regime, \tqAF\ is biased by the color of the mid-IR continuum \citep[\cf\ Figure~1 of ][]{galliano08a}.

Overall, we find clear correlations in both panels of \reffig{fig:scalAF}, consistent with past studies.
However, the relation appears more scattered with \tUav, than with \tZOH.
The correlation coefficient of panel~\textit{(a)} is only $\rho=-0.434_{-0.028}^{+0.038}$ with $\CR{\rho}=[-0.49,-0.35]$, while it is $\rho=0.762_{-0.030}^{+0.018}$ with $\CR{\rho}=[0.70,0.79]$ for panel~\textit{(b)}.
We could argue that panel~\textit{(b)} of \reffig{fig:scalAF} contains only the 376 sources with metallicity measurements (\reftab{tab:ancillary}), while panel~\textit{(a)} contains the 798 sources with photometric fluxes, including very noisy ETG SEDs.
If we consider only the subsample of panel~\textit{(a)} with metallicity measurements, the correlation coefficient does not significantly improve:
$\rho=-0.486_{-0.026}^{+0.034}$ with $\CR{\rho}=[-0.54,-0.42]$.
If we also exclude the two biased estimates of \izw\ and \sbs{0335-052}, the correlation coefficients do not change much:
$\rho=-0.432_{-0.029}^{+0.040}$ with $\CR{\rho}=[-0.49,-0.35]$ for panel~\textit{(a)} and 
$\rho=0.768_{-0.031}^{+0.018}$ with $\CR{\rho}=[0.70,0.80]$ for panel~\textit{(b)}.
 
We could also question the accuracy of \tUav\ as an ISRF tracer.
Indeed, \tUav\ is a mass-averaged ISRF intensity, giving a large weight to the coldest regions within the beam.
\citet{remy-ruyer15} and \citet{nersesian19} used sSFR, which is luminosity weighted, in place of \tUav\ and found a good negative correlation with $q_\sms{PAH}$ and $q_\sms{AF}$, respectively.
In our sample, the correlation coefficient between $\ln\qAF$ and $\ln\textnormal{sSFR}$ is not improved: $\rho=-0.488_{-0.024}^{+0.028}$ with $\CR{\rho}=[-0.53,-0.43]$.

Therefore, it appears that \tqAF\ correlates much more robustly with metallicity than ISRF indicators, in our sample.
This result is worth noting, especially since several studies focussing on a narrower metallicity range concluded the opposite 
\citep[\eg\ ][]{gordon08,wu11}.
It probably relies on the fact the metallicities we have adopted (\refsec{sec:metal}) correspond to well-sampled galaxy averages, while in the past a single metallicity, often central, was available and may have not been representative of the entire galaxy.
This result suggests that photodestruction, although real at the scale of star-forming regions, might not be the dominant mechanism at galaxy-wide scales and that one needs to invoke one of the inhibited formation processes discussed in \refsec{sec:qAFscenario}.

    \subsection{Approximate Analytical Metallicity Trends}
  
\begin{figure}[htbp]
  \includegraphics[width=\linewidth]{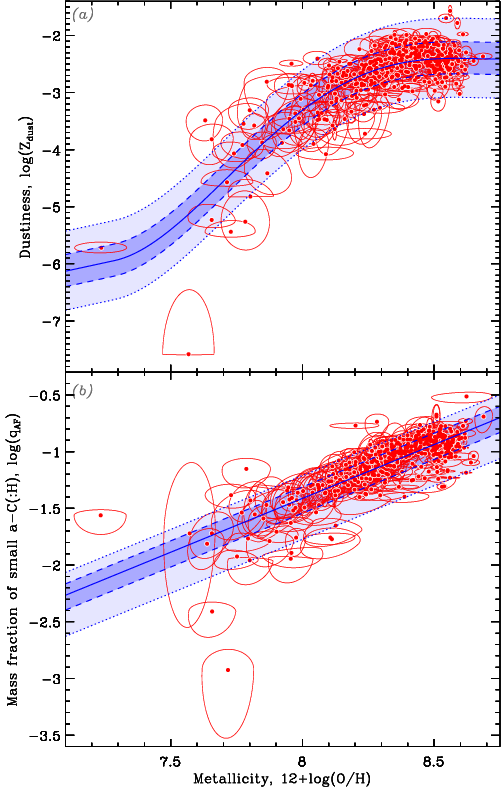}
  \caption{\textsl{Analytical fit of the scaling relations.}
           The red SUEs show the data of panel~\textit{(d)} of 
           \reffig{fig:scaldust} and panel~\textit{(b)} of \reffig{fig:scalAF}.
           The blue curve in panel~\textit{(a)} shows the analytical fit of 
           \refeq{eq:analfit1} modified by \refeq{eq:analfit2}.
           The blue curve in panel~\textit{(b)} shows the analytical fit of 
           \refeq{eq:analfit3}.
           In both panels, the dashed lines display the envelope encompassing 
           $68\,\%$ of the sources, and the dotted lines, the envelope 
           encompassing $95\,\%$ of the sources. 
           We show here the \textit{decimal} logarithm of the \tZd\ and \tqAF.}
  \label{fig:analfit}
\end{figure}
It can be interesting to have a simple analytical approximation describing how \tZd\ and \tqAF\ vary as a function of \tmet.
To that purpose, we have performed polynomial fits of the relations in panel~\textit{(d)} of \reffig{fig:scaldust} and panel~\textit{(b)} of \reffig{fig:scalAF}.
These fits are displayed in \reffig{fig:analfit}.

A 4$^\sms{th}$ degree polynomial fit of the dustiness-metallicity relation gives, posing
$x=\met$:
\begin{equation}
  \begin{aligned}
  \log\Zd\simeq11471.808-5669.5959x+1045.9713x^2 \\
    -85.434332x^3+2.6078774x^4.
  \end{aligned}
  \label{eq:analfit1}
\end{equation}
This equation provides a good fit of our trend, for the metallicity range covered by our sources.
However, it gives a rising dustiness trend with decreasing metallicity below \izw.
To improve this relation, we can assume that the dustiness will be proportional to \tmet\ at extremelly low-metallicity, as we will show in \refsec{sec:dustvol_par} that dust evolution in this regime is dominated by \snii\ production.
We therefore modify \refeq{eq:analfit1}, as:
\begin{equation}
  \log\Zd\simeq-13.230+x \;\;\;\mbox{ for }x<7.3.
  \label{eq:analfit2}
\end{equation}
This fit splits our data in two equal size samples above and below.
To quantify its uncertainty, we can compute the envelopes corresponding to encompassing $68\,\%$ and $95\,\%$ of the sources, adding $[-0.27,+0.29]$ and $[-0.68,+0.70]$ respectively to the fit of $\log\Zd$.

In the case of \tqAF, a 1$^\sms{st}$ degree polynomial gives a satisfactory fit as a function of \tmet:
\begin{equation}
  \log\qAF\simeq-9.001+0.9486x.
  \label{eq:analfit3}
\end{equation}
Its $68\,\%$ and $95\,\%$ envelopes are obtained adding $[-0.13,+0.10]$ and $[-0.36,+0.20]$, respectively.

\section{EMPIRICAL QUANTIFICATION OF KEY DUST 
         EVOLUTION PROCESSES}
\label{sec:dustvol}

We now analyze the trends of \reffig{fig:scaldust}, in a more quantitative way, with the help of a grain evolution model.
The goal here is two-fold:
\begin{inlinelist}
  \item by fitting theoretical tracks to our sample, we intend to test
    if it is possible to account for the variation of the main galaxy 
    parameters, with only a few evolutionary processes;
  \item we aim to give a self-consistent empirical quantification of the main 
    \textit{ad hoc} tuning parameters controlling these evolutionary 
    processes.
\end{inlinelist}

  \subsection{The Cosmic Dust Evolution Model}
  \label{sec:dustvolmod}

Cosmic dust evolution models compute the variation of the dust content\footnote{they can also eventually compute the variation of the grain size distribution and composition with time \citep[\eg\ ][]{hirashita15}.} of a galaxy as a function of time.
\citet{dwek80} presented the first model of this kind, developed as an extension of models computing the elemental enrichment of the ISM \citep[][ for an early review]{audouze76}.
Subsequent attempts, with various degrees of complexity, followed
\citep[\eg\ ][]{dwek98,lisenfeld98,hirashita99,morgan03,inoue03,dwek07,galliano08a,calura08,valiante09,mattsson12b,asano13,zhukovska14,feldmann15,de-looze20,nanni20}.
Recently, such models have been used to post-process numerical simulations in order to provide a more comprehensive understanding of galaxy evolution \citep[\eg\ ][]{aoyama17}.

    \subsubsection{Model Hypothesis}

We adopt the one-zone dust evolution model of \citet{rowlands14} updated by \citet{de-vis17}\footnote{We started from the public Python code provided by the authors at \href{https://github.com/zemogle/chemevol}{https://github.com/zemogle/chemevol}. We have rewritten it in Fortran for numerical efficiency, since we needed to generate large grids of models. We have also implemented an adaptative temporal grid to ensure numerical precision at later stages of evolution.}.
It solves the coupled differential equations accounting for the time evolution of the mass of the four following quantities.
\begin{description}
  \item[Stars] are made out of the gas, which is partially returned to 
    the ISM at the end of their lifetime.
    The model tracks their evolution as a function of their mass, which 
    determines their lifetime and their elemental and dust yields.
    The role of this component therefore relies greatly on the form of the 
    assumed IMF \citep[\eg\ ][]{salpeter55,chabrier03}.
  \item[Gas] is depleted by astration and outflow, and is 
    replenished by stellar feedback and by inflow of metal- and dust-free gas.
  \item[Heavy elements] are injected in the ISM by stars at the end of their 
    lifetime.
    A fraction of them is recycled into stars through astration and lost 
    \textit{via} outflow.
  \item[Dust] is produced by three main processes:
    \begin{inlinelist}
      \item condensation in \expression{Low- and Intemediate-Mass Star} (LIMS) 
        ejecta 
        ($m_\star<8\eMsun$)\footnote{following \citet{de-vis17}, we assume that
         LIMSs condense $15\,\%$ of their heavy elements into dust.};
      \item condensation in \snii\ ejecta ($m_\star\ge8\eMsun$);
      \item grain growth in the ISM, by accretion of elements onto grain seeds.
    \end{inlinelist}
    Dust is removed from the ISM by:
    \begin{inlinelist}
      \item destruction by \snii\ blast waves;
      \item outflow;
      \item astration.
    \end{inlinelist}
\end{description}

The drivers of the evolution of these quantities are:
\begin{inlinelist}
  \item the assumed SFH (\ie\ the SFR as a function of time);
  \item the assumed inflow and outflow rates.
\end{inlinelist}

    \subsubsection{The Tuning Parameters}

The efficiencies of individual dust evolution processes are poorly known \citep[see reviews by][]{dwek05,draine09,jones16a,jones16b,jones16c,galliano18}.
From a theoretical point of view, these efficiencies depend on the most elusive detailed microscopic dust properties: chemical constitution, structure (crystalline, amorphous, aggregate, \etc) and size.
Consequently, theoretical estimates usually span several orders of magnitude.
From an observational point of view, unambiguous constraints of the evolution rates are also problematic, because it is nearly impossible to isolate a dust source or sink in a telescope beam.
Dust evolution models therefore use simple parametric efficiencies controlled by tuning parameters.

\paragraph{Dust condensation in SN-II ejecta.}
Dust formed by massive stars ($m_\star\ge8\eMsun$) is thought to dominate the dust production at early stages, below the \expression{critical metallicity} (\cf~\refsec{sec:D2M}).
The dust evolution model we are using assumes the theoretical dust yields, $m_\sms{dust}^\sms{SN}$, of \citet{todini01}, modified by a general tuning parameter, $\delta_\sms{SN}$ (\reftab{tab:SNcond}).
These dust yields can also be integrated over the IMF, $\phi(m_\star)$:
\begin{equation}
  \langle Y_\sms{SN}\rangle \equiv 
  \frac{\displaystyle\int_{8\eMsun}^{40\eMsun}\phi(m_\star)\times 
        m_\sms{dust}^\sms{SN}(m_\star)\ddiff m_\star}
       {\displaystyle\int_{8\eMsun}^{40\eMsun}\phi(m_\star)\ddiff m_\star},
  \label{eq:cond}
\end{equation}
to provide a single averaged dust yield per \snii.
For both \citet{salpeter55} and \citet{chabrier03} IMFs, it is:
$\langle Y_\sms{SN}\rangle\simeq0.35\times\delta_\sms{SN}\eMsun/\textnormal{SN}$.
The dust condensation timescale, $\tau_\sms{cond}$, can be expressed as a function of the \snii\ rate, $R_\sms{SN}$:
\begin{equation}
  \frac{1}{\tau_\sms{cond}(t)} 
    = \langle Y_\sms{SN}\rangle\frac{R_\sms{SN}(t)}{\Md(t)}.
\end{equation}
\begin{table}[htbp]
  \caption{\textsl{SN~II dust yields.} These values are the \citet{todini01} 
           yields, compiled by \citet{rowlands14}.
           They are multiplied by the tuning parameter $\delta_\sms{SN}$, used
           by \citet{de-vis17}.}
  \label{tab:SNcond}
  \centering
  \begin{tabular}{lr}
    \hline\hline
      Individual star mass, $m_\star$ & Dust yield, $m_\sms{dust}^\sms{SN}$ \\
    \hline
      $8.5\eMsun$ & $0\eMsun\times\delta_\sms{SN}$ \\
      $9\eMsun$  & $0.17\eMsun\times\delta_\sms{SN}$ \\
      $12\eMsun$ & $0.2\eMsun\times\delta_\sms{SN}$ \\
      $15\eMsun$ & $0.5\eMsun\times\delta_\sms{SN}$ \\
      $20\eMsun$ & $0.5\eMsun\times\delta_\sms{SN}$ \\
      $22\eMsun$ & $0.8\eMsun\times\delta_\sms{SN}$ \\
      $25\eMsun$ & $1.0\eMsun\times\delta_\sms{SN}$ \\
      $30\eMsun$ & $1.0\eMsun\times\delta_\sms{SN}$ \\
      $35\eMsun$ & $0.6\eMsun\times\delta_\sms{SN}$ \\
      $40\eMsun$ & $0.4\eMsun\times\delta_\sms{SN}$ \\
    \hline
  \end{tabular}
\end{table}

\paragraph{Grain growth in the ISM.}
The dust build-up by accretion of gas atoms onto pre-existing dust seeds is a potentially dominant production process at late stages, above the \expression{critical metallicity} (\refsec{sec:D2M}).
We adopt the parametrization of \citet{mattsson12a}, where the grain growth timescale, $\tau_\sms{grow}$, is parametrized by a tuning parameter, $\epsilon_\sms{grow}$:
\begin{equation}
  \frac{1}{\tau_\sms{grow}(t)} = 
    \epsilon_\sms{grow}\frac{\psi(t)}{\Mg(t)}\left(Z(t)-Z_\sms{dust}(t)\right),
  \label{eq:grow}
\end{equation}
where $\psi(t)$ is the SFR as a function of time $t$.
This tuning parameter encompasses our uncertainty about grain sizes, sticking coefficients and the fraction of cold clouds where dust growth occurs\footnote{We did not implement the parameter $f_\sms{c}$, the fraction of cold clouds, introduced by \citet{de-vis17}, as its effect can be encompassed in $\epsilon_\sms{grow}$ and $m_\sms{gas}^\sms{dest}$ \refeqp{eq:dest}.}.

\paragraph{Dust destruction by SN~II blast waves.}
\snii\ shock waves destroy dust by kinetic sputtering and grain-grain collisional vaporization \citep[\eg\ ][]{jones94}.
We adopt the dust destruction timescale, $\tau_\sms{dest}$, of \citet{dwek80}:
\begin{equation}
  \frac{1}{\tau_\sms{dest}(t)} = 
    \frac{m_\sms{gas}^\sms{dest}R_\sms{SN}(t)}{\Mg(t)},
  \label{eq:dest}
\end{equation}
where $m_\sms{gas}^\sms{dest}$ is a tuning parameter quantifying the effective mass of ISM swept by a single \snii\ blast wave within which all the grains are destroyed.

  \subsection{Modelling the Evolution of Individual Galaxies}

We use the dust evolution model of \refsec{sec:dustvolmod} to fit, for each galaxy: \tMd, \tMg, \tMs, SFR and $Z$.
Dust destruction in X-ray emitting gas is not taken into account in this model. We therefore exclude the ETGs ($T\le0$) which would not be consistently fitted (\cf\ \refsec{sec:Lx}).
The sample we model here contains 556 sources.

    \subsubsection{The Model Grid}

We have generated a large grid of models that we interpolate to find the best evolution tracks for each of our sources.
We assume a \expression{delayed} SFH \citep{lee10sk}:
\begin{equation}
  \psi(t)=\psi_0   
    \frac{t}{\tau_\sms{SFH}}\exp\left(-\frac{t}{\tau_\sms{SFH}}\right),
  \label{eq:SFH}
\end{equation}
parametrized by the SFH timescale, $\tau_\sms{SFH}$, and a scaling factor, $\psi_0$.
We assume a \citet{salpeter55} IMF in order to be consistent with the observed SFR and stellar mass estimates (\refsec{sec:ancillary}).
Exchanges of matter between galaxies and their environment can have a significant role in regulating the global dustiness at all redshifts \citep[\eg\ ][]{jones18,ohyama19,sanders20,burgarella20}.
We account for inflow and outflow, assuming their rates are proportional to SFR:
$R_\sms{in/out}(t)=\delta_\sms{in/out}\times\psi(t)$.
\reftab{tab:dustmod} gives the parameters of our dust evolution model grid.
We perform log-log interpolation of this grid in order to estimate the tracks corresponding to any arbitrary combination of parameters.
\begin{table}[htbp]
  \caption{\textsl{Dust evolution parameter grid.}
           \expression{Individual galaxy parameters} control the particular SFH 
           of each galaxy.
           \expression{Common dust evolution parameters} are fitted to the 
           whole sample, assuming their values are the same for each galaxy.
           The grid step (4$^{th}$ column) is logarithmic for all parameters 
           except the time grid.
           The SFR scale is normalized to the total initial mass of the galaxy, 
           $M_0$.
           All extensive quantities computed by the model are proportional to 
           $M_0$, while intensive quantities are independent of 
           $M_0$.
           In practice, we generate a grid for an arbitrary 
           $M_0=4\E{10}\eMsun$.
           This parameter cancels out in the process, as we are fitting ratios
           (\refsec{sec:HBdustvol}).}
  \label{tab:dustmod}
  \begin{tabularx}{\linewidth}{Xrrrr}
    \hline\hline
      Parameter             & Notation            & Range       & Step\\
    \hline
      \multicolumn{4}{c}{Individual galaxy parameters} \\
    \hline
      Age [Gyr]             & $t$                 & $[0.001,15]$& 0.15     \\
      SF timescale [Gyr]    & $\tau_\sms{SFH}$     & $[0.1,30]$  & 0.57 (ln) \\
      SFR scale [Gyr$^{-1}$] & $\psi_0/M_0$ & $[0.0125,12.5]$    & 0.26 (ln)\\
      Inflow rate / SFR     & $\delta_\sms{in}$    & $[0.05,4]$  & 0.37 (ln) \\
      Outflow rate / SFR    & $\delta_\sms{out}$   & $[0.05,4]$  & 0.13 (ln) \\
    \hline
      \multicolumn{4}{c}{Common dust evolution parameters} \\
    \hline
      SN condensation       & $\delta_\sms{SN}$    & $[0.001,1]$  & 0.37 (ln) \\
      Grain growth          & $\epsilon_\sms{grow}$& $[100,10000]$& 0.17 (ln) \\
      SN dest. [\tMsun]     & $m_\sms{gas}^\sms{dest}$& $[50,1500]$  &0.34 (ln)\\
    \hline
  \end{tabularx}
\end{table}

The SFH we have adopted here, to compute the chemical evolution, is different from the SFH used by \citet{nersesian19} to model the SEDs and estimate the observed SFR and \tMs\ (\cf~\refsec{sec:Mstar}).
Ideally, adopting the same SFH would be more consistent.
However, adding a second functional form to \refeq{eq:SFH}, accounting for a potential recent burst, would push this model beyond our current computational capabilities, by increasing the number of free parameters.
At the same time, the elemental and dust enrichment by this recent burst would probably be negligible.
Alternatively, we could have also modelled the SED, using solely \refeq{eq:SFH}.
The problem, in this case, is that the use of a single stellar population biases the estimate of \tMs\ (\cf~\refapp{app:Mstar}).

    \subsubsection{Hierarchical Bayesian Dust Evolution Inference}
    \label{sec:HBdustvol}

To constrain our dust evolution model, we consider the posterior inference of our \expression{reference} run (\refsec{sec:ref}) as a set of observational constraints.
We fit the four independent, intensive quantities: \tsMd, \tsMg, sSFR and $Z$.
The complex PDF, including the intricate parameter correlations, is preserved in this process.
We make the restrictive assumption that the three tuning parameters, $\delta_\sms{SN}$, $\epsilon_\sms{grow}$ and $m_\sms{gas}^\sms{dest}$ are universal and are therefore identical in every galaxy.
Consequently, we assume that the difference between galaxies is solely the result of their particular individual SFH.
We vary the age of the galaxy, $t$, the two SFH parameters, $\tau_\sms{SFH}$ and $\psi_0$, and the inflow and outflow rates, $\delta_\sms{in}$ and $\delta_\sms{out}$.

\paragraph{On the universality of the tuning parameters.}
As we have seen, the three tuning parameters hide effects of the dust constitution, size distribution, fraction of cold clouds, \etc\ 
These quantities could vary across different environments.
However, as we will show in \refsec{sec:dustvol_comind}, the SN~II dust condensation dominates in the low-metallicity regime, while grain growth and SN destruction are important above the critical metallicity.
Thus, the parameters that we will constrain are representative of the regime where they dominate the dust budget.
Exploring their variation with the environment is probably premature.
We will however be able to infer the variation of the timescales across galaxies, from \refeqs{eq:cond}{eq:dest}.
In addition, if we were to infer one set of tuning parameters per galaxy, it would imply that they vary as a function of the galaxy's global parameters.
In order to be consistent, we would then need to implement these variations of the tuning parameters in the dust evolution model, and change their value at each time step, accordingly.

\paragraph{The model prior.}
We have built a HB model to infer these parameters.
The prior of the five SFH-related parameters is a multivariate Student's $t$ distribution controlled by hyperparameters, similar to the treatment in \HB.
We have assumed wide log-normal priors for the three common tuning parameters.
These priors are centered at $\ln0.1$, $\ln1000$ and $\ln320\,\Msun$ with standard-deviations, $0.8$, $0.8$ and $0.4$, for $\delta_\sms{SN}$, $\epsilon_\sms{grow}$ and $m_\sms{gas}^\sms{dest}$, respectively.
These priors were designed so that their $\pm3\sigma$ range roughly corresponds to the extent of the values reported in the literature: $1\,\%\lesssim\delta_\sms{SN}\lesssim100\,\%$, $100\lesssim\epsilon_\sms{grow}\lesssim10^4$ and $100\,\Msun\lesssim m_\sms{gas}^\sms{dest}\lesssim1000\,\Msun$.
It does not mean that these parameters can not exceed these limits, but it will be \textit{a priori} improbable.
The reason to adopt weakly informative priors is to avoid unrealistic degeneracies.
For instance, there is a well-known degeneracy between grain growth and grain destruction \citep[\eg\ ][]{mattsson12b,de-vis17,de-looze20}.
Indeed, the ratio of the two timescales, $\tau_\sms{dest}(t)/\tau_\sms{grow}(t)\propto\epsilon_\sms{grow}/m_\sms{gas}^\sms{dest}\times(Z(t)-Z_\sms{dust}(t))$ is not constant, but the metallicity dependence is quite mild in the narrow range above the critical metallicity.
This is enough to allow the MCMC to explore unrealistically high values of $\epsilon_\sms{grow}$ and $m_\sms{gas}^\sms{dest}$, if we assume a flat prior.

\paragraph{HB run specifications.}
Accounting for the missing ancillary data, we have a total of 1913  observational constraints, for our 556 sources.
One may note that we are fitting a model with $5\times556+3=2783$ parameters 
and $5+5+C_{5}^{2}=20$ hyperparameters.
This is not an issue in the Bayesian approach \citep[\eg\ ][]{hogg10b}.
Model parameters are rarely completely independent.
Furthermore, the hyperprior is the dominant constraint for very scattered parameters \citepalias{galliano18a}.
Finally, by sampling the full joint PDF, a Bayesian model clearly delineates the various degeneracies, provided the MCMC has converged towards the stationary posterior.
We have run 12 parallel MCMCs, starting from uniformly distributed random initial conditions within the parameter ranges in \reftab{tab:dustmod}, in order to ensure the uniqueness of our posterior.
The results discussed below are from a $10^5$ length MCMC, removing the first $10^4$ steps to account for burn-in.
The longest integrated autocorrelation time is 6200.

    \subsubsection{The Fitted Dust Evolution Tracks}
    \label{sec:tracks}

\begin{figure*}[htbp]
  \includegraphics[width=\textwidth]{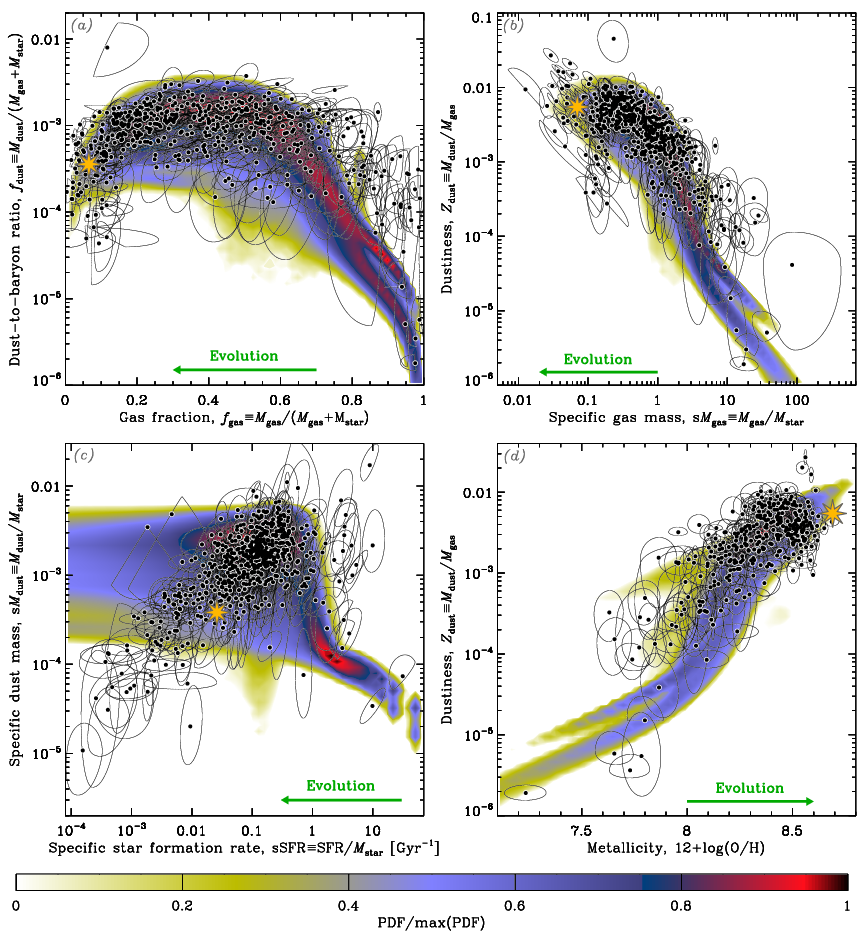}
  \caption{\textsl{Fitted dust evolution tracks,} 
           assuming a \citet{salpeter55} IMF.
           The four panels represent the same quantities as 
           \reffig{fig:scaldust}.
           The black SUEs represent the 556 galaxies of our subsample. 
           The yellow star is the Milky Way.
           The colored density contours represent the posterior PDF
           of dust evolution tracks, marginalizing over the individual SFH
           of each galaxy.}
  \label{fig:tracks}
\end{figure*}
The four panels of \reffig{fig:tracks} present the same relations as in \reffig{fig:scaldust}, for the subsample of 556 sources.
The posterior PDF of the dust evolution tracks is displayed as colored density contours, marginalizing over the SFH of individual galaxies.

\paragraph{Qualitative inspection.}
At first glance, we can see that the tracks of \reffig{fig:tracks} reproduce the data quite well.
Apart from a few outliers, there are however two notable systematic discrepancies.
\begin{enumerate}
  \item The \tsMd-sSFR trend of panel~\textit{(c)} is poorly reproduced.
    The initial rise of \tsMd\ at 
    $\textnormal{sSFR}\gtrsim0.3\;\textnormal{Gyr}^{-1}$ undershoots several of
    the starbursting dwarf galaxies.
    Overall, this discrepancy is quite mild, as most of the SUEs in this area
    are less than 2$\sigma$ away from the trend.
    The recent burst of star formation, that is not accounted for by the 
    chemical evolution model, is likely responsible for the enhanced observed 
    SFR.
    On the contrary, a starburst occuring in more evolved objects, on the left 
    of this trend, might simply contribute to the general scatter of the 
    relation and go unnoticed.
    Below $\textnormal{sSFR}\lesssim0.1\;\textnormal{Gyr}^{-1}$, there is a 
    general decreasing trend of \tsMd\ with decreasing sSFR, but our model is 
    essentially flat.
    This is the most troublesome discrepancy of our analysis.
    We will discuss it fully below.
  \item In panel~\textit{(a)}, several high-gas-fraction sources are undershot  
    by the model, although most of them are only less than 2$\sigma$ away from 
    the tracks.
    Those are the irregular galaxies around the critical metallicity regime.
    They are in the stage where the dustiness changes rapidly, due to the 
    increased contribution of grain growth.
    These outliers can be seen in panel~\textit{(b)}, around $\sMg\simeq10$,
    in panel~\textit{(c)}, around 
    $\textnormal{sSFR}\simeq3\;\textnormal{Gyr}^{-1}$,
     and in panel~\textit{(d)}, around $\met\simeq8$.
    The model poorly reproduces the rapid dustiness rise around the 
    critical metallicity, in panel~\textit{(d)}.
\end{enumerate}

\begin{figure}[htbp]
  \includegraphics[width=\linewidth]{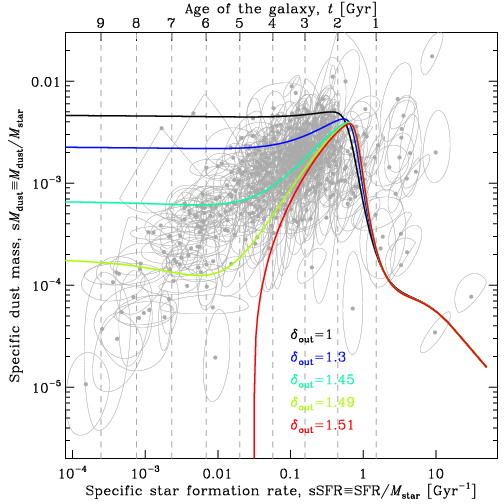}
  \caption{\textsl{Effect of outflow} on the \tsMd-sSFR relation.
           The data is identical to panel~\textit{(c)} of \reffig{fig:tracks}.
           The colored lines represent our dust evolution model.
           We have fixed all the parameters close to their maximum 
           \textit{a posterior} values, except the outflow rate, 
           $\delta_\sms{out}$: 
           $\tau_\sms{SFH}=0.8$~Gyr, $\psi_0=40\eMsun/\textnormal{yr}$,
           $\delta_\sms{in}=1.0$, $\delta_\sms{SN}=0.01$, 
           $\epsilon_\sms{grow}=4500$ and 
           $m_\sms{gas}^\sms{dest}=1200\eMsun/\textnormal{SN}$.
           The different lines correspond to different values of 
           $\delta_\sms{out}$.
           The top axis displays the age of the galaxy corresponding to the
           particular SFH of the model run.
           We have used a \citet{salpeter55} IMF.}
  \label{fig:outflow}
\end{figure}

\paragraph{On the \tsMd-sSFR relation.}
The quasi-linear trend of \tsMd\ with sSFR, for $\textnormal{sSFR}\lesssim0.1\;\textnormal{Gyr}^{-1}$ (panel~\textit{(c)} of \reffig{fig:tracks}), is not accounted for by our model.
\citetalias{nanni20} argue that outflows is responsible for this trend\footnote{We note that \citetalias{nanni20} do not have sources below $\sMd\lesssim10^{-4}$, while it is where our fit gets the most problematic.}.
Indeed, outflow depletes the dust content proportionally to SFR, without affecting the stellar content.
It seems like a natural explanation.
The outflow rate, $\delta_\sms{out}$, is a free parameter in our model.
However, it does not produce a linear \tsMd-sSFR relation.
\reffig{fig:outflow} shows the same data as in panel~\textit{(c)} of \reffig{fig:tracks}.
We have overlaid several dust evolution tracks corresponding to our maximum \textit{a posteriori} parameters, varying $\delta_\sms{out}$.
This figure is similar to Fig.~6 of \citetalias{nanni20}.
We note the following.
\begin{enumerate}
  \item We can see that no single track can reproduce the trend.
    It could be a satisfactory explanation if galaxies were rapidly 
    evolving off the main trend, staying only a small fraction of their 
    lifetime on the horizontal branch.
    However, looking at the top axis of \reffig{fig:outflow}, we realize that
    a galaxy should spend about half of its lifetime on the horizontal branch.
    If our model was correct, there should have been, statistically, a 
    significant number of sources deviating from the main trend.
  \item The range of $\delta_\sms{out}$ values that can account for most 
    of the trend is quite narrow (\reffig{fig:outflow}). 
    It would mean that the outflow has to be precisely regulated.
    It seems unlikely.
  \item If this trend was solely due to outflow, the specific gas mass would 
    follow it closely, which is not the case.
    This point has also been discussed by \citet{cortese12}.
\end{enumerate}
In summary, it appears our failure at reproducing the \tsMd-sSFR relation is rather an incapacity of our model to produce a realistic trend, rather than a fitting issue.
As we will see in \refsec{sec:IMF}, this is not an IMF-related issue.
Since the other trends are acceptably reproduced, the problem might lie in the only quantity represented in panel~\textit{(c)} of \reffig{fig:tracks} not appearing in the other panels: the SFR.
Our adopted SFH \refeqp{eq:SFH} and our inflow and outflow prescriptions might be too simplistic to account for the wealth of data we have here 
\citep[\cf\ \eg\ ][ for a discussion on the limitations of the delayed SFH]{leja19}.

\paragraph{The importance of fitting dust evolution models.}
We stress the importance of actually fitting, in a consistent way, dust evolution models, rather than simply performing visual comparisons.
Here, we have attempted to fit \tMs, \tMg, \tMd, Z and SFR for each galaxy.
This rigorous process highlights the model limitations.
Most past studies have merely overlaid tracks on their data, producing a convincing but inconsistent interpretation.
For instance, \citet{de-vis17}, who used the same dust evolution model as we use, and compared it to a similar sample as ours, were able to produce tracks accounting for most of the observations in the panels~\textit{(a)} and \textit{(d)} of our \reffig{fig:tracks}.
However, two quantities of a given galaxy, such as the dust and stellar masses, were usually explained with different values of the dust evolution parameters, at different ages.
On the contrary, our approach allows us to avoid mutually inconsistent explanations of different trends and correlations.
Overall, performing a rigorous fit does not help getting better solutions, but it definitely helps avoiding bad ones.

  \subsection{The Inferred Dust Evolution Parameters}
  \label{sec:dustvol_par}

We now discuss the parameters inferred from the fit of \refsec{sec:tracks}.

    \subsubsection{Parameter Distribution}
    \label{sec:dustvol_comind}

The fits of \reffig{fig:tracks} allow us to infer the common dust evolution tuning parameters, as well as the individual SFH-related parameters (\reftab{tab:dustmod}).

\begin{figure}[htbp]
  \includegraphics[width=\linewidth]{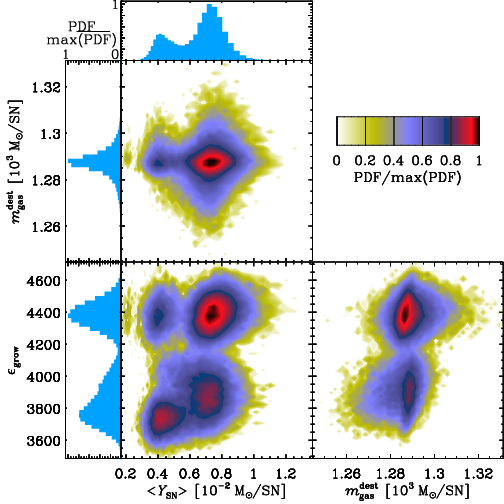}
  \caption{\textsl{Posterior distribution of the tuning parameters,}
           assuming a \citet{salpeter55} IMF.
           The three central panels with colored contours display the 
           bidimensional posterior PDF of pairs of parameters, marginalizing 
           over all the other ones.
           The three margin plots show the posterior PDF of each tuning 
           parameter.
           The PDFs are scaled (divided by their maximum).
           The displayed ranges encompass every single parameter draw after 
           burn-in.}
  \label{fig:corrcom}
\end{figure}

\paragraph{The dust evolution tuning parameters.}
The PDF of the three common dust evolution tuning parameters is displayed in \reffig{fig:corrcom}.
We infer the following values:
\begin{itemize}
  \item $\langle Y_\sms{SN}\rangle\simeq7.3_{-0.3}^{+0.2}\E{-3}\eMsun/\textnormal{SN}$;
  \item $\epsilon_\sms{grow}\simeq4045_{-354}^{404}$;
  \item $m_\sms{gas}^\sms{dest}\simeq1288_{-8}^{+7}\eMsun/\textnormal{SN}$.
\end{itemize}
The relatively small uncertainties on these parameters reflect the fitting uncertainties.
They indicate the values we infer are not ambiguous.
However, they do not include the assumption-dependent uncertainties.
In particular, the precise value of $\langle Y_\sms{SN}\rangle$ relies mainly on our estimate of the dustiness of the few ELMGs in our sample.
In \refsec{sec:D2M}, we have extensively discussed the different systematic effects that could have biased these estimates.
If some of these effects are, at some point, proven to be relevant, our inference of $\langle Y_\sms{SN}\rangle$ would have to be revised accordingly. 
To first order, the SN dust yield is proportional to the dustiness of the lowest metallicity objects:
\begin{equation}
  \langle Y_\sms{SN}\rangle\simeq0.007\eMsun/\textnormal{SN}
    \times\frac{Z_\sms{dust}(\textnormal{\izw})}{1.9\E{-6}}.
\end{equation}
We also notice that $\delta_\sms{SN}$ and $m_\sms{gas}^\sms{dest}$ are peaking at the tail of their prior.
It means our data carries a large weight evidence. 
It also means the values we infer are probably an upper limit for $\delta_\sms{SN}$ and a lower limit for $m_\sms{gas}^\sms{dest}$.
Since $m_\sms{gas}^\sms{dest}$ and $\epsilon_\sms{grow}$ are correlated, our inference of $\epsilon_\sms{grow}$ is also probably a lower limit.
We have estimated in \refsec{sec:D2M} that the dustiness of ELMGs could have been underestimated by a factor of at most $\simeq4.25$, the conservative conclusions we can draw from our analysis are thus that:
\begin{itemize}
  \item $\langle Y_\sms{SN}\rangle\lesssim0.03\eMsun/\textnormal{SN}$;
  \item $\epsilon_\sms{grow}\gtrsim3000$;
  \item $m_\sms{gas}^\sms{dest}\gtrsim1200\eMsun/\textnormal{SN}$.
\end{itemize}

\begin{figure*}[htbp]
  \includegraphics[width=\textwidth]{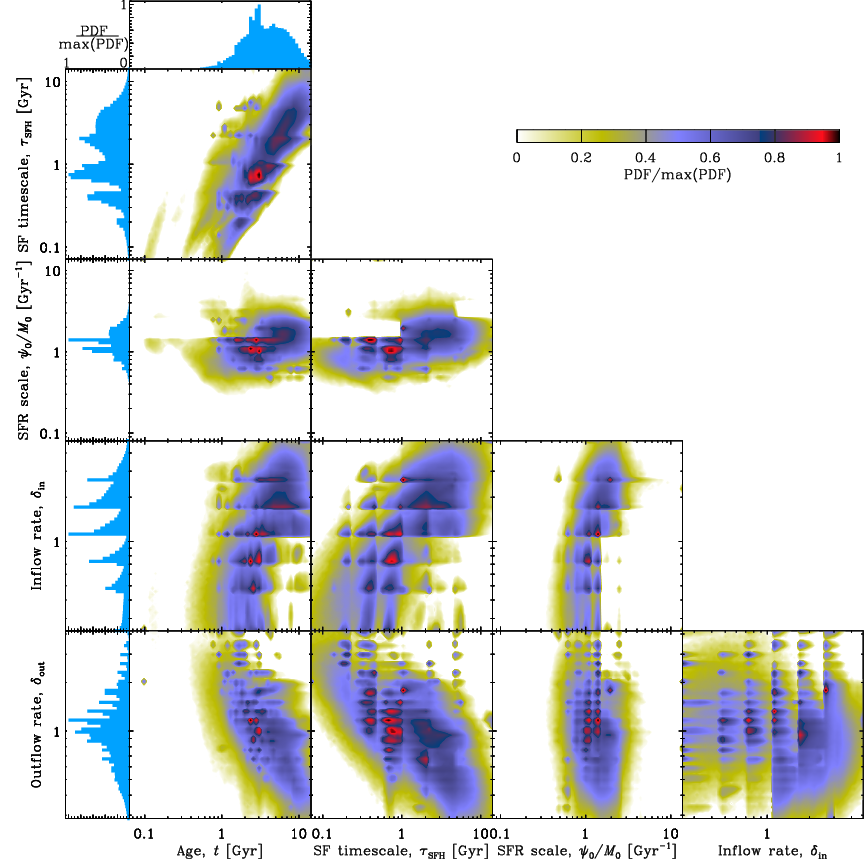}
  \caption{\textsl{Posterior distribution of the SFH-related parameters.}
           The plotting conventions are identical to \reffig{fig:corrcom}.
           The displayed PDF is the distribution of the individual parameters
           of every galaxy, marginalizing over the dust evolution tuning 
           parameters.}
  \label{fig:corrind}
\end{figure*}

\paragraph{The SFH-related parameters.}
\reffig{fig:corrind} displays the corner plot of the five parameters controlling the individual SFH of each galaxy.
The displayed PDF is the posterior of the parameters of every galaxy.
There is a relatively large scatter of these parameters, implying that our different galaxies have different SFHs.
Note that the interpolation between models does not produce a perfectly smooth distribution. 
The edges of our model grid are visible in this corner plot.
This does not however impact the results as our grid samples well enough the 
PDF.

    \subsubsection{The Dust Evolution Timescales}
    \label{sec:dustvol_tau}

\begin{figure}[htbp]
  \includegraphics[width=\linewidth]{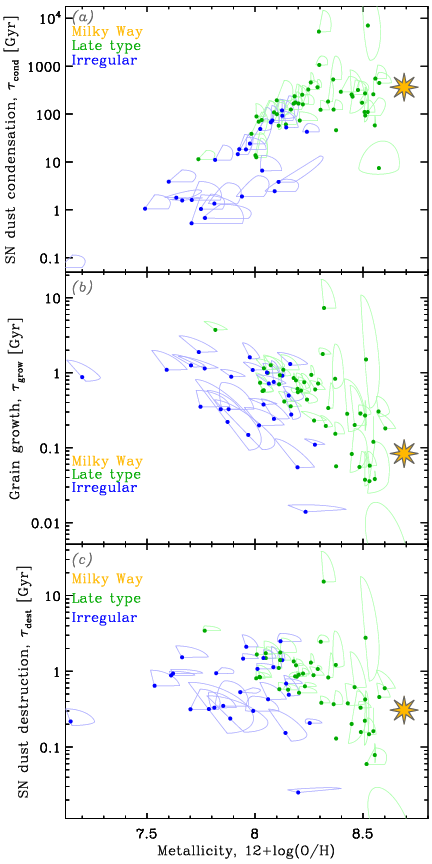}
  \caption{\textsl{Dust evolution timescales,} assuming a \citet{salpeter55}
           IMF.
           The three panels display the posterior PDF of the
           three dust evolution timescales.
           The SUEs represent the value of these timescales as a function of
           metallicity, for each galaxy.
           The yellow star represents the Milky Way, at the maximum a posteriori
           of the three tuning parameters.}
  \label{fig:tau}
\end{figure}
It is possible to estimate the posterior PDF of the dust evolution timescales of each galaxy, from the inferred parameters of \reffigs{fig:corrcom}{fig:corrind}, using \refeqs{eq:cond}{eq:dest}.
\reffig{fig:tau} displays these timescales as a function of metallicity.
Although there is some scatter due to the different SFH and age of galaxies in a given metallicity bin, we note that these timescales evolve.
\begin{description}
  \item[The SN~II dust condensation timescale] (panel~\textit{(a)} of 
    \reffig{fig:tau}) is around $\tau_\sms{cond}\simeq100$~Myr for ELMGs, 
    implying 
    that dust can be dominated by stardust in this regime.
    It rises up to $\tau_\sms{cond}\simeq 1000$~Gyr around solar metallicity,
    indicating this process is not sufficient to account for all ISM dust
    in evolved systems.
  \item[The grain growth timescale] (panel~\textit{(b)} of \reffig{fig:tau})
    is quite scattered.
    It starts around $\tau_\sms{grow}\simeq1$~Gyr in the ELMG regime 
    and decreases down to $\tau_\sms{grow}\simeq50$~Myr around solar metallicity.
    The average value of our sample at $\met\ge8.5$ is 
    $\tau_\sms{grow}\simeq45$~Myr.
    It is another way to show that dust formation is dominated by grain growth
    around solar metallicity.
  \item[The SN~II dust destruction timescale] (panel~\textit{(c)} of 
    \reffig{fig:tau}) is also scattered but stays around 
    $\tau_\sms{dest}\simeq300$~Myr across our metallicity range.
\end{description}
The Milky Way value displayed in the three panels is consistent with the cloud of points in the highest metallicity domain.

There is a common misconception that dust destruction by SN~II blast waves is unimportant at early-stages, because the dustiness is so low that few grains are destroyed by a single SN~II.
However, we show here that this is not the case as:
\begin{inlinelist}
  \item the SN~II rate is on average higher at low-metallicity;
  \item the fraction of grains destroyed by a single SN~II is 
    dustiness-independent.
\end{inlinelist}

    \subsubsection{Sensitivity to the IMF}
    \label{sec:IMF}

As we discussed in \refsecs{sec:Mgas}{sec:SFR}, the main uncertainty on our estimated \tMs\ and SFR comes from our IMF assumptions.
Those were derived with a \citet{salpeter55} IMF.
Here, we explore the sensitivity of our results assuming a \citet{chabrier03} IMF.
To be consistent, we first need to correct our estimated \tMs\ and SFR.
According to \citet[][ Sect.~3.1]{madau14}, we need to multiply SFR by 0.63 and \tMs\ by 0.61.
We have then generated another grid of models similar to \reftab{tab:dustmod} 
with the \citet{chabrier03} IMF, and performed the fitting process of \refsec{sec:tracks}.
The full results are given in \refapp{app:Chab}.
The upshot is that the inferred tuning parameters are roughly consistent with our estimates in \refsec{sec:dustvol_comind}:
\begin{itemize}
  \item $\langle Y_\sms{SN}\rangle\simeq2.5_{-0.3}^{+0.2}\E{-2}\eMsun/\textnormal{SN}$;
  \item $\epsilon_\sms{grow}\simeq4485_{-17}^{+11}$;
  \item $m_\sms{gas}^\sms{dest}\simeq1289_{-7}^{+6}\eMsun/\textnormal{SN}$.
\end{itemize}
The differences can be explained by the different relative contribution from LIMSs.
The only value that differs sensibly is $\langle Y_\sms{SN}\rangle$.
This difference is however simply due to the fact the low-metallicity dustiness is overshot by the model with the \citet{chabrier03} IMF.

This agreement was expected. 
The reason is that, to first order, LIMSs represent a dead weight in both the SED and the chemical enrichment.
The total power emitted by stellar populations is dominated by massive stars.
This is the reason why SFR and \tMs\ estimates are so dependent on the IMF assumption.
From the dust evolution modelling, the elemental and stardust enrichment is also dominated by massive stars.
The only quantity for which we infer a different value is logically $\psi_0$.
It is $\simeq50\,\%$ higher to generate the same number of massive stars, compensating for the 0.63 correction factor.

  \subsection{Discussion}
  \label{sec:disc}

    \subsubsection{On the Low Inferred SN~II Dust Yield}

Our SN~II dust yield is lower than the most recent estimates \textit{in situ}.
Measuring the dust mass produced by a single SN~II is quite difficult, as it implies disentangling the freshly-formed dust from the surrounding ISM.
It also carries the usual uncertainty about dust optical properties.
A decade ago, the largest dust yield ever measured was $Y_\sms{SN}\simeq0.02\eMsun$ \citep[in SN~2003gd;][]{sugerman06}.
The \hersc\ space telescope has been instrumental in estimating the cold mass of \expression{SuperNova Remnants} (SNR).
The yields of the three most well-studied SNRs are now an order of magnitude higher:
\begin{description}
  \item[Cassiopeia A:] $Y_\sms{SN}\simeq0.04-1.1\eMsun$ 
    \citep{barlow10,arendt14,de-looze17,bevan17,priestley19};
  \item[The Crab nebula:] $Y_\sms{SN}\simeq0.03-0.23\eMsun$ 
    \citep{gomez12,temim13,de-looze19};
  \item[SN~1987A:] $Y_\sms{SN}\simeq0.45-0.8\eMsun$ \citep{dwek15,matsuura15}.
\end{description}
In all these cases, the newly-formed grains have not yet experienced the reverse shock \citep{bocchio16b}.
The net yield is thus expected to be significantly lower.

Even if $\simeq10-20\,\%$ of the dust condensed in an SN~II ejecta survives its reverse shock \citep[\eg\ ][]{nozawa06,micelotta16,bocchio16b}, we have to also consider the fact that massive stars are born in clusters.
The freshly-formed dust injected by a particular SN~II, having survived the reverse shock, will thus be exposed to the forward shock waves of nearby SNe~II \citep[\eg\ ][]{martinez-gonzalez18}. 
This effect is not accounted for by \refeq{eq:dest}, as it does not account for clustering, nor does it account for the excess dustiness around these stars due to the recent grain production.
Our estimate of $\langle Y_\sms{SN}\rangle$ is therefore an effective empirical yield, that probably accounts for this effect.

    \subsubsection{The Relevance of Local Low-Metallicity Galaxies}

Our analysis confirms the long-lasting consensus that Milky Way dust is essentially grown in the ISM (\refsec{sec:intro}).
The apparently paradoxical fact here is that we have drawn this conclusion from the low-metallicity domain.
It is because dust production is dominated by SN~II condensation below the critical metallicity that we could constrain its efficiency and show it is unimportant at solar metallicity.
The relevance of dwarf galaxies here is not necessarily that they can be considered as analogs of primordial distant galaxies, but that they sample a particular, key, dust production regime.

    \subsubsection{Implications for High-Redshift Systems}

High-redshift ($z\gtrsim6$) objects exhibiting copious amounts of dust ($\simeq10^7-10^8\eMsun$), close to the reionization era, have been challenging grain formation scenarios \citep[\eg\ ][]{dwek07,valiante09,dwek14,laporte17}.
These objects are indeed only a few 100~Myr old, but have a roughly Galactic dustiness.
SN~II dust condensation would need to have a high efficiency ($\simeq1\eMsun/\textnormal{SN}$) to account for the observed dust mass \citep{dwek07,dwek14}.
AGB star yield can explain this dust content for $z\simeq6$ objects \citep{valiante09}, but not at $z\simeq8$ \citep{laporte17}.

Our results imply that grain growth should be the dominant dust formation mechanism in these galaxies.
The dustiness of these massive objects being roughly Galactic, their metallicity should thus be Galactic too.
The grain growth timescale should therefore be shorter than $\simeq100$~Myr
(panel~\textit{(b)} of \reffig{fig:tau}), well below the age of these systems.
Consequently, these very distant galaxies may not be the best laboratories to constrain the SN~II dust yield.

  \subsubsection{Comparison with Recent Studies}

As stated in \refsecs{sec:intro}{} and \ref{sec:tracks}, past studies have not been rigorously fitting cosmic dust evolution models to observations of galaxies.
Recently, \citetalias{nanni20} and \citet[][ hereafter \citetalias{de-looze20}]{de-looze20} have addressed this issue.
\citetalias{nanni20} have adopted a relatively coarse dust evolution model grid with a frequentist fitting approach.
Although interesting, their approach does not allow them to rigorously quantify parameter degeneracies and uncertainties.
The comparison with their results is therefore limited.
We have discussed the main discrepancies between the present work and theirs in \refsecs{sec:DGS}{} and \refsec{sec:tracks}.

\citetalias{de-looze20} have analyzed the JINGLE galaxy sample, following a methodology similar to ours, fitting the IR SED of each galaxy, and subsequently fitting one-zone dust evolution tracks to their derived \tMs, \tMg, \tMd, $Z$ and SFR.
They adopted a non-hierarchical Bayesian approach.
To our knowledge, this is the first article, similar to ours, to perform a rigorous fit of dust evolution models, with clearly quantified parameter uncertainties.
Their results however qualitatively differ from ours.
\citetalias{de-looze20} find an overall high SN~II dust yield, and a low grain growth efficiency (the tuning parameters are not assumed universal in their study).
In our opinion, the discrepancy between our results and those of \citetalias{de-looze20} comes from the two following points.
\begin{enumerate}
  \item \citetalias{de-looze20}'s metallicity coverage is more limited than 
    ours.
    They have only one source below $\met=8.0$, and none below $\met=7.8$
    (see \eg\ their figure 7).
    Consequently, they do not sample the stardust regime, below the critical 
    metallicity.
    Disentangling SN~II dust condensation, grain 
    growth and shock destruction, with their data, is therefore rendered 
    difficult.
  \item The posterior distribution of the dust evolution parameters inferred by
    \citetalias{de-looze20}, as seen in their figures G.7 to G.12 (they 
    focus their analysis on their galaxy bins 3 and 4), rarely goes 
    down to zero probability within the range allowed by their uniform prior.
    It means the weight of evidence provided by their data is relatively mild.
    Our inference of $\epsilon_\sms{grow}$ and $m_\sms{gas}^\sms{dest}$ are 
    consistent with their PDF, as they fall in a high probability domain in all 
    their models.
    We simply have a smaller uncertainty, thanks to our low-metallicity 
    coverage.
    This is more pronounced for their distribution of $\delta_\sms{SN}$
    (their $f_\sms{survival}$).
    \citetalias{de-looze20} would have also benefitted from extending their 
    prior range down to smaller values, as they do not consider yields below 
    $\delta_\sms{SN}=10\,\%$, while we
    infer a value around $\delta_\sms{SN}\simeq1\,\%$.
\end{enumerate}

  \section{SUMMARY AND CONCLUSION}
  \label{sec:summary}

This article has presented an observational study aimed at constraining the timescales of the main processes controlling the evolution of interstellar dust in galaxies.
The principal highlights are the following.
\begin{enumerate}
  \item We have gathered the 3-\tmic-to-1-mm photometry and ancillary 
    data of 798 nearby galaxies
    from the DustPedia \citep{davies17} and DGS \citep{madden13} surveys.
    We have attempted to create the most conservative, homogeneous sample, by 
    controlling the factors that could lead to systematic biases 
    (\refsec{sec:data}):
    \begin{enumerate}
      \item the DustPedia and DGS IR data reduction and photometry have been
        performed consistently;
      \item the stellar mass and SFR have been estimated using the same 
        IMF;
      \item the metallicities have been estimated using one uniform calibration;
      \item the total gas masses have been derived from \hiline\ and \COio\ 
        observations, when available.
        When CO data was not available, the molecular gas mass was estimated
        from a scaling relation.
      \item Resolved interferometric \hiline\ observations of 20 of the
        lowest metallicity objects were used in order to extract the gas mass
        corresponding exactly to the IR photometric aperture.
    \end{enumerate}
  \item We have performed a hierarchical Bayesian dust SED fit of the 798 
    galaxies, using the code \HB\ \citepalias{galliano18a} with the \THE\
    grain properties \citep{jones17}.
    \begin{enumerate}
      \item This allowed us to infer the dust mass, \tMd, mean starlight 
        intensity, \tUav, and
        the mass fraction of aromatic-feature-emitting grains, \tqAF, in each 
        galaxy (\refsec{sec:ref}).
        The inferred parameters are given in \reftab{tab:resref}.
      \item We have compared our inferred parameters to a series of additional 
        runs, as well as to independent literature studies, in order to 
        demonstrate the influence of the different assumptions of our model 
        and assess the robustness of our results
        (\refsecs{sec:robust}{sec:sophiasco}).
    \end{enumerate}
  \item We have displayed several well-known scaling relations involving \tMs, 
    \tMg, \tMd, \tmet, SFR, \tqAF\ and \tUav, for our sample 
    (\refsec{sec:trends}).
    \begin{enumerate}
      \item We have shown that there is a drastic evolution with metallicity
        of the dust-to-metal mass ratio (by two orders of magnitude).
        We have extensively discussed the different biases that could 
        artificially produce such a trend, concluding they were unlikely 
        (\refsec{sec:D2M}).
      \item We have noticed that early-type galaxies have a 
        systematically lower dust-to-gas mass ratio than other types in the 
        same gas-to-stellar mass ratio range.
        We have investigated the possibility that this was resulting from 
        the enhanced dust destruction due to thermal sputtering in the hot X-ray
        emitting gas permeating these objects.
        This scenario is supported by a rough negative correlation between
        the dust-to-star mass ratio and the X-ray photon rate per dust grain
        (\refsec{sec:Lx}).
      \item We have displayed the well-known trends of \tqAF\ with \tmet\ and
        \tUav.
        Our data indicate the correlation with \tmet\ is significantly
        better (\refsec{sec:qAFev}).
        It implies that, at the scale of a galaxy, the overall abundance of
        small a-C(:H) grains might be principally controlled by the efficiency
        of their formation (stardust production and/or shattering of larger 
        carbon grains).
        The photodestruction of small a-C(:H) might overall be circumscribed 
        around star forming regions.
    \end{enumerate}
  \item We have performed a hierarchical Bayesian fit of a one-zone dust
    evolution model to the derived \tMs, \tMg, \tMd, \tmet\ and SFR of
    a subsample of 556 late-type and irregular objects (\refsec{sec:dustvol}).
    \begin{enumerate}
      \item We have inferred the efficiency of the three main dust evolution
        tuning parameters (\refsec{sec:dustvol_par}):
        \begin{inlinelist}
          \item the IMF-averaged SN~II dust yield is 
            $\langle Y_\sms{SN}\rangle\lesssim0.03\eMsun/\textnormal{SN}$;
          \item the grain growth efficiency parameter \citep{mattsson12a}
            is $\epsilon_\sms{grow}\gtrsim3000$;
          \item the average gas mass cleared of dust by a single SN~II shock 
            wave is $m_\sms{gas}^\sms{dest}\gtrsim1200\eMsun/\textnormal{SN}$.
        \end{inlinelist}
        Our results therefore imply that dust production is dominated
        by grain growth in the ISM above a critical metallicity of 
        $\met\simeq8.0$.
        They also suggest that the massive amounts of dust detected in high 
        redshift systems ($z\gtrsim6$) likely grew in the ISM.
      \item We have shown that ELMGs were crucial
        in constraining these parameters, as they sample a regime where dust
        production is dominated by SN~II condensation.
        A steep, strongly non-linear, dustiness-metallicity relation, such as
        the one we have found, is the unambiguous evidence that stardust can
        not dominate the content of solar metallicity systems.
      \item We have shown and explained why these conclusions were, to
        first order, independent of our IMF assumption (\refsec{sec:IMF}).
      \item Our model fails at reproducing the relation between the sSFR and 
        the dust-to-star mass ratio.
        We suspect this is due to the oversimplicity of our SFH, inflow and 
        outflow prescriptions.
    \end{enumerate}
\end{enumerate}

Several of the limitations of our study could be addressed by spatially resolving star forming regions.
Performing a similar analysis on $\simeq100$~pc scales within galaxies would allow us to access another important parameter: the density of the ISM.
This quantity drives mantle growth and coagulation.
Local variations of the dust-to-metal mass ratio are good indications of grain growth and can help us break the degeneracy with SN~II destruction (\refsec{sec:HBdustvol}).
In addition, spatial resolution would be a way to address the origin of the trend of \tqAF\ with metallicity (\refsec{sec:qAFev}), as well as resolving ETGs to quantify the grain sputtering timescale (\refsec{sec:Lx}).
Numerous spatially resolved dust studies have already been published \citep[\eg\ ][]{galliano11,mattsson12b,draine13,hunt15,roman-duval17,aniano20}.
We are currently preparing several papers of a resolved subsample, with the same consistent approach as employed in the present manuscript \citepprep{roychowdhury20,casasola20b}.
The main observational challenge in order to address these degeneracies is however to obtain spatial resolution in ELMGs, at far-IR wavelengths.
The data are currently very limited, because the sensitivity of \hersc\ was not high enough.
\spica\ \citep{vandertak18} was the only observatory on the horizon to have the sensitivity to produce resolved far-IR maps of ELMGs.
Its sudden cancellation by ESA might leave the important open questions raised in this paper unanswered for several decades.

  \begin{acknowledgements}
    We thank Marc-Antoine \familyname{Miville-Desch\^enes} for useful 
    discussions about SED fitting, and about \planck\ and \iras\ data, Albrecht 
    \familyname{Poglitsch} for discussions about PACS calibration, and 
    Joana \familyname{Frontera-Pons} for her expertise about M-estimators.
    We also thank Vianney \familyname{Lebouteiller} and Julia 
    \familyname{Roman-Duval} for insightful interpretations of absorption line 
    measurements in DLA systems.
    Finally, we thank the referee, Denis \familyname{Burgarella}, for 
    his numerous comments that helped clarify the text of this article and 
    strengthen the reliability of our stellar mass estimates.
    DustPedia is a collaborative focused research project supported by the 
    European Union under the Seventh Framework Programme (2007–2013) call 
    (proposal no.~606847, PI J.~I.~Davies). 
    The data used in this work is publicly available at 
    \href{http://dustpedia.astro.noa.gr}{http://dustpedia.astro.noa.gr}.
    This work was supported by the \expression{Programme National “Physique et 
    Chimie du Milieu Interstellaire” (PCMI)} of CNRS/INSU with INC/INP 
    co-funded by CEA and CNES (France).
    It has also been supported by the \expression{Programme National de
    Cosmologie et Galaxies} (PNCG) CNRS/INSU with INC/IN2P3 co-funded by CEA 
    and CNES (France).
    Simone \familyname{Bianchi} and Viviana \familyname{Casasola} acknowledge 
    funding from the INAF mainstream 2018 program “Gas-DustPedia: A definitive 
    view of the ISM in the Local Universe".
    Ilse \familyname{De Looze} acknowledges support from European Research 
    Council (ERC) Starting Grant 851622 DustOrigin.
    Maud \familyname{Galametz}  has received funding from the European Research 
    Council (ERC) under the European Union Horizon 2020 research and innovation 
    programme (MagneticYSOs project, grant agreement No 679937, PI: Maury).
    Aleksandr \familyname{Mosenkov} acknowledges financial support from the 
    Russian Science Foundation (grant no.~20-72-10052).
    This work has made extensive use of the HDF5 library, developed by The HDF 
    Group and by the National Center for Supercomputing Applications at the 
    University of Illinois at Urbana‐Champaign.
    This research has made use of the NASA/IPAC Extragalactic Database (NED),
    which is operated by the Jet Propulsion Laboratory, California Institute of
    Technology, under contract with the National Aeronautics and Space 
    Administration.
    This research has made use of the VizieR catalogue access tool, CDS, 
    Strasbourg, France (DOI: 10.26093/cds/vizier). 
    The original description of the VizieR service was published in 
    \citet{ochsenbein00}. 
  \end{acknowledgements}

  \bibliographystyle{bib_notes}
  \bibliography{references}

\begin{thebibliography}{249}
\expandafter\ifx\csname natexlab\endcsname\relax\def\natexlab#1{#1}\fi

\bibitem[{{Akylas} \& {Georgantopoulos}(2009)}]{akylas09}
\normalfont
 {Akylas}, A. \& {Georgantopoulos}, I.
  2009\href{https://ui.adsabs.harvard.edu/abs/2009A&A...500..999A}{, \aap, 500,
  999}

\bibitem[{{Aloisi} {et~al.}(2007){Aloisi}, {Clementini}, {Tosi}, {Annibali},
  {Contreras}, {Fiorentino}, {Mack}, {Marconi}, {Musella}, {Saha}, {Sirianni},
  \& {van der Marel}}]{aloisi07}
\normalfont
 {Aloisi}, A., {Clementini}, G., {Tosi}, M., {et~al.}
  2007\href{http://adsabs.harvard.edu/abs/2007ApJ...667L.151A}{, \apjl, 667,
  L151}

\bibitem[{{Andersson} {et~al.}(2015){Andersson}, {Lazarian}, \&
  {Vaillancourt}}]{andersson15}
\normalfont
 {Andersson}, B.-G., {Lazarian}, A., \& {Vaillancourt}, J.~E.
  2015\href{http://cdsads.u-strasbg.fr/abs/2015ARA%26A..53..501A}{, \araa, 53,
  501}

\bibitem[{Aniano {et~al.}(2020)Aniano, Draine, Hunt, Sandstrom, Calzetti,
  Kennicutt, Dale, Galametz, Gordon, Leroy, Smith, Roussel, Sauvage, Walter,
  Armus, Bolatto, Boquien, Crocker, {De Looze}, {Donovan Meyer}, Helou, Hinz,
  Johnson, Koda, Miller, Montiel, Murphy, Rela&ntilde;o, Rix, Schinnerer,
  Skibba, Wolfire, \& Engelbracht}]{aniano20}
\normalfont
 Aniano, G., Draine, B., Hunt, L., {et~al.}
  2020\href{https://ui.adsabs.harvard.edu/abs/2020ApJ...889..150A/abstract}{,
  Astrophysical Journal, 889}

\bibitem[{{Aoyama} {et~al.}(2020){Aoyama}, {Hirashita}, \&
  {Nagamine}}]{aoyama20}
\normalfont
 {Aoyama}, S., {Hirashita}, H., \& {Nagamine}, K.
  2020\href{https://ui.adsabs.harvard.edu/abs/2020MNRAS.491.3844A}{, \mnras,
  491, 3844}

\bibitem[{{Aoyama} {et~al.}(2017){Aoyama}, {Hou}, {Shimizu}, {Hirashita},
  {Todoroki}, {Choi}, \& {Nagamine}}]{aoyama17}
\normalfont
 {Aoyama}, S., {Hou}, K.-C., {Shimizu}, I., {et~al.}
  2017\href{https://ui.adsabs.harvard.edu/abs/2017MNRAS.466..105A}{, \mnras,
  466, 105}

\bibitem[{{Arendt} {et~al.}(2014){Arendt}, {Dwek}, {Kober}, {Rho}, \&
  {Hwang}}]{arendt14}
\normalfont
 {Arendt}, R.~G., {Dwek}, E., {Kober}, G., {Rho}, J., \& {Hwang}, U.
  2014\href{https://ui.adsabs.harvard.edu/abs/2014ApJ...786...55A}{, \apj, 786,
  55}

\bibitem[{{Asano} {et~al.}(2013){Asano}, {Takeuchi}, {Hirashita}, \&
  {Inoue}}]{asano13}
\normalfont
 {Asano}, R.~S., {Takeuchi}, T.~T., {Hirashita}, H., \& {Inoue}, A.~K.
  2013\href{http://cdsads.u-strasbg.fr/abs/2013EP%26S...65..213A}{, Earth,
  Planets, and Space, 65, 213}

\bibitem[{{Asplund} {et~al.}(2009){Asplund}, {Grevesse}, {Sauval}, \&
  {Scott}}]{asplund09}
\normalfont
 {Asplund}, M., {Grevesse}, N., {Sauval}, A.~J., \& {Scott}, P.
  2009\href{http://cdsads.u-strasbg.fr/abs/2009ARA%26A..47..481A}{, \araa, 47,
  481}

\bibitem[{{Assef} {et~al.}(2018){Assef}, {Stern}, {Noirot}, {Jun}, {Cutri}, \&
  {Eisenhardt}}]{assef18}
\normalfont
 {Assef}, R.~J., {Stern}, D., {Noirot}, G., {et~al.}
  2018\href{https://ui.adsabs.harvard.edu/abs/2018ApJS..234...23A}{, \apjs,
  234, 23}

\bibitem[{{Astropy Collaboration} {et~al.}(2018){Astropy Collaboration},
  {Price-Whelan}, {Sip{\H{o}}cz}, {G{\"u}nther}, {Lim}, {Crawford}, {Conseil},
  {Shupe}, {Craig}, {Dencheva}, {Ginsburg}, {Vand erPlas}, {Bradley},
  {P{\'e}rez-Su{\'a}rez}, {de Val-Borro}, {Aldcroft}, {Cruz}, {Robitaille},
  {Tollerud}, {Ardelean}, {Babej}, {Bach}, {Bachetti}, {Bakanov}, {Bamford},
  {Barentsen}, {Barmby}, {Baumbach}, {Berry}, {Biscani}, {Boquien}, {Bostroem},
  {Bouma}, {Brammer}, {Bray}, {Breytenbach}, {Buddelmeijer}, {Burke},
  {Calderone}, {Cano Rodr{\'\i}guez}, {Cara}, {Cardoso}, {Cheedella}, {Copin},
  {Corrales}, {Crichton}, {D'Avella}, {Deil}, {Depagne}, {Dietrich}, {Donath},
  {Droettboom}, {Earl}, {Erben}, {Fabbro}, {Ferreira}, {Finethy}, {Fox},
  {Garrison}, {Gibbons}, {Goldstein}, {Gommers}, {Greco}, {Greenfield},
  {Groener}, {Grollier}, {Hagen}, {Hirst}, {Homeier}, {Horton}, {Hosseinzadeh},
  {Hu}, {Hunkeler}, {Ivezi{\'c}}, {Jain}, {Jenness}, {Kanarek}, {Kendrew},
  {Kern}, {Kerzendorf}, {Khvalko}, {King}, {Kirkby}, {Kulkarni}, {Kumar},
  {Lee}, {Lenz}, {Littlefair}, {Ma}, {Macleod}, {Mastropietro}, {McCully},
  {Montagnac}, {Morris}, {Mueller}, {Mumford}, {Muna}, {Murphy}, {Nelson},
  {Nguyen}, {Ninan}, {N{\"o}the}, {Ogaz}, {Oh}, {Parejko}, {Parley}, {Pascual},
  {Patil}, {Patil}, {Plunkett}, {Prochaska}, {Rastogi}, {Reddy Janga},
  {Sabater}, {Sakurikar}, {Seifert}, {Sherbert}, {Sherwood-Taylor}, {Shih},
  {Sick}, {Silbiger}, {Singanamalla}, {Singer}, {Sladen}, {Sooley},
  {Sornarajah}, {Streicher}, {Teuben}, {Thomas}, {Tremblay}, {Turner},
  {Terr{\'o}n}, {van Kerkwijk}, {de la Vega}, {Watkins}, {Weaver}, {Whitmore},
  {Woillez}, {Zabalza}, \& {Astropy Contributors}}]{astropy2}
\normalfont
 {Astropy Collaboration}, {Price-Whelan}, A.~M., {Sip{\H{o}}cz}, B.~M., {et~al.}
  2018\href{https://ui.adsabs.harvard.edu/abs/2018AJ....156..123A}{, \aj, 156,
  123}

\bibitem[{{Astropy Collaboration} {et~al.}(2013){Astropy Collaboration},
  {Robitaille}, {Tollerud}, {Greenfield}, {Droettboom}, {Bray}, {Aldcroft},
  {Davis}, {Ginsburg}, {Price-Whelan}, {Kerzendorf}, {Conley}, {Crighton},
  {Barbary}, {Muna}, {Ferguson}, {Grollier}, {Parikh}, {Nair}, {Unther},
  {Deil}, {Woillez}, {Conseil}, {Kramer}, {Turner}, {Singer}, {Fox}, {Weaver},
  {Zabalza}, {Edwards}, {Azalee Bostroem}, {Burke}, {Casey}, {Crawford},
  {Dencheva}, {Ely}, {Jenness}, {Labrie}, {Lim}, {Pierfederici}, {Pontzen},
  {Ptak}, {Refsdal}, {Servillat}, \& {Streicher}}]{astropy1}
\normalfont
 {Astropy Collaboration}, {Robitaille}, T.~P., {Tollerud}, E.~J., {et~al.}
  2013\href{https://ui.adsabs.harvard.edu/abs/2013A&A...558A..33A}{, \aap, 558,
  A33}

\bibitem[{{Audouze} \& {Tinsley}(1976)}]{audouze76}
\normalfont
 {Audouze}, J. \& {Tinsley}, B.~M.
  1976\href{https://ui.adsabs.harvard.edu/abs/1976ARA&A..14...43A}{, \araa, 14,
  43}

\bibitem[{{Baes} \& {Dejonghe}(2001)}]{baes01}
\normalfont
 {Baes}, M. \& {Dejonghe}, H.
  2001\href{https://ui.adsabs.harvard.edu/abs/2001MNRAS.326..733B}{, \mnras,
  326, 733}

\bibitem[{{Balog} {et~al.}(2014){Balog}, {M{\"u}ller}, {Nielbock}, {Altieri},
  {Klaas}, {Blommaert}, {Linz}, {Lutz}, {Mo{\'o}r}, {Billot}, {Sauvage}, \&
  {Okumura}}]{balog14}
\normalfont
 {Balog}, Z., {M{\"u}ller}, T., {Nielbock}, M., {et~al.}
  2014\href{https://ui.adsabs.harvard.edu/abs/2014ExA....37..129B}{,
  Experimental Astronomy, 37, 129}

\bibitem[{{Barlow} {et~al.}(2010){Barlow}, {Krause}, {Swinyard}, {Sibthorpe},
  {Besel}, {Wesson}, {Ivison}, {Dunne}, {Gear}, {Gomez}, {Hargrave}, {Henning},
  {Leeks}, {Lim}, {Olofsson}, \& {Polehampton}}]{barlow10}
\normalfont
 {Barlow}, M.~J., {Krause}, O., {Swinyard}, B.~M., {et~al.}
  2010\href{https://ui.adsabs.harvard.edu/abs/2010A&A...518L.138B}{, \aap, 518,
  L138}

\bibitem[{Barnard {et~al.}(2000)Barnard, McCulloch, \& Meng}]{barnard00}
\normalfont
 Barnard, J., McCulloch, R., \& Meng, X.-L.
  2000\href{http://www3.stat.sinica.edu.tw/statistica/j10n4/j10n416/j10n416.htm}{,
  Statistica Sinica, 10, 1281}

\bibitem[{{Beers} {et~al.}(1990){Beers}, {Flynn}, \& {Gebhardt}}]{beers90}
\normalfont
 {Beers}, T.~C., {Flynn}, K., \& {Gebhardt}, K.
  1990\href{https://ui.adsabs.harvard.edu/abs/1990AJ....100...32B}{, \aj, 100,
  32}

\bibitem[{{Begum} {et~al.}(2008){Begum}, {Chengalur}, {Karachentsev},
  {Sharina}, \& {Kaisin}}]{begum08}
\normalfont
 {Begum}, A., {Chengalur}, J.~N., {Karachentsev}, I.~D., {Sharina}, M.~E., \&
  {Kaisin}, S.~S.
  2008\href{https://ui.adsabs.harvard.edu/abs/2008MNRAS.386.1667B}{, \mnras,
  386, 1667}

\bibitem[{{Bell} {et~al.}(2019){Bell}, {Onaka}, {Galliano}, {Wu}, {Doi},
  {Kaneda}, {Ishihara}, \& {Giard}}]{bell19}
\normalfont
 {Bell}, A.~C., {Onaka}, T., {Galliano}, F., {et~al.}
  2019\href{https://ui.adsabs.harvard.edu/abs/2019PASJ...71..123B}{, \pasj, 71,
  123}

\bibitem[{{Bendo} {et~al.}(2006){Bendo}, {Buckalew}, {Dale}, {Draine},
  {Joseph}, {Kennicutt}, {Sheth}, {Smith}, {Walter}, {Calzetti}, {Cannon},
  {Engelbracht}, {Gordon}, {Helou}, {Hollenbach}, {Murphy}, \&
  {Roussel}}]{bendo06}
\normalfont
 {Bendo}, G.~J., {Buckalew}, B.~A., {Dale}, D.~A., {et~al.}
  2006\href{http://cdsads.u-strasbg.fr/abs/2006ApJ...645..134B}{, \apj, 645,
  134}

\bibitem[{{Bevan} {et~al.}(2017){Bevan}, {Barlow}, \&
  {Milisavljevic}}]{bevan17}
\normalfont
 {Bevan}, A., {Barlow}, M.~J., \& {Milisavljevic}, D.
  2017\href{https://ui.adsabs.harvard.edu/abs/2017MNRAS.465.4044B}{, \mnras,
  465, 4044}

\bibitem[{{Bianchi} {et~al.}(2019){Bianchi}, {Casasola}, {Baes}, {Clark},
  {Corbelli}, {Davies}, {De Looze}, {De Vis}, {Dobbels}, {Galametz},
  {Galliano}, {Jones}, {Madden}, {Magrini}, {Mosenkov}, {Nersesian}, {Viaene},
  {Xilouris}, \& {Ysard}}]{bianchi19}
\normalfont
 {Bianchi}, S., {Casasola}, V., {Baes}, M., {et~al.}
  2019\href{https://ui.adsabs.harvard.edu/abs/2019A&A...631A.102B}{, \aap, 631,
  A102}

\bibitem[{{Bianchi} {et~al.}(2018){Bianchi}, {De Vis}, {Viaene}, {Nersesian},
  {Mosenkov}, {Xilouris}, {Baes}, {Casasola}, {Cassar{\`a}}, {Clark}, {Davies},
  {De Looze}, {Dobbels}, {Galametz}, {Galliano}, {Jones}, {Lianou}, {Madden},
  \& {Tr{\v{c}}ka}}]{bianchi18}
\normalfont
 {Bianchi}, S., {De Vis}, P., {Viaene}, S., {et~al.}
  2018\href{https://ui.adsabs.harvard.edu/abs/2018A&A...620A.112B}{, \aap, 620,
  A112}

\bibitem[{{Bianchi} \& {Schneider}(2007)}]{bianchi07b}
\normalfont
 {Bianchi}, S. \& {Schneider}, R.
  2007\href{https://ui.adsabs.harvard.edu/abs/2007MNRAS.378..973B}{, \mnras,
  378, 973}

\bibitem[{{Bocchio} {et~al.}(2016){Bocchio}, {Marassi}, {Schneider}, {Bianchi},
  {Limongi}, \& {Chieffi}}]{bocchio16b}
\normalfont
 {Bocchio}, M., {Marassi}, S., {Schneider}, R., {et~al.}
  2016\href{https://ui.adsabs.harvard.edu/abs/2016A&A...587A.157B}{, \aap, 587,
  A157}

\bibitem[{{Bocchio} {et~al.}(2012){Bocchio}, {Micelotta}, {Gautier}, \&
  {Jones}}]{bocchio12}
\normalfont
 {Bocchio}, M., {Micelotta}, E.~R., {Gautier}, A.-L., \& {Jones}, A.~P.
  2012\href{http://cdsads.u-strasbg.fr/abs/2012A%26A...545A.124B}{, \aap, 545,
  A124}

\bibitem[{{Bolatto} {et~al.}(2013){Bolatto}, {Wolfire}, \& {Leroy}}]{bolatto13}
\normalfont
 {Bolatto}, A.~D., {Wolfire}, M., \& {Leroy}, A.~K.
  2013\href{http://cdsads.u-strasbg.fr/abs/2013ARA%26A..51..207B}{, \araa, 51,
  207}

\bibitem[{{Boquien} {et~al.}(2019){Boquien}, {Burgarella}, {Roehlly}, {Buat},
  {Ciesla}, {Corre}, {Inoue}, \& {Salas}}]{boquien19}
\normalfont
 {Boquien}, M., {Burgarella}, D., {Roehlly}, Y., {et~al.}
  2019\href{https://ui.adsabs.harvard.edu/abs/2019A&A...622A.103B}{, \aap, 622,
  A103}

\bibitem[{{Bot} {et~al.}(2010){Bot}, {Ysard}, {Paradis}, {Bernard}, {Lagache},
  {Israel}, \& {Wall}}]{bot10}
\normalfont
 {Bot}, C., {Ysard}, N., {Paradis}, D., {et~al.}
  2010\href{http://cdsads.u-strasbg.fr/abs/2010A%26A...523A..20B}{, \aap, 523,
  A20+}

\bibitem[{{Boulanger} {et~al.}(1998){Boulanger}, {Boisssel}, {Cesarsky}, \&
  {Ryter}}]{boulanger98}
\normalfont
 {Boulanger}, F., {Boisssel}, P., {Cesarsky}, D., \& {Ryter}, C.
  1998\href{http://adsabs.harvard.edu/cgi-bin/nph-bib_query?bibcode=1998A%26A...339..194B&db_key=AST}{,
  \aap, 339, 194}

\bibitem[{{Brauher} {et~al.}(2008){Brauher}, {Dale}, \& {Helou}}]{brauher08}
\normalfont
 {Brauher}, J.~R., {Dale}, D.~A., \& {Helou}, G.
  2008\href{https://ui.adsabs.harvard.edu/abs/2008ApJS..178..280B}{, \apjs,
  178, 280}

\bibitem[{{Brightman} \& {Nandra}(2011)}]{brightman11}
\normalfont
 {Brightman}, M. \& {Nandra}, K.
  2011\href{https://ui.adsabs.harvard.edu/abs/2011MNRAS.413.1206B}{, \mnras,
  413, 1206}

\bibitem[{{Brinkmann} {et~al.}(1994){Brinkmann}, {Siebert}, \&
  {Boller}}]{brinkman94}
\normalfont
 {Brinkmann}, W., {Siebert}, J., \& {Boller}, T.
  1994\href{https://ui.adsabs.harvard.edu/abs/1994A&A...281..355B}{, \aap, 281,
  355}

\bibitem[{{Bron} {et~al.}(2014){Bron}, {Le Bourlot}, \& {Le Petit}}]{bron14}
\normalfont
 {Bron}, E., {Le Bourlot}, J., \& {Le Petit}, F.
  2014\href{http://cdsads.u-strasbg.fr/abs/2014A%26A...569A.100B}{, \aap, 569,
  A100}

\bibitem[{{Brown} {et~al.}(2014){Brown}, {Jarrett}, \& {Cluver}}]{brown14}
\normalfont
 {Brown}, M.~J.~I., {Jarrett}, T.~H., \& {Cluver}, M.~E.
  2014\href{https://ui.adsabs.harvard.edu/\#abs/2014PASA...31...49B}{,
  Publications of the Astronomical Society of Australia, 31, e049}

\bibitem[{{Buat} {et~al.}(2014){Buat}, {Heinis}, {Boquien}, {Burgarella},
  {Charmandaris}, {Boissier}, {Boselli}, {Le Borgne}, \& {Morrison}}]{buat14}
\normalfont
 {Buat}, V., {Heinis}, S., {Boquien}, M., {et~al.}
  2014\href{https://ui.adsabs.harvard.edu/abs/2014A&A...561A..39B}{, \aap, 561,
  A39}

\bibitem[{{Burgarella} {et~al.}(2020){Burgarella}, {Nanni}, {Hirashita},
  {Theul{\'e}}, {Inoue}, \& {Takeuchi}}]{burgarella20}
\normalfont
 {Burgarella}, D., {Nanni}, A., {Hirashita}, H., {et~al.}
  2020\href{https://ui.adsabs.harvard.edu/abs/2020A&A...637A..32B}{, \aap, 637,
  A32}

\bibitem[{{Calura} {et~al.}(2008){Calura}, {Pipino}, \& {Matteucci}}]{calura08}
\normalfont
 {Calura}, F., {Pipino}, A., \& {Matteucci}, F.
  2008\href{http://adsabs.harvard.edu/abs/2008A%26A...479..669C}{, \aap, 479,
  669}

\bibitem[{{Camps} {et~al.}(2015){Camps}, {Misselt}, {Bianchi}, {Lunttila},
  {Pinte}, {Natale}, {Juvela}, {Fischera}, {Fitzgerald}, {Gordon}, {Baes}, \&
  {Steinacker}}]{camps15b}
\normalfont
 {Camps}, P., {Misselt}, K., {Bianchi}, S., {et~al.}
  2015\href{https://ui.adsabs.harvard.edu/abs/2015A&A...580A..87C}{, \aap, 580,
  A87}

\bibitem[{{Camps} {et~al.}(2016){Camps}, {Trayford}, {Baes}, {Theuns},
  {Schaller}, \& {Schaye}}]{camps16}
\normalfont
 {Camps}, P., {Trayford}, J.~W., {Baes}, M., {et~al.}
  2016\href{https://ui.adsabs.harvard.edu/abs/2016MNRAS.462.1057C}{, \mnras,
  462, 1057}

\bibitem[{{Camps} {et~al.}(2018){Camps}, {Tr{\v{c}}ka}, {Trayford}, {Baes},
  {Theuns}, {Crain}, {McAlpine}, {Schaller}, \& {Schaye}}]{camps18}
\normalfont
 {Camps}, P., {Tr{\v{c}}ka}, A., {Trayford}, J., {et~al.}
  2018\href{https://ui.adsabs.harvard.edu/abs/2018ApJS..234...20C}{, \apjs,
  234, 20}

\bibitem[{{Cappi} {et~al.}(2006){Cappi}, {Panessa}, {Bassani}, {Dadina}, {Di
  Cocco}, {Comastri}, {della Ceca}, {Filippenko}, {Gianotti}, {Ho}, {Malaguti},
  {Mulchaey}, {Palumbo}, {Piconcelli}, {Sargent}, {Stephen}, {Trifoglio}, \&
  {Weaver}}]{cappi06}
\normalfont
 {Cappi}, M., {Panessa}, F., {Bassani}, L., {et~al.}
  2006\href{https://ui.adsabs.harvard.edu/abs/2006A&A...446..459C}{, \aap, 446,
  459}

\bibitem[{{Cardelli} {et~al.}(1989){Cardelli}, {Clayton}, \&
  {Mathis}}]{cardelli89}
\normalfont
 {Cardelli}, J.~A., {Clayton}, G.~C., \& {Mathis}, J.~S.
  1989\href{http://cdsads.u-strasbg.fr/abs/1989ApJ...345..245C}{, \apj, 345,
  245}

\bibitem[{{Cartledge} {et~al.}(2005){Cartledge}, {Clayton}, {Gordon},
  {Rachford}, {Draine}, {Martin}, {Mathis}, {Misselt}, {Sofia}, {Whittet}, \&
  {Wolff}}]{cartledge05}
\normalfont
 {Cartledge}, S.~I.~B., {Clayton}, G.~C., {Gordon}, K.~D., {et~al.}
  2005\href{http://cdsads.u-strasbg.fr/abs/2005ApJ...630..355C}{, \apj, 630,
  355}

\bibitem[{{Casasola} {et~al.}(2020{\natexlab{a}}){Casasola}, {Bianchi}, {De
  Vis}, {Magrini}, {Corbelli}, {Clark}, {Fritz}, {Nersesian}, {Viaene}, {Baes},
  {Cassar{\`a}}, {Davies}, {De Looze}, {Dobbels}, {Galametz}, {Galliano},
  {Jones}, {Madden}, {Mosenkov}, {Tr{\v{c}}ka}, \& {Xilouris}}]{casasola20}
\normalfont
 {Casasola}, V., {Bianchi}, S., {De Vis}, P., {et~al.}
  2020{\natexlab{a}}\href{https://ui.adsabs.harvard.edu/abs/2020A&A...633A.100C}{,
  \aap, 633, A100}

\bibitem[{{Casasola} {et~al.}(2020{\natexlab{b}}){Casasola}, {co-I1}, {co-I2},
  {co-I3}, {co-I4}, {co-I5}, \& {co-I6}}]{casasola20b}
\normalfont
 {Casasola}, V., {co-I1}, {co-I2}, {et~al.}
  2020{\natexlab{b}}\href{http://http://cdsads.u-strasbg.fr/abstract_service.html}{,
  {\it in prep.}}

\bibitem[{{Chabrier}(2003)}]{chabrier03}
\normalfont
 {Chabrier}, G.
  2003\href{https://ui.adsabs.harvard.edu/abs/2003PASP..115..763C}{, \pasp,
  115, 763}

\bibitem[{{Chastenet} {et~al.}(2019){Chastenet}, {Sandstrom}, {Chiang},
  {Leroy}, {Utomo}, {Bot}, {Gordon}, {Draine}, {Fukui}, {Onishi}, \&
  {Tsuge}}]{chastenet19}
\normalfont
 {Chastenet}, J., {Sandstrom}, K., {Chiang}, I.~D., {et~al.}
  2019\href{https://ui.adsabs.harvard.edu/abs/2019ApJ...876...62C}{, \apj, 876,
  62}

\bibitem[{{Clark} {et~al.}(2019){Clark}, {De Vis}, {Baes}, {Bianchi},
  {Casasola}, {Cassar{\`a}}, {Davies}, {Dobbels}, {Lianou}, {De Looze},
  {Evans}, {Galametz}, {Galliano}, {Jones}, {Madden}, {Mosenkov}, {Verstocken},
  {Viaene}, {Xilouris}, \& {Ysard}}]{clark19}
\normalfont
 {Clark}, C.~J.~R., {De Vis}, P., {Baes}, M., {et~al.}
  2019\href{https://ui.adsabs.harvard.edu/abs/2019MNRAS.489.5256C}{, \mnras,
  489, 5256}

\bibitem[{{Clark} {et~al.}(2015){Clark}, {Dunne}, {Gomez}, {Maddox}, {De Vis},
  {Smith}, {Eales}, {Baes}, {Bendo}, {Bourne}, {Driver}, {Dye}, {Furlanetto},
  {Grootes}, {Ivison}, {Schofield}, {Robotham}, {Rowlands}, {Valiante},
  {Vlahakis}, {van der Werf}, {Wright}, \& {de Zotti}}]{clark15}
\normalfont
 {Clark}, C.~J.~R., {Dunne}, L., {Gomez}, H.~L., {et~al.}
  2015\href{http://cdsads.u-strasbg.fr/abs/2015MNRAS.452..397C}{, \mnras, 452,
  397}

\bibitem[{{Clark} {et~al.}(2018){Clark}, {Verstocken}, {Bianchi}, {Fritz},
  {Viaene}, {Smith}, {Baes}, {Casasola}, {Cassara}, {Davies}, {De Looze}, {De
  Vis}, {Evans}, {Galametz}, {Jones}, {Lianou}, {Madden}, {Mosenkov}, \&
  {Xilouris}}]{clark18}
\normalfont
 {Clark}, C.~J.~R., {Verstocken}, S., {Bianchi}, S., {et~al.}
  2018\href{http://cdsads.u-strasbg.fr/abs/2018A%26A...609A..37C}{, \aap, 609,
  A37}
\normalfont{\textcolor{purple}{(C18)}}
\bibitem[{{Clayton} {et~al.}(2003){Clayton}, {Gordon}, {Salama}, {Allamandola},
  {Martin}, {Snow}, {Whittet}, {Witt}, \& {Wolff}}]{clayton03}
\normalfont
 {Clayton}, G.~C., {Gordon}, K.~D., {Salama}, F., {et~al.}
  2003\href{http://cdsads.u-strasbg.fr/abs/2003ApJ...592..947C}{, \apj, 592,
  947}

\bibitem[{{Cohen} {et~al.}(2003){Cohen}, {Megeath}, {Hammersley},
  {Mart{\'\i}n-Luis}, \& {Stauffer}}]{cohen03}
\normalfont
 {Cohen}, M., {Megeath}, S.~T., {Hammersley}, P.~L., {Mart{\'\i}n-Luis}, F., \&
  {Stauffer}, J.
  2003\href{https://ui.adsabs.harvard.edu/abs/2003AJ....125.2645C}{, \aj, 125,
  2645}

\bibitem[{{Compi{\`e}gne} {et~al.}(2011){Compi{\`e}gne}, {Verstraete}, {Jones},
  {Bernard}, {Boulanger}, {Flagey}, {Le Bourlot}, {Paradis}, \&
  {Ysard}}]{compiegne11}
\normalfont
 {Compi{\`e}gne}, M., {Verstraete}, L., {Jones}, A., {et~al.}
  2011\href{http://cdsads.u-strasbg.fr/abs/2011A%26A...525A.103C}{, \aap, 525,
  A103+}

\bibitem[{{Cormier} {et~al.}(2019){Cormier}, {Abel}, {Hony}, {Lebouteiller},
  {Madden}, {Polles}, {Galliano}, {De Looze}, {Galametz}, \&
  {Lambert-Huyghe}}]{cormier19}
\normalfont
 {Cormier}, D., {Abel}, N.~P., {Hony}, S., {et~al.}
  2019\href{https://ui.adsabs.harvard.edu/abs/2019A&A...626A..23C}{, \aap, 626,
  A23}

\bibitem[{{Cormier} {et~al.}(2015){Cormier}, {Madden}, {Lebouteiller}, {Abel},
  {Hony}, {Galliano}, {R{\'e}my-Ruyer}, {Bigiel}, {Baes}, {Boselli},
  {Chevance}, {Cooray}, {De Looze}, {Doublier}, {Galametz}, {Hughes},
  {Karczewski}, {Lee}, {Lu}, \& {Spinoglio}}]{cormier15}
\normalfont
 {Cormier}, D., {Madden}, S.~C., {Lebouteiller}, V., {et~al.}
  2015\href{http://cdsads.u-strasbg.fr/abs/2015A%26A...578A..53C}{, \aap, 578,
  A53}

\bibitem[{{Cortese} {et~al.}(2012){Cortese}, {Ciesla}, {Boselli}, {Bianchi},
  {Gomez}, {Smith}, {Bendo}, {Eales}, {Pohlen}, {Baes}, {Corbelli}, {Davies},
  {Hughes}, {Hunt}, {Madden}, {Pierini}, {di Serego Alighieri}, {Zibetti},
  {Boquien}, {Clements}, {Cooray}, {Galametz}, {Magrini}, {Pappalardo},
  {Spinoglio}, \& {Vlahakis}}]{cortese12}
\normalfont
 {Cortese}, L., {Ciesla}, L., {Boselli}, A., {et~al.}
  2012\href{http://cdsads.u-strasbg.fr/abs/2012A%26A...540A..52C}{, \aap, 540,
  A52}

\bibitem[{{Crinklaw} {et~al.}(1994){Crinklaw}, {Federman}, \&
  {Joseph}}]{crinklaw94}
\normalfont
 {Crinklaw}, G., {Federman}, S.~R., \& {Joseph}, C.~L.
  1994\href{http://cdsads.u-strasbg.fr/abs/1994ApJ...424..748C}{, \apj, 424,
  748}

\bibitem[{{Dale} {et~al.}(2017){Dale}, {Cook}, {Roussel}, {Turner}, {Armus},
  {Bolatto}, {Boquien}, {Brown}, {Calzetti}, {De Looze}, {Galametz}, {Gordon},
  {Groves}, {Jarrett}, {Helou}, {Herrera-Camus}, {Hinz}, {Hunt}, {Kennicutt},
  {Murphy}, {Rest}, {Sandstrom}, {Smith}, {Tabatabaei}, \& {Wilson}}]{dale17}
\normalfont
 {Dale}, D.~A., {Cook}, D.~O., {Roussel}, H., {et~al.}
  2017\href{http://cdsads.u-strasbg.fr/abs/2017ApJ...837...90D}{, \apj, 837,
  90}

\bibitem[{{Dale} {et~al.}(2001){Dale}, {Helou}, {Contursi}, {Silbermann}, \&
  {Kolhatkar}}]{dale01}
\normalfont
 {Dale}, D.~A., {Helou}, G., {Contursi}, A., {Silbermann}, N.~A., \&
  {Kolhatkar}, S.
  2001\href{http://adsabs.harvard.edu/cgi-bin/nph-bib_query?bibcode=2001ApJ...549..215D&db_key=AST}{,
  \apj, 549, 215}

\bibitem[{{David} {et~al.}(2006){David}, {Jones}, {Forman}, {Vargas}, \&
  {Nulsen}}]{david06}
\normalfont
 {David}, L.~P., {Jones}, C., {Forman}, W., {Vargas}, I.~M., \& {Nulsen}, P.
  2006\href{https://ui.adsabs.harvard.edu/abs/2006ApJ...653..207D}{, \apj, 653,
  207}

\bibitem[{{Davies} {et~al.}(2017){Davies}, {Baes}, {Bianchi}, {Jones},
  {Madden}, {Xilouris}, {Bocchio}, {Casasola}, {Cassara}, {Clark}, {De Looze},
  {Evans}, {Fritz}, {Galametz}, {Galliano}, {Lianou}, {Mosenkov}, {Smith},
  {Verstocken}, {Viaene}, {Vika}, {Wagle}, \& {Ysard}}]{davies17}
\normalfont
 {Davies}, J.~I., {Baes}, M., {Bianchi}, S., {et~al.}
  2017\href{http://cdsads.u-strasbg.fr/abs/2017PASP..129d4102D}{, \pasp, 129,
  044102}

\bibitem[{{Davies} {et~al.}(2019){Davies}, {Nersesian}, {Baes}, {Bianchi},
  {Casasola}, {Cassar{\`a}}, {Clark}, {De Looze}, {De Vis}, {Evans}, {Fritz},
  {Galametz}, {Galliano}, {Jones}, {Lianou}, {Madden}, {Mosenkov}, {Smith},
  {Verstocken}, {Viaene}, {Vika}, {Xilouris}, \& {Ysard}}]{davies19}
\normalfont
 {Davies}, J.~I., {Nersesian}, A., {Baes}, M., {et~al.}
  2019\href{https://ui.adsabs.harvard.edu/abs/2019A&A...626A..63D}{, \aap, 626,
  A63}

\bibitem[{{De Cia} {et~al.}(2016){De Cia}, {Ledoux}, {Mattsson}, {Petitjean},
  {Srianand}, {Gavignaud}, \& {Jenkins}}]{de-cia16}
\normalfont
 {De Cia}, A., {Ledoux}, C., {Mattsson}, L., {et~al.}
  2016\href{http://cdsads.u-strasbg.fr/abs/2016A%26A...596A..97D}{, \aap, 596,
  A97}

\bibitem[{{De Looze} {et~al.}(2019){De Looze}, {Barlow}, {Bandiera}, {Bevan},
  {Bietenholz}, {Chawner}, {Gomez}, {Matsuura}, {Priestley}, \&
  {Wesson}}]{de-looze19}
\normalfont
 {De Looze}, I., {Barlow}, M.~J., {Bandiera}, R., {et~al.}
  2019\href{https://ui.adsabs.harvard.edu/abs/2019MNRAS.488..164D}{, \mnras,
  488, 164}

\bibitem[{{De Looze} {et~al.}(2017){De Looze}, {Barlow}, {Swinyard}, {Rho},
  {Gomez}, {Matsuura}, \& {Wesson}}]{de-looze17}
\normalfont
 {De Looze}, I., {Barlow}, M.~J., {Swinyard}, B.~M., {et~al.}
  2017\href{https://ui.adsabs.harvard.edu/abs/2017MNRAS.465.3309D}{, \mnras,
  465, 3309}

\bibitem[{{De Looze} {et~al.}(2020){De Looze}, {Lamperti}, {Saintonge},
  {Rela{\~n}o}, {Smith}, {Clark}, {Wilson}, {Decleir}, {Jones}, {Kennicutt},
  {Accurso}, {Brinks}, {Bureau}, {Cigan}, {Clements}, {De Vis}, {Fanciullo},
  {Gao}, {Gear}, {Ho}, {Hwang}, {Micha{\l}owski}, {Lee}, {Li}, {Lin}, {Liu},
  {Lomaeva}, {Pan}, {Sargent}, {Williams}, {Xiao}, \& {Zhu}}]{de-looze20}
\normalfont
 {De Looze}, I., {Lamperti}, I., {Saintonge}, A., {et~al.}
  2020\href{https://ui.adsabs.harvard.edu/abs/2020MNRAS.496.3668D}{, \mnras,
  496, 3668}
\normalfont{\textcolor{purple}{(DL20)}}
\bibitem[{{De Vis} {et~al.}(2017{\natexlab{a}}){De Vis}, {Dunne}, {Maddox},
  {Gomez}, {Clark}, {Bauer}, {Viaene}, {Schofield}, {Baes}, {Baker}, {Bourne},
  {Driver}, {Dye}, {Eales}, {Furlanetto}, {Ivison}, {Robotham}, {Rowlands},
  {Smith}, {Smith}, {Valiante}, \& {Wright}}]{de-vis17a}
\normalfont
 {De Vis}, P., {Dunne}, L., {Maddox}, S., {et~al.}
  2017{\natexlab{a}}\href{http://cdsads.u-strasbg.fr/abs/2017MNRAS.464.4680D}{,
  \mnras, 464, 4680}

\bibitem[{{De Vis} {et~al.}(2017{\natexlab{b}}){De Vis}, {Gomez}, {Schofield},
  {Maddox}, {Dunne}, {Baes}, {Cigan}, {Clark}, {Gomez}, {Lara-L{\'o}pez}, \&
  {Owers}}]{de-vis17}
\normalfont
 {De Vis}, P., {Gomez}, H.~L., {Schofield}, S.~P., {et~al.}
  2017{\natexlab{b}}\href{http://cdsads.u-strasbg.fr/abs/2017MNRAS.471.1743D}{,
  \mnras, 471, 1743}

\bibitem[{{De Vis} {et~al.}(2019){De Vis}, {Jones}, {Viaene}, {Casasola},
  {Clark}, {Baes}, {Bianchi}, {Cassara}, {Davies}, {De Looze}, {Galametz},
  {Galliano}, {Lianou}, {Madden}, {Manilla-Robles}, {Mosenkov}, {Nersesian},
  {Roychowdhury}, {Xilouris}, \& {Ysard}}]{de-vis19}
\normalfont
 {De Vis}, P., {Jones}, A., {Viaene}, S., {et~al.}
  2019\href{https://ui.adsabs.harvard.edu/abs/2019A&A...623A...5D}{, \aap, 623,
  A5}

\bibitem[{{Demyk} {et~al.}(2017{\natexlab{a}}){Demyk}, {Meny}, {Leroux},
  {Depecker}, {Brubach}, {Roy}, {Nayral}, {Ojo}, \& {Delpech}}]{demyk17b}
\normalfont
 {Demyk}, K., {Meny}, C., {Leroux}, H., {et~al.}
  2017{\natexlab{a}}\href{https://ui.adsabs.harvard.edu/abs/2017A&A...606A..50D}{,
  \aap, 606, A50}

\bibitem[{{Demyk} {et~al.}(2017{\natexlab{b}}){Demyk}, {Meny}, {Lu},
  {Papatheodorou}, {Toplis}, {Leroux}, {Depecker}, {Brubach}, {Roy}, {Nayral},
  {Ojo}, {Delpech}, {Paradis}, \& {Gromov}}]{demyk17a}
\normalfont
 {Demyk}, K., {Meny}, C., {Lu}, X.~H., {et~al.}
  2017{\natexlab{b}}\href{https://ui.adsabs.harvard.edu/abs/2017A&A...600A.123D}{,
  \aap, 600, A123}

\bibitem[{{Diehl} \& {Statler}(2007)}]{diehl07}
\normalfont
 {Diehl}, S. \& {Statler}, T.~S.
  2007\href{https://ui.adsabs.harvard.edu/abs/2007ApJ...668..150D}{, \apj, 668,
  150}

\bibitem[{{Draine}(1978)}]{draine78}
\normalfont
 {Draine}, B.~T. 1978\href{http://cdsads.u-strasbg.fr/abs/1978ApJS...36..595D}{,
  \apjs, 36, 595}

\bibitem[{{Draine}(2009)}]{draine09}
\normalfont
 {Draine}, B.~T. 2009, in
  \href{http://cdsads.u-strasbg.fr/abs/2009ASPC..414..453D}{Astronomical
  Society of the Pacific Conference Series, Vol. 414, Astronomical Society of
  the Pacific Conference Series, ed. {T.~Henning, E.~Gr{\"u}n, \&
  J.~Steinacker}}, 453--+

\bibitem[{{Draine}(2011)}]{draine11}
\normalfont
 {Draine}, B.~T. 2011, in
  \href{http://cdsads.u-strasbg.fr/abs/2011EAS....46...29D}{EAS Publications
  Series, Vol.~46, EAS Publications Series, ed. C.~{Joblin} \& A.~G.~G.~M.
  {Tielens}}, 29--42

\bibitem[{{Draine} {et~al.}(2007){Draine}, {Dale}, {Bendo}, {Gordon}, {Smith},
  {Armus}, {Engelbracht}, {Helou}, {Kennicutt}, {Li}, {Roussel}, {Walter},
  {Calzetti}, {Moustakas}, {Murphy}, {Rieke}, {Bot}, {Hollenbach}, {Sheth}, \&
  {Teplitz}}]{draine07b}
\normalfont
 {Draine}, B.~T., {Dale}, D.~A., {Bendo}, G., {et~al.}
  2007\href{http://adsabs.harvard.edu/abs/2007ApJ...663..866D}{, \apj, 663,
  866}

\bibitem[{{Draine} \& {Hensley}(2012)}]{draine12}
\normalfont
 {Draine}, B.~T. \& {Hensley}, B.
  2012\href{http://cdsads.u-strasbg.fr/abs/2012ApJ...757..103D}{, \apj, 757,
  103}

\bibitem[{{Draine} \& {Hensley}(2013)}]{draine13}
\normalfont
 {Draine}, B.~T. \& {Hensley}, B.
  2013\href{http://cdsads.u-strasbg.fr/abs/2013ApJ...765..159D}{, \apj, 765,
  159}

\bibitem[{{Draine} \& {Li}(2007)}]{draine07}
\normalfont
 {Draine}, B.~T. \& {Li}, A.
  2007\href{http://adsabs.harvard.edu/abs/2007ApJ...657..810D}{, \apj, 657,
  810}
\normalfont{\textcolor{purple}{(DL07)}}
\bibitem[{{Draine} \& {Salpeter}(1979)}]{draine79}
\normalfont
 {Draine}, B.~T. \& {Salpeter}, E.~E.
  1979\href{http://adsabs.harvard.edu/cgi-bin/nph-bib_query?bibcode=1979ApJ...231..438D&db_key=AST}{,
  \apj, 231, 438}

\bibitem[{{Dumke} {et~al.}(2004){Dumke}, {Krause}, \& {Wielebinski}}]{dumke04}
\normalfont
 {Dumke}, M., {Krause}, M., \& {Wielebinski}, R.
  2004\href{http://cdsads.u-strasbg.fr/abs/2004A%26A...414..475D}{, \aap, 414,
  475}

\bibitem[{{Dupac} {et~al.}(2003){Dupac}, {Bernard}, {Boudet}, {Giard},
  {Lamarre}, {M{\'e}ny}, {Pajot}, {Ristorcelli}, {Serra}, {Stepnik}, \&
  {Torre}}]{dupac03}
\normalfont
 {Dupac}, X., {Bernard}, J.-P., {Boudet}, N., {et~al.}
  2003\href{http://adsabs.harvard.edu/abs/2003A%26A...404L..11D}{, \aap, 404,
  L11}

\bibitem[{{Dwek}(1998)}]{dwek98}
\normalfont
 {Dwek}, E.
  1998\href{http://adsabs.harvard.edu/cgi-bin/nph-bib_query?bibcode=1998ApJ...501..643D&db_key=AST}{,
  \apj, 501, 643}

\bibitem[{{Dwek}(2005)}]{dwek05}
\normalfont
 {Dwek}, E. 2005, in
  \href{http://adsabs.harvard.edu/cgi-bin/nph-bib_query?bibcode=2005AIPC..761..103D&db_key=AST}{AIP
  Conf. Proc. 761: The Spectral Energy Distributions of Gas-Rich Galaxies:
  Confronting Models with Data, ed. C.~C. {Popescu} \& R.~J. {Tuffs}}, 103

\bibitem[{{Dwek} \& {Arendt}(2015)}]{dwek15}
\normalfont
 {Dwek}, E. \& {Arendt}, R.~G.
  2015\href{https://ui.adsabs.harvard.edu/abs/2015ApJ...810...75D}{, \apj, 810,
  75}

\bibitem[{{Dwek} {et~al.}(2007){Dwek}, {Galliano}, \& {Jones}}]{dwek07}
\normalfont
 {Dwek}, E., {Galliano}, F., \& {Jones}, A.~P.
  2007\href{http://adsabs.harvard.edu/abs/2007ApJ...662..927D}{, \apj, 662,
  927}

\bibitem[{{Dwek} \& {Scalo}(1980)}]{dwek80}
\normalfont
 {Dwek}, E. \& {Scalo}, J.~M.
  1980\href{http://adsabs.harvard.edu/cgi-bin/nph-bib_query?bibcode=1980ApJ...239..193D&db_key=AST}{,
  \apj, 239, 193}

\bibitem[{{Dwek} {et~al.}(2014){Dwek}, {Staguhn}, {Arendt}, {Kovacks}, {Su}, \&
  {Benford}}]{dwek14}
\normalfont
 {Dwek}, E., {Staguhn}, J., {Arendt}, R.~G., {et~al.}
  2014\href{https://ui.adsabs.harvard.edu/abs/2014ApJ...788L..30D}{, \apjl,
  788, L30}

\bibitem[{{Efron} \& {Morris}(1977)}]{efron77}
\normalfont
 {Efron}, B. \& {Morris}, C.
  1977\href{https://ui.adsabs.harvard.edu/abs/1977SciAm.236e.119E}{, Scientific
  American, 236, 119}

\bibitem[{{Engelbracht} {et~al.}(2007){Engelbracht}, {Blaylock}, {Su}, {Rho},
  {Rieke}, {Muzerolle}, {Padgett}, {Hines}, {Gordon}, {Fadda},
  {Noriega-Crespo}, {Kelly}, {Latter}, {Hinz}, {Misselt}, {Morrison},
  {Stansberry}, {Shupe}, {Stolovy}, {Wheaton}, {Young}, {Neugebauer},
  {Wachter}, {P{\'e}rez-Gonz{\'a}lez}, {Frayer}, \& {Marleau}}]{engelbracht07}
\normalfont
 {Engelbracht}, C.~W., {Blaylock}, M., {Su}, K.~Y.~L., {et~al.}
  2007\href{http://cdsads.u-strasbg.fr/abs/2007PASP..119..994E}{, \pasp, 119,
  994}

\bibitem[{{Engelbracht} {et~al.}(2005){Engelbracht}, {Gordon}, {Rieke},
  {Werner}, {Dale}, \& {Latter}}]{engelbracht05}
\normalfont
 {Engelbracht}, C.~W., {Gordon}, K.~D., {Rieke}, G.~H., {et~al.}
  2005\href{http://adsabs.harvard.edu/cgi-bin/nph-bib_query?bibcode=2005ApJ...628L..29E&db_key=AST}{,
  \apjl, 628, L29}

\bibitem[{{Ercolano} {et~al.}(2007){Ercolano}, {Barlow}, \&
  {Sugerman}}]{ercolano07}
\normalfont
 {Ercolano}, B., {Barlow}, M.~J., \& {Sugerman}, B.~E.~K.
  2007\href{https://ui.adsabs.harvard.edu/abs/2007MNRAS.375..753E}{, \mnras,
  375, 753}

\bibitem[{{Eskew} {et~al.}(2012){Eskew}, {Zaritsky}, \& {Meidt}}]{eskew12}
\normalfont
 {Eskew}, M., {Zaritsky}, D., \& {Meidt}, S.
  2012\href{https://ui.adsabs.harvard.edu/\#abs/2012AJ....143..139E}{, \aj,
  143, 139}

\bibitem[{{Fabbiano} {et~al.}(1992){Fabbiano}, {Kim}, \&
  {Trinchieri}}]{fabbiano92}
\normalfont
 {Fabbiano}, G., {Kim}, D.~W., \& {Trinchieri}, G.
  1992\href{https://ui.adsabs.harvard.edu/abs/1992ApJS...80..531F}{, \apjs, 80,
  531}

\bibitem[{{Fanciullo} {et~al.}(2017){Fanciullo}, {Guillet}, {Boulanger}, \&
  {Jones}}]{fanciullo17}
\normalfont
 {Fanciullo}, L., {Guillet}, V., {Boulanger}, F., \& {Jones}, A.~P.
  2017\href{https://ui.adsabs.harvard.edu/abs/2017A&A...602A...7F}{, \aap, 602,
  A7}

\bibitem[{{Feldmann}(2015)}]{feldmann15}
\normalfont
 {Feldmann}, R.
  2015\href{https://ui.adsabs.harvard.edu/abs/2015MNRAS.449.3274F}{, \mnras,
  449, 3274}

\bibitem[{{Fitzpatrick}(1986)}]{fitzpatrick86}
\normalfont
 {Fitzpatrick}, E.~L.
  1986\href{http://cdsads.u-strasbg.fr/abs/1986AJ.....92.1068F}{, \aj, 92,
  1068}

\bibitem[{{Galametz} {et~al.}(2014){Galametz}, {Albrecht}, {Kennicutt},
  {Aniano}, {Bertoldi}, {Calzetti}, {Croxall}, {Dale}, {Draine}, {Engelbracht},
  {Gordon}, {Hinz}, {Hunt}, {Kirkpatrick}, {Murphy}, {Roussel}, {Skibba},
  {Walter}, {Weiss}, \& {Wilson}}]{galametz14}
\normalfont
 {Galametz}, M., {Albrecht}, M., {Kennicutt}, R., {et~al.}
  2014\href{https://ui.adsabs.harvard.edu/abs/2014MNRAS.439.2542G}{, \mnras,
  439, 2542}

\bibitem[{{Galametz} {et~al.}(2013){Galametz}, {Hony}, {Galliano}, {Madden},
  {Albrecht}, {Bot}, {Cormier}, {Engelbracht}, {Fukui}, {Israel}, {Kawamura},
  {Lebouteiller}, {Li}, {Meixner}, {Misselt}, {Montiel}, {Okumura}, {Panuzzo},
  {Roman-Duval}, {Rubio}, {Sauvage}, {Seale}, {Sewi{\l}o}, \& {van
  Loon}}]{galametz13}
\normalfont
 {Galametz}, M., {Hony}, S., {Galliano}, F., {et~al.}
  2013\href{http://cdsads.u-strasbg.fr/abs/2013MNRAS.431.1596G}{, \mnras, 431,
  1596}

\bibitem[{{Galametz} {et~al.}(2009){Galametz}, {Madden}, {Galliano}, {Hony},
  {Schuller}, {Beelen}, {Bendo}, {Sauvage}, {Lundgren}, \&
  {Billot}}]{galametz09}
\normalfont
 {Galametz}, M., {Madden}, S., {Galliano}, F., {et~al.}
  2009\href{http://cdsads.u-strasbg.fr/abs/2009A%26A...508..645G}{, \aap, 508,
  645}

\bibitem[{{Galametz} {et~al.}(2011){Galametz}, {Madden}, {Galliano}, {Hony},
  {Bendo}, \& {Sauvage}}]{galametz11}
\normalfont
 {Galametz}, M., {Madden}, S.~C., {Galliano}, F., {et~al.}
  2011\href{http://cdsads.u-strasbg.fr/abs/2011A%26A...532A..56G}{, \aap, 532,
  A56}

\bibitem[{{Galliano}(2018)}]{galliano18a}
\normalfont
 {Galliano}, F. 2018\href{http://cdsads.u-strasbg.fr/abs/2018MNRAS.476.1445G}{,
  \mnras, 476, 1445}
\normalfont{\textcolor{purple}{(G18)}}
\bibitem[{{Galliano} {et~al.}(2008{\natexlab{a}}){Galliano}, {Dwek}, \&
  {Chanial}}]{galliano08a}
\normalfont
 {Galliano}, F., {Dwek}, E., \& {Chanial}, P.
  2008{\natexlab{a}}\href{http://adsabs.harvard.edu/abs/2008ApJ...672..214G}{,
  \apj, 672, 214}

\bibitem[{{Galliano} {et~al.}(2018){Galliano}, {Galametz}, \&
  {Jones}}]{galliano18}
\normalfont
 {Galliano}, F., {Galametz}, M., \& {Jones}, A.~P.
  2018\href{https://ui.adsabs.harvard.edu/#abs/2018ARA&A..56..673G}{, \araa,
  56, 673}

\bibitem[{{Galliano} {et~al.}(2011){Galliano}, {Hony}, {Bernard}, {Bot},
  {Madden}, {Roman-Duval}, {Galametz}, {Li}, {Meixner}, {Engelbracht},
  {Lebouteiller}, {Misselt}, {Montiel}, {Panuzzo}, {Reach}, \&
  {Skibba}}]{galliano11}
\normalfont
 {Galliano}, F., {Hony}, S., {Bernard}, J.-P., {et~al.}
  2011\href{http://cdsads.u-strasbg.fr/abs/2011A%26A...536A..88G}{, \aap, 536,
  A88}

\bibitem[{{Galliano} {et~al.}(2005){Galliano}, {Madden}, {Jones}, {Wilson}, \&
  {Bernard}}]{galliano05}
\normalfont
 {Galliano}, F., {Madden}, S.~C., {Jones}, A.~P., {Wilson}, C.~D., \& {Bernard},
  J.-P.
  2005\href{http://adsabs.harvard.edu/cgi-bin/nph-bib_query?bibcode=2005A%26A...434..867G&db_key=AST}{,
  \aap, 434, 867}

\bibitem[{{Galliano} {et~al.}(2003){Galliano}, {Madden}, {Jones}, {Wilson},
  {Bernard}, \& {Le Peintre}}]{galliano03}
\normalfont
 {Galliano}, F., {Madden}, S.~C., {Jones}, A.~P., {et~al.}
  2003\href{http://adsabs.harvard.edu/cgi-bin/nph-bib_query?bibcode=2003A%26A...407..159G&db_key=AST}{,
  \aap, 407, 159}

\bibitem[{{Galliano} {et~al.}(2008{\natexlab{b}}){Galliano}, {Madden},
  {Tielens}, {Peeters}, \& {Jones}}]{galliano08b}
\normalfont
 {Galliano}, F., {Madden}, S.~C., {Tielens}, A.~G.~G.~M., {Peeters}, E., \&
  {Jones}, A.~P.
  2008{\natexlab{b}}\href{http://adsabs.harvard.edu/abs/2008ApJ...679..310G}{,
  \apj, 679, 310}

\bibitem[{{Garnett} {et~al.}(1995){Garnett}, {Skillman}, {Dufour}, {Peimbert},
  {Torres-Peimbert}, {Terlevich}, {Terlevich}, \& {Shields}}]{garnett95}
\normalfont
 {Garnett}, D.~R., {Skillman}, E.~D., {Dufour}, R.~J., {et~al.}
  1995\href{http://cdsads.u-strasbg.fr/abs/1995ApJ...443...64G}{, \apj, 443,
  64}

\bibitem[{{Gelman} {et~al.}(2004){Gelman}, {Carlin}, {Stern}, \&
  {Rubin}}]{gelman04}
\normalfont
 {Gelman}, A., {Carlin}, J., {Stern}, H., \& {Rubin}, D. 2004, Bayesian Data
  Analysis (Chapman \&\ Hall)

\bibitem[{Geman \& Geman(1984)}]{geman84}
\normalfont
 Geman, S. \& Geman, D. 1984\href{http://dl.acm.org/citation.cfm?id=2286617}{,
  IEEE Trans. Pattern Anal. Mach. Intell., 6, 721}

\bibitem[{{Gomez} {et~al.}(2012){Gomez}, {Krause}, {Barlow}, {Swinyard},
  {Owen}, {Clark}, {Matsuura}, {Gomez}, {Rho}, {Besel}, {Bouwman}, {Gear},
  {Henning}, {Ivison}, {Polehampton}, \& {Sibthorpe}}]{gomez12}
\normalfont
 {Gomez}, H.~L., {Krause}, O., {Barlow}, M.~J., {et~al.}
  2012\href{https://ui.adsabs.harvard.edu/abs/2012ApJ...760...96G}{, \apj, 760,
  96}

\bibitem[{{Gonz{\'a}lez-Mart{\'\i}n} {et~al.}(2009){Gonz{\'a}lez-Mart{\'\i}n},
  {Masegosa}, {M{\'a}rquez}, {Guainazzi}, \&
  {Jim{\'e}nez-Bail{\'o}n}}]{gonzalez-martin09}
\normalfont
 {Gonz{\'a}lez-Mart{\'\i}n}, O., {Masegosa}, J., {M{\'a}rquez}, I., {Guainazzi},
  M., \& {Jim{\'e}nez-Bail{\'o}n}, E.
  2009\href{https://ui.adsabs.harvard.edu/abs/2009A&A...506.1107G}{, \aap, 506,
  1107}

\bibitem[{{Gordon} {et~al.}(2003){Gordon}, {Clayton}, {Misselt}, {Landolt}, \&
  {Wolff}}]{gordon03}
\normalfont
 {Gordon}, K.~D., {Clayton}, G.~C., {Misselt}, K.~A., {Landolt}, A.~U., \&
  {Wolff}, M.~J.
  2003\href{http://adsabs.harvard.edu/cgi-bin/nph-bib_query?bibcode=2003ApJ...594..279G&db_key=AST}{,
  \apj, 594, 279}

\bibitem[{{Gordon} {et~al.}(2007){Gordon}, {Engelbracht}, {Fadda},
  {Stansberry}, {Wachter}, {Frayer}, {Rieke}, {Noriega-Crespo}, {Latter},
  {Young}, {Neugebauer}, {Balog}, {Beeman}, {Dole}, {Egami}, {Haller}, {Hines},
  {Kelly}, {Marleau}, {Misselt}, {Morrison}, {P{\'e}rez-Gonz{\'a}lez}, {Rho},
  \& {Wheaton}}]{gordon07}
\normalfont
 {Gordon}, K.~D., {Engelbracht}, C.~W., {Fadda}, D., {et~al.}
  2007\href{http://cdsads.u-strasbg.fr/abs/2007PASP..119.1019G}{, \pasp, 119,
  1019}

\bibitem[{{Gordon} {et~al.}(2008){Gordon}, {Engelbracht}, {Rieke}, {Misselt},
  {Smith}, \& {Kennicutt}}]{gordon08}
\normalfont
 {Gordon}, K.~D., {Engelbracht}, C.~W., {Rieke}, G.~H., {et~al.}
  2008\href{http://cdsads.u-strasbg.fr/abs/2008ApJ...682..336G}{, \apj, 682,
  336}

\bibitem[{{Gordon} {et~al.}(2014){Gordon}, {Roman-Duval}, {Bot}, {Meixner},
  {Babler}, {Bernard}, {Bolatto}, {Boyer}, {Clayton}, {Engelbracht}, {Fukui},
  {Galametz}, {Galliano}, {Hony}, {Hughes}, {Indebetouw}, {Israel}, {Jameson},
  {Kawamura}, {Lebouteiller}, {Li}, {Madden}, {Matsuura}, {Misselt}, {Montiel},
  {Okumura}, {Onishi}, {Panuzzo}, {Paradis}, {Rubio}, {Sandstrom}, {Sauvage},
  {Seale}, {Sewi{\l}o}, {Tchernyshyov}, \& {Skibba}}]{gordon14}
\normalfont
 {Gordon}, K.~D., {Roman-Duval}, J., {Bot}, C., {et~al.}
  2014\href{http://cdsads.u-strasbg.fr/abs/2014ApJ...797...85G}{, \apj, 797,
  85}

\bibitem[{{Gould} \& {Salpeter}(1963)}]{gould63}
\normalfont
 {Gould}, R.~J. \& {Salpeter}, E.~E.
  1963\href{http://cdsads.u-strasbg.fr/abs/1963ApJ...138..393G}{, \apj, 138,
  393}

\bibitem[{{Grebel}(1999)}]{grebel99}
\normalfont
 {Grebel}, E.~K. 1999, in
  \href{https://ui.adsabs.harvard.edu/abs/1999IAUS..192...17G}{The Stellar
  Content of Local Group Galaxies, ed. P.~{Whitelock} \& R.~{Cannon}, Vol.
  192}, 17

\bibitem[{{Greenberg} {et~al.}(2000){Greenberg}, {Gillette}, {Mu{\~n}oz Caro},
  {Mahajan}, {Zare}, {Li}, {Schutte}, {de Groot}, \&
  {Mendoza-G{\'o}mez}}]{greenberg00}
\normalfont
 {Greenberg}, J.~M., {Gillette}, J.~S., {Mu{\~n}oz Caro}, G.~M., {et~al.}
  2000\href{https://ui.adsabs.harvard.edu/abs/2000ApJ...531L..71G}{, \apjl,
  531, L71}

\bibitem[{{Grier} {et~al.}(2011){Grier}, {Mathur}, {Ghosh}, \&
  {Ferrarese}}]{grier11}
\normalfont
 {Grier}, C.~J., {Mathur}, S., {Ghosh}, H., \& {Ferrarese}, L.
  2011\href{https://ui.adsabs.harvard.edu/abs/2011ApJ...731...60G}{, \apj, 731,
  60}

\bibitem[{{Griffin} {et~al.}(2013){Griffin}, {North}, {Schulz},
  {Amaral-Rogers}, {Bendo}, {Bock}, {Conversi}, {Conley}, {Dowell}, {Ferlet},
  {Glenn}, {Lim}, {Pearson}, {Pohlen}, {Sibthorpe}, {Spencer}, {Swinyard}, \&
  {Valtchanov}}]{griffin13}
\normalfont
 {Griffin}, M.~J., {North}, C.~E., {Schulz}, B., {et~al.}
  2013\href{http://cdsads.u-strasbg.fr/abs/2013MNRAS.434..992G}{, \mnras, 434,
  992}

\bibitem[{{Hahn}(2005)}]{hahn05}
\normalfont
 {Hahn}, R. 2005, Pierre Simon Laplace, 1749-1827: A Determined Scientist
  (Harvard University Press)

\bibitem[{{Hamanowicz} {et~al.}(2020){Hamanowicz}, {P{\'e}roux}, {Zwaan},
  {Rahmani}, {Pettini}, {York}, {Klitsch}, {Augustin}, {Krogager}, {Kulkarni},
  {Fresco}, {Biggs}, {Milliard}, \& {Vernet}}]{hamanowicz20}
\normalfont
 {Hamanowicz}, A., {P{\'e}roux}, C., {Zwaan}, M.~A., {et~al.}
  2020\href{https://ui.adsabs.harvard.edu/abs/2020MNRAS.492.2347H}{, \mnras,
  492, 2347}

\bibitem[{{Hirashita}(1999)}]{hirashita99}
\normalfont
 {Hirashita}, H.
  1999\href{https://ui.adsabs.harvard.edu/abs/1999ApJ...510L..99H}{, \apjl,
  510, L99}

\bibitem[{{Hirashita}(2012)}]{hirashita12}
\normalfont
 {Hirashita}, H.
  2012\href{https://ui.adsabs.harvard.edu/abs/2012MNRAS.422.1263H}{, \mnras,
  422, 1263}

\bibitem[{{Hirashita} \& {Aoyama}(2019)}]{hirashita19}
\normalfont
 {Hirashita}, H. \& {Aoyama}, S.
  2019\href{https://ui.adsabs.harvard.edu/abs/2019MNRAS.482.2555H}{, \mnras,
  482, 2555}

\bibitem[{{Hirashita} \& {Murga}(2020)}]{hirashita20}
\normalfont
 {Hirashita}, H. \& {Murga}, M.~S.
  2020\href{https://ui.adsabs.harvard.edu/abs/2020MNRAS.492.3779H}{, \mnras,
  492, 3779}

\bibitem[{{Hirashita} {et~al.}(2015){Hirashita}, {Nozawa}, {Villaume}, \&
  {Srinivasan}}]{hirashita15}
\normalfont
 {Hirashita}, H., {Nozawa}, T., {Villaume}, A., \& {Srinivasan}, S.
  2015\href{http://adsabs.harvard.edu/abs/2015MNRAS.454.1620H}{, \mnras, 454,
  1620}

\bibitem[{{Hogg} {et~al.}(2010){Hogg}, {Bovy}, \& {Lang}}]{hogg10b}
\normalfont
 {Hogg}, D.~W., {Bovy}, J., \& {Lang}, D.
  2010\href{https://ui.adsabs.harvard.edu/abs/2010arXiv1008.4686H}{, arXiv
  e-prints, arXiv:1008.4686}

\bibitem[{{Hou} {et~al.}(2017){Hou}, {Hirashita}, {Nagamine}, {Aoyama}, \&
  {Shimizu}}]{hou17}
\normalfont
 {Hou}, K.-C., {Hirashita}, H., {Nagamine}, K., {Aoyama}, S., \& {Shimizu}, I.
  2017\href{http://cdsads.u-strasbg.fr/abs/2017MNRAS.469..870H}{, \mnras, 469,
  870}

\bibitem[{{Houck} {et~al.}(2004){Houck}, {Charmandaris}, {Brandl}, {Weedman},
  {Herter}, {Armus}, {Soifer}, {Bernard-Salas}, {Spoon}, {Devost}, \&
  {Uchida}}]{houck04}
\normalfont
 {Houck}, J.~R., {Charmandaris}, V., {Brandl}, B.~R., {et~al.}
  2004\href{http://adsabs.harvard.edu/cgi-bin/nph-bib_query?bibcode=2004ApJS..154..211H&db_key=AST}{,
  \apjs, 154, 211}

\bibitem[{{Hunt} {et~al.}(2015){Hunt}, {Draine}, {Bianchi}, {Gordon}, {Aniano},
  {Calzetti}, {Dale}, {Helou}, {Hinz}, {Kennicutt}, {Roussel}, {Wilson},
  {Bolatto}, {Boquien}, {Croxall}, {Galametz}, {Gil de Paz}, {Koda},
  {Mu{\~n}oz-Mateos}, {Sandstrom}, {Sauvage}, {Vigroux}, \& {Zibetti}}]{hunt15}
\normalfont
 {Hunt}, L.~K., {Draine}, B.~T., {Bianchi}, S., {et~al.}
  2015\href{http://adsabs.harvard.edu/abs/2015A%26A...576A..33H}{, \aap, 576,
  A33}

\bibitem[{{Hunter} \& {Gallagher}(1989)}]{hunter89b}
\normalfont
 {Hunter}, D.~A. \& {Gallagher}, J.~S., I.
  1989\href{https://ui.adsabs.harvard.edu/abs/1989Sci...243.1557H}{, Science,
  243, 1557}

\bibitem[{{Inoue}(2003)}]{inoue03}
\normalfont
 {Inoue}, A.~K.
  2003\href{https://ui.adsabs.harvard.edu/abs/2003PASJ...55..901I}{, \pasj, 55,
  901}

\bibitem[{{James} {et~al.}(2002){James}, {Dunne}, {Eales}, \&
  {Edmunds}}]{james02}
\normalfont
 {James}, A., {Dunne}, L., {Eales}, S., \& {Edmunds}, M.~G.
  2002\href{http://adsabs.harvard.edu/cgi-bin/nph-bib_query?bibcode=2002MNRAS.335..753J&db_key=AST}{,
  \mnras, 335, 753}

\bibitem[{{Janowiecki} {et~al.}(2017){Janowiecki}, {Salzer}, {van Zee},
  {Rosenberg}, \& {Skillman}}]{janowiecki17}
\normalfont
 {Janowiecki}, S., {Salzer}, J.~J., {van Zee}, L., {Rosenberg}, J.~L., \&
  {Skillman}, E.
  2017\href{https://ui.adsabs.harvard.edu/abs/2017ApJ...836..128J}{, \apj, 836,
  128}

\bibitem[{{Jarrett} {et~al.}(2011){Jarrett}, {Cohen}, {Masci}, {Wright},
  {Stern}, {Benford}, {Blain}, {Carey}, {Cutri}, {Eisenhardt}, {Lonsdale},
  {Mainzer}, {Marsh}, {Padgett}, {Petty}, {Ressler}, {Skrutskie}, {Stanford},
  {Surace}, {Tsai}, {Wheelock}, \& {Yan}}]{jarrett11}
\normalfont
 {Jarrett}, T.~H., {Cohen}, M., {Masci}, F., {et~al.}
  2011\href{http://adsabs.harvard.edu/abs/2011ApJ...735..112J}{, \apj, 735,
  112}

\bibitem[{Jaynes(1976)}]{jaynes76}
\normalfont
 Jaynes, E.~T. 1976, Confidence Intervals vs Bayesian Intervals (W. L. Harper
  and C. A. Hooker), 175

\bibitem[{{Jenkins}(2009)}]{jenkins09}
\normalfont
 {Jenkins}, E.~B.
  2009\href{http://cdsads.u-strasbg.fr/abs/2009ApJ...700.1299J}{, \apj, 700,
  1299}

\bibitem[{{Joblin} {et~al.}(1992){Joblin}, {Leger}, \& {Martin}}]{joblin92}
\normalfont
 {Joblin}, C., {Leger}, A., \& {Martin}, P.
  1992\href{http://adsabs.harvard.edu/cgi-bin/nph-bib_query?bibcode=1992ApJ...393L..79J&db_key=AST}{,
  \apjl, 393, L79}

\bibitem[{{Jones}(2016{\natexlab{a}})}]{jones16a}
\normalfont
 {Jones}, A.~P.
  2016{\natexlab{a}}\href{http://cdsads.u-strasbg.fr/abs/2016RSOS....360221J}{,
  Royal Society Open Science, 3, 160221}

\bibitem[{{Jones}(2016{\natexlab{b}})}]{jones16b}
\normalfont
 {Jones}, A.~P.
  2016{\natexlab{b}}\href{http://cdsads.u-strasbg.fr/abs/2016RSOS....360223J}{,
  Royal Society Open Science, 3, 160223}

\bibitem[{{Jones}(2016{\natexlab{c}})}]{jones16c}
\normalfont
 {Jones}, A.~P.
  2016{\natexlab{c}}\href{http://cdsads.u-strasbg.fr/abs/2016RSOS....360224J}{,
  Royal Society Open Science, 3, 160224}

\bibitem[{{Jones} {et~al.}(2013){Jones}, {Fanciullo}, {K{\"o}hler},
  {Verstraete}, {Guillet}, {Bocchio}, \& {Ysard}}]{jones13}
\normalfont
 {Jones}, A.~P., {Fanciullo}, L., {K{\"o}hler}, M., {et~al.}
  2013\href{http://cdsads.u-strasbg.fr/abs/2013A%26A...558A..62J}{, \aap, 558,
  A62}

\bibitem[{{Jones} {et~al.}(2017){Jones}, {K{\"o}hler}, {Ysard}, {Bocchio}, \&
  {Verstraete}}]{jones17}
\normalfont
 {Jones}, A.~P., {K{\"o}hler}, M., {Ysard}, N., {Bocchio}, M., \& {Verstraete},
  L. 2017\href{http://cdsads.u-strasbg.fr/abs/2017A%26A...602A..46J}{, \aap,
  602, A46}

\bibitem[{{Jones} {et~al.}(1994){Jones}, {Tielens}, {Hollenbach}, \&
  {McKee}}]{jones94}
\normalfont
 {Jones}, A.~P., {Tielens}, A.~G.~G.~M., {Hollenbach}, D.~J., \& {McKee}, C.~F.
  1994\href{http://cdsads.u-strasbg.fr/abs/1994ApJ...433..797J}{, \apj, 433,
  797}

\bibitem[{{Jones} {et~al.}(2018){Jones}, {Stark}, \& {Ellis}}]{jones18}
\normalfont
 {Jones}, T., {Stark}, D.~P., \& {Ellis}, R.~S.
  2018\href{https://ui.adsabs.harvard.edu/abs/2018ApJ...863..191J}{, \apj, 863,
  191}

\bibitem[{{Kelly} {et~al.}(2012){Kelly}, {Shetty}, {Stutz}, {Kauffmann},
  {Goodman}, \& {Launhardt}}]{kelly12}
\normalfont
 {Kelly}, B.~C., {Shetty}, R., {Stutz}, A.~M., {et~al.}
  2012\href{http://cdsads.u-strasbg.fr/abs/2012ApJ...752...55K}{, \apj, 752,
  55}

\bibitem[{{Khramtsova} {et~al.}(2013){Khramtsova}, {Wiebe}, {Boley}, \&
  {Pavlyuchenkov}}]{khramtsova13}
\normalfont
 {Khramtsova}, M.~S., {Wiebe}, D.~S., {Boley}, P.~A., \& {Pavlyuchenkov}, Y.~N.
  2013\href{http://cdsads.u-strasbg.fr/abs/2013MNRAS.431.2006K}{, \mnras, 431,
  2006}

\bibitem[{{Kim} {et~al.}(2019){Kim}, {Anderson}, {Burke}, {D'Abrusco},
  {Fabbiano}, {Fruscione}, {Lauer}, {McCollough}, {Morgan}, {Mossman},
  {O'Sullivan}, {Paggi}, {Vrtilek}, \& {Trinchieri}}]{kim19}
\normalfont
 {Kim}, D.-W., {Anderson}, C., {Burke}, D., {et~al.}
  2019\href{https://ui.adsabs.harvard.edu/abs/2019ApJS..241...36K}{, \apjs,
  241, 36}

\bibitem[{{Kim} {et~al.}(1994){Kim}, {Martin}, \& {Hendry}}]{kim94}
\normalfont
 {Kim}, S.-H., {Martin}, P.~G., \& {Hendry}, P.~D.
  1994\href{https://ui.adsabs.harvard.edu/abs/1994ApJ...422..164K}{, \apj, 422,
  164}

\bibitem[{{Kimura}(2016)}]{kimura16}
\normalfont
 {Kimura}, H. 2016\href{http://cdsads.u-strasbg.fr/abs/2016MNRAS.459.2751K}{,
  \mnras, 459, 2751}

\bibitem[{{Kirchschlager} {et~al.}(2019){Kirchschlager}, {Schmidt}, {Barlow},
  {Fogerty}, {Bevan}, \& {Priestley}}]{kirchschlager19}
\normalfont
 {Kirchschlager}, F., {Schmidt}, F.~D., {Barlow}, M.~J., {et~al.}
  2019\href{https://ui.adsabs.harvard.edu/abs/2019MNRAS.489.4465K}{, \mnras,
  489, 4465}

\bibitem[{{K{\"o}hler} {et~al.}(2014){K{\"o}hler}, {Jones}, \&
  {Ysard}}]{kohler14}
\normalfont
 {K{\"o}hler}, M., {Jones}, A., \& {Ysard}, N.
  2014\href{http://cdsads.u-strasbg.fr/abs/2014A%26A...565L...9K}{, \aap, 565,
  L9}

\bibitem[{{K{\"o}hler} {et~al.}(2015){K{\"o}hler}, {Ysard}, \&
  {Jones}}]{kohler15}
\normalfont
 {K{\"o}hler}, M., {Ysard}, N., \& {Jones}, A.~P.
  2015\href{http://cdsads.u-strasbg.fr/abs/2015A%26A...579A..15K}{, \aap, 579,
  A15}

\bibitem[{{Kunth} \& {{\"O}stlin}(2000)}]{kunth00}
\normalfont
 {Kunth}, D. \& {{\"O}stlin}, G.
  2000\href{http://adsabs.harvard.edu/cgi-bin/nph-bib_query?bibcode=2000A%26ARv..10....1K&db_key=AST}{,
  \aapr, 10, 1}

\bibitem[{{Laigle} {et~al.}(2019){Laigle}, {Davidzon}, {Ilbert}, {Devriendt},
  {Kashino}, {Pichon}, {Capak}, {Arnouts}, {de la Torre}, {Dubois},
  {Gozaliasl}, {Le Borgne}, {Lilly}, {McCracken}, {Salvato}, \&
  {Slyz}}]{laigle19}
\normalfont
 {Laigle}, C., {Davidzon}, I., {Ilbert}, O., {et~al.}
  2019\href{https://ui.adsabs.harvard.edu/abs/2019MNRAS.486.5104L}{, \mnras,
  486, 5104}

\bibitem[{{Lamperti} {et~al.}(2019){Lamperti}, {Saintonge}, {De Looze},
  {Accurso}, {Clark}, {Smith}, {Wilson}, {Brinks}, {Brown}, {Bureau},
  {Clements}, {Eales}, {Glass}, {Hwang}, {Lee}, {Lin}, {Michalowski},
  {Sargent}, {Williams}, {Xiao}, \& {Yang}}]{lamperti19}
\normalfont
 {Lamperti}, I., {Saintonge}, A., {De Looze}, I., {et~al.}
  2019\href{https://ui.adsabs.harvard.edu/abs/2019MNRAS.489.4389L}{, \mnras,
  489, 4389}

\bibitem[{{Laporte} {et~al.}(2017){Laporte}, {Ellis}, {Boone}, {Bauer},
  {Qu{\'e}nard}, {Roberts-Borsani}, {Pell{\'o}}, {P{\'e}rez-Fournon}, \&
  {Streblyanska}}]{laporte17}
\normalfont
 {Laporte}, N., {Ellis}, R.~S., {Boone}, F., {et~al.}
  2017\href{https://ui.adsabs.harvard.edu/abs/2017ApJ...837L..21L}{, \apjl,
  837, L21}

\bibitem[{Lax(1975)}]{lax75}
\normalfont
 Lax, D. 1975, An Interim Report of a Monte Carlo Study of Robust Estimators of
  Width (Department of Statistics, Princeton University)

\bibitem[{{Lee} {et~al.}(2010){Lee}, {Ferguson}, {Somerville}, {Wiklind}, \&
  {Giavalisco}}]{lee10sk}
\normalfont
 {Lee}, S.-K., {Ferguson}, H.~C., {Somerville}, R.~S., {Wiklind}, T., \&
  {Giavalisco}, M.
  2010\href{https://ui.adsabs.harvard.edu/abs/2010ApJ...725.1644L}{, \apj, 725,
  1644}

\bibitem[{{Leja} {et~al.}(2019){Leja}, {Carnall}, {Johnson}, {Conroy}, \&
  {Speagle}}]{leja19}
\normalfont
 {Leja}, J., {Carnall}, A.~C., {Johnson}, B.~D., {Conroy}, C., \& {Speagle},
  J.~S. 2019\href{https://ui.adsabs.harvard.edu/abs/2019ApJ...876....3L}{,
  \apj, 876, 3}

\bibitem[{{Lianou} {et~al.}(2019){Lianou}, {Barmby}, {Mosenkov}, {Lehnert}, \&
  {Karczewski}}]{lianou19}
\normalfont
 {Lianou}, S., {Barmby}, P., {Mosenkov}, A.~A., {Lehnert}, M., \& {Karczewski},
  O. 2019\href{https://ui.adsabs.harvard.edu/abs/2019A&A...631A..38L}{, \aap,
  631, A38}
\normalfont{\textcolor{purple}{(L19)}}
\bibitem[{{Lisenfeld} \& {Ferrara}(1998)}]{lisenfeld98}
\normalfont
 {Lisenfeld}, U. \& {Ferrara}, A.
  1998\href{http://adsabs.harvard.edu/cgi-bin/nph-bib_query?bibcode=1998ApJ...496..145L&db_key=AST}{,
  \apj, 496, 145}

\bibitem[{{Liu} \& {Zhang}(2002)}]{liu02}
\normalfont
 {Liu}, F.~K. \& {Zhang}, Y.~H.
  2002\href{https://ui.adsabs.harvard.edu/abs/2002A&A...381..757L}{, \aap, 381,
  757}

\bibitem[{{Liu}(2011)}]{liu11}
\normalfont
 {Liu}, J. 2011\href{https://ui.adsabs.harvard.edu/abs/2011ApJS..192...10L}{,
  \apjs, 192, 10}

\bibitem[{{Lopes} {et~al.}(2020){Lopes}, {Telles}, \& {Melnick}}]{lopes20}
\normalfont
 {Lopes}, A.~R., {Telles}, E., \& {Melnick}, J.
  2020\href{https://ui.adsabs.harvard.edu/abs/2020MNRAS.tmp.3115L}{, \mnras}

\bibitem[{{Madau} \& {Dickinson}(2014)}]{madau14}
\normalfont
 {Madau}, P. \& {Dickinson}, M.
  2014\href{https://ui.adsabs.harvard.edu/abs/2014ARA&A..52..415M}{, \araa, 52,
  415}

\bibitem[{{Madden} {et~al.}(2006){Madden}, {Galliano}, {Jones}, \&
  {Sauvage}}]{madden06}
\normalfont
 {Madden}, S.~C., {Galliano}, F., {Jones}, A.~P., \& {Sauvage}, M.
  2006\href{http://adsabs.harvard.edu/cgi-bin/nph-bib_query?bibcode=2006A%26A...446..877M&db_key=AST}{,
  \aap, 446, 877}

\bibitem[{{Madden} {et~al.}(2013){Madden}, {R{\'e}my-Ruyer}, {Galametz},
  {Cormier}, {Lebouteiller}, {Galliano}, {Hony}, {Bendo}, {Smith}, {Pohlen},
  {Roussel}, {Sauvage}, {Wu}, {Sturm}, {Poglitsch}, {Contursi}, {Doublier},
  {Baes}, {Barlow}, {Boselli}, {Boquien}, {Carlson}, {Ciesla}, {Cooray},
  {Cortese}, {de Looze}, {Irwin}, {Isaak}, {Kamenetzky}, {Karczewski}, {Lu},
  {MacHattie}, {O''Halloran}, {Parkin}, {Rangwala}, {Schirm}, {Schulz},
  {Spinoglio}, {Vaccari}, {Wilson}, \& {Wozniak}}]{madden13}
\normalfont
 {Madden}, S.~C., {R{\'e}my-Ruyer}, A., {Galametz}, M., {et~al.}
  2013\href{http://cdsads.u-strasbg.fr/abs/2013PASP..125..600M}{, \pasp, 125,
  600}

\bibitem[{{Madden} {et~al.}(2014){Madden}, {R{\'e}my-Ruyer}, {Galametz},
  {Cormier}, {Lebouteiller}, {Galliano}, {Hony}, {Bendo}, {Smith}, {Pohlen},
  {Roussel}, {Sauvage}, {Wu}, {Sturm}, {Poglitsch}, {Contursi}, {Doublier},
  {Baes}, {Barlow}, {Boselli}, {Boquien}, {Carlson}, {Ciesla}, {Cooray},
  {Cortese}, {De Looze}, {Irwin}, {Isaak}, {Kamenetzky}, {Karczewski}, {Lu},
  {MacHattie}, {O'Halloran}, {Parkin}, {Rangwala}, {Schirm}, {Schulz},
  {Spinoglio}, {Vaccari}, {Wilson}, \& {Wozniak}}]{madden14}
\normalfont
 {Madden}, S.~C., {R{\'e}my-Ruyer}, A., {Galametz}, M., {et~al.}
  2014\href{https://ui.adsabs.harvard.edu/abs/2014PASP..126.1079M}{, \pasp,
  126, 1079}

\bibitem[{{Makarov} {et~al.}(2014){Makarov}, {Prugniel}, {Terekhova},
  {Courtois}, \& {Vauglin}}]{makarov14}
\normalfont
 {Makarov}, D., {Prugniel}, P., {Terekhova}, N., {Courtois}, H., \& {Vauglin},
  I. 2014\href{http://adsabs.harvard.edu/abs/2014A%26A...570A..13M}{, \aap,
  570, A13}

\bibitem[{{Malhotra} {et~al.}(1997){Malhotra}, {Helou}, {Stacey}, {Hollenbach},
  {Lord}, {Beichman}, {Dinerstein}, {Hunter}, {Lo}, {Lu}, {Rubin},
  {Silbermann}, {Thronson}, \& {Werner}}]{malhotra97}
\normalfont
 {Malhotra}, S., {Helou}, G., {Stacey}, G., {et~al.}
  1997\href{https://ui.adsabs.harvard.edu/abs/1997ApJ...491L..27M}{, \apjl,
  491, L27}

\bibitem[{{Malhotra} {et~al.}(2001){Malhotra}, {Kaufman}, {Hollenbach},
  {Helou}, {Rubin}, {Brauher}, {Dale}, {Lu}, {Lord}, {Stacey}, {Contursi},
  {Hunter}, \& {Dinerstein}}]{malhotra01}
\normalfont
 {Malhotra}, S., {Kaufman}, M.~J., {Hollenbach}, D., {et~al.}
  2001\href{https://ui.adsabs.harvard.edu/abs/2001ApJ...561..766M}{, \apj, 561,
  766}

\bibitem[{{Marassi} {et~al.}(2019){Marassi}, {Schneider}, {Limongi}, {Chieffi},
  {Graziani}, \& {Bianchi}}]{marassi19}
\normalfont
 {Marassi}, S., {Schneider}, R., {Limongi}, M., {et~al.}
  2019\href{https://ui.adsabs.harvard.edu/abs/2019MNRAS.484.2587M}{, \mnras,
  484, 2587}

\bibitem[{{Mart{\'\i}nez-Gonz{\'a}lez}
  {et~al.}(2018){Mart{\'\i}nez-Gonz{\'a}lez}, {W{\"u}nsch}, {Palou{\v{s}}},
  {Mu{\~n}oz-Tu{\~n}{\'o}n}, {Silich}, \&
  {Tenorio-Tagle}}]{martinez-gonzalez18}
\normalfont
 {Mart{\'\i}nez-Gonz{\'a}lez}, S., {W{\"u}nsch}, R., {Palou{\v{s}}}, J.,
  {et~al.} 2018\href{https://ui.adsabs.harvard.edu/abs/2018ApJ...866...40M}{,
  \apj, 866, 40}

\bibitem[{{Mathews} \& {Brighenti}(2003)}]{mathews03}
\normalfont
 {Mathews}, W.~G. \& {Brighenti}, F.
  2003\href{https://ui.adsabs.harvard.edu/abs/2003ARA&A..41..191M}{, \araa, 41,
  191}

\bibitem[{{Mathis} {et~al.}(1983){Mathis}, {Mezger}, \& {Panagia}}]{mathis83}
\normalfont
 {Mathis}, J.~S., {Mezger}, P.~G., \& {Panagia}, N.
  1983\href{http://adsabs.harvard.edu/cgi-bin/nph-bib_query?bibcode=1983A%26A...128..212M&db_key=AST}{,
  \aap, 128, 212}

\bibitem[{{Matsuura} {et~al.}(2015){Matsuura}, {Dwek}, {Barlow}, {Babler},
  {Baes}, {Meixner}, {Cernicharo}, {Clayton}, {Dunne}, {Fransson}, {Fritz},
  {Gear}, {Gomez}, {Groenewegen}, {Indebetouw}, {Ivison}, {Jerkstrand},
  {Lebouteiller}, {Lim}, {Lundqvist}, {Pearson}, {Roman-Duval}, {Royer},
  {Staveley-Smith}, {Swinyard}, {van Hoof}, {van Loon}, {Verstappen}, {Wesson},
  {Zanardo}, {Blommaert}, {Decin}, {Reach}, {Sonneborn}, {Van de Steene}, \&
  {Yates}}]{matsuura15}
\normalfont
 {Matsuura}, M., {Dwek}, E., {Barlow}, M.~J., {et~al.}
  2015\href{http://cdsads.u-strasbg.fr/abs/2015ApJ...800...50M}{, \apj, 800,
  50}

\bibitem[{{Mattsson} \& {Andersen}(2012)}]{mattsson12b}
\normalfont
 {Mattsson}, L. \& {Andersen}, A.~C.
  2012\href{https://ui.adsabs.harvard.edu/abs/2012MNRAS.423...38M}{, \mnras,
  423, 38}

\bibitem[{{Mattsson} {et~al.}(2012){Mattsson}, {Andersen}, \&
  {Munkhammar}}]{mattsson12a}
\normalfont
 {Mattsson}, L., {Andersen}, A.~C., \& {Munkhammar}, J.~D.
  2012\href{https://ui.adsabs.harvard.edu/abs/2012MNRAS.423...26M}{, \mnras,
  423, 26}

\bibitem[{McGrayne(2011)}]{mcgrayne11}
\normalfont
 McGrayne, S. 2011, The Theory That Would Not Die: How Bayes' Rule Cracked the
  Enigma Code, Hunted Down Russian Submarines, and Emerged Triumphant from Two
  Centuries of Controversy (Yale University Press)

\bibitem[{{Meny} {et~al.}(2007){Meny}, {Gromov}, {Boudet}, {Bernard},
  {Paradis}, \& {Nayral}}]{meny07}
\normalfont
 {Meny}, C., {Gromov}, V., {Boudet}, N., {et~al.}
  2007\href{http://cdsads.u-strasbg.fr/abs/2007A%26A...468..171M}{, \aap, 468,
  171}

\bibitem[{{Micelotta} {et~al.}(2016){Micelotta}, {Dwek}, \&
  {Slavin}}]{micelotta16}
\normalfont
 {Micelotta}, E.~R., {Dwek}, E., \& {Slavin}, J.~D.
  2016\href{https://ui.adsabs.harvard.edu/abs/2016A&A...590A..65M}{, \aap, 590,
  A65}

\bibitem[{{Mitchell} {et~al.}(2013){Mitchell}, {Lacey}, {Baugh}, \&
  {Cole}}]{mitchell13}
\normalfont
 {Mitchell}, P.~D., {Lacey}, C.~G., {Baugh}, C.~M., \& {Cole}, S.
  2013\href{https://ui.adsabs.harvard.edu/abs/2013MNRAS.435...87M}{, \mnras,
  435, 87}

\bibitem[{{Morgan} \& {Edmunds}(2003)}]{morgan03}
\normalfont
 {Morgan}, H.~L. \& {Edmunds}, M.~G.
  2003\href{http://adsabs.harvard.edu/cgi-bin/nph-bib_query?bibcode=2003MNRAS.343..427M&db_key=AST}{,
  \mnras, 343, 427}

\bibitem[{{Mosteller} \& {Tukey}(1977)}]{mosteller77}
\normalfont
 {Mosteller}, F. \& {Tukey}, J.~W. 1977, {Data analysis and regression. A second
  course in statistics} (Addison-Wesley Series in Behavioral Science:
  Quantitative Methods, Reading, Mass.: Addison-Wesley, 1977)

\bibitem[{{Nanni} {et~al.}(2020){Nanni}, {Burgarella}, {Theul{\'e}},
  {C{\^o}t{\'e}}, \& {Hirashita}}]{nanni20}
\normalfont
 {Nanni}, A., {Burgarella}, D., {Theul{\'e}}, P., {C{\^o}t{\'e}}, B., \&
  {Hirashita}, H.
  2020\href{https://ui.adsabs.harvard.edu/abs/2020A&A...641A.168N}{, \aap, 641,
  A168}
\normalfont{\textcolor{purple}{(N20)}}
\bibitem[{{Nersesian} {et~al.}(2019){Nersesian}, {Xilouris}, {Bianchi},
  {Galliano}, {Jones}, {Baes}, {Casasola}, {Cassar{\`a}}, {Clark}, {Davies},
  {Decleir}, {Dobbels}, {De Looze}, {De Vis}, {Fritz}, {Galametz}, {Madden},
  {Mosenkov}, {Tr{\v{c}}ka}, {Verstocken}, {Viaene}, \& {Lianou}}]{nersesian19}
\normalfont
 {Nersesian}, A., {Xilouris}, E.~M., {Bianchi}, S., {et~al.}
  2019\href{https://ui.adsabs.harvard.edu/abs/2019A&A...624A..80N}{, \aap, 624,
  A80}

\bibitem[{{Nozawa} {et~al.}(2006){Nozawa}, {Kozasa}, \& {Habe}}]{nozawa06}
\normalfont
 {Nozawa}, T., {Kozasa}, T., \& {Habe}, A.
  2006\href{https://ui.adsabs.harvard.edu/abs/2006ApJ...648..435N}{, \apj, 648,
  435}

\bibitem[{{Ochsenbein} {et~al.}(2000){Ochsenbein}, {Bauer}, \&
  {Marcout}}]{ochsenbein00}
\normalfont
 {Ochsenbein}, F., {Bauer}, P., \& {Marcout}, J.
  2000\href{https://ui.adsabs.harvard.edu/abs/2000A&AS..143...23O}{, \aaps,
  143, 23}

\bibitem[{{Ohyama} {et~al.}(2019){Ohyama}, {Sakamoto}, {Aalto}, \&
  {Gallagher}}]{ohyama19}
\normalfont
 {Ohyama}, Y., {Sakamoto}, K., {Aalto}, S., \& {Gallagher}, John~S., I.
  2019\href{https://ui.adsabs.harvard.edu/abs/2019ApJ...871..191O}{, \apj, 871,
  191}

\bibitem[{{O'Sullivan} {et~al.}(2001){O'Sullivan}, {Forbes}, \&
  {Ponman}}]{osullivan01}
\normalfont
 {O'Sullivan}, E., {Forbes}, D.~A., \& {Ponman}, T.~J.
  2001\href{https://ui.adsabs.harvard.edu/abs/2001MNRAS.328..461O}{, \mnras,
  328, 461}

\bibitem[{{Parvathi} {et~al.}(2012){Parvathi}, {Sofia}, {Murthy}, \&
  {Babu}}]{parvathi12}
\normalfont
 {Parvathi}, V.~S., {Sofia}, U.~J., {Murthy}, J., \& {Babu}, B.~R.~S.
  2012\href{https://ui.adsabs.harvard.edu/abs/2012ApJ...760...36P}{, \apj, 760,
  36}

\bibitem[{{Pei}(1992)}]{pei92}
\normalfont
 {Pei}, Y.~C.
  1992\href{http://adsabs.harvard.edu/cgi-bin/nph-bib_query?bibcode=1992ApJ...395..130P&db_key=AST}{,
  \apj, 395, 130}

\bibitem[{{Pilyugin} \& {Grebel}(2016)}]{pilyugin16}
\normalfont
 {Pilyugin}, L.~S. \& {Grebel}, E.~K.
  2016\href{https://ui.adsabs.harvard.edu/abs/2016MNRAS.457.3678P}{, \mnras,
  457, 3678}

\bibitem[{{Planck Collaboration} {et~al.}(2014){Planck Collaboration},
  {Abergel}, {Ade}, {Aghanim}, {Alves}, {Aniano}, {Armitage-Caplan}, {Arnaud},
  {Ashdown}, {Atrio-Barandela}, \& et~al.}]{planck-collaboration14c}
\normalfont
 {Planck Collaboration}, {Abergel}, A., {Ade}, P.~A.~R., {et~al.}
  2014\href{http://adsabs.harvard.edu/abs/2014A%26A...571A..11P}{, \aap, 571,
  A11}

\bibitem[{{Planck Collaboration} {et~al.}(2016){Planck Collaboration}, {Adam},
  {Ade}, {Aghanim}, {Arnaud}, {Ashdown}, {Aumont}, {Baccigalupi}, {Banday},
  {Barreiro}, {Bartolo}, {Battaner}, {Benabed}, {Beno{\^\i}t},
  {Benoit-L{\'e}vy}, {Bernard}, {Bersanelli}, {Bertincourt}, {Bielewicz},
  {Bock}, {Bonavera}, {Bond}, {Borrill}, {Bouchet}, {Boulanger}, {Bucher},
  {Burigana}, {Calabrese}, {Cardoso}, {Catalano}, {Challinor}, {Chamballu},
  {Chiang}, {Christensen}, {Clements}, {Colombi}, {Colombo}, {Combet},
  {Couchot}, {Coulais}, {Crill}, {Curto}, {Cuttaia}, {Danese}, {Davies},
  {Davis}, {de Bernardis}, {de Rosa}, {de Zotti}, {Delabrouille}, {Delouis},
  {D{\'e}sert}, {Diego}, {Dole}, {Donzelli}, {Dor{\'e}}, {Douspis}, {Ducout},
  {Dupac}, {Efstathiou}, {Elsner}, {En{\ss}lin}, {Eriksen}, {Falgarone},
  {Fergusson}, {Finelli}, {Forni}, {Frailis}, {Fraisse}, {Franceschi},
  {Frejsel}, {Galeotta}, {Galli}, {Ganga}, {Ghosh}, {Giard},
  {Giraud-H{\'e}raud}, {Gjerl{\o}w}, {Gonz{\'a}lez-Nuevo}, {G{\'o}rski},
  {Gratton}, {Gruppuso}, {Gudmundsson}, {Hansen}, {Hanson}, {Harrison},
  {Henrot-Versill{\'e}}, {Herranz}, {Hildebrandt}, {Hivon}, {Hobson}, {Holmes},
  {Hornstrup}, {Hovest}, {Huffenberger}, {Hurier}, {Jaffe}, {Jaffe}, {Jones},
  {Juvela}, {Keih{\"a}nen}, {Keskitalo}, {Kisner}, {Kneissl}, {Knoche}, {Kunz},
  {Kurki-Suonio}, {Lagache}, {Lamarre}, {Lasenby}, {Lattanzi}, {Lawrence}, {Le
  Jeune}, {Leahy}, {Lellouch}, {Leonardi}, {Lesgourgues}, {Levrier}, {Liguori},
  {Lilje}, {Linden-V{\o}rnle}, {L{\'o}pez-Caniego}, {Lubin},
  {Mac{\'\i}as-P{\'e}rez}, {Maggio}, {Maino}, {Mandolesi}, {Mangilli}, {Maris},
  {Martin}, {Mart{\'\i}nez-Gonz{\'a}lez}, {Masi}, {Matarrese}, {McGehee},
  {Melchiorri}, {Mendes}, {Mennella}, {Migliaccio}, {Mitra},
  {Miville-Desch{\^e}nes}, {Moneti}, {Montier}, {Moreno}, {Morgante},
  {Mortlock}, {Moss}, {Mottet}, {Munshi}, {Murphy}, {Naselsky}, {Nati},
  {Natoli}, {Netterfield}, {N{\o}rgaard-Nielsen}, {Noviello}, {Novikov},
  {Novikov}, {Oxborrow}, {Paci}, {Pagano}, {Pajot}, {Paoletti}, {Pasian},
  {Patanchon}, {Pearson}, {Perdereau}, {Perotto}, {Perrotta}, {Pettorino},
  {Piacentini}, {Piat}, {Pierpaoli}, {Pietrobon}, {Plaszczynski},
  {Pointecouteau}, {Polenta}, {Pratt}, {Pr{\'e}zeau}, {Prunet}, {Puget},
  {Rachen}, {Reinecke}, {Remazeilles}, {Renault}, {Renzi}, {Ristorcelli},
  {Rocha}, {Rosset}, {Rossetti}, {Roudier}, {Rusholme}, {Sandri}, {Santos},
  {Sauv{\'e}}, {Savelainen}, {Savini}, {Scott}, {Seiffert}, {Shellard},
  {Spencer}, {Stolyarov}, {Stompor}, {Sudiwala}, {Sutton}, {Suur-Uski},
  {Sygnet}, {Tauber}, {Terenzi}, {Toffolatti}, {Tomasi}, {Tristram}, {Tucci},
  {Tuovinen}, {Valenziano}, {Valiviita}, {Van Tent}, {Vibert}, {Vielva},
  {Villa}, {Wade}, {Wandelt}, {Watson}, {Wehus}, {Yvon}, {Zacchei}, \&
  {Zonca}}]{planck-collaboration16d}
\normalfont
 {Planck Collaboration}, {Adam}, R., {Ade}, P.~A.~R., {et~al.}
  2016\href{https://ui.adsabs.harvard.edu/abs/2016A&A...594A...8P}{, \aap, 594,
  A8}

\bibitem[{{Priestley} {et~al.}(2019){Priestley}, {Barlow}, \& {De
  Looze}}]{priestley19}
\normalfont
 {Priestley}, F.~D., {Barlow}, M.~J., \& {De Looze}, I.
  2019\href{https://ui.adsabs.harvard.edu/abs/2019MNRAS.485..440P}{, \mnras,
  485, 440}

\bibitem[{{Rau} {et~al.}(2019){Rau}, {Hirashita}, \& {Murga}}]{rau19}
\normalfont
 {Rau}, S.-J., {Hirashita}, H., \& {Murga}, M.
  2019\href{https://ui.adsabs.harvard.edu/abs/2019MNRAS.489.5218R}{, \mnras,
  489, 5218}

\bibitem[{{Reach} {et~al.}(2005){Reach}, {Megeath}, {Cohen}, {Hora}, {Carey},
  {Surace}, {Willner}, {Barmby}, {Wilson}, {Glaccum}, {Lowrance}, {Marengo}, \&
  {Fazio}}]{reach05}
\normalfont
 {Reach}, W.~T., {Megeath}, S.~T., {Cohen}, M., {et~al.}
  2005\href{http://cdsads.u-strasbg.fr/abs/2005PASP..117..978R}{, \pasp, 117,
  978}

\bibitem[{{R{\'e}my-Ruyer} {et~al.}(2014){R{\'e}my-Ruyer}, {Madden},
  {Galliano}, {Galametz}, {Takeuchi}, {Asano}, {Zhukovska}, {Lebouteiller},
  {Cormier}, {Jones}, {Bocchio}, {Baes}, {Bendo}, {Boquien}, {Boselli},
  {DeLooze}, {Doublier-Pritchard}, {Hughes}, {Karczewski}, \&
  {Spinoglio}}]{remy-ruyer14}
\normalfont
 {R{\'e}my-Ruyer}, A., {Madden}, S.~C., {Galliano}, F., {et~al.}
  2014\href{http://cdsads.u-strasbg.fr/abs/2014A%26A...563A..31R}{, \aap, 563,
  A31}

\bibitem[{{R{\'e}my-Ruyer} {et~al.}(2013){R{\'e}my-Ruyer}, {Madden},
  {Galliano}, {Hony}, {Sauvage}, {Bendo}, {Roussel}, {Pohlen}, {Smith},
  {Galametz}, {Cormier}, {Lebouteiller}, {Wu}, {Baes}, {Barlow}, {Boquien},
  {Boselli}, {Ciesla}, {De Looze}, {Karczewski}, {Panuzzo}, {Spinoglio},
  {Vaccari}, \& {Wilson}}]{remy-ruyer13}
\normalfont
 {R{\'e}my-Ruyer}, A., {Madden}, S.~C., {Galliano}, F., {et~al.}
  2013\href{http://cdsads.u-strasbg.fr/abs/2013A%26A...557A..95R}{, \aap, 557,
  A95}

\bibitem[{{R{\'e}my-Ruyer} {et~al.}(2015){R{\'e}my-Ruyer}, {Madden},
  {Galliano}, {Lebouteiller}, {Baes}, {Bendo}, {Boselli}, {Ciesla}, {Cormier},
  {Cooray}, {Cortese}, {De Looze}, {Doublier-Pritchard}, {Galametz}, {Jones},
  {Karczewski}, {Lu}, \& {Spinoglio}}]{remy-ruyer15}
\normalfont
 {R{\'e}my-Ruyer}, A., {Madden}, S.~C., {Galliano}, F., {et~al.}
  2015\href{http://cdsads.u-strasbg.fr/abs/2015A%26A...582A.121R}{, \aap, 582,
  A121}

\bibitem[{{Rieke} {et~al.}(1985){Rieke}, {Lebofsky}, \& {Low}}]{rieke85}
\normalfont
 {Rieke}, G.~H., {Lebofsky}, M.~J., \& {Low}, F.~J.
  1985\href{https://ui.adsabs.harvard.edu/abs/1985AJ.....90..900R}{, \aj, 90,
  900}

\bibitem[{{Rodriguez-Gomez} {et~al.}(2019){Rodriguez-Gomez}, {Snyder}, {Lotz},
  {Nelson}, {Pillepich}, {Springel}, {Genel}, {Weinberger}, {Tacchella},
  {Pakmor}, {Torrey}, {Marinacci}, {Vogelsberger}, {Hernquist}, \&
  {Thilker}}]{rodriguez-gomez19}
\normalfont
 {Rodriguez-Gomez}, V., {Snyder}, G.~F., {Lotz}, J.~M., {et~al.}
  2019\href{https://ui.adsabs.harvard.edu/abs/2019MNRAS.483.4140R}{, \mnras,
  483, 4140}

\bibitem[{{R{\"o}llig} {et~al.}(2007){R{\"o}llig}, {Abel}, {Bell}, {Bensch},
  {Black}, {Ferland}, {Jonkheid}, {Kamp}, {Kaufman}, {Le Bourlot}, {Le Petit},
  {Meijerink}, {Morata}, {Ossenkopf}, {Roueff}, {Shaw}, {Spaans}, {Sternberg},
  {Stutzki}, {Thi}, {van Dishoeck}, {van Hoof}, {Viti}, \&
  {Wolfire}}]{rollig07}
\normalfont
 {R{\"o}llig}, M., {Abel}, N.~P., {Bell}, T., {et~al.}
  2007\href{http://cdsads.u-strasbg.fr/abs/2007A%26A...467..187R}{, \aap, 467,
  187}

\bibitem[{{Roman-Duval} {et~al.}(2017){Roman-Duval}, {Bot}, {Chastenet}, \&
  {Gordon}}]{roman-duval17}
\normalfont
 {Roman-Duval}, J., {Bot}, C., {Chastenet}, J., \& {Gordon}, K.
  2017\href{http://cdsads.u-strasbg.fr/abs/2017ApJ...841...72R}{, \apj, 841,
  72}

\bibitem[{{Rosa Gonz{\'a}lez} {et~al.}(2009){Rosa Gonz{\'a}lez}, {Terlevich},
  {Jim{\'e}nez Bail{\'o}n}, {Terlevich}, {Ranalli}, {Comastri}, {Laird}, \&
  {Nandra}}]{rosa-gonzalez09}
\normalfont
 {Rosa Gonz{\'a}lez}, D., {Terlevich}, E., {Jim{\'e}nez Bail{\'o}n}, E.,
  {et~al.} 2009\href{https://ui.adsabs.harvard.edu/abs/2009MNRAS.399..487R}{,
  \mnras, 399, 487}

\bibitem[{{Rowlands} {et~al.}(2014){Rowlands}, {Gomez}, {Dunne},
  {Arag{\'o}n-Salamanca}, {Dye}, {Maddox}, {da Cunha}, \& {van der
  Werf}}]{rowlands14}
\normalfont
 {Rowlands}, K., {Gomez}, H.~L., {Dunne}, L., {et~al.}
  2014\href{https://ui.adsabs.harvard.edu/abs/2014MNRAS.441.1040R}{, \mnras,
  441, 1040}

\bibitem[{{Roychowdhury} {et~al.}(2020){Roychowdhury}, {Galliano},
  {Roychowdhury}, {Jones}, {Ysard}, {co-I5}, \& {co-I6}}]{roychowdhury20}
\normalfont
 {Roychowdhury}, S., {Galliano}, F., {Roychowdhury}, S., {et~al.}
  2020\href{http://http://cdsads.u-strasbg.fr/abstract_service.html}{, {\it in
  prep.}}

\bibitem[{{Salpeter}(1955)}]{salpeter55}
\normalfont
 {Salpeter}, E.~E.
  1955\href{https://ui.adsabs.harvard.edu/abs/1955ApJ...121..161S}{, \apj, 121,
  161}

\bibitem[{{Sanders} {et~al.}(2020){Sanders}, {Shapley}, {Jones}, {Reddy},
  {Kriek}, {Siana}, {Coil}, {Mobasher}, {Shivaei}, {Dav{\'e}}, {Azadi},
  {Price}, {Leung}, {Freeman}, {Fetherolf}, {de Groot}, {Zick}, \&
  {Barro}}]{sanders20}
\normalfont
 {Sanders}, R.~L., {Shapley}, A.~E., {Jones}, T., {et~al.}
  2020\href{https://ui.adsabs.harvard.edu/abs/2020arXiv200907292S}{, arXiv
  e-prints, arXiv:2009.07292}

\bibitem[{{Sandstrom} {et~al.}(2010){Sandstrom}, {Bolatto}, {Draine}, {Bot}, \&
  {Stanimirovi{\'c}}}]{sandstrom10}
\normalfont
 {Sandstrom}, K.~M., {Bolatto}, A.~D., {Draine}, B.~T., {Bot}, C., \&
  {Stanimirovi{\'c}}, S.
  2010\href{http://cdsads.u-strasbg.fr/abs/2010ApJ...715..701S}{, \apj, 715,
  701}

\bibitem[{{Savage} \& {Mathis}(1979)}]{savage79}
\normalfont
 {Savage}, B.~D. \& {Mathis}, J.~S.
  1979\href{http://adsabs.harvard.edu/cgi-bin/nph-bib_query?bibcode=1979ARA%26A..17...73S&db_key=AST}{,
  \araa, 17, 73}

\bibitem[{{Schirmer} {et~al.}(2020){Schirmer}, {Abergel}, {Verstraete},
  {Ysard}, {Juvela}, {Jones}, \& {Habart}}]{schirmer20}
\normalfont
 {Schirmer}, T., {Abergel}, A., {Verstraete}, L., {et~al.}
  2020\href{https://ui.adsabs.harvard.edu/abs/2020arXiv200305902S}{, arXiv
  e-prints, arXiv:2003.05902}

\bibitem[{{Seok} {et~al.}(2014){Seok}, {Hirashita}, \& {Asano}}]{seok14}
\normalfont
 {Seok}, J.~Y., {Hirashita}, H., \& {Asano}, R.~S.
  2014\href{http://cdsads.u-strasbg.fr/abs/2014MNRAS.439.2186S}{, \mnras, 439,
  2186}

\bibitem[{{Shetty} {et~al.}(2009){Shetty}, {Kauffmann}, {Schnee}, \&
  {Goodman}}]{shetty09}
\normalfont
 {Shetty}, R., {Kauffmann}, J., {Schnee}, S., \& {Goodman}, A.~A.
  2009\href{http://cdsads.u-strasbg.fr/abs/2009ApJ...696..676S}{, \apj, 696,
  676}

\bibitem[{{Slavin} {et~al.}(2015){Slavin}, {Dwek}, \& {Jones}}]{slavin15}
\normalfont
 {Slavin}, J.~D., {Dwek}, E., \& {Jones}, A.~P.
  2015\href{http://cdsads.u-strasbg.fr/abs/2015ApJ...803....7S}{, \apj, 803, 7}

\bibitem[{{Smith} {et~al.}(2012){Smith}, {Gomez}, {Eales}, {Ciesla}, {Boselli},
  {Cortese}, {Bendo}, {Baes}, {Bianchi}, {Clemens}, {Clements}, {Cooray},
  {Davies}, {De Looze}, {di Serego Alighieri}, {Fritz}, {Gavazzi}, {Gear},
  {Madden}, {Mentuch}, {Panuzzo}, {Pohlen}, {Spinoglio}, {Verstappen},
  {Vlahakis}, {Wilson}, \& {Xilouris}}]{smith12}
\normalfont
 {Smith}, M.~W.~L., {Gomez}, H.~L., {Eales}, S.~A., {et~al.}
  2012\href{http://adsabs.harvard.edu/abs/2012ApJ...748..123S}{, \apj, 748,
  123}

\bibitem[{{Stansberry} {et~al.}(2007){Stansberry}, {Gordon}, {Bhattacharya},
  {Engelbracht}, {Rieke}, {Marleau}, {Fadda}, {Frayer}, {Noriega-Crespo},
  {Wachter}, {Young}, {M{\"u}ller}, {Kelly}, {Blaylock}, {Henderson},
  {Neugebauer}, {Beeman}, \& {Haller}}]{stansberry07}
\normalfont
 {Stansberry}, J.~A., {Gordon}, K.~D., {Bhattacharya}, B., {et~al.}
  2007\href{http://cdsads.u-strasbg.fr/abs/2007PASP..119.1038S}{, \pasp, 119,
  1038}

\bibitem[{Stein(1956)}]{stein56}
\normalfont
 Stein, C. 1956, in
  \href{https://projecteuclid.org/euclid.bsmsp/1200501656}{Proceedings of the
  Third Berkeley Symposium on Mathematical Statistics and Probability, Volume
  1: Contributions to the Theory of Statistics} (Berkeley, Calif.: University
  of California Press), 197--206

\bibitem[{{Stepnik} {et~al.}(2003){Stepnik}, {Abergel}, {Bernard}, {Boulanger},
  {Cambr{\'e}sy}, {Giard}, {Jones}, {Lagache}, {Lamarre}, {Meny}, {Pajot}, {Le
  Peintre}, {Ristorcelli}, {Serra}, \& {Torre}}]{stepnik03}
\normalfont
 {Stepnik}, B., {Abergel}, A., {Bernard}, J., {et~al.}
  2003\href{http://cdsads.u-strasbg.fr/abs/2003A%26A...398..551S}{, \aap, 398,
  551}

\bibitem[{{Sugerman} {et~al.}(2006){Sugerman}, {Ercolano}, {Barlow}, {Tielens},
  {Clayton}, {Zijlstra}, {Meixner}, {Speck}, {Gledhill}, {Panagia}, {Cohen},
  {Gordon}, {Meyer}, {Fabbri}, {Bowey}, {Welch}, {Regan}, \&
  {Kennicutt}}]{sugerman06}
\normalfont
 {Sugerman}, B.~E.~K., {Ercolano}, B., {Barlow}, M.~J., {et~al.}
  2006\href{http://adsabs.harvard.edu/cgi-bin/nph-bib_query?bibcode=2006Sci...313..196S&db_key=AST}{,
  Science, 313, 196}

\bibitem[{{Tajer} {et~al.}(2005){Tajer}, {Trinchieri}, {Wolter}, {Campana},
  {Moretti}, \& {Tagliaferri}}]{tajer05}
\normalfont
 {Tajer}, M., {Trinchieri}, G., {Wolter}, A., {et~al.}
  2005\href{https://ui.adsabs.harvard.edu/abs/2005A&A...435..799T}{, \aap, 435,
  799}

\bibitem[{{Temim} \& {Dwek}(2013)}]{temim13}
\normalfont
 {Temim}, T. \& {Dwek}, E.
  2013\href{https://ui.adsabs.harvard.edu/abs/2013ApJ...774....8T}{, \apj, 774,
  8}

\bibitem[{{Temim} {et~al.}(2017){Temim}, {Dwek}, {Arendt}, {Borkowski},
  {Reynolds}, {Slane}, {Gelfand}, \& {Raymond}}]{temim17}
\normalfont
 {Temim}, T., {Dwek}, E., {Arendt}, R.~G., {et~al.}
  2017\href{https://ui.adsabs.harvard.edu/abs/2017ApJ...836..129T}{, \apj, 836,
  129}

\bibitem[{{Tielens}(1998)}]{tielens98}
\normalfont
 {Tielens}, A.~G.~G.~M.
  1998\href{http://adsabs.harvard.edu/cgi-bin/nph-bib_query?bibcode=1998ApJ...499..267T&db_key=AST}{,
  \apj, 499, 267}

\bibitem[{{Todini} \& {Ferrara}(2001)}]{todini01}
\normalfont
 {Todini}, P. \& {Ferrara}, A.
  2001\href{https://ui.adsabs.harvard.edu/abs/2001MNRAS.325..726T}{, \mnras,
  325, 726}

\bibitem[{{Trayford} {et~al.}(2017){Trayford}, {Camps}, {Theuns}, {Baes},
  {Bower}, {Crain}, {Gunawardhana}, {Schaller}, {Schaye}, \&
  {Frenk}}]{trayford17}
\normalfont
 {Trayford}, J.~W., {Camps}, P., {Theuns}, T., {et~al.}
  2017\href{https://ui.adsabs.harvard.edu/abs/2017MNRAS.470..771T}{, \mnras,
  470, 771}

\bibitem[{{Tr{\v{c}}ka} {et~al.}(2020){Tr{\v{c}}ka}, {Baes}, {Camps}, {Meidt},
  {Trayford}, {Bianchi}, {Casasola}, {Cassar{\`a}}, {De Looze}, {De Vis},
  {Dobbels}, {Fritz}, {Galametz}, {Galliano}, {Katsianis}, {Madden},
  {Mosenkov}, {Nersesian}, {Viaene}, \& {Xilouris}}]{trcka20}
\normalfont
 {Tr{\v{c}}ka}, A., {Baes}, M., {Camps}, P., {et~al.}
  2020\href{https://ui.adsabs.harvard.edu/abs/2020MNRAS.494.2823T}{, \mnras,
  494, 2823}

\bibitem[{{Valiante} {et~al.}(2009){Valiante}, {Schneider}, {Bianchi}, \&
  {Andersen}}]{valiante09}
\normalfont
 {Valiante}, R., {Schneider}, R., {Bianchi}, S., \& {Andersen}, A.~C.
  2009\href{https://ui.adsabs.harvard.edu/abs/2009MNRAS.397.1661V}{, \mnras,
  397, 1661}

\bibitem[{{van der Tak} {et~al.}(2018){van der Tak}, {Madden}, {Roelfsema},
  {Armus}, {Baes}, {Bernard-Salas}, {Bolatto}, {Bontemps}, {Bot}, {Bradford},
  {Braine}, {Ciesla}, {Clements}, {Cormier}, {Fern{\'a}ndez-Ontiveros},
  {Galliano}, {Giard}, {Gomez}, {Gonz{\'a}lez-Alfonso}, {Herpin}, {Johnstone},
  {Jones}, {Kaneda}, {Kemper}, {Lebouteiller}, {De Looze}, {Matsuura},
  {Nakagawa}, {Onaka}, {P{\'e}rez-Gonz{\'a}lez}, {Shipman}, \&
  {Spinoglio}}]{vandertak18}
\normalfont
 {van der Tak}, F.~F.~S., {Madden}, S.~C., {Roelfsema}, P., {et~al.}
  2018\href{http://cdsads.u-strasbg.fr/abs/2018PASA...35....2V}{, \pasa, 35,
  e002}

\bibitem[{{Viaene} {et~al.}(2014){Viaene}, {Fritz}, {Baes}, {Bendo},
  {Blommaert}, {Boquien}, {Boselli}, {Ciesla}, {Cortese}, {De Looze}, {Gear},
  {Gentile}, {Hughes}, {Jarrett}, {Karczewski}, {Smith}, {Spinoglio}, {Tamm},
  {Tempel}, {Thilker}, \& {Verstappen}}]{viaene14}
\normalfont
 {Viaene}, S., {Fritz}, J., {Baes}, M., {et~al.}
  2014\href{https://ui.adsabs.harvard.edu/abs/2014A&A...567A..71V}{, \aap, 567,
  A71}

\bibitem[{Villani \& Larsson(2006)}]{villani06}
\normalfont
 Villani, M. \& Larsson, R.
  2006\href{http://www.ingentaconnect.com/content/tandf/lsta/2006/00000035/00000006/art00015}{,
  Communications in Statistics-Theory and Methods, 35, 1123}

\bibitem[{{Wagenmakers} {et~al.}(2008){Wagenmakers}, {Lee}, {Lodewyckx}, \&
  {Iverson}}]{wagenmakers08}
\normalfont
 {Wagenmakers}, E.-J., {Lee}, M., {Lodewyckx}, T., \& {Iverson}, G.~J. 2008,
  Bayesian Versus Frequentist Inference (New York, NY: Springer New York),
  181--207

\bibitem[{{Walter} {et~al.}(2007){Walter}, {Cannon}, {Roussel}, {Bendo},
  {Calzetti}, {Dale}, {Draine}, {Helou}, {Kennicutt}, {Moustakas}, {Rieke},
  {Armus}, {Engelbracht}, {Gordon}, {Hollenbach}, {Lee}, {Li}, {Meyer},
  {Murphy}, {Regan}, {Smith}, {Brinks}, {de Blok}, {Bigiel}, \&
  {Thornley}}]{walter07}
\normalfont
 {Walter}, F., {Cannon}, J.~M., {Roussel}, H., {et~al.}
  2007\href{http://cdsads.u-strasbg.fr/abs/2007ApJ...661..102W}{, \apj, 661,
  102}

\bibitem[{{Wheelock} {et~al.}(1994){Wheelock}, {Gautier}, {Chillemi}, {Kester},
  {McCallon}, {Oken}, {White}, {Gregorich}, {Boulanger}, \&
  {Good}}]{wheelock94}
\normalfont
 {Wheelock}, S.~L., {Gautier}, T.~N., {Chillemi}, J., {et~al.} 1994, {IRAS sky
  survey atlas: Explanatory supplement}, Tech.
  rep.\href{https://irsa.ipac.caltech.edu/IRASdocs/exp.sup/toc.html}{, IRSA}

\bibitem[{{Witt} \& {Gordon}(2000)}]{witt00}
\normalfont
 {Witt}, A.~N. \& {Gordon}, K.~D.
  2000\href{https://ui.adsabs.harvard.edu/abs/2000ApJ...528..799W}{, \apj, 528,
  799}

\bibitem[{{Witt} {et~al.}(1992){Witt}, {Thronson}, \& {Capuano}}]{witt92}
\normalfont
 {Witt}, A.~N., {Thronson}, Harley~A., J., \& {Capuano}, John~M., J.
  1992\href{https://ui.adsabs.harvard.edu/abs/1992ApJ...393..611W}{, \apj, 393,
  611}

\bibitem[{{Wu} {et~al.}(2011){Wu}, {Hogg}, \& {Moustakas}}]{wu11}
\normalfont
 {Wu}, R., {Hogg}, D.~W., \& {Moustakas}, J.
  2011\href{http://cdsads.u-strasbg.fr/abs/2011ApJ...730..111W}{, \apj, 730,
  111}

\bibitem[{{Wu} {et~al.}(2007){Wu}, {Charmandaris}, {Hunt}, {Bernard-Salas},
  {Brandl}, {Marshall}, {Lebouteiller}, {Hao}, \& {Houck}}]{wu07}
\normalfont
 {Wu}, Y., {Charmandaris}, V., {Hunt}, L.~K., {et~al.}
  2007\href{http://cdsads.u-strasbg.fr/abs/2007ApJ...662..952W}{, \apj, 662,
  952}

\bibitem[{{Ysard} {et~al.}(2016){Ysard}, {K{\"o}hler}, {Jones}, {Dartois},
  {Godard}, \& {Gavilan}}]{ysard16}
\normalfont
 {Ysard}, N., {K{\"o}hler}, M., {Jones}, A., {et~al.}
  2016\href{https://ui.adsabs.harvard.edu/abs/2016A&A...588A..44Y}{, \aap, 588,
  A44}

\bibitem[{{Ysard} {et~al.}(2015){Ysard}, {K{\"o}hler}, {Jones},
  {Miville-Desch{\^e}nes}, {Abergel}, \& {Fanciullo}}]{ysard15}
\normalfont
 {Ysard}, N., {K{\"o}hler}, M., {Jones}, A., {et~al.}
  2015\href{http://cdsads.u-strasbg.fr/abs/2015A%26A...577A.110Y}{, \aap, 577,
  A110}

\bibitem[{{Zhukovska}(2014)}]{zhukovska14}
\normalfont
 {Zhukovska}, S.
  2014\href{http://cdsads.u-strasbg.fr/abs/2014A%26A...562A..76Z}{, \aap, 562,
  A76}

\bibitem[{{Zubko} {et~al.}(2004){Zubko}, {Dwek}, \& {Arendt}}]{zubko04}
\normalfont
 {Zubko}, V., {Dwek}, E., \& {Arendt}, R.~G.
  2004\href{http://adsabs.harvard.edu/cgi-bin/nph-bib_query?bibcode=2004ApJS..152..211Z&db_key=AST}{,
  \apjs, 152, 211}

\end{thebibliography}

\appendix

  \section{CALIBRATION UNCERTAINTIES}
  \label{app:calunc}

The HB approach is one of the only methods allowing a rigorous treatment of the photometric calibration uncertainties.
These uncertainties are indeed systematic effects, with both spectral and spatial correlations.
Some studies account for their spectral correlations, but ignore their spatial
correlations \citep[\eg\ ][]{gordon14}.
In the present paper, following \citetalias{galliano18a}, we assume that calibration uncertainties are perfectly correlated between sources, and partially correlated between wavelengths.
In this section, we discuss the values of these uncertainties and of their correlation coefficients.

The DustPedia photometric fluxes, given in the archive\footnote{\href{http://dustpedia.astro.noa.gr/Photometry}{http://dustpedia.astro.noa.gr/Photometry}}, $F_\nu(\lambda_0)$ at wavelength $\lambda_0$, are accompanied with a total uncertainty being the quadratic sum of noise and calibration errors: 
\begin{equation}
  \sigma_\nu^\sms{C18}(\lambda_0) 
  = \sqrt{\left(\sigma_\nu^\sms{noise}(\lambda_0)\right)^2 
         +\left(\delta_\sms{cal}^\sms{C18}(\lambda_0)\times F_\nu(\lambda_0)\right)^2},
\end{equation}
where $\delta_\sms{cal}^\sms{C18}(\lambda_0)$ is the calibration uncertainty \citepalias[Table~1 of][]{clark18}.
\HB\ treats the noise and calibration uncertainties separately.
Therefore, we keep the noise uncertainty, $\sigma_\nu^\sms{noise}(\lambda_0)$, but we build a covariance matrix of the absolute calibration uncertainties, $V_\sms{cal}$, to replace the $\delta_\sms{cal}^\sms{C18}$ coefficients.


    \subsection{The Full Covariance Matrix}

Using the \expression{separation strategy} \citep{barnard00}, we decompose the covariance matrix of the absolute calibration uncertainties as:
\begin{equation}
  V_\sms{cal} = S_\sms{cal}R_\sms{cal}S_\sms{cal},
  \label{eq:Vcal}
\end{equation}
where $S_\sms{cal}$ is the diagonal matrix of standard-deviations and $R_\sms{cal}$ the correlation matrix.
The non-trivial elements of $S_\sms{cal}$ and $R_\sms{cal}$ are given in \reftabs{tab:Scal}{tab:Rcal}.
We describe, in the following sections~\ref{app:IRAC} to \ref{app:HFI}, how we obtained these values from the literature.
\begin{table}[htbp]
  \centering
  \caption{Absolute calibration uncertainties (diagonal of 
           $S_\sms{cal}$; \refeqnp{eq:Vcal}).
           See \reftab{tab:photometry} for waveband naming 
           conventions.}
  \label{tab:Scal}
  \begin{tabular}{lr}
    \hline\hline
      Waveband label & Calibration uncertainty \\
    \hline
IRAC1 & $10.2\,\%$ \\
IRAC2 & $10.2\,\%$ \\
IRAC3 & $10.2\,\%$ \\
IRAC4 & $10.2\,\%$ \\
    \hline
MIPS1 & $4.0\,\%$ \\
MIPS2 & $5.0\,\%$ \\
MIPS3 & $11.6\,\%$ \\
    \hline
PACS1 & $5.4\,\%$ \\
PACS2 & $5.4\,\%$ \\
PACS3 & $5.4\,\%$ \\
    \hline
SPIRE1 & $5.9\,\%$ \\
SPIRE2 & $5.9\,\%$ \\
SPIRE3 & $5.9\,\%$ \\
    \hline
WISE1 & $3.2\,\%$ \\
WISE2 & $3.5\,\%$ \\
WISE3 & $5.0\,\%$ \\
WISE4 & $7.0\,\%$ \\
    \hline
IRAS1 & $7.6\,\%$ \\
IRAS2 & $11.0\,\%$ \\
IRAS3 & $14.6\,\%$ \\
IRAS4 & $20.2\,\%$ \\
    \hline
HFI1 & $4.3\,\%$ \\
HFI2 & $4.2\,\%$ \\
HFI3 & $0.90\,\%$ \\
    \hline
  \end{tabular}
\end{table}

\begin{sidewaystable*}[htbp]
  \caption{Correlations of the absolute calibration uncertainties 
           ($R_\sms{cal}$ matrix; \refeqnp{eq:Vcal}).
           See \reftab{tab:photometry} for waveband naming 
           conventions.}
  \label{tab:Rcal}
{\small
  \begin{tabularx}{\linewidth}{X|rrrr|rrr|rrr|rrr|rrrr|rrrr|rrr}
    \hline\hline
 & \multicolumn{4}{c|}{IRAC}  & \multicolumn{3}{c|}{MIPS}  & \multicolumn{3}{c|}{PACS}  & \multicolumn{3}{c|}{SPIRE}  & \multicolumn{4}{c|}{WISE}  & \multicolumn{4}{c|}{IRAS}  & \multicolumn{3}{c}{HFI} \\
    \hline
IRAC1 & 1 & 0.02 & 0.02 & 0.02 & 0 & 0 & 0 & 0 & 0 & 0 & 0 & 0 & 0 & 0.13 & 0.06 & 0.04 & 0 & 0 & 0 & 0 & 0 & 0 & 0 & 0 \\
IRAC2 & 0.02 & 1 & 0.02 & 0.02 & 0 & 0 & 0 & 0 & 0 & 0 & 0 & 0 & 0 & 0.07 & 0.12 & 0.04 & 0 & 0 & 0 & 0 & 0 & 0 & 0 & 0 \\
IRAC3 & 0.02 & 0.02 & 1 & 0.02 & 0 & 0 & 0 & 0 & 0 & 0 & 0 & 0 & 0 & 0.07 & 0.06 & 0.04 & 0 & 0 & 0 & 0 & 0 & 0 & 0 & 0 \\
IRAC4 & 0.02 & 0.02 & 0.02 & 1 & 0 & 0 & 0 & 0 & 0 & 0 & 0 & 0 & 0 & 0.07 & 0.06 & 0.09 & 0 & 0 & 0 & 0 & 0 & 0 & 0 & 0 \\
    \hline
MIPS1 & 0 & 0 & 0 & 0 & 1 & 0 & 0.34 & 0 & 0 & 0 & 0 & 0 & 0 & 0 & 0 & 0 & 0.57 & 0 & 0 & 0 & 0 & 0 & 0 & 0 \\
MIPS2 & 0 & 0 & 0 & 0 & 0 & 1 & 0.43 & 0 & 0 & 0 & 0 & 0 & 0 & 0 & 0 & 0 & 0 & 0 & 0 & 0 & 0 & 0 & 0 & 0 \\
MIPS3 & 0 & 0 & 0 & 0 & 0.34 & 0.43 & 1 & 0 & 0 & 0 & 0 & 0 & 0 & 0 & 0 & 0 & 0.20 & 0 & 0 & 0 & 0 & 0 & 0 & 0 \\
    \hline
PACS1 & 0 & 0 & 0 & 0 & 0 & 0 & 0 & 1 & 0.86 & 0.86 & 0 & 0 & 0 & 0 & 0 & 0 & 0 & 0 & 0 & 0 & 0 & 0 & 0 & 0 \\
PACS2 & 0 & 0 & 0 & 0 & 0 & 0 & 0 & 0.86 & 1 & 0.86 & 0 & 0 & 0 & 0 & 0 & 0 & 0 & 0 & 0 & 0 & 0 & 0 & 0 & 0 \\
PACS3 & 0 & 0 & 0 & 0 & 0 & 0 & 0 & 0.86 & 0.86 & 1 & 0 & 0 & 0 & 0 & 0 & 0 & 0 & 0 & 0 & 0 & 0 & 0 & 0 & 0 \\
    \hline
SPIRE1 & 0 & 0 & 0 & 0 & 0 & 0 & 0 & 0 & 0 & 0 & 1 & 0.47 & 0.47 & 0 & 0 & 0 & 0 & 0 & 0 & 0 & 0 & 0.40 & 0.41 & 0 \\
SPIRE2 & 0 & 0 & 0 & 0 & 0 & 0 & 0 & 0 & 0 & 0 & 0.47 & 1 & 0.47 & 0 & 0 & 0 & 0 & 0 & 0 & 0 & 0 & 0.40 & 0.41 & 0 \\
SPIRE3 & 0 & 0 & 0 & 0 & 0 & 0 & 0 & 0 & 0 & 0 & 0.47 & 0.47 & 1 & 0 & 0 & 0 & 0 & 0 & 0 & 0 & 0 & 0.40 & 0.41 & 0 \\
    \hline
WISE1 & 0.13 & 0.07 & 0.07 & 0.07 & 0 & 0 & 0 & 0 & 0 & 0 & 0 & 0 & 0 & 1 & 0.21 & 0.14 & 0 & 0 & 0 & 0 & 0 & 0 & 0 & 0 \\
WISE2 & 0.06 & 0.12 & 0.06 & 0.06 & 0 & 0 & 0 & 0 & 0 & 0 & 0 & 0 & 0 & 0.21 & 1 & 0.13 & 0 & 0 & 0 & 0 & 0 & 0 & 0 & 0 \\
WISE3 & 0.04 & 0.04 & 0.04 & 0.09 & 0 & 0 & 0 & 0 & 0 & 0 & 0 & 0 & 0 & 0.14 & 0.13 & 1 & 0 & 0 & 0 & 0 & 0 & 0 & 0 & 0 \\
WISE4 & 0 & 0 & 0 & 0 & 0.57 & 0 & 0.20 & 0 & 0 & 0 & 0 & 0 & 0 & 0 & 0 & 0 & 1 & 0 & 0 & 0 & 0 & 0 & 0 & 0 \\
    \hline
IRAS1 & 0 & 0 & 0 & 0 & 0 & 0 & 0 & 0 & 0 & 0 & 0 & 0 & 0 & 0 & 0 & 0 & 0 & 1 & 0.69 & 0.52 & 0.38 & 0 & 0 & 0 \\
IRAS2 & 0 & 0 & 0 & 0 & 0 & 0 & 0 & 0 & 0 & 0 & 0 & 0 & 0 & 0 & 0 & 0 & 0 & 0.69 & 1 & 0.36 & 0.26 & 0 & 0 & 0 \\
IRAS3 & 0 & 0 & 0 & 0 & 0 & 0 & 0 & 0 & 0 & 0 & 0 & 0 & 0 & 0 & 0 & 0 & 0 & 0.52 & 0.36 & 1 & 0.72 & 0 & 0 & 0 \\
IRAS4 & 0 & 0 & 0 & 0 & 0 & 0 & 0 & 0 & 0 & 0 & 0 & 0 & 0 & 0 & 0 & 0 & 0 & 0.38 & 0.26 & 0.72 & 1 & 0 & 0 & 0 \\
    \hline
HFI1 & 0 & 0 & 0 & 0 & 0 & 0 & 0 & 0 & 0 & 0 & 0.40 & 0.40 & 0.40 & 0 & 0 & 0 & 0 & 0 & 0 & 0 & 0 & 1 & 0.69 & 0 \\
HFI2 & 0 & 0 & 0 & 0 & 0 & 0 & 0 & 0 & 0 & 0 & 0.41 & 0.41 & 0.41 & 0 & 0 & 0 & 0 & 0 & 0 & 0 & 0 & 0.69 & 1 & 0 \\
HFI3 & 0 & 0 & 0 & 0 & 0 & 0 & 0 & 0 & 0 & 0 & 0 & 0 & 0 & 0 & 0 & 0 & 0 & 0 & 0 & 0 & 0 & 0 & 0 & 1 \\
    \hline
  \end{tabularx}}
\end{sidewaystable*}

    \subsection{\spitz/IRAC}
    \label{app:IRAC}

The calibration of IRAC data is presented by \citet{reach05}.
It is performed using a subsample of the stellar catalog of \citet{cohen03}.
The components entering in the uncertainty on the absolute calibration are given by Eq.~(13) of \citet{reach05} and the corresponding values in their Table~7.
The first term is the correlated uncertainty, $\sigma_\sms{abs}=1.5\,\%$, corresponding to the uncertainty in the predicted fluxes of Vega and Sirius.
The second term is the uncorrelated uncertainty on the calibrator fluxes:
$\sqrt{(\sigma_\sms{groud}^2-\sigma_\sms{abs}^2)/n}=0.87\,\%$, with $\sigma_\sms{ground}=2.3\,\%$ and $n=4$.
Finally, there is an uncorrelated term introduced by the dispersion of the IRAC observations of the calibrators: $\sigma_\sms{rms}/\sqrt{n}$, where
$\sigma_\sms{rms}=2.0\,\%$, $1.9\,\%$, $2.1\,\%$ and $2.1\,\%$ for IRAC1, IRAC2, IRAC3 and IRAC4, respectively.

In addition, one must multiply the point source calibrated fluxes by aperture correction factors.
According to the IRAC instrument handbook\footnote{\href{https://irsa.ipac.caltech.edu/data/SPITZER/docs/irac/iracinstrumenthandbook/}{https://irsa.ipac.caltech.edu/\-data/\-SPITZER/\-docs/\-irac/\-irac\-ins\-tru\-ment\-hand\-book/}} (Sect.~4.11), these factors have a $10\,\%$ uncertainty.
We include this extra source of uncertainty, assuming it is uncorrelated.
These correction factors have been applied by \citetalias{clark18} for all the DustPedia galaxies and by \citet{remy-ruyer15} for half of the DGS galaxies, the other half being nearly point sources.
The IRAC calibration uncertainties will thus be overestimated for about 15 of our sources.

    \subsection{\spitz/MIPS}

MIPS calibration has been independently performed for each of its three wavebands.
The 24~\tmic\ calibration is presented by \citet{engelbracht07}. 
It is primarily based on observations of A stars.
The authors recommend adopting a net calibration uncertainty of $4\,\%$.
The 70~\tmic\ calibration is presented by \citet{gordon07}.
It is primarily based on observations of B and M stars.
The authors recommend using a calibration uncertainty of $5\,\%$, dominated by repeatability scatter.
Finally, the 160~\tmic\ calibration, presented by \citet{stansberry07} is performed on observations of asteroids, simulatenously in the three MIPS bands.
This 160~\tmic\ calibration uncertainty is therefore tied to 24 and 70~\tmic.
The authors estimate a $7\,\%$ uncertainty tied to these bands, for a total uncertainty of $11.6\,\%$.
To simplify, we assume that the calibration uncertainty is the quadratical sum of $4\,\%$ correlated with MIPS1, $5\,\%$ correlated with MIPS2, and the remaining $9.7\,\%$ uncorrelated.

    \subsection{\hersc/PACS}

PACS calibration is performed on five stars, used as primary calibrators \citep{balog14}.
The absolute calibration uncertainty for point sources appears to be dominated by the stellar model uncertainty of $5\,\%$, correlated between bands.
In addition, there is a $2\,\%$ repeatability uncertainty, uncorrelated between bands.
The extended source calibration uncertainty is not quantified by the instrument team.

    \subsection{\hersc/SPIRE}
    \label{app:SPIRE}

SPIRE calibration, presented by \citet{griffin13} is performed on Neptune.
The uncertainty comes from three sources.
There is a $4\,\%$ correlated uncertainty on Neptune's model.
There is also a $1.5\,\%$ uncorrelated uncertainty due to the noise in Neptune's observations.
Finally, there is a $4\,\%$ uncorrelated uncertainty on SPIRE's beam area.

    \subsection{\wise}

\wise\ data have been calibrated by \citet{jarrett11} using observations of a network of stars in both ecliptic poles \citep{cohen03} and an additional red source, the galaxy \ngc{6552}.
These calibrators have been observed with \spitz/IRAC and \spitz/MIPS, and compared to \wise.
It appears that the relative \wise\ calibration, tied to \spitz, is accurate to $2.4\,\%$, $2.8\,\%$, $4.5\,\%$ and $5.7\,\%$ (for WISE1, WISE2, WISE3 and WISE4, respectively).
These uncertainties come from the scatter of individual calibrators and can therefore be considered independent.
The red calibrator (\ngc{6552}) is reported to be too bright by $\simeq6\,\%$
by \citet{jarrett11}.
However, \citet{brown14} revised the central wavelength of the WISE4 band to correct systematic discrepancies observed with red sources.

The absolute calibration accuracy of \wise\ is not discussed by \citet{jarrett11}.
We can simply assume that each \wise\ band will be tied to the closest \spitz\ band: WISE1 with IRAC1; WISE2 with IRAC2; WISE3 with IRAC4; WISE4 with MIPS1.
We can thus quadratically add the \spitz\ calibration uncertainties to these bands and assume that this term will be correlated with the \spitz\ bands.
However, we do not include the IRAC uncertainty on the extended source correction (\refapp{app:IRAC}), as these calibrators are point sources.

    \subsection{\iras}

The \iras\ flux calibration is presented by \citet[][ Sect.~VI.C.2]{wheelock94}\footnote{\href{https://irsa.ipac.caltech.edu/IRASdocs/exp.sup/toc.html}{https://irsa.ipac.caltech.edu/IRASdocs/exp.sup/toc.html}}.
The 12~\tmic\ flux is calibrated on observations of the star $\alpha$-Tau, 
assuming the absolute flux value derived from the ground-based 10~\tmic\ photometry by \citet{rieke85}, accurate within $3\,\%$.
An additional source of error comes from the uncertainty of $2\,\%$ on the model used to extrapolate the flux at 12~\tmic.
The 25 and 60~\tmic\ fluxes are calibrated by extrapolating the 12~\tmic\ flux, using stellar models, normalized to the Sun.
The uncertainty in this extrapolation, based on the scatter of individual stars, is estimated to be $4\,\%$ and $5.3\,\%$, respectively \citep[Table VI..C.1 of][]{wheelock94}.
These uncertainties, which can be assumed to be independent, have to be quadratically added to the 12~\tmic\ uncertainty, which is a fully correlated term.
The 100~\tmic\ calibration is based on observations of asteroids, where the 60~\tmic\ flux is extrapolated to 100~\tmic, assuming that the 25/60 and 60/100 color temperatures are identical.
\citet{wheelock94} estimate that the total relative calibration uncertainty at 100~\tmic\ is $10\,\%$, due to the model uncertainty and scatter of the asteroid observations. 
This uncertainty is correlated with the 60~\tmic.

For extended sources, uncertainties in the frequency response and the effective detector area have to be considered \citep[][ Sect.~VI.C.4]{wheelock94}.
The uncertainty on the frequency response is not fully quantified, except that it should be $7\,\%$/\tmic\ of the central bandwidth, with an uncertainty of 0.3~\tmic\ on the bandwidth \citep[Sect.~VI.C.3 of][]{wheelock94}.
It is also specified that the 100~\tmic\ uncertainty can be \textit{significant} for objects cooler than 30~K, which is our case.
The uncertainty on the detector areas is given by Table~IV.A.1 of \citet{wheelock94} showing the flux dispersion of each individual detector, during a cross-scan of the planetary nebula \ngc{6543}.
The average dispersions of $6.4\,\%$, $6.5\,\%$, $11\,\%$ and $10.9\,\%$ (for IRAS1, IRAS2, IRAS3 and IRAS4, respectively) can be considered as an additional, independent source of uncertainty.

    \subsection{\planck/HFI}
    \label{app:HFI}

The calibration of the HFI data used by \citetalias{clark18} is presented by \citet{planck-collaboration16d}.
The two short wavelength bands (350 and 550~\tmic) are calibrated on Neptune and Uranus.
The uncertainty on the model is $5\,\%$ including $2\,\%$ spectrally independent.
The statistical errors on the observations of the calibrators are $1.4\,\%$ and $1.1\,\%$, respectively, which can be considered independent.
The models of Neptune used here and for SPIRE (\refapp{app:SPIRE}) are the ESA3 and ESA4 models (\href{ftp://ftp.sciops.esa.int/pub/hsc-calibration/PlanetaryModels/}{ftp://ftp.sciops.esa.int/pub/hsc-calibration/PlanetaryModels/}), respectively.
Both models differ only by 0.3--0.6$\,\%$ in the SPIRE and HFI bands.
They will thus induce correlation.
To simplify, we assume that the HFI calibration factors derived from Neptune (correlated with SPIRE) and Uranus (not correlated with SPIRE) are averaged.
We obtain calibration uncertainties $\simeq30\,\%$ lower than \citet{planck-collaboration16d}, as they have opted for the more conservative, linear addition of independent errors.
We choose here the more rigorous quadratrical addition.

The four long wavelength bands of HFI are calibrated on the CMB dipole.
The first source of uncertainty is the scatter of individual bolometers: $0.78\,\%$, $0.16\,\%$, $0.07\,\%$ and $0.09\,\%$ at 850~\tmic, 1.38, 2.1 and 3.0~mm.
The second source of uncertainty comes from the gain variations: $0.03\,\%$, $0.04\,\%$, $0.05\,\%$ and $0.05\,\%$, respectively.
These different uncertainties can be considered independent.

  \section{HOMOGENEITY OF THE SAMPLE}

    \subsection{DustPedia and DGS Photometry}
    \label{app:phot}

\begin{figure}[htbp]
  \includegraphics[width=\linewidth]{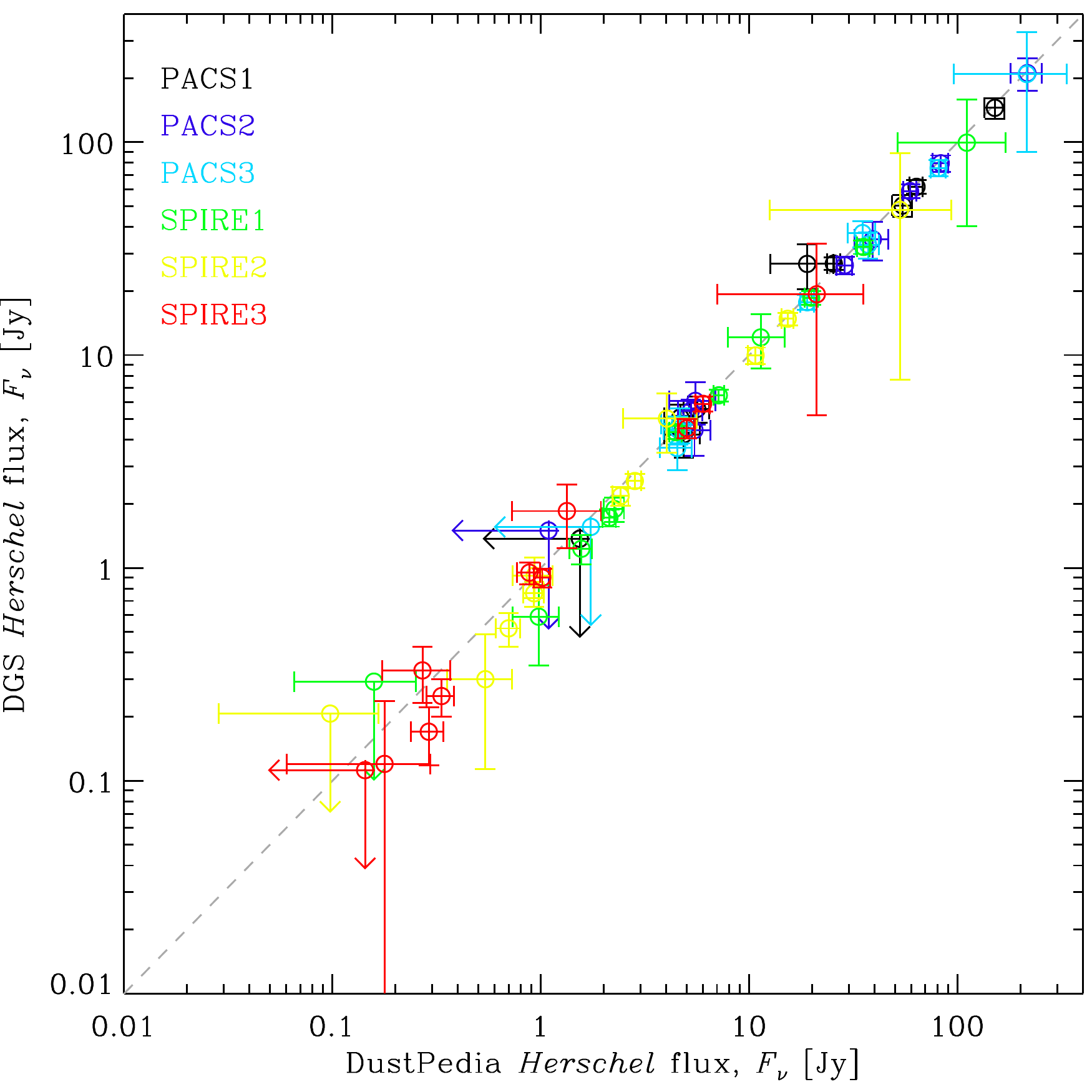}
  \caption{\textsl{Comparison between the DustPedia and DGS photometry.}
           The 13 sources of the DGS which are part of the DustPedia catalog
           are compared in the 3 PACS and 3 SPIRE bands (one color per filter).
           The $x$ axis shows the photometry estimated by \citetalias{clark18},
           while the $y$ axis shows the photometry estimated by 
           \citet{remy-ruyer15}.
           The dashed line represents the 1:1 relation.}
  \label{fig:compphot}
\end{figure}
\reffig{fig:compphot} displays the comparison, in the 6 \hersc\ bands, of the photometry of the sources of the DGS that are part of the DustPedia sample.
The DustPedia photometry has been estimated by \citetalias{clark18} (\cf~\refsec{sec:DustPedia}), and the DGS photometry, by 
\citet[][ \cf~\refsec{sec:DGS}]{remy-ruyer15}.
Both are in very good agreement.

    \subsection{Derivation of the Stellar Mass}
    \label{app:Mstar}

\begin{figure}[htbp]
  \includegraphics[width=\linewidth]{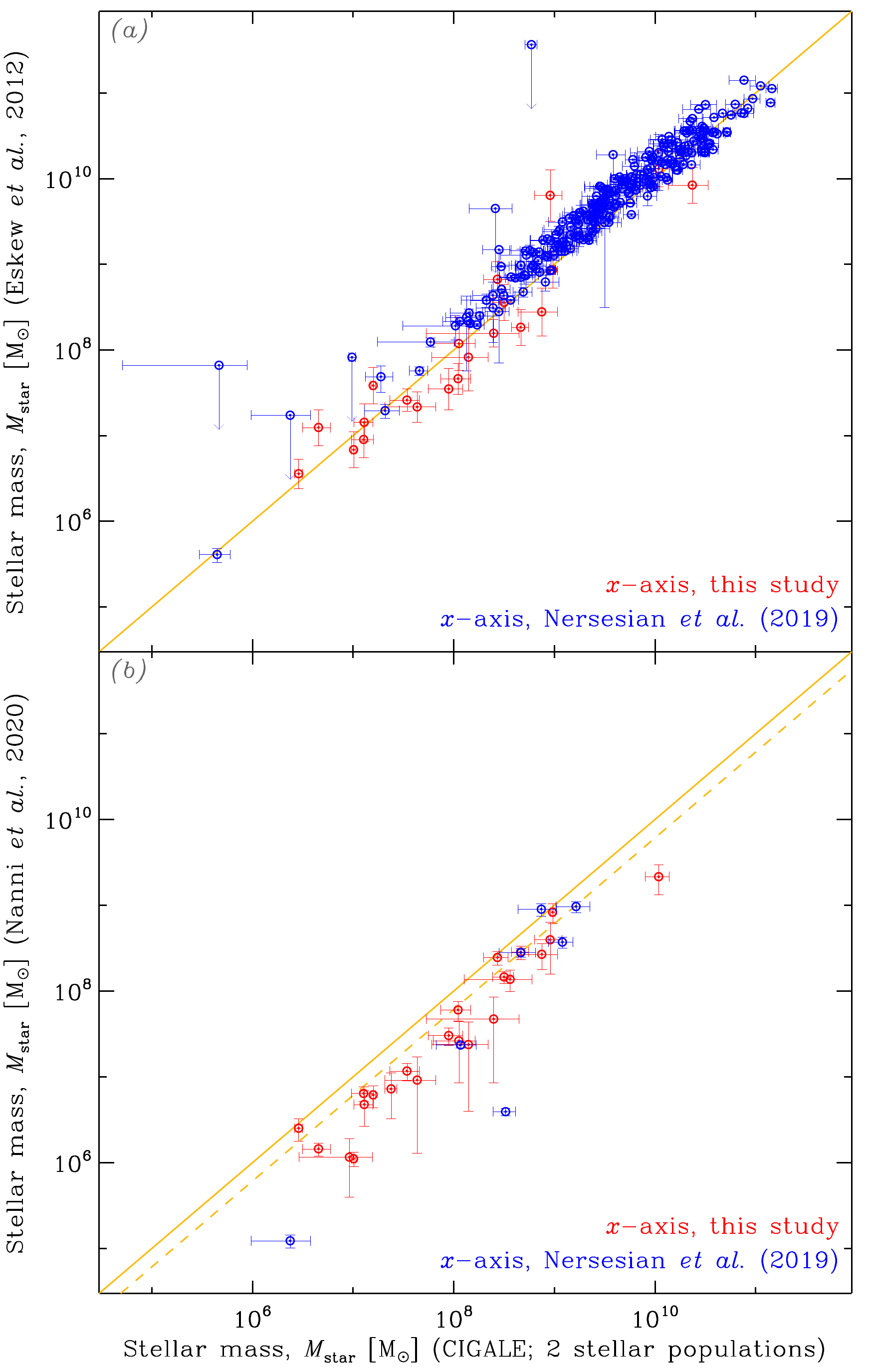}
  \caption{\textsl{Comparison of stellar mass estimators.}
           Panel~\textit{(a)} compares the \CIG\ estimates of \tMs, using two 
           stellar populations ($x$-axis), with the 
           \citet[][ $y$-axis]{eskew12} empirical 
           approximation, both assuming a \citet{salpeter55} IMF.
           Panel~\textit{(b)} compares the same $x$-axis as panel~\textit{(a)}
           to the \CIG\ estimates, using a single stellar population and a 
           \citet{chabrier03} IMF, by \citetalias{nanni20}.
           In both panels, the blue symbols correspond to the values presented
           in \citet{nersesian19}, and the red symbols correspond to the values
           estimated for this study, with the same \CIG\ settings.
           The solid yellow line represents the 1:1 relation.
           The dashed yellow line in panel~\textit{(b)} represents the 
           1:0.61 relation that accounts for the difference in IMFs.}
  \label{fig:eskew}
\end{figure}
To further investigate the discrepancy between the \CIG\ fits of \citetalias{nanni20} and the \citet{eskew12} approximation used by \citet{madden14}, discussed in \refsec{sec:Mstar}, we have compared the two estimators on our sample.
To close the controversy, we have modelled the UV-to-mm SED of the 35 DGS galaxies that were not in DustPedia, with \CIG, using the same settings as \citet[][ \ie\ using two stellar populations]{nersesian19}.
We therefore have \CIG-derived \tMs\ for our whole sample.
This is shown in panel~\textit{(a)} of \reffig{fig:eskew}.
The stellar mass derived using \CIG\ with two stellar populations are in very good agreement with the \citet{eskew12} approximation.

Panel~\textit{(b)} of \reffig{fig:eskew} compares the same \CIG-two-population estimates with the values of \tMs\ reported by \citetalias{nanni20}.
The main differences are the following, noting that the stellar masses quoted by \citetalias{nanni20} come from the SED modelling presented by \citet{burgarella20}.
\begin{enumerate}
  \item \citet{nersesian19} used the UV-optical photometry of \citet{clark18},
    which have been homogenized (\cf~\refsec{sec:photometry}), whereas
    \citet{burgarella20} gathered UV-optical data from the 
    \expression{NASA/IPAC Extragalactic Database} (NED).
    The latter is a compilation of data from the literature that can be 
    inhomogeneous in terms of photometric aperture, extinction correction 
    and removal of foreground stars.
  \item \citet{nersesian19} used \CIG\ with two stellar populations:
    \begin{inlinelist}
      \item an old exponentially decreasing SFH;
      \item a young burst.
    \end{inlinelist}
    In contrast, \citet{burgarella20} used a single delayed SFH, which can lead 
    to underestimating the mass, because of the outshining effect of the recent 
    star formation \citep[\eg\ ][]{lopes20}.
    For instance, \citet{buat14} conducted a large comparison of stellar mass
    estimates and concluded that models using two stellar populations were 
    providing the best fits.
    This can explain why the IRAC1 and IRAC2 fluxes are systematically 
    underestimated by the mean model of \citetalias{nanni20} (\cf~their figure 
    A.1), while it is not the case for \citet[][     
    \href{http://dustpedia.astro.noa.gr/Cigale}%
         {http://dustpedia.astro.noa.gr/Cigale}]{nersesian19}.
    Yet, these bands sample the spectral range where the bulk of the stellar 
    mass dominates the emission.
    In addition, most of the galaxies of the DGS are \expression{Blue Compact 
    Dwarf} galaxies (BCD).
    These galaxies are known to be actively forming stars and have an old 
    underlying stellar population 
    \citep[\eg\ ][]{hunter89b,grebel99,kunth00,aloisi07,janowiecki17}.
  \item \citet{nersesian19} used a \citet{salpeter55} IMF, whereas
    \citet{burgarella20} used a \citet{chabrier03} IMF leading to a systematic 
    scaling by a factor 0.61.
\end{enumerate}
The comparison of these two estimates, in panel~\textit{(b)} of 
\reffig{fig:eskew}, shows that the stellar masses of \citetalias{nanni20} are most of the time systematically lower than those of \citet{nersesian19}, sometimes drastically.

  \section{ADDITIONAL DISCUSSION ON THE 
            PARAMETRIZATION OF \THE}
  \label{app:themis}

We have presented the way we parametrize the \THE\ size distribution
in \refsec{sec:themis}. 
It consists in linearly scaling the fractions of a-C(:H) smaller than 7~\AA, and between 7 and 15~\AA.
We call it the \expression{linear} parametrization.
It is controlled by the two parameters \tqAF\ and $f_\sms{VSAC}$, introduced in \refsec{sec:themis}.
In the present appendix, we demonstrate, on an example, that this parametrization is flexible enough to mimic the original parametrization of \THE.
The \THE\ size distribution of small a-C(:H) is a power-law, with minimum cut-off size $a_\sms{min}=4\;\AA$ and index $\alpha=5$ \citep[Table~2 of][]{jones13}.
We call it the \expression{non-linear} parametrization.
Varying $a_\sms{min}$ and $\alpha$ is probably more physical than our \expression{linear} method, as it preserves the continuity of the size distribution.
However, it produces SEDs with very similar shapes.

\begin{figure}[htbp]
  \includegraphics[width=\linewidth]{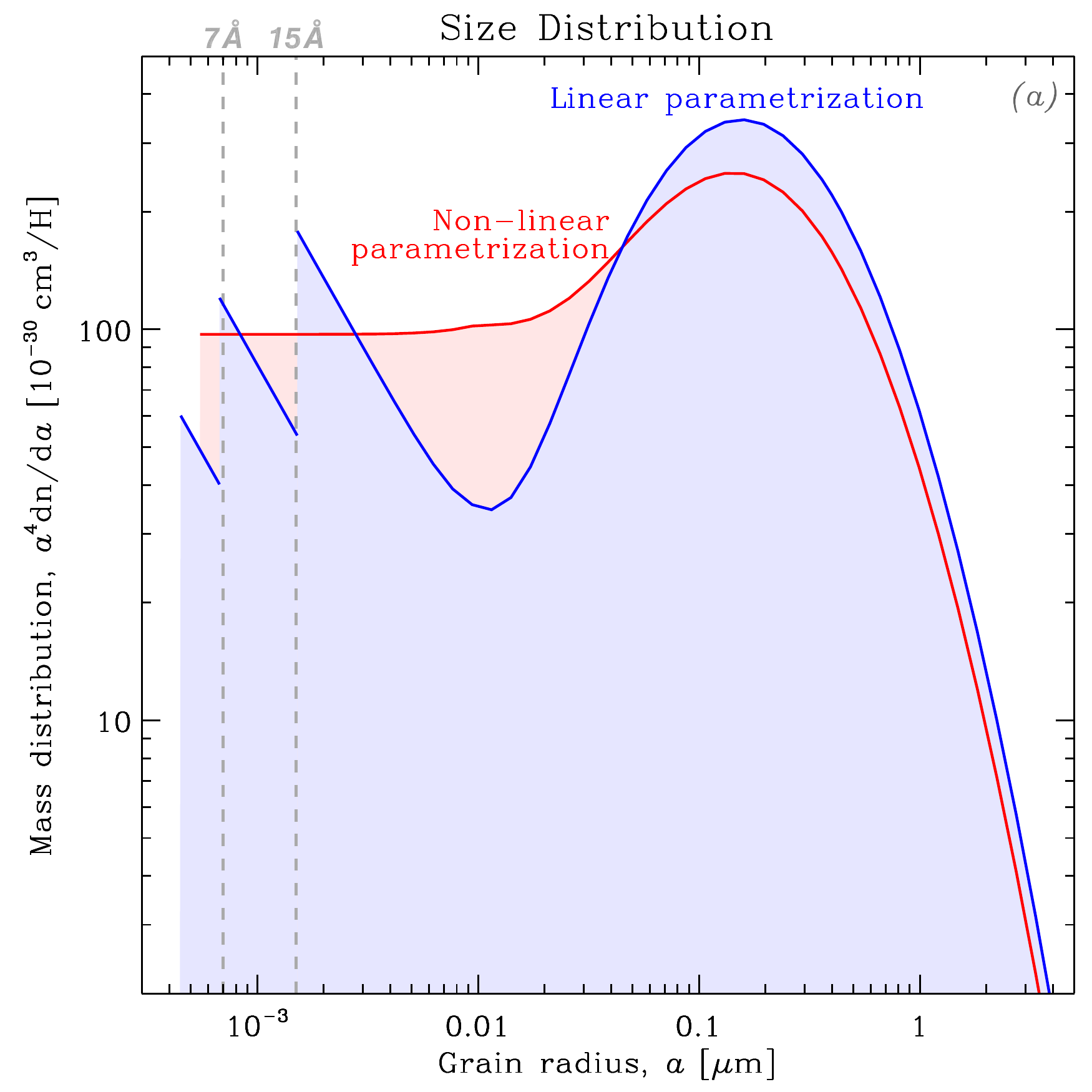} \\
  \includegraphics[width=\linewidth]{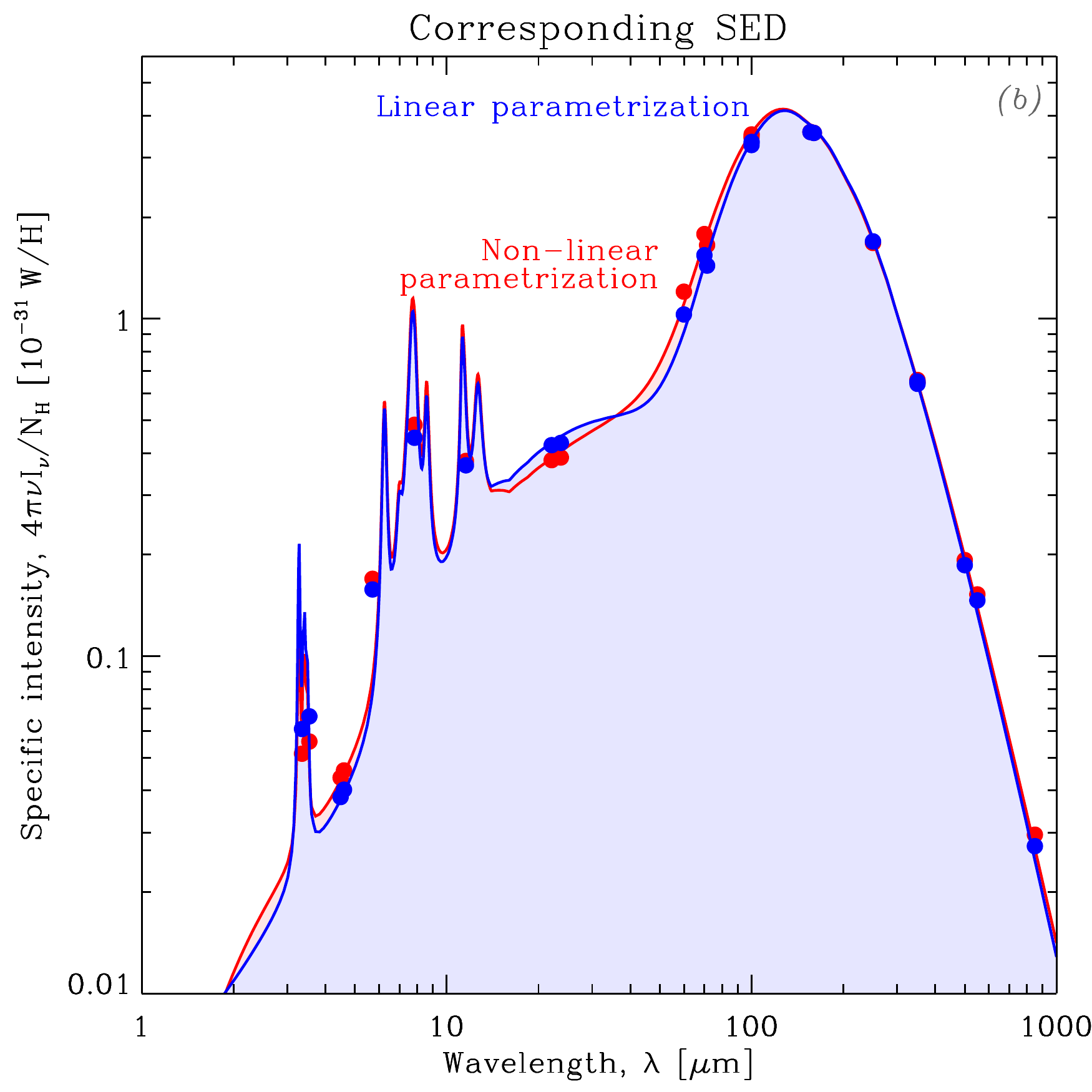}
  \caption{\textsl{\THE\ parametrization comparison.}
           The two panels display the same quantities as
           \reffig{fig:themis}.
           On each panel, the red curve shows \THE, 
           \expression{non-linearly} parametrized, \ie\ varying the minimum 
           cut-off radius, $a_\sms{min}$, and the index of the power-law1 size 
           distribution, $\alpha$.
           The blue curve shows \THE, \expression{linearly} parametrized, 
           with the method described in \refsec{sec:themis}.
           The dots on the two SED curves (panel \textit{b}) represent the 
           synthetic photometry, \ie\ the model integrated in the filters of 
           \reftab{tab:photometry}.
          }
  \label{fig:themis_parm}
\end{figure}
This is demonstrated on \reffig{fig:themis_parm}.
This figure displays the same quantities as \reffig{fig:themis}, but compares the two parametrizations.
The red curves show the size distribution (panel \textit{a}) and the corresponding SED (panel \textit{b}), for the \expression{non-linear} method, with $a_\sms{min}=5\;\AA$ and $\alpha=4$.
The blue curves show the same quantities, with the \expression{linear} method.
We see that the broadband fluxes (dots in panel \textit{b}) produced by these two parametrizations are very similar. 
The fact that they are not exactly the same is not important here.
What is relevant is that the two methods can alter the SED with the same dynamical range.

  \section{POSTERIOR PREDICTIVE P-VALUE}
  \label{app:ppp}

We have estimated the \expression{posterior predictive $p$-value} \citep[PPP; \eg][]{gelman04} of our reference run (\refsec{sec:ref}).
PPPs are the Bayesian equivalent to a chi-squared test.
They allow us to quantify the goodness of our fit.
PPPs are estimated by generating sets of replicated observables, noted $D_\sms{rep}$, from our posterior distribution, conditional on the actual data (\refsec{sec:data}), noted $D$:
\begin{equation}
  p(D_\sms{rep}|D)=\int p(D_\sms{rep}|X)p(X|D)\ddiff X,
\end{equation}
where $X$ are the model parameters.
In practice, these replicated sets are simply estimated by computing the SED model for samples of the parameters drawn from the MCMC.
The comparison to the data requires the assumption of a test statistic, $T(D,X)$.
We adopt the commonly-used $\chi^2$ discrepancy quantity: 
\begin{equation}
  T(D,X) \equiv \sum_i\frac{\left[D_i-\mu(D_i|X)\right]^2}{\sigma(D_i|X)^2},
\end{equation}
where the index $i$ represents every observable of every galaxy, and the quantities $\mu(D_i|X)$ and $\sigma(D_i|X)$ are the mean and standard-deviation of the replicated data.
We then need to estimate the probability:
\begin{equation}
  p_B\equiv P\left(T(D_\sms{rep}|X)\ge T(D|X)\,|\,D\right).
  \label{eq:ppp}
\end{equation}
If the difference between the replicated set and the data is solely due to statistical fluctuations, this probability should be on average $50\,\%$.
A model passes the test, at the $98\,\%$ credence level, if $1\,\%<p_B<99\,\%$.

\begin{figure}[htbp]
  \includegraphics[width=\linewidth]{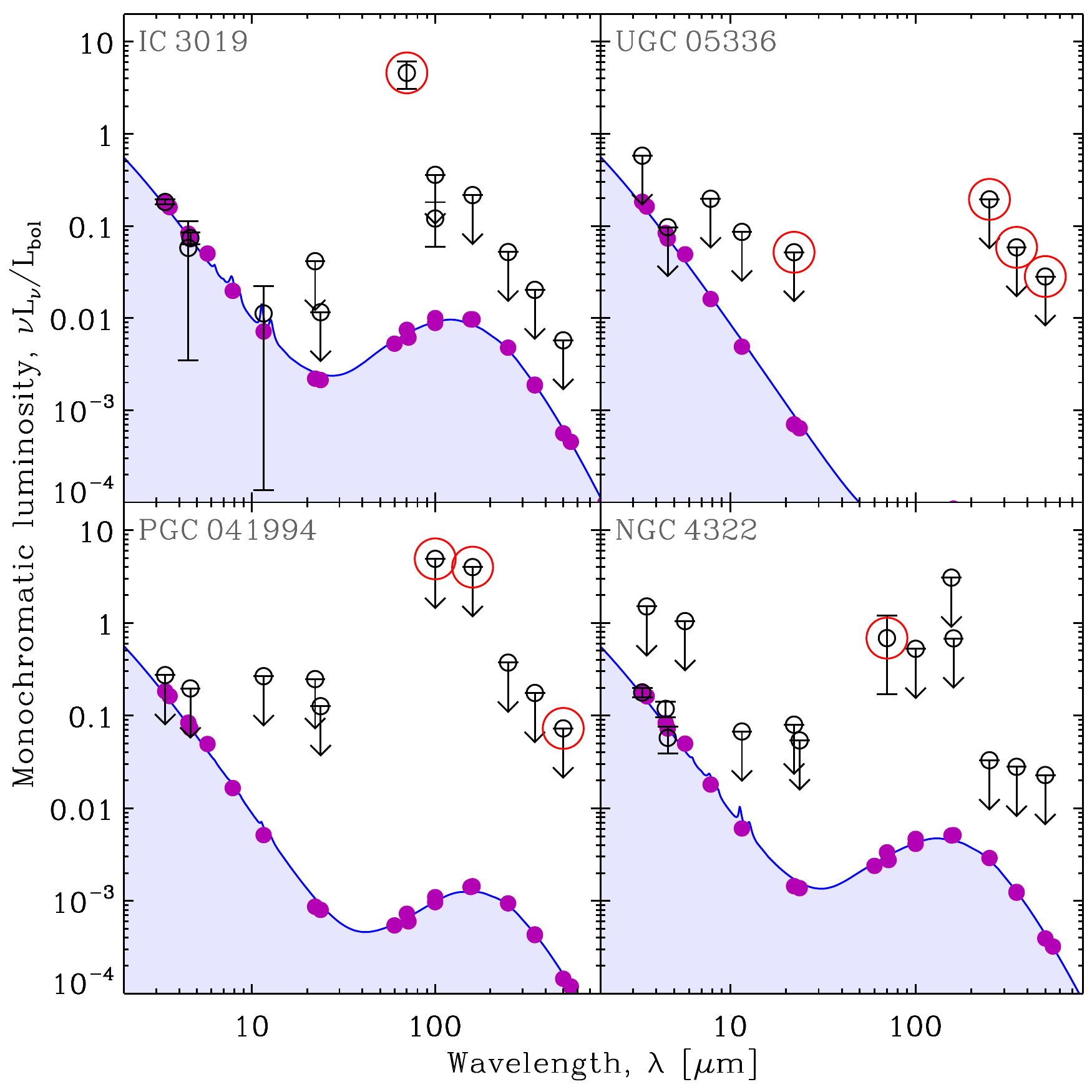}
  \caption{\textsl{SEDs of the 4 galaxies causing the maximum deviations of the 
                   PPP.}
           The black circles with error bars are the observables.
           Most of them are upper limits.
           The blue line is the maximum \textit{a posteriori} model, and
           the purple dots are the synthetic photometry.
           The problematic fluxes are identified by a red circle.}
  \label{fig:ppp}
\end{figure}
In our case, we get $p_B=0.68\,\%$, indicating our model fails this test (it passes the $99\,\%$ credence level, though).
Investigating the cause of this low $p$-value, we notice that a few observations are responsible for large deviations.
\reffig{fig:ppp} displays the SEDs containing the 9 most deviant fluxes.
\begin{description}
  \item[\IC{0319} and \ngc{4322}:] the problematic flux is PACS1.
    For \IC{0319}, it clearly lies above the rest of the observations, but has 
    a small error bar. 
    There is likely a problem with this particular measure.
    The problem is more difficult to assess for \ngc{4322}.
    However, the model is pulled well below this flux by the three SPIRE 
    upper limits.
  \item[\ugc{05336} and \pgc{041994}:]
    the problems come from several far-IR upper limits.
    We display $3\sigma$ upper limits when the value of the measured flux
    is smaller than its $1\sigma$ noise level.
    In the case of the highlighted bands, the actual value of the fluxes are 
    negative, either because of statistical fluctuations, or background 
    over-subtraction.
    Since the modelled flux can not be negative, we have  
    $T(D_\sms{rep}|X)\ll T(D|X)$ for all replicated sets, pulling $p_B$ to the 
    tail of the distribution.
\end{description}
In summary, we are not in the case where the model would provide a poor fit on average.
A few discrepant data, that are treated as outliers by the model (\ie\ which have a very limited impact on the results), are responsible for failing the PPP test.
Excluding these 9 observables, which represent only $0.07\,\%$ of our sample, we get $p_B=1.3\,\%$, making our model pass the test.
In practice, such a test is quite strict, knowing the complexity of our hierarchical model and of the data.
Indeed, the frequentist equivalent would be to test the chi-squared of all our SED fits at once.
Accounting for missing data, the model for our 798 galaxies would have 3655 degrees of freedom.
For a $98\,\%$ confidence level, the reduced chi-squared would be required to lie in the narrow range $0.95<\bar\chi^2<1.05$.
This is because the more data we have, the smaller the statistical fluctuations should be.

  \section{DERIVED TEMPERATURE-EMISSIVITY RELATION}
  \label{app:MBB}

\begin{figure}[htbp]
  \includegraphics[width=\linewidth]{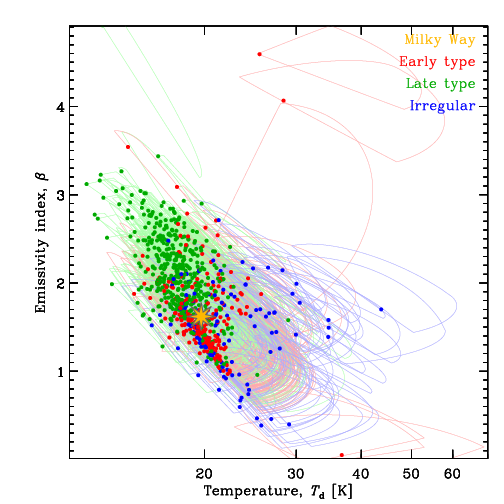}
  \caption{\textsl{Modified black body results.}
           The SUEs show the relation between the temperature, $T_\sms{d}$, and 
           the emissivity index, $\beta$, derived from the $\beta$-free MBB run 
           presented in \refsec{sec:robustmod}.
           Galaxies are color-coded according to their type 
           (\cf~\refsec{sec:robustmod}).
           The diffuse ISM of the Milky Way 
           \citep[$T_\sms{d}^\sms{MW}=20$~K,
           $\beta_\sms{MW}=1.6$;][]{planck-collaboration14c} is displayed as a 
           yellow star.}
  \label{fig:beta_T}
\end{figure}
\reffig{fig:beta_T} shows the $\beta-T_\sms{d}$ relation derived from the HB MBB fit to the sample of \refsec{sec:data} (\cf~\refsec{sec:robustmod}).
We can see that most sources are clustered around
 $(T_\sms{d};\beta) \simeq \left(18.7_{-0.2}^{+5.6}\,\textnormal{K};
2.18_{-0.57}^{+0.00}\right)$,
 with $\pm1\sigma$ distribution widths $\Delta T_\sms{d}=\left[15.7_{--1.0}^{+0.8}\,\textnormal{K},22.2_{--1.0}^{+1.1}\,\textnormal{K}\right]$ and 
 $\Delta \beta=\left[1.68_{-0.06}^{+0.02},2.68_{-0.06}^{+0.02}\right]$.
The correlation coefficient is $\rho(\ln T_\sms{d},\beta)\simeq-0.54_{-0.01}^{+0.18}$, and 
$\CR{\rho(\ln T_\sms{d},\beta)}=[-0.56,-0.13]$.

There are two clear outliers to this trend: the elliptical galaxies \ngc{4268} and \ngc{5507} (top-right of the panel).
The fit of both galaxies is constrained only with an \irasiv\ detection and three SPIRE upper limits.

This trend is consistent with the Galactic diffuse ISM (yellow star).
This relation appears intrinsically scattered, with irregulars (in blue) being significantly hotter than ETGs (in red), themselves hotter than LTGs (in green).
There is a significant $\beta-T_\sms{d}$ negative correlation \citep[such as shown by \eg][]{dupac03}.
We show this plot as a reference for comparison to other studies, since the MBB is the most commonly used model.
Nonetheless, we emphasize that its physical interpretation is rendered difficult by the fact that $\beta$ is degenerate with the mixing of physical conditions \citep[\cf\ Sect.~2.3.1 of ][ for a review]{galliano18}.

  \section{UNCERTAINTY REPRESENTATION}
  \label{app:unc}

We detail here the way we display uncertainties in correlation plots throughout this article.

  \subsection{Skewed Uncertainty Ellipses}
  \label{app:SUE}

\begin{figure}[htbp]
  \includegraphics[width=\linewidth]{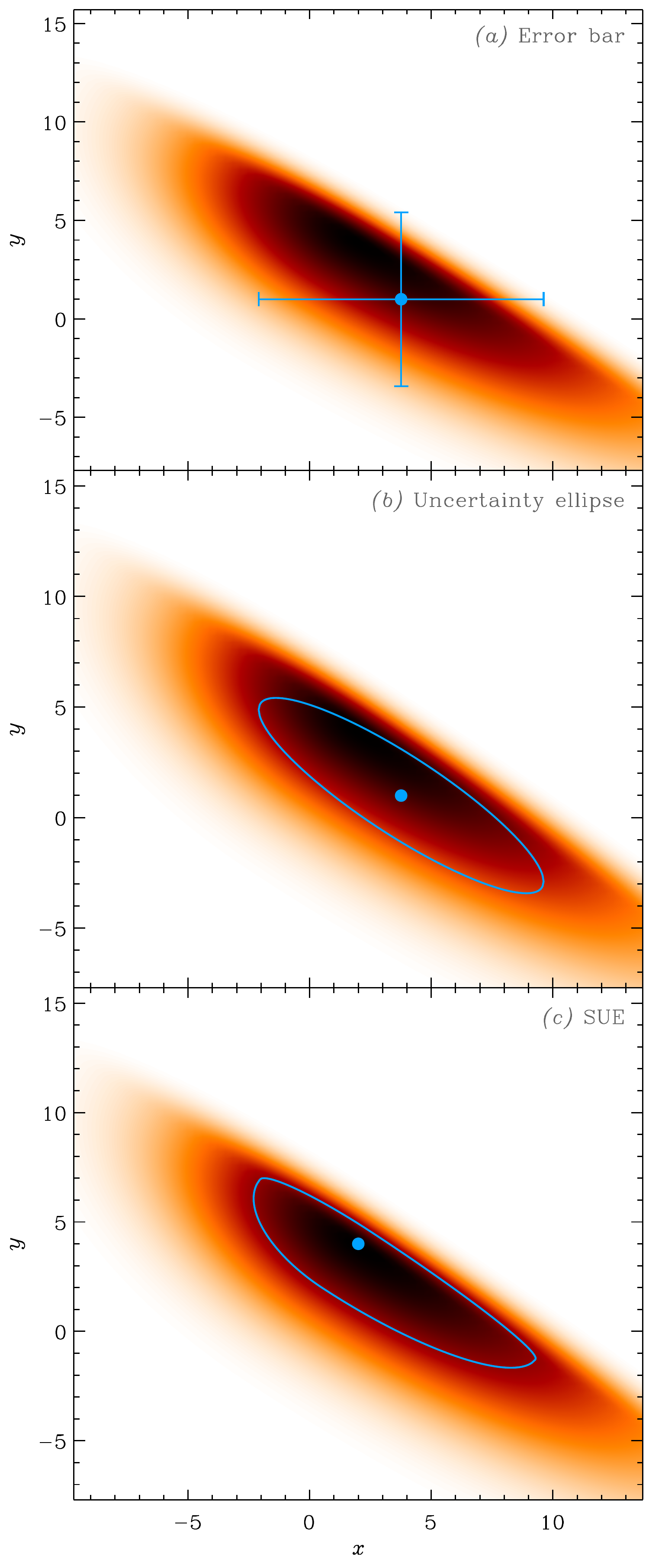}
  \caption{\textsl{Uncertainty display.} 
           The orange density contours in the three panels represent an
           arbitrary bivariate PDF of two variables $x$ and $y$.
           Panel~\textit{(a)} shows the corresponding traditional error bar:
           the dot is the mean and the bars show the $\pm1\sigma$ extent along 
           both axes.
           Panel~\textit{(b)} shows the widely-used uncertainty ellipse, which 
           can be viewed as the mode and $1\sigma$ contour of a bivariate 
           normal distribution having the same means, standard-deviations and 
           correlation coefficient as the PDF.
           Panel~\textit{(c)} shows the concept of SUE introduced in 
           \refapp{app:SUE}.}
  \label{fig:SUE}
\end{figure}

Our HB model computes the full posterior distribution of the parameters of each source.
Such a distribution is usually asymmetric and correlations between parameters can be strong.
Displaying this marginalized posterior distribution as density contours, for several hundreds of sources, is visually ineffective.
In 2D correlation plots, uncertainty ellipses are a widely-used device to display the extent of the posterior and show the correlation between parameters.
However, it does not account for its skewness.
To address this issue, we display each posterior as a \expression{Skewed Uncertainty Ellipse} (SUE), which is the 1$\sigma$ contour of a \expression{Bivariate Split-Normal distribution} \citep[BSN;][]{villani06}, having the same means, variances, skewnesses and correlation coefficient as the posterior.
We add a central dot corresponding to the mode of this BSN.
\reffig{fig:SUE} demonstrates the different ways of displaying uncertainties, for a typical PDF (orange density). 
Panel~\textit{(c)} shows the corresponding SUE.
It has the advantage of retaining a lot of information from the posterior, with only one dot and one contour.

Displaying a SUE is not straightforward.
Indeed, a BSN is parametrized by its position vector, $\vect{X}_0=(x_0,y_0)$, its scale vector, $\vect{\Lambda}=(\lambda_x,\lambda_y)$, its shape vector, $\vect{T}=(\tau_x,\tau_y)$, and its rotation angle, $-\pi/2<\theta<\pi/2$.
If we call $\vect{X}=(x,y)$ our original coordinates and $\vect{X^\prime}=(x^\prime,y^\prime)$ the coordinates in the centered, rotated, reference frame ($\mat{R}$ being the rotation matrix):
\begin{eqnarray}
  \vect{X^\prime} & = &
  \begin{array}{ccc}
    \left(\begin{array}{c}
      x^\prime(x,y) \\
      y^\prime(x,y)
    \end{array}\right)
    & = &
    \mat{R}^T(\vect{X}-\vect{X}_0) 
  \end{array}
  \nonumber\\
  & = &
  \left(\begin{array}{c}
    (x-x_0)\cos\theta+(y-y_0)\sin\theta \\
    -(x-x_0)\sin\theta+(y-y_0)\cos\theta
  \end{array}\right),
\end{eqnarray}
the BSN PDF can be expressed as:
\begin{equation}
  p(x,y) = \mathcal{N} \times
  \left\{
  \begin{array}{l}
    \displaystyle
    \exp\left[-\frac{1}{2}\left(\frac{x^\prime(x,y)}{\lambda_x}\right)^2
              -\frac{1}{2}\left(\frac{y^\prime(x,y)}{\lambda_y}\right)^2
              \right] \\
    \mbox{ if } x^\prime(x,y) < 0 \mbox{ and } y^\prime(x,y) < 0 \\ \\
    \displaystyle
    \exp\left[-\frac{1}{2}\left(\frac{x^\prime(x,y)}{\lambda_x\tau_x}\right)^2
              -\frac{1}{2}\left(\frac{y^\prime(x,y)}{\lambda_y}\right)^2
              \right] \\
    \mbox{ if } x^\prime(x,y) \ge 0 \mbox{ and } y^\prime(x,y) < 0 \\ \\
    \displaystyle
    \exp\left[-\frac{1}{2}\left(\frac{x^\prime(x,y)}{\lambda_x}\right)^2
              -\frac{1}{2}\left(\frac{y^\prime(x,y)}{\lambda_y\tau_y}\right)^2
              \right] \\
    \mbox{ if } x^\prime(x,y) < 0 \mbox{ and } y^\prime(x,y) \ge 0 \\ \\
    \displaystyle
    \exp\left[-\frac{1}{2}\left(\frac{x^\prime(x,y)}{\lambda_x\tau_x}\right)^2
              -\frac{1}{2}\left(\frac{y^\prime(x,y)}{\lambda_y\tau_y}\right)^2
              \right] \\
    \mbox{ if } x^\prime(x,y) \ge 0 \mbox{ and } y^\prime(x,y) \ge 0,
  \end{array}
  \right.
  \label{eq:BSN}
\end{equation}
where the normalization constant is:
\begin{equation}
\mathcal{N} = \frac{2}{\pi\lambda_x\lambda_y(1+\tau_x)(1+\tau_y)}.
\end{equation}

We start by estimating the means ($\langle x\rangle, \langle y\rangle$), standard-deviations ($\sigma_x,\sigma_y$), skewnesses ($\gamma_x,\gamma_y$) and the correlation coefficient ($\rho$) of the posterior.
We detail how we estimate these moments in \refapp{app:Mest}.
These moments can be expressed as a function of the BSN's parameters ($x_0$, $y_0$, $\lambda_x$, $\lambda_y$, $\tau_x$, $\tau_y$, $\theta$):
\begin{eqnarray}
  \langle x\rangle
    & = & \sqrt{\frac{2}{\pi}}\left[\lambda_x(\tau_x-1)\cos\theta
                                   -\lambda_y(\tau_y-1)\sin\theta\right]+x_0
  \label{eq:mombeg}
  \\
  \langle y\rangle
    & = & \sqrt{\frac{2}{\pi}}\left[\lambda_x(\tau_x-1)\sin\theta
                                   +\lambda_y(\tau_y-1)\cos\theta\right]+y_0
  \\
  \sigma_x^2
    & = & \lambda_x^2B(\tau_x)\cos^2\theta + \lambda_y^2B(\tau_y)\sin^2\theta
  \\
  \sigma_y^2
    & = & \lambda_x^2B(\tau_x)\sin^2\theta + \lambda_y^2B(\tau_y)\cos^2\theta
  \\
  \rho\sigma_x\sigma_y
    & = & \left(\lambda_x^2B(\tau_x)-\lambda_y^2B(\tau_y)\right)
          \cos\theta\sin\theta
  \\
  \gamma_x\sigma_x^3
    & = & \sqrt{\frac{2}{\pi}}\left(\lambda_x^3C(\tau_x)\cos^3\theta
                                   -\lambda_y^3C(\tau_y)\sin^3\theta\right)
  \\
  \gamma_y\sigma_y^3
    & = & \sqrt{\frac{2}{\pi}}\left(\lambda_x^3C(\tau_x)\sin^3\theta
                                   +\lambda_y^3C(\tau_y)\cos^3\theta\right),
  \label{eq:momend}
\end{eqnarray}
with:
\begin{eqnarray}
  B(\tau) & = & \left(1-\frac{2}{\pi}\right)(\tau-1)^2+\tau
  \\
  C(\tau) & = & \left[\left(\frac{4}{\pi}-1\right)\tau^2
                 +\left(3-\frac{8}{\pi}\right)\tau
                 +\frac{4}{\pi}-1\right] (\tau-1).
\end{eqnarray}
We then simply need to solve the system of \refeqs{eq:mombeg}{eq:momend}.
To do that, we first solve $\theta$:
\begin{equation}
  \theta = \frac{1}{2}\arctan\left(\frac{2\rho\sigma_x\sigma_y}
                                        {\sigma_x^2-\sigma_y^2}\right).
\end{equation}
We then solve numerically for $\tau_x$ and $\tau_y$, independently, from the
two equations:
\begin{eqnarray}
  \frac{C(\tau_x)}{B(\tau_x)^{3/2}} & = &
    \sqrt{\frac{\pi}{2}}
    \frac{\gamma_x\sigma_x^3\cos^3\theta+\gamma_x\sigma_x^3\sin^3\theta}
         {\cos^6\theta+\sin^6\theta} \nonumber\\
    &\times&
         \left(\frac{\cos^2\theta-\sin^2\theta}
              {\sigma_x^2\cos^2\theta-\sigma_y^2\sin^2\theta}\right)^{3/2} \\
  \frac{C(\tau_y)}{B(\tau_y)^{3/2}} & = &
    \sqrt{\frac{\pi}{2}}
    \frac{\gamma_y\sigma_y^3\cos^3\theta-\gamma_x\sigma_x^3\sin^3\theta}
         {\cos^6\theta+\sin^6\theta} \nonumber\\
    &\times&
         \left(\frac{\cos^2\theta-\sin^2\theta}
              {\sigma_y^2\cos^2\theta-\sigma_x^2\sin^2\theta}\right)^{3/2}.
\end{eqnarray}
We derive the remaining parameters, using the following equations:
\begin{eqnarray}
  \lambda_x & = & \sqrt{\frac{1}{B(\tau_x)}\frac{\sigma_x^2\cos^2\theta
                                     -\sigma_y^2\sin^2\theta}
                                     {\cos^2\theta-\sin^2\theta}} \\
  \lambda_y & = & \sqrt{\frac{1}{B(\tau_y)}\frac{\sigma_y^2\cos^2\theta
                                     -\sigma_x^2\sin^2\theta}
                                     {\cos^2\theta-\sin^2\theta}} \\
  x_0 & = & \langle x\rangle
          - \sqrt{\frac{2}{\pi}}\left[\lambda_x(\tau_x-1)\cos\theta
                                     -\lambda_y(\tau_y-1)\sin\theta\right] \\
  y_0 & = & \langle y\rangle
          -  \sqrt{\frac{2}{\pi}}\left[\lambda_x(\tau_x-1)\sin\theta
                                   +\lambda_y(\tau_y-1)\cos\theta\right].
\end{eqnarray}

  \subsection{Robust Estimate of the Posterior Distribution  
                 Moments}
  \label{app:Mest}

We estimate the moments of \refeqs{eq:mombeg}{eq:momend} numerically, from the marginalized posterior.
Some outliers can be present due to incomplete burn-in removal, requiring robust estimators.
We choose \expression{M-estimators} \citep{mosteller77}, which are a generalization of maximum-likelihood estimators.
They have been popularized in astrophysics by \citet{beers90}.
Location and scale M-estimators, using \expression{Tukey's biweight} loss function were presented by \citet{lax75}.
Posing $u_i=(x_i-l_x)/(c\times s_x)$, where $l_x=\textnormal{med}(X)$ is the median of the sample $X=\{x_i\}_{i=1\ldots N}$, $s_x=1.4826\times\textnormal{med}(|X-\textnormal{med}(X)|)$ is the \expression{Median Absolute Deviation} (MAD) of $X$, and $c$ is a tuning parameter, the M-estimator of the mean is:
\begin{equation}
  \hat\mu(X) = l_x + \frac{\sum_{|u_i|<1}(x_i-l_x)(1-u_i^2)^2}
                      {\sum_{|u_i|<1}(1-u_i^2)^2},
  \label{eq:Mest_mu}
\end{equation}
and for the variance:
\begin{equation}
  \hat{V}(X) = \frac{N\times\sum_{|u_i|<1}(x_i-l_x)^2(1-u_i^2)^4}
                       {\left[\sum_{|u_i|<1}(1-u_i^2)(1-5u_i^2)\right]
                         \left[\sum_{|u_i|<1}(1-u_i^2)(1-5u_i^2)-1\right]}.
  \label{eq:Mest_stdev}
\end{equation}
Considering a second sample $Y=\{y_i\}_{i=1\ldots N}$, an M-estimator of the covariance is presented by \citet{mosteller77}.
Posing $v_i=(y_i-l_y)/(c\times s_y)$, the covariance can be estimated as:
\begin{equation}
  \hat V(X,Y) 
  = \frac{N\times\sum_{|u_i|<1,|v_i|<1}(x_i-l_x)(1-u_i^2)^2(y_i-l_y)(1-v_i^2)^2}
                       {\left[\sum_{|u_i|<1}(1-u_i^2)(1-5u_i^2)\right]
                        \left[\sum_{|v_i|<1}(1-v_i^2)(1-5v_i^2)\right]}.
  \label{eq:Mest_corr}
\end{equation}

These estimators are implemented in the Python \ncode{astropy} module \citep{astropy1,astropy2}. 
We have designed our own skewness W-estimator:
\begin{equation}
  \hat\gamma(X) = \frac{\sum_{|u_i|<1}u_i^3(1-u_i^2)^2}{\sum_{|u_i|<1}(1-u_i^2)^2}.
  \label{eq:Mest_skew}
\end{equation}
We have also slightly improved \ncode{astropy}'s implementation by iterating on \refeqs{eq:Mest_mu}{eq:Mest_skew}, replacing $l_x$, $l_y$, $s_x$ and $s_y$ with $\hat{\mu}(X)$, $\hat{\mu}(Y)$, $\sqrt{\hat{V}(X)}$ and $\sqrt{\hat{V}(Y)}$, respectively, until a $10^{-5}$ relative accuracy is reached.
The tuning parameter is usually taken as $c=6$ for the mean and $c=9$ for the variance, covariance and skewness.

  \section{DUST EVOLUTION RESULTS ASSUMING A     
            CHABRIER IMF}
  \label{app:Chab}

\begin{figure*}[htbp]
  \includegraphics[width=\textwidth]{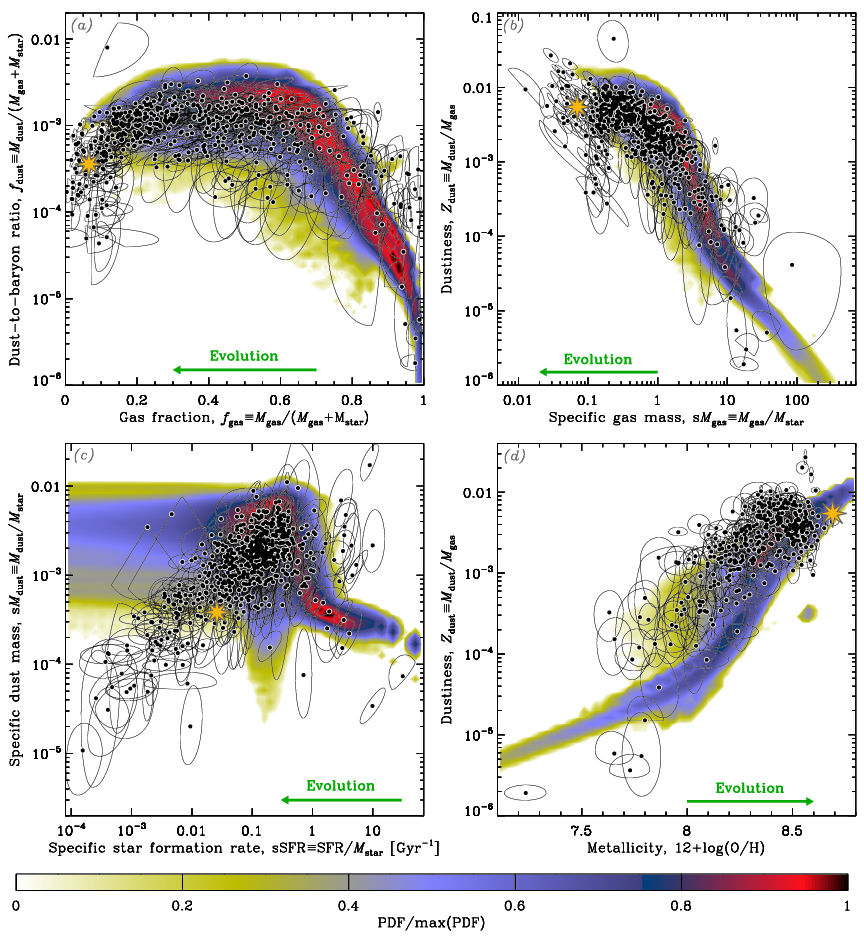}
  \caption{\textsl{Fitted dust evolution tracks,} assuming a \citet{chabrier03} 
           IMF.
           This is the equivalent of \reffig{fig:tracks}.}
  \label{fig:tracks2}
\end{figure*}
\begin{figure}[htbp]
  \includegraphics[width=\linewidth]{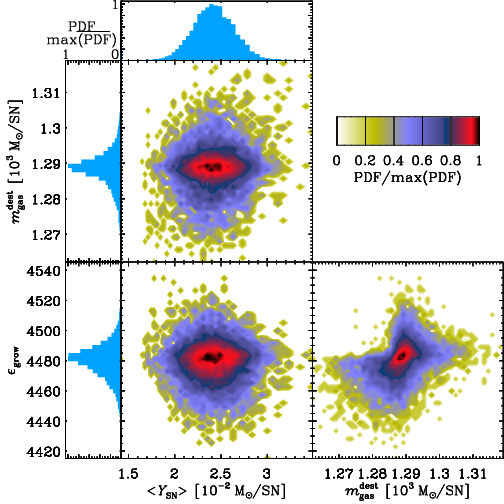}
  \caption{\textsl{Posterior distribution of the tuning parameters,}
           assuming a \citet{chabrier03} IMF.
           This is the equivalent of \reffig{fig:corrcom}.}
  \label{fig:corrcom2}
\end{figure}
\begin{figure*}[htbp]
  \includegraphics[width=\textwidth]{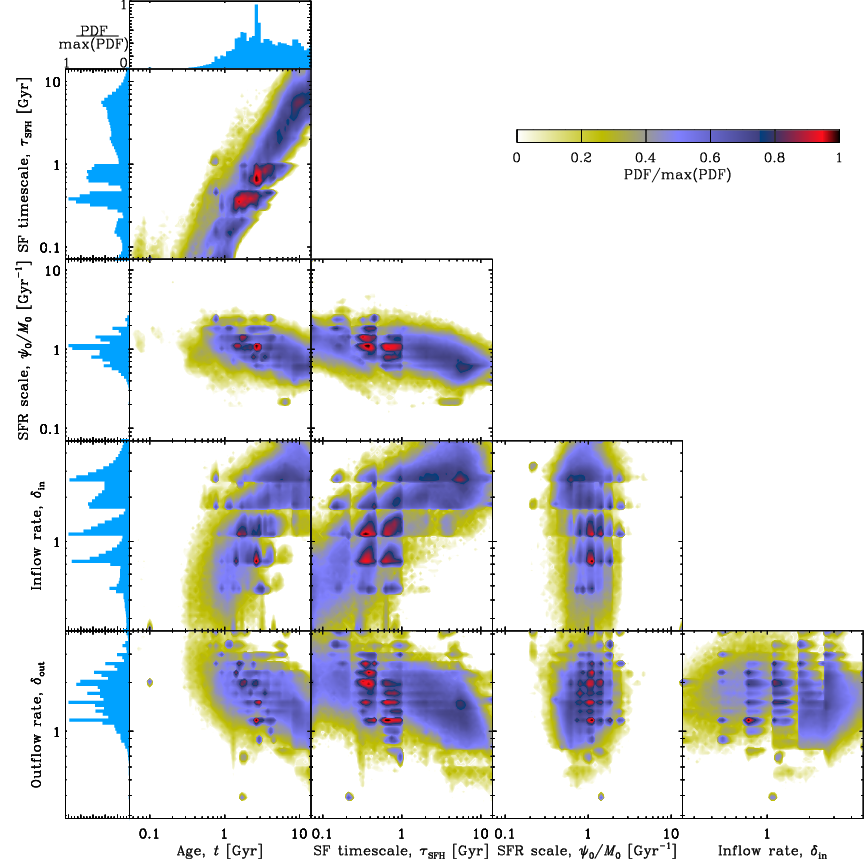}
  \caption{\textsl{Posterior distribution of the SFH-related parameters,}
           assuming a \citet{chabrier03} IMF.
           This is the equivalent of \reffig{fig:corrind}.}
  \label{fig:corrind2}
\end{figure*}
\begin{figure}[htbp]
  \includegraphics[width=\linewidth]{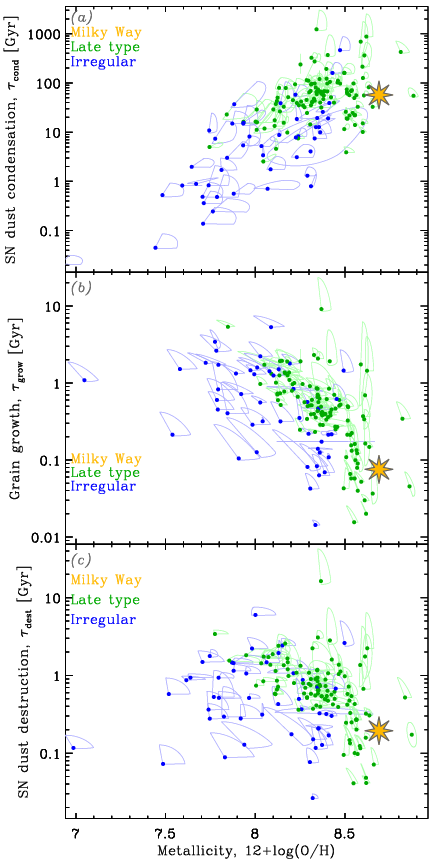}
  \caption{\textsl{Posterior distribution of the tuning parameters,}
           assuming a \citet{chabrier03} IMF.
           This is the equivalent of \reffig{fig:tau}.}
  \label{fig:tau2}
\end{figure}
We have performed the modelling of \refsec{sec:HBdustvol}, assuming a \citet{chabrier03} IMF.
The data have been corrected accordingly: \tMs\ has been multiplied by 0.61 and SFR by 0.63 \citep{madau14}, as these two quantities were derived assuming a \citet{salpeter55} IMF.
\reffig{fig:tracks2} shows the fit of the scaling relations; it is the equivalent of \reffig{fig:tracks}.
\reffigs{fig:corrcom2}{fig:corrind2} display the PDF of the dust evolution and SFH-related parameters, respectively; they are the equivalent of \reffigs{fig:corrcom}{fig:corrind}.
The inferred timescales are displayed in \reffig{fig:tau2}; it is the equivalent of \reffig{fig:tau}.

  \section{REFERENCE RUN PARAMETERS}
  
\reftab{tab:resref} gives the mean and standard-deviation of the most relevant parameters derived with the \expression{reference} run (\cf~\refsec{sec:ref}), for each galaxy of the sample presented in \refsec{sec:data}.
\begin{landscape}
\begin{longtable}[c]{l|*{3}{r}|*{5}{r}}
  \caption{Main inferred quantities for each galaxy, with the 
           \textit{reference} run.
           For each quantity, we quote the mean $\pm$ the 
           standard-deviation of the posterior PDF.}
  \label{tab:resref}\\
  \hline\hline
{\scriptsize Galaxy} & {\scriptsize $\log M_\sms{dust}/M_\odot$} & {\scriptsize $\log\langle U\rangle$} & {\scriptsize $\log q_\sms{AF}$} & {\scriptsize $\log M_\sms{gas}/M_\odot$} & {\scriptsize $\log f_\sms{\hmol}$} & {\scriptsize $\log M_\star/M_\odot$} & {\scriptsize $\log\textnormal{sSFR}/\textnormal{yr}^{-1}$} & {\scriptsize $12+\log(\textnormal{O/H})$} \\
  \hline
  \endfirsthead
  \caption{continued.} \\
  \hline\hline
{\scriptsize Galaxy} & {\scriptsize $\log M_\sms{dust}/M_\odot$} & {\scriptsize $\log\langle U\rangle$} & {\scriptsize $\log q_\sms{AF}$} & {\scriptsize $\log M_\sms{gas}/M_\odot$} & {\scriptsize $\log f_\sms{\hmol}$} & {\scriptsize $\log M_\star/M_\odot$} & {\scriptsize $\log\textnormal{sSFR}/\textnormal{yr}^{-1}$} & {\scriptsize $12+\log(\textnormal{O/H})$} \\
  \hline
  \endhead
{\scriptsize ESO\,149-013} & {\scriptsize $5.4 \pm 0.5$} & {\scriptsize $-0.4 \pm 0.5$} & {\scriptsize $-1.10 \pm 0.21$} & {\scriptsize $8.864 \pm 0.023$} & {\scriptsize \ldots} & {\scriptsize $8.42 \pm 0.19$} & {\scriptsize \ldots} & {\scriptsize \ldots}\\
{\scriptsize NGC\,0007} & {\scriptsize $6.20 \pm 0.17$} & {\scriptsize $0.13 \pm 0.18$} & {\scriptsize $-1.22 \pm 0.11$} & {\scriptsize $9.31 \pm 0.03$} & {\scriptsize \ldots} & {\scriptsize $9.06 \pm 0.10$} & {\scriptsize $-9.75 \pm 0.11$} & {\scriptsize \ldots}\\
{\scriptsize ESO\,410-005} & {\scriptsize $3.5 \pm 0.5$} & {\scriptsize $0.4 \pm 0.8$} & {\scriptsize $-1.4 \pm 0.4$} & {\scriptsize \ldots} & {\scriptsize \ldots} & {\scriptsize $6.28 \pm 0.14$} & {\scriptsize \ldots} & {\scriptsize \ldots}\\
{\scriptsize IC\,0010} & {\scriptsize $5.61 \pm 0.20$} & {\scriptsize $0.47 \pm 0.19$} & {\scriptsize $-1.25 \pm 0.16$} & {\scriptsize $8.08 \pm 0.06$} & {\scriptsize \ldots} & {\scriptsize $8.66 \pm 0.12$} & {\scriptsize $-9.99 \pm 0.31$} & {\scriptsize $8.18 \pm 0.09$}\\
{\scriptsize NGC\,0115} & {\scriptsize $6.80 \pm 0.12$} & {\scriptsize $0.02 \pm 0.13$} & {\scriptsize $-1.22 \pm 0.08$} & {\scriptsize $9.67 \pm 0.04$} & {\scriptsize \ldots} & {\scriptsize $9.48 \pm 0.10$} & {\scriptsize $-9.78 \pm 0.10$} & {\scriptsize \ldots}\\
{\scriptsize ESO\,410-012} & {\scriptsize $5.4 \pm 0.5$} & {\scriptsize $-0.3 \pm 0.5$} & {\scriptsize $-1.11 \pm 0.21$} & {\scriptsize $8.915 \pm 0.029$} & {\scriptsize \ldots} & {\scriptsize $8.13 \pm 0.13$} & {\scriptsize $-9.73 \pm 0.14$} & {\scriptsize \ldots}\\
{\scriptsize NGC\,0131} & {\scriptsize $6.63 \pm 0.07$} & {\scriptsize $0.12 \pm 0.07$} & {\scriptsize $-0.99 \pm 0.04$} & {\scriptsize $9.408 \pm 0.005$} & {\scriptsize \ldots} & {\scriptsize $9.30 \pm 0.09$} & {\scriptsize $-10.01 \pm 0.11$} & {\scriptsize \ldots}\\
\hline
\end{longtable}
\end{landscape}

\end{document}